\newif\ifsubmit

\documentclass[11pt
]{llncs}
\usepackage{geometry}
\usepackage{Xiao-llncs}

\newcommand{\msf}{\mathsf}
\newcommand{\mcal}{\mathcal}

\newcommand{\mbf}{\mathbf}
\newcommand{\bs}[1]{{\boldsymbol #1}}

\newcommand{\Adv}{\ensuremath{\mathcal{A}}\xspace}
\newcommand{\A}{\ensuremath{\mathcal{A}}\xspace}

\newcommand{\acc}{\ensuremath{\mathsf{Acc}}\xspace}

\newcommand{\aux}{\ensuremath{\reg{aux}}\xspace}

\newcommand{\bp}{\mathbf{p}}
\renewcommand{\bar}{\overline}

\newcommand{\bits}{\ensuremath{\{0,1\}}\xspace}

\newcommand{\BPP}{\ensuremath{\mathbf{BPP}}\xspace}

\newcommand{\bother}{\mathbf{other}}
\newcommand{\BQP}{\ensuremath{\mathbf{BQP}}\xspace}

\newcommand{\cH}{\ensuremath{\mathcal{H}}}

\newcommand{\cM}{\ensuremath{\mathcal{M}}}

\newcommand{\calX}{\ensuremath{\mathcal{X}}}
\newcommand{\calY}{\ensuremath{\mathcal{Y}}}

\newcommand{\cind}{\ensuremath{\stackrel{\text{c}}{\approx}}\xspace}

\newcommand{\Cont}{\ensuremath{\mathbf{Cont}}}

\renewcommand{\epsilon}{\varepsilon}

\newcommand{\func}{\ensuremath{\mathcal{F}}\xspace}

\newcommand{\Hil}{\ensuremath{\mathcal{H}}\xspace}

\renewcommand{\hat}{\widehat}

\newcommand{\keyspace}{\ensuremath{\mathcal{K}}\xspace}

\newcommand{\Lang}{\ensuremath{\mathcal{L}}\xspace}
\newcommand{\la}{\ensuremath{\leftarrow}\xspace}

\newcommand{\Naturals}{\ensuremath{\mathbb{N}}\xspace}

\newcommand{\negl}{\ensuremath{\mathsf{negl}}\xspace}

\newcommand{\NP}{\ensuremath{\mathbf{NP}}\xspace}

\newcommand{\oracle}{\mathcal{O}}
\newcommand{\ora}{\mathcal{O}}

\renewcommand{\paragraph}{\para}

\newcommand{\pick}{\ensuremath{\xleftarrow{\$}}\xspace}

\newcommand{\Prover}{\ensuremath{\mathcal{P}}\xspace}

\newcommand{\poly}{\ensuremath{\mathsf{poly}}\xspace}

\newcommand{\QMA}{\mathbf{QMA}}

\newcommand{\ra}{\ensuremath{\rightarrow}\xspace}

\newcommand{\red}[1]{{\color{red} #1}}
\newcommand{\Relation}{\ensuremath{\mathcal{R}}\xspace}

\newcommand{\reg}{\mathtt}
\newcommand{\reprogram}{\mathsf{Reprogram}}

\newcommand{\sample}{\gets}

\newcommand{\Sexp}{\ensuremath{S_{\mathsf{exp}}}\xspace}

\newcommand{\secpar}{\ensuremath{\lambda}\xspace}

\newcommand{\Set}[1]{\ensuremath{\{#1\}}\xspace}

\newcommand{\Sim}{\ensuremath{\mathcal{S}}\xspace}

\newcommand{\tensor}{\ensuremath{\otimes}\xspace}

\renewcommand{\tilde}{\widetilde}

\newcommand{\vecone}{\mathbf{1}}

\newcommand{\Verifier}{\ensuremath{\mathcal{V}}\xspace}
\newcommand{\Vexp}{\ensuremath{{V_{\mathsf{exp}}}}\xspace}
\newcommand{\Vhon}{\ensuremath{{V_{\mathsf{hon}}}}\xspace}

\usepackage{bm}
\usepackage{mdframed}

\makeatletter
\def\@fnsymbol#1{\ensuremath{\ifcase#1\or *\or \dagger\or \ddagger\or
   \mathsection\or \mathparagraph\or \|\or **\or \dagger\dagger
   \or \ddagger\ddagger \else\@ctrerr\fi}}
\makeatother

\begin{document}

\ifdraft{
}{}

\ifdraft{
    \listoffixmes
    \addcontentsline{toc}{section}{List of Corrections}
    \newpage
}{}

\title{The Black-Box Simulation Barrier Persists in a Fully Quantum World}

\author{Nai-Hui Chia\inst{1}
 \and
Kai-Min Chung\inst{2}
 \and
 Xiao Liang\inst{3}
 \and 
 Jiahui Liu\inst{4}
 }
 \institute{
 Rice University, USA\\ \email{nc67@rice.edu}
 \and
 Academia Sinica, Taiwan\\ \email{kmchung@iis.sinica.edu.tw}
  \and
 The Chinese University of Hong Kong, Hong Kong\\ \email{xiaoliang@cuhk.edu.hk}
 \and
 Massachusetts Institute of Technology, USA\\ \email{jiahuiliu@csail.mit.edu}
 }

\let\oldaddcontentsline\addcontentsline
\def\addcontentsline#1#2#3{}
\maketitle
\def\addcontentsline#1#2#3{\oldaddcontentsline{#1}{#2}{#3}}

\begin{abstract}\addcontentsline{toc}{section}{Abstract}

Zero-Knowledge (ZK) protocols have been a subject of intensive study due to their fundamental importance and versatility in modern cryptography. However, the inherently different nature of quantum information significantly alters the landscape, necessitating a re-examination of ZK designs.

\vspace{0.5em}
\hspace{1em} A crucial aspect of ZK protocols is their round complexity, intricately linked to {\em simulation}, which forms the foundation of their formal definition and security proofs. In the {\em post-quantum} setting, where honest parties and their communication channels are all classical but the adversaries could be quantum, Chia, Chung, Liu, and Yamakawa [FOCS'21] demonstrated the non-existence of constant-round {\em black-box-simulatable} ZK arguments (BBZK) for $\NP$ unless $\NP \subseteq \BQP$. However, this problem remains widely open in the full-fledged quantum future that will eventually arrive, where all parties (including the honest ones) and their communication are naturally quantum.

\vspace{0.5em}
\hspace{1em} Indeed, this problem is of interest to the broader theory of quantum computing. It has been an important theme to investigate how quantum power fundamentally alters traditional computational tasks, such as the {\em unconditional} security of Quantum Key Distribution and the incorporation of Oblivious Transfers in MiniQCrypt. Moreover, quantum communication has led to round compression for commitments and interactive arguments. Along this line, the above problem is of great significance in understanding whether quantum computing could also change the nature of ZK protocols in some fundamentally manner.

\vspace{0.5em}
\hspace{1em} We resolved this problem by proving that only languages in $\BQP$ admit constant-round {\em fully-quantum} BBZK. This result holds significant implications. Firstly, it illuminates the nature of quantum zero-knowledge and provides valuable insights for designing future protocols in the quantum realm. Secondly, it relates ZK round complexity with the intriguing problem of $\BQP$ vs $\QMA$, which is out of the reach of previous analogue impossibility results in the classical or post-quantum setting. Lastly, it justifies the need for the {\em non-black-box} simulation techniques or the relaxed security notions employed in existing constant-round fully-quantum BBZK protocols.

\end{abstract}

	\pagenumbering{roman}
\setcounter{tocdepth}{4} 
\tableofcontents
\addcontentsline{toc}{section}{Table of Contents}
\clearpage

\pagenumbering{arabic}

\section{Introduction}
\label{sec:intro}

{\em Zero-Knowledge} (ZK) protocols, pioneered by Goldwasser, Micali, and Rackoff \cite{STOC:GolMicRac85}, enable a prover in an interactive proof system to demonstrate the truth of a statement (e.g., one in an $\NP$ language) without disclosing additional information (e.g., the $\NP$ witness) beyond the statement's truthfulness. Since their inception, ZK protocols have garnered considerable research interest and emerged as a cornerstone of cryptography. Apart from serving as versatile primitives, ZK protocols are integral to {\em secure multi-party computation} (MPC), another cornerstone of modern cryptography. Furthermore, exploration of ZK protocols has unveiled intriguing aspects of computational complexity theory, as evidenced by various studies \cite{C:ImpYun87,FOCS:SahVad97,DBLP:conf/dimacs/SahaiV97,C:BGGHKMR88,STOC:BelMicOst90b,FOCS:GriSloYue19,DBLP:conf/stoc/GhosalIKKLS23} etc. Thus, investigating the nature of ZK protocols is of significant importance.

\para{Round Complexity and Black-Box Simulation.} A crucial aspect of ZK protocol research is understanding their round complexity, which is intricately linked to the concept of {\em simulation}, upon which the formal definition of ZK protocols (and consequently their security proofs) relies.

The formal definition of ZK mandates the presence of a {\em simulator} $\Sim$ that interacts with a malicious verifier $\Verifier^*$ and convinces $\Verifier^*$ that it is communicating with the honest prover. Importantly, $\Sim$ possesses no secret input akin to that of the honest prover. Thus, the success of $\Sim$ in convincing $\Verifier^*$ encapsulates the intuitive essence of ZK protocols --- $\Verifier^*$ gains no extra information (other than the statement's truthfulness) from the interaction with the honest prover.




Typically, the simulator $\Sim$ interacts with $\Verifier^*$ in a {\em black-box} manner, meaning that $\Sim$ solely leverages the Input/Output behavior of $\Verifier^*$ and treats it as an oracle. {Such a simulator is allowed to \emph{rewind} the verifier: that is, unlike the prover who can only interact with the verifier in a straight-line, sequential manner, $\Sim$ can send, say, a fourth message to $\Verifier^*$ in one query, and in the next query, send a fresh first message with the hope to see a new completion of starting there. This capability of rewinding essentially allows  $\Sim$   to simulate an accepting execution without a witness.}
This black-box approach is arguably the most natural choice for simulation, often being 
simpler and more modular compared to non-black-box methods. Indeed, black-box simulation is widely employed in most positive results for ZK (and MPC) protocols, including notable works such as \cite{FOCS:GolMicWig86,C:FeiSha89,JC:GolKah96,brassard1991constant,EC:BelJakYun97}, among others (exceptions are discussed in \Cref{sec:more-related}).

Strong impossibility results have been established for ZK arguments\footnote{This refers to ZK protocols with {\em computational} soundness guarantee and negligible soundness error.} employing black-box simulation (dubbed BBZK henceforth). Goldreich and Krawczyk \cite{goldreich1996composition} demonstrated that no three-round BBZK protocols or public-coin constant-round BBZK protocols for $\NP$ exist unless $\NP \subseteq \BPP$. Barak and Lindell \cite{STOC:BarLin02} further established that constant-round BBZK protocols with {\em strict polynomial-time} simulation are non-existent unless $\NP \subseteq \BPP$. Here, strict polynomial-time simulation indicates that the simulator always runs in a fixed polynomial time, unlike an {\em expected polynomial-time} simulator whose runtime is polynomial time only in expectation. Notably, as per \cite{STOC:BarLin02}, all existing constant-round BBZK protocols for $\NP$ rely on expected polynomial-time simulation. 

These impossibility results significantly advance our comprehension of zero-knowledge and offer valuable guidance for positive results (i.e., constructions).

\para{BBZK in the Quantum Era.} All the above results are in the classical setting. However, it is known that quantum information behaves in a fundamentally different manner. This poses an intriguing question: {\em What can we say about the round complexity of ZK protocols in the quantum setting?} Toward answering this question, two models need consideration --- the {\em post-quantum} model and the {\em fully quantum} model.

\subpara{Post-Quantum ZK.} This is the model where honest parties and their communication channels are all classical, but the adversary could be a quantum machine. This model is particularly interesting in the near future, where adversaries may gain early access to quantum computing capabilities while honest parties are not required to catch up to remain protected against them\footnote{Note that to properly define post-quantum zero knowledge, the simulator $\Sim$ has to possess quantum power as well in order to black-boxly interact with a quantum malicious verifier. Therefore, the lower bound against classical malicious verifiers in \cite{STOC:BarLin02} does not trivially give us a lower bound for post-quantum setting.}.

Exciting progress has been made in this model. The recent breakthrough by Bitansky and Shmueli \cite{STOC:BitShm20} presents a constant-round post-quantum ZK argument for $\NP$ using {\em non-black-box simulation}. On the other hand, Chia, Chung, Liu, and Yamakawa \cite{FOCS:CCLY21} show that there does not exist a constant-round post-quantum BBZK for $\NP$ unless $\NP \subseteq \BQP$. It is worth noting that the result by \cite{FOCS:CCLY21} holds even for {\em expected} quantum-polynomial-time (QPT) simulation. Thus, it is {\em not} merely a analogue of \cite{STOC:BarLin02}, but rather represents a qualitatively stronger impossibility result in the post-quantum setting. These two results together essentially provide a complete characterization of the round complexity of BBZK in the post-quantum setting.

It is also worth mentioning that the recent work by Lombardi, Ma, and Spooner \cite{lombardi2022post} introduced a new notion of simulation called {\em coherent-runtime} expected QPT simulation. This provides an alternative framework that helps circumvent the impossibility result of \cite{FOCS:CCLY21}. For further details, we refer the reader to \Cref{sec:more-related}.

\subpara{Fully-Quantum ZK.} This model represents the full-fledged quantum scenario meant to capture the future quantum landscape, where all parties (including the honest ones) and communication channels are allowed to be quantum. Unlike the post-quantum model, the fully-quantum capabilities across all parties in this model opens up new possibilities. One example is the existence of ZK protocols for $\QMA$, which is infeasible in the post-quantum setting because classical honest provers cannot utilize quantum $\QMA$ witnesses.

In contrast to the post-quantum model, our understanding of the round complexity of {\em fully-quantum} BBZK is rather limited. While there exist fully-quantum BBZK protocols for $\NP$ and even $\QMA$ (e.g., \cite{FOCS:BJSW16,FOCS:BroGri20}, etc.), they require at least super-constant rounds. Regarding impossibility results, due to the quantum power of the honest parties, it is unclear if the aforementioned results in the post-quantum setting still hold. The only marginally relevant result is the work of \cite{jain2009parallel}, which demonstrates the nonexistence of constant-round public-coin (or three-round but potentially private-coin) post-quantum BBZK proofs for $\NP$ unless $\NP \subseteq \BQP$, which holds even if the last message in the protocol is quantum. In summary, the central problem in the area remains widely open:
\begin{quote}
{\bf \hypertarget{link:question}{Question}:} {\em Do there exist constant-round fully-quantum BBZK protocols for $\NP$ (or $\QMA$) with strict (or expected) QPT simulation?}
\end{quote}


\subsection{Controversial Clues Toward an Answer}	
\label{sec:contravercial-clues}

Interestingly, both positive and negative indications have emerged in response to the above \hyperlink{link:question}{Question}, particularly in light of recent advances in quantum cryptography.

On the one hand, a central theme in quantum computing has been to investigate how quantum capabilities fundamentally alter traditional computational tasks. Recent progress along this line has produced a growing number of examples demonstrating how classical impossibilities can (delightfully) be surpassed in the {\em fully-quantum} regime. Just to name a few categories of such results:
\begin{itemize}
 \item 
 Quantum communication has been shown to relax the assumptions required for certain cryptographic tasks. Notable examples include the {\em unconditional} security of Quantum Key Distribution \cite{bennet1984quantum}, and the incorporation of Oblivious Transfer  within MiniQCrypt \cite{EC:GLSV21,C:BCKM21b}. Quantum communication has also enabled round compression for commitments \cite{DBLP:conf/asiacrypt/Yan22} and for interactive arguments \cite{STOC:KitWat00,DBLP:journals/cc/KempeKMV09,bostanci2023efficient}.

\item 
An active line of recent work \cite{C:JiLiuSon18,DBLP:conf/tqc/Kretschmer21,C:AnaQiaYue22,C:MorYam22,DBLP:conf/stoc/KretschmerQST23,KhuTom-STOC-25}, among others, has made significant progress in instantiating cryptographic primitives with the honest parties leveraging {\em fully-quantum} capabilities. Remarkably, these results serve as strong evidence that various cryptographic tasks may remain feasible even if $\mathbf{P} = \mathbf{NP}$, a regime in which all classical/post-quantum cryptographic tools become fundamentally inapplicable.
\end{itemize} 
In light of these exciting results, it is quite natural to expect that such ``quantum supremacy'' might also be observed in the context of ZK protocols, leading to an encouraging ``YES'' as the eventual answer to the above \hyperlink{link:question}{Question}.

On the other hand, the recent success of \cite{FOCS:CCLY21} in establishing a post-quantum impossibility  seems to suggest a ``NO'' answer in the fully quantum world as well. However, this impression largely rests on the hope that the techniques from \cite{STOC:BarLin02,FOCS:CCLY21} could be successfully extended to the fully-quantum setting. Upon closer examination, this hope appears far from being promising: 

The techniques in \cite{STOC:BarLin02,FOCS:CCLY21} rely heavily on the classical nature of the honest verifier in classical/post-quantum protocols.\footnote{Although these works focus on impossibility results, the malicious verifier they construct is essentially the {\em classical} honest verifier augmented with a (classical) random-aborting behavior.} So far, it remains entirely unclear how to adapt these techniques to the setting of a genuinely {\em fully-quantum} verifier. In such a setting, the advantages present in the post-quantum case --- such as classical communication channels and the ability to ``de-randomize'' the verifier to a deterministic one --- become obstacles that must be overcome when dealing with a genuinely quantum adversarial verifier. It is, a priori, unclear whether these obstacles can even be overcome at all! (
See \Cref{sec:tech-overview:CCLY-review} for more details.)

Indeed, similar difficulties in extending classical impossibility results have been observed even in the so-called ``QCCC'' setting, {\em where all quantum computation is performed locally by the parties who communicate only via classical messages}, let alone in the fully-quantum setting. For example, it is notoriously difficult to mirror in the QCCC setting the basic (classical) impossibility of constructing key agreement from one-way functions \cite{C:ACCFLM22,DBLP:conf/crypto/LiLLL24}.

These concrete, technical obstacles seem to resonate well with the ``quantum supremacy'' view described earlier, further reinforcing hesitation toward attempting to establish a ``NO'' answer to the question.

The contrasting ``YES vs NO'' perspectives further elevate the significance of the \hyperlink{link:question}{Question}, attesting to its scientific value. Any definitive answer would be highly interesting: a ``YES'' would reveal new, intrinsically quantum effects that fundamentally distinguish the quantum world from the classical/post-quantum; a ``NO'' would require the development of novel mathematical tools capable of overcoming the aforementioned technical obstacles and formalizing the impossibility in the fully-quantum setting. This represents a compelling {\em win-win scenario} for both quantum computing and cryptography research.

Beyond ultimately resolving the controversial issues discussed above, answers to this \hyperlink{link:question}{Question} hold significance for several additional reasons:
\begin{enumerate}
\item
Similar to the impossibility results in the classical and post-quantum settings, answers to this \hyperlink{link:question}{Question} would help further our understanding of zero-knowledge and serve as valuable guidance for protocol design in the full-fledged quantum future.

\item \label{item:question:value:2}
Answers to this \hyperlink{link:question}{Question} would relate zero-knowledge with the intriguing problem of $\BQP$ vs $\QMA$, while post-quantum protocols could at most relate to $\BQP$ vs $\NP$, as explained above.

\item \label{item:question:value:3}
Recent works (e.g., \cite{C:ChiChuYam21,C:CCLY22}) have achieved constant-round constructions for post-quantum ZK and even 2PC w.r.t.\ a relaxed security notion called {\em $\epsilon$-simulation}. These results, along with the line of work \cite{STOC:BitShm20,EC:ABGKM21,C:BCKM21a,DBLP:conf/focs/0014PY23} that utilizes non-black-box simulation, all extend to the fully-quantum setting\footnote{It means that they achieve quantum ZK for $\QMA$ or quantum MPC for {\em quantum} functionalities.}. Without any answers to the above \hyperlink{link:question}{Question}, it remains unclear whether these results, when viewed in the fully-quantum setting, are truly optimal (in the sense that the relaxation to $\epsilon$-simulation or the use of non-black-box simulation is really necessary).
\end{enumerate}

\xiao{Older version hidden here.}

\subsection{Our Results}
\label{sec:intro:results}

We answer the above \hyperlink{link:question}{Question} by establishing the following impossibility result. 
\begin{theorem}\label{thm:main:intro}
For any language $\Lang$, if there exists a constant-round fully-quantum BBZK protocol with expected QPT simulation, then it holds that $\Lang \in \BQP$.
\end{theorem}
It is worth noticing that \Cref{thm:main:intro} rules out {\em expected} QPT simulation, and thus it can be viewed as the exact analogue of the \cite{FOCS:CCLY21} impossibility in the fully-quantum context. However, our techniques to establishing \Cref{thm:main:intro} diverge significantly from \cite{FOCS:CCLY21} due to the fundamentally different nature inherent in fully-quantum protocols. Further discussion on this matter will be provided in \Cref{sec:tech-overview}.

As mentioned earlier in \Cref{item:question:value:2}, since \Cref{thm:main:intro} is about quantum protocols, it allows us to connect zero-knowledge with the quantum complexity class $\QMA$, which we state as the following corollary.
\begin{corollary}[of \Cref{thm:main:intro}]
There does not exist any constant-round fully-quantum BBZK protocol for $\QMA$ (resp.\ $\NP$) unless $\QMA \subseteq \BQP$ (resp.\ $\NP \subseteq \BQP$).
\end{corollary}
As previously mentioned in \Cref{item:question:value:3}, \Cref{thm:main:intro} justifies the necessity and tightness of the aforementioned positive results with relaxed security notions, {\em in the fully-quantum context}, of (1) the use of non-black-box simulation in \cite{STOC:BitShm20}, (2) the relaxation to $\epsilon$-simulation in \cite{C:ChiChuYam21,C:CCLY22}, and (3) the non-constant round complexity observed in existing BBZK or secure {\em quantum} computation protocols with black-box simulation, e.g., \cite{C:DupNieSal12,FOCS:BJSW16,FOCS:BroGri20,EC:DGJMS20,EC:GLSV21,C:AnaChuLaP21} etc.


\subsection{Future Prospects of Our Techniques}
\label{sec:future-prospects}

The key to establishing \Cref{thm:main:intro} is to mathematically formalize the intuition that ``rewinding does not help'' against a randomly aborting verifier in the {\em fully-quantum} setting. As our main technical contribution, we develop a set of novel tools to achieve this. As explained earlier, it was far from clear how to do so, even with the techniques from \cite{FOCS:CCLY21}, which succeeded only in the {\em post-quantum} regime. With this understanding in place, it is not hard to see that our new tools hold promise for broader applications:

Understanding the round complexity of ZK protocols is a central research theme, and establishing impossibility results has long been a fruitful and inspiring direction \cite{goldreich1996composition,STOC:CKPR01,STOC:BarLin02,TCC:Katz08a,TCC:ChuPasTse12,C:KalRotRot17,EC:HazPasVen20,FOCS:CCLY21}. From this perspective, the techniques we develop for stand-alone quantum ZK are, in fact, quite general: they provide a useful toolkit for future work on lower bounds in {\em fully-quantum} interactions (or even reductions), particularly in scenarios where rewinding is a concern.

For instance, we are optimistic that our techniques can contribute to future investigations into lower bounds for concurrent quantum ZK protocols,  an important and extensively studied topic in the classical setting, e.g., \cite{STOC:CKPR01,TCC:ChuPasTse12}. Notably, both of these concurrent impossibility results rely heavily on techniques originally developed for the stand-alone setting. While concurrent settings require new ``scheduling-level'' insights and observations to leverage concurrency constraints, they still fundamentally depend on the mathematical tools used to formally ``nullify'' the power of rewinding --- namely, the construction of a randomly aborting malicious verifier $\Verifier^*$ and the accompanying analysis.

In particular, without techniques that formalize the intuition that ``rewinding does not help'' against a randomly aborting $\Verifier^*$ in the stand-alone setting, it is difficult to envision how such concurrent lower bounds \cite{STOC:CKPR01,TCC:ChuPasTse12} could be rigorously established.

We believe our techniques will play a similarly foundational role as the field progresses toward understanding concurrent ZK in the {\em fully-quantum} setting.

\subsection{More Related Work}
\label{sec:more-related}

In the classical setting, there exist constant-round constructions of ZK (and even MPC) that utilize non-black-box simulation to bypass the \cite{STOC:BarLin02} lower bound, e.g., \cite{FOCS:Barak01,EPRINT:BarGol01,STOC:Lindell03,STOC:Pass04,FOCS:BitPan12,STOC:Goyal13,STOC:ChuPasSet13,TCC:PanPraSah15,STOC:GOSV14}, etc. But compared with black-box simulation, our knowledge of non-black-box simulation is still limited, with a small number of non-black-box simulation techniques known. 


We summarize additional impossibility results that are not directly related to the current work. \cite{TCC:Katz08a} proved that there does not exist a four-round ZK {\em proof} with black-box simulation for $\NP$ unless $\NP \subseteq \mbf{coMA}$. \cite{EC:HazPasVen20} showed that only languages in $\mbf{coMA}$ admit a four-round {\em fully}\footnote{This means that not only the simulation but also the construction is black-box (see \cite{TCC:ReiTreVad04}).} black-box ZK {\em argument} based on one-way functions. \cite{C:KalRotRot17} demonstrated that there does not exist constant-round public-coin ZK proofs for $\NP$ even with non-black-box simulation under certain assumptions on obfuscation. \cite{EC:FleGoyJai18} showed that there does not exist three-round ZK proofs for $\NP$ even with non-black-box simulation under the same assumptions.

The recent work by Lombardi, Ma, and Spooner \cite{lombardi2022post} introduced a new notion for simulation called {\em coherent-runtime expected QPT} simulation. This notion allows the simulator to execute expected QPT procedures coherently, resulting in a superposition over computations with different runtimes, and subsequently ``revert'' the runtime by executing the same computation in reverse. Utilizing this simulation notion, \cite{lombardi2022post} achieved a set of interesting results that bypass the lower bounds established by \cite{FOCS:CCLY21}.


We note that the current work focuses exclusively on the notion of expected QPT as adapted by \cite{FOCS:CCLY21}. For a detailed discussion on the provenance of (expected) quantum polynomial-time Turing machines, please refer to Section 1.2 of \cite{lombardi2021postquantumzeroknowledgerevisited}, particularly Footnote 2 therein. 






\section{Technical Overview}
\label{sec:tech-overview}

At a high level, our approach to \Cref{thm:main:intro} aligns with the paradigm established by \cite{STOC:BarLin02,FOCS:CCLY21} in the post-quantum setting. However, the {\em fully} quantum nature of the protocol introduces fundamentally different challenges.


In the following, we begin by reviewing the approach of \cite{FOCS:CCLY21} in \Cref{sec:tech-overview:CCLY-review}, where we also clarify the challenges faced in generalizing their techniques to a fully quantum setting. Next, we present our ideas, focusing on how to overcome these obstacles. To help navigate the technical complexities, we first provide a high-level summary of our approach in \Cref{sec:birds-eye}. We then describe our strategies in greater detail in \Cref{sec:overview:NPE,sec:overview:message-order,sec:overview:sim-rewinding-q,sec:overview:rug,sec:overview:putting-together}.

\subsection{On the \cite{FOCS:CCLY21} Approach}
\label{sec:tech-overview:CCLY-review}

\cite{FOCS:CCLY21} establishes that only languages in $\BQP$ admit constant-round post-quantum BBZK (PQ-BBZK) with expected QPT simulation. At a high level, their approach operates in two main steps:
\begin{itemize}
\item 
{\bf Step 1:} Initially, for any language $\Lang$, the method begins by assuming the existence of a constant-round PQ-BBZK protocol $\langle \Prover, \Verifier \rangle$ with a {\em strict} QPT simulator $\Sim$. Next, it constructs a specialized malicious verifier $\tilde{\Verifier}$ based on the honest $\Verifier$. Then, it builds a QPT decider for $\Lang$ using this $\tilde{\Verifier}$ and the simulator $\Sim$, thereby placing $\Lang$ in $\BQP$.

\item 
{\bf Step 2:} This step extends the above result concerning {\em strict} QPT simulation to {\em expected} QPT simulation. To achieve this, the authors create another malicious verifier $\tilde{\Verifier}'$, which effectively runs the honest $\Verifier$ and the aforementioned $\tilde{\Verifier}$ in superposition. They then prove that if an expected QPT simulator $\Sim$ exists, simulating the view of $\tilde{\Verifier}'$, it can be truncated into a strict QPT simulator $\Sim'$ while still providing sufficient simulation guarantees. Specifically, $\Sim'$ and $\tilde{\Verifier}'$ can be leveraged to construct a $\BQP$ decider using a similar argument as in {\bf Step 1}, thereby ruling out expected QPT simulation as well. This step crucially relies on the quantum nature of $\tilde{\Verifier}'$ and $\Sim'$, a characteristic absent in the {\em classical setting}, where prior work such as \cite{STOC:BarLin02} fails to extend to expected polynomial-time simulation.

\end{itemize}
Throughout this technical overview, our sole focus will be on {\bf Step 1}. This is because our approach in the fully quantum setting mirrors the two-step structure outlined above, and in particular, our {\bf Step 2} can be done in a similar manner as in \cite{FOCS:CCLY21}. Consequently, in the subsequent discussion, we will exclusively recapitulate the techniques pertaining to {\bf Step 1}, as it is the pertinent component for understanding our new ideas later.

The main idea behind {\bf Step 1}, originating from \cite{STOC:BarLin02}
builds upon the intuition that we can design a specific malicious verifier $\tilde{\Verifier}$ such that when dealing with this verifier, the simulator $\Sim$'s \emph{rewinding capability 
becomes useless}. In other words, $\Sim$ essentially acts like a prover (but without a witness) in a straight-line fashion when interacting with $\tilde{\Verifier}$.
Therefore, the interaction between $\Sim$ and $\tilde{\Verifier}$ can be treated as an efficient decider for BPP (or in \cite{FOCS:CCLY21}, BQP respectively) languages in the following sense:
\begin{itemize}
\item 
On a true statement $x$, $\Sim^{\tilde{\Verifier}}(x)$ is guaranteed to emulate an accepting execution just as in the real world. This follows from the zero-knowledge property of the protocol.

\item 
On a false statement $x$, $\Sim^{\tilde{\Verifier}}(x)$ is bound to emulate a rejecting execution. This is a consequence of the protocol's soundness property, as well as the straight-line interaction manner between $\Sim$ and  $\tilde{\Verifier}$, as discussed earlier.
\end{itemize}

Formalizing the above intuition proves to be quite challenging. First, a specific malicious verifier $\tilde{\Verifier}$ is constructed as a random-aborting variant of the honest $\Verifier$. At each round $k \in [K]$, where $K$ denotes the constant round complexity of the protocol, $\tilde{\Verifier}$ operates as follows: it initially applies a function $H_\epsilon$ (explained shortly) to the prover's messages $(p_1, \ldots, p_k)$ received up to that point (referred to as the {\em prefix}). If $H_{\epsilon}(p_1, \ldots, p_k) = 0$, $\tilde{\Verifier}$ halts immediately and outputs a rejection symbol $\bot$; otherwise (i.e., $H_{\epsilon}(p_1, \ldots, p_k) = 1$), $\tilde{\Verifier}$ behaves identically to the honest $\Verifier$. Here, $H_{\epsilon}$ is a random function\footnote{Although the use of such a random function may render $\tilde{V}$ inefficient, this can be rectified by replacing $H_\epsilon$ with a $q$-wise independent hash in the final $\BQP$ decider. Hence, for the sake of clarity, we will continue to employ $H_{\epsilon}$ throughout this overview without loss of generality.} that takes input of variable length $k \in [K]$ and maps each $(p_1, \ldots, p_k)$ to $1$ (or $0$) with a properly chosen probability $\epsilon$ (or $1-\epsilon$). Subsequently, a $\BQP$ decider $\mathcal{B}$ can be constructed as follows: given input $x$, $\mathcal{B}(x)$ simply executes the simulator $\Sim^{\tilde{\Verifier}(x)}(x)$ with black-box access to $\tilde{\Verifier}(x)$ and outputs the result produced by $\Sim^{\tilde{\Verifier}(x)}(x)$. To demonstrate that $\mathcal{B}$ indeed functions as a valid $\BQP$ decider, it suffices to establish the following:
\begin{itemize}
\item 
{\bf Completeness:} For any $x \in \Lang$, $\mcal{B}(x)$ outputs acceptance with inverse-polynomial probability $1/\poly(|x|)$;
\item 
{\bf Soundness:} For any $x \notin \Lang$, $\mcal{B}(x)$ outputs acceptance with negligible probability $\negl(|x|)$.
\end{itemize}

To establish completeness, we observe that during the real execution between $\tilde{\Verifier}(x)$ and the honest prover $\Prover(x, w)$ with witness $w$ for $x$, $\tilde{\Verifier}(x)$ will output acceptance with a probability of at least $\epsilon^K$. This probability arises because $\tilde{\Verifier}(x)$ behaves identically to the honest verifier, {\em conditioned on the function $H_\epsilon$ outputting $1$ in each of the $K$ rounds}. Subsequently, owing to the ZK property, $\Sim^{\tilde{\Verifier}(x)}(x)$ will output acceptance with a probability of at least $\epsilon^K - \negl(|x|)$, which can be lower-bounded by an inverse polynomial. This is achievable due to the fact that $K$ is a constant, coupled with a suitable choice of $\epsilon$.

The proof of soundness is rather intricate. Let us first outline it in the classical context in \cite{STOC:BarLin02}, where the idea originated. Assuming for contradiction that $\mathcal{B}(x)$ outputs acceptance with some non-negligible probability $\delta(|x|)$ for some $x \notin \Lang$, \cite{STOC:BarLin02} illustrates how to transform $\mathcal{B}$ into a malicious prover $\tilde{\Prover}$ that induces the honest verifier $\Verifier$ to accept $x$ with probability $\delta(|x|)$, thereby compromising the soundness of the original ZK protocol. The approach involves $\tilde{\Prover}(x)$ internally running the $\Sim^{\tilde{\Verifier}(x)}(x)$ as $\mathcal{B}(x)$, while externally interacting with the honest verifier $\Verifier(x)$. For each $k \in [K]$, when $\Sim$ queries its oracle with input $(p_1, \ldots, p_k)$ to determine the next verifier's message $v_k$, $\tilde{\Prover}(x)$ sends $p_k$ to the external verifier $\Verifier(x)$ to obtain $v_k$, which it then forwards to $\Sim$ as the response.

Initially, it is unclear whether such a $\tilde{\Prover}(x)$ would be effective due to a significant discrepancy: the internal $\Sim$ might rewind its oracle $\tilde{\Verifier}$ to obtain the next message for two distinct history transcripts $(p_1, \ldots, p_k) \neq (p'_1, \ldots, p'_k)$, while the external verifier $\Verifier(x)$ anticipates a ``straight-line'' interaction with $\tilde{\Prover}(x)$. \cite{STOC:BarLin02} resolves this issue using the following strategy: through a careful selection of $\epsilon$, it is very unlikely that $\Sim$ could find $(p_1, \ldots, p_k) \neq (p'_1, \ldots, p'_k)$ for some $k \in [K]$ so that $H_\epsilon(p_1, \ldots, p_k) = H_\epsilon(p'_1, \ldots, p'_k) = 1$. Consequently, if $\Sim$ has already acquired a message $v_k$ pertaining to a $(p_1, \ldots, p_k)$ that satisfies $H_\epsilon$, we can confidently presume that all subsequent queries by $\Sim$ in the form of $(p'_1, \ldots, p'_k)$ do not fulfill the condition imposed by $H_\epsilon$. According to the definition of $\tilde{\Verifier}$, responses to such queries do not necessitate $\tilde{\Prover}$ to engage with the external $\Verifier$, as the response is simply the symbol $\bot$. Hence, all messages relayed by $\tilde{\Prover}$ between the internal $\Sim$ and the external $\Verifier$ indeed align with a ``straight-line'' execution.

\para{The Measure-and-Reprogram (MnR) Technique.} The post-quantum setting considered in \cite{FOCS:CCLY21} introduces a fundamental departure from the classical framework described in \cite{STOC:BarLin02}. A quantum simulator $\Sim$ has the capability to execute quantum queries to its oracle $\tilde{\Verifier}$, which in turn prompts quantum queries to the random function $H_\epsilon$. Notably, a single quantum query to $H_\epsilon$ enables $\Sim$ to acquire the responses $v_k$ for multiple prefixes $(p_1, \ldots, p_k)$ {\em in superposition}. Thus, it becomes unclear if the aforementioned idea from \cite{STOC:BarLin02} could work in this setting.

To address this challenge, \cite{FOCS:CCLY21} utilized the {\em measure-and-reprogram} (MnR) technique, originally developed in \cite{C:DFMS19,C:DonFehMaj20} to establish post-quantum Fiat-Shamir transform. First, note that the operation of $\Sim^{\tilde{\Verifier}(x)}(x)$ can be conceptualized as that of an oracle machine $\msf{SIM}^{H_\epsilon}(x)$, with $H_\epsilon$ acting as the quantum oracle. Let $H_0$ denote a null oracle that outputs $0$ for all inputs. \cite{FOCS:CCLY21} propose replacing $\msf{SIM}^{H_\epsilon}(x)$ in the description of $\mathcal{B}(x)$ with a ``MnR version'' of a new game $\msf{SIM}^{H_0}(x)$, which is almost identical to $\msf{SIM}^{H_\epsilon}(x)$, except for the following distinctions:
\begin{enumerate}
\item 
	It initializes the oracle as $H_0$ (instead of $H_\epsilon$).

\item \label{item:overview:MnR:game:2}
Assuming $\msf{SIM}$ makes $q$ quantum queries to its oracle in total, the game begins by randomly selecting $K$ queries out of all these $q$ queries. These selected queries are intended to be measured (see the next step).

\item \label{item:overview:MnR:game:3}
Each time $\msf{SIM}$ makes a query intended for measurement, the game measures this query to determine a classical prefix $(p_1, \ldots, p_k)$. The oracle is then ``reprogrammed'' to output $1$ for input $(p_1, \ldots, p_k)$. Subsequent queries are answered using this updated oracle (until it gets reprogrammed again).
\end{enumerate}
By utilizing this revised definition of $\mathcal{B}(x)$, \cite{FOCS:CCLY21} successfully adapts the arguments from \cite{STOC:BarLin02} to their post-quantum setting as follows:

For completeness, it relies on a so-called {\em MnR lemma} (established in \cite{C:DFMS19,C:DonFehMaj20,EC:YamZha21}), which asserts that the output of the ``MnR version'' of $\msf{SIM}^{H_0}$ does not differ much from the game $\msf{SIM}^{H_0}$. Furthermore, $\msf{SIM}^{H_0}$ does not differ much from the original game $\msf{SIM}^{H_\epsilon}$ due to the sparsity of the random function $H_\epsilon$. As previously argued, $\msf{SIM}^{H_\epsilon}$ outputs acceptances for $x\in \Lang$ with a ``good'' probability due to the ZK guarantee; then, so does the new $\mcal{B}(x)$ that utilizes the ``MnR version'' of $\msf{SIM}^{H_0}$.

For soundness, first note that the MnR version of $\msf{SIM}^{H_0}$ initiates with the null oracle $H_0$. As the execution progresses, it undergoes reprogramming to output $1$ solely for the {\em measured, classical} (instead of super-position) queries made by $\msf{SIM}$. This essentially facilitates the recovery of the \cite{STOC:BarLin02} arguments to establish that the messages relayed by $\tilde{\Prover}$ between the internal MnR version of $\msf{SIM}^{H_0}$ and the external verifier $\Verifier$ indeed constitute a ``straight-line'' execution. This eventually completes the reduction from the soundness of $\mathcal{B}(x)$ to that of the original ZK protocol. (Further discussion on this matter will be provided shortly.)

\jiahui{Comment: I'm using one of Luowen's suggestions. it's better to emphasize our "barriers" to overcome than saying the previous works is "easy". This will suit most readers' mindset more naturally. }

\para{Barriers in the Fully Quantum Setting (i.e. Benefits of the Classical Nature of the Protocol in \cite{FOCS:CCLY21}).} In the upcoming discussion, we will emphasize the barriers we face when proving the result in the fully quantum setting, which correspond to aspects (or benefits) in the {\em classical nature} of post-quantum ZK protocols and underpin the \cite{FOCS:CCLY21} techniques discussed above. It is worth noting that while some of these benefits were only implicit in \cite{FOCS:CCLY21}, we need to articulate them more explicitly as they represent the obstacles in our concerned setting, where the honest parties are {\em fully quantum}.

\jiahui{Comment: Adding a short title to each benefit so that when the readers refer back, they don't have to reread entire paragraph}

\hypertarget{benefit1}{\subpara{Barrier 1: Quantum communication of the honest fully-quantum protocol.}} The most apparent benefit of analyzing a classical protocol's post-quantum security lies in the classical nature of the prover's messages. This is pivotal for adapting the random-aborting technique from \cite{STOC:BarLin02} to the post-quantum setting, where the $H_\epsilon$ is evaluated on the classical prover messages at each round, determining whether $\tilde{\Verifier}$ needs to prematurely abort. However, defining the random-aborting verifier using a hash function over quantum transcripts can be challenging. Even worse, it is unclear what "transcripts" mean in the fully quantum scenario as the entire protocol can be seen as an evolution over a single quantum state over the joint systems of the prover and the verifier.

\hypertarget{benefit2}{\subpara{Barrier 2: Inherently randomized behaviors of the quantum verifier.}}  Note that $\tilde{\Verifier}$ in \cite{FOCS:CCLY21} is constructed from the classical honest $\Verifier$, thus also maintaining a classical nature. Once we fix the random tap $r$ of $\tilde{\Verifier}$, the machine $\tilde{\Verifier}_r$ becomes deterministic. Specifically, since $\tilde{\Verifier}_r$ is deterministic, when fed with the same prover's messages $(p_1, \ldots, p_k)$, it will consistently output the same next message $v_k$. This point is crucial for the final construction of the malicious prover $\tilde{\Prover}$.

Recall that $\tilde{\Prover}$ internally runs the simulator, which may attempt to rewind the execution. At a certain round $k$, $\Sim$ may ask to view the verifier's message multiple times via rewinding. While the MnR technique with the $H_0$ oracle could limit $\Sim$ to learning the value $v_k$ for only one prefix $(p_1, \ldots, p_k)$, it cannot prevent $\Sim$ from requesting to view the response to the same $(p_1, \ldots, p_k)$ again. Now, since $\tilde{\Verifier}_r$ is deterministic, the response must be $v_k$ again. This implies that $\tilde{\Prover}$ only needs to learn the value $v_k$ from the external verifier {\em only once} to accommodate the internal $\Sim$.

To further illustrate this advantage, let us momentarily assume that the verifier is not deterministic, meaning that when queried on $(p_1, \ldots, p_k)$, it may provide different $v_k$ values. In such a scenario, to perfectly simulate the environment for the internal $\Sim$, $\tilde{\Prover}$ would need to forward $(p_1, \ldots, p_k)$ externally to learn the (potentially different) response $v_k$ when requested by $\Sim$. However, note that in this soundness reduction, $\tilde{\Prover}$ interacts with the external verifier in a {\em straight-line} manner. Consequently, $\tilde{\Prover}$ is not permitted to query the external verifier twice on the same prefix.

However, we do not know of any techniques to fix a classical randomness tape of a quantum algorithm due to the inherent randomness in quantum computation (which would be surprising if possible). Therefore, we cannot apply the above analysis to a quantum honest verifier's behaviors.

\hypertarget{benefit3}{\subpara{Barrier 3: No guarantee on the ordering of the measured messages.}} The last benefit of the classical protocol in \cite{FOCS:CCLY21} also pertains to the determinism of $\tilde{\Verifier}_r$, albeit in a more subtle manner. It relates to an inherent feature of the MnR technique that was not explicitly mentioned in the previous discussion. To grasp this issue, let us revisit the steps outlined in \Cref{item:overview:MnR:game:2,item:overview:MnR:game:3}, where $K$ queries are selected to be measured. Although it can be argued\footnote{The argument is not immediately evident and requires a close examination of the MnR technique. However, we choose to omit the related intricacies as they are less pertinent to the current discussion.} that these measured queries would ultimately form a complete transcript, {\em there is no guarantee that the prover's messages will appear in the desired order}. 

To delve into this further, consider the scenario where the $k$-th ($k\in [K]$) selected query to be measured from the oracle corresponds exactly to some $(p_1, \ldots, p_k)$, enabling $\tilde{\Prover}$ to straightforwardly forward $p_k$ to learn the response $v_k$. However, this ideal situation does not always occur. For instance, imagine the case where the constructed $\tilde{\Prover}$ has yet to send any message to the external $\Verifier$. Ideally, the subsequent move would involve $\tilde{\Prover}$ forwarding the first prover message $p_1$, obtained by measuring the first selected query from the internal $\msf{SIM}^{H_0}$, to learn $\Verifier$'s response $v_1$. However, it is indeed possible that the first measurement results in a query format of $(p_1, p_2)$ instead. Intuitively, this represents a discrepancy where the internal execution of $\msf{SIM}^{H_0}$ advances to round 2 while the external execution between $\tilde{\Prover}$ and $\Verifier$ has not even commenced.

Indeed, this is a known issue in the literature, including the original MnR paper (see \cite[Section 4.1]{C:DonFehMaj20}). Typically, it poses no significant problem for post-quantum applications. For instance, \cite{FOCS:CCLY21} tackled it by leveraging the determinism of the verifier (once the random tap is fixed) as follows: in the scenario described earlier, $\tilde{\Prover}$ can pause the internal execution and ``synchronize'' the external execution as follows: first, $\tilde{\Prover}$ sends $p_1$ (in the measurement outcome $(p_1, p_2)$) to the external $\Verifier$ to learn and store the response $v_1$. Then, it sends the $p_2$ to the external $\Verifier$ to learn $v_2$. Now that $\tilde{\Prover}$ has obtained the message $v_2$ required for subsequent execution, it can resume the internal execution. Additionally, if a future selected query from $\msf{SIM}^{H_0}$ measures to $p_1$, $\tilde{\Prover}$ can respond using the $v_1$ obtained earlier. {\em Note that this strategy works due to the deterministic nature of the verifier}. Specifically, when the random tap is fixed, the response to $p_1$ is always $v_1, $ irrespective of whether $\Verifier_r$ has previously been invoked on $(p_1, p_2)$ or not.

We next shift our focus to our proposals in the fully quantum setting, specifically addressing how to handle the aforementioned dependency on the classical nature of the honest parties (or the protocol).

\subsection{A Bird's Eye View on Our Techniques}
\label{sec:birds-eye}
\xiao{xiao's version}

The above three benefits of a classical honest protocol exploited in \cite{FOCS:CCLY21}, in turn, represent obstacles in the fully quantum setting we concern. In this subsection, we provide a high-level overview of our ideas to tackle them.

Regarding the aforementioned \hyperlink{benefit1}{\underline{Barrier 1}}, an immediate obstacle in our fully quantum setting is that we can no longer define the random-aborting verifier using a hash function over transcripts, as the exchanged messages are now quantum. It is even challenging to define what ``transcripts'' mean in this context.

Our first idea is to ``dequantize'' the communication by transforming any fully quantum honest protocol into a protocol where verifier and prover pre-share EPR pairs (distributed by some trusted third party) and send each other classical messages via quantum teleportation. When it comes to simulation, the simulator $\Sim$ is required to honestly use the (honest prover's share of) EPRs when interacting with the malicious verifier. That is, for each query, $\Sim$ will use its EPR pairs for teleportation purpose only, just as an honest prover will do, and then query $\Verifier$ on the message obtained from the teleportation measurements. Other than this constraint, $\Sim $ can query $\Verifier$ in superposition and rewind it as usual. We refer to this model as the \underline{N}on-\underline{P}rogrammable \underline{E}PR (NPE) model. 
We first demonstrate that any protocol secure in this NPE model will imply a protocol secure in the fully quantum model. It then suffices to show an impossibility result in this intermediate model. In this way, we could recover $\tilde{\Verifier}$'s aborting behavior by evaluating the prover messages, which are classical now, with its random oracle.

However, we want to emphasize that this is more of a ``hard fix'' that {\em superficially} enforces a well-defined random-aborting behavior. It does not address the real issue --- the inherently random nature of quantum verifiers, as demonstrated in  \hyperlink{benefit2}{\underline{Barrier 2}} and \hyperlink{benefit3}{\underline{Barrier 3}} in a fully quantum setting. {\em Worse, we may even have complicated the task further by additionally introducing pre-shared entanglement into the honest protocol.}

To deal with these issues, we develop new analytical tools to characterize the evolution of $\tilde{\Verifier}$'s state throughout its interaction with $\Sim$, {\em even in the presence of the pre-shared entanglement}. Roughly speaking, we show that the overall state is always in a superposition of a ``good'' branch and a ``bad'' branch. The good branch corresponds to the messages and internal verifier state that an honest verifier will execute on to decide whether to accept the input $x$. The bad branch consists of ``garbage'' states which result from pre-maturely aborted messages and $\Sim$'s malicious rewinding behaviors, and we have to handle them carefully.

As a key observation, we found that those garbage states maintain certain structures during ZK simulation, and we managed to accurately characterize the that structure in the recursive evolution of the garbage states.
This key observation allows us to construct a dummy verifier operator on the bad branch when $\Sim$ queries $\tilde{\Verifier}$ or rewinds it. This is the main observation that brought us the initial hope to implement the ``rewinding does not help'' intuition in the fully quantum setting {\em without any need to de-randomize the verifier}.

While inspiring, the above observations do not resolve the ``out of synchronization'' issue described in \hyperlink{benefit3}{\underline{Barrier 3}}, which again arises from the verifier's inherent quantumness. To address this, we design the reduction to use two counters which, with careful construction, achieve the following effects:
\begin{itemize}
\item 
The first counter is {\em global}. It increments (resp.\ decrements) each time the simulator $\Sim$ queries $\tilde{\Verifier}$ (resp.\ $\tilde{\Verifier}^\dagger$).

\item 
The second counter is {\em local}. It mirrors the behavior of the global counter only on the good branch, while effectively ``deadlocking'' the bad branch whenever the bad-branch execution lags behind the good branch. Additionally, each time $\Sim$ rewinds the good branch to align with the depth of the bad branch, the two branches merge, and the local counter resumes mirroring the global counter.
\end{itemize}
We show that this design serves as an effective alternative to \cite{FOCS:CCLY21} for inducing the desired query order during measure-and-reprogram, {\em without requiring derandomization of the verifier}.

Eventually, the above ideas together guarantee that anything meaningful $\Sim$ can do appears only in the good branch, which has analogous behaviors to a non-aborting, ``straight-line'' interaction in the post-quantum setting of \cite{FOCS:CCLY21}. We can then complete the proof by ``gluing'' these ideas together.  However, we note that the above description is a significant oversimplification; implementing the approach requires non-trivial additional ideas. In the sequel, we provide an overview in greater detail.

\subsection{Non-Programmable EPR Model with Classical Prover Messages}
\label{sec:overview:NPE}

To address the reliance on the prover's classical messages, as discussed in \hyperlink{benefit1}{\underline{Barrier 1}}, we introduce an intermediate model for zero-knowledge protocols that partially ``de-quantizes'' a fully quantum protocol. This model, which we will refer to as the {\em Non-Programmable EPR} (NPE for short) model, operates as follows. In the NPE model, a trusted third party prepares a polynomial number of EPR pairs. At the start of the protocol, the prover and verifier each possess half of these EPR pairs. Throughout the protocol, they are permitted to utilize these EPR pairs. When defining the zero-knowledge property, the simulator holds the prover's shares of the EPR pairs-- half the trusted EPR pairs are given to the simulator as part of the input and the other half are given through black box access to the verifier. 
It is crucial to emphasize that in the definition of zero-knowledge, the simulator and verifier's EPR shares are still prepared and distributed by the trusted party. This stands in contrast to the widely-used Common Reference String (CRS) model, where the simulator is responsible for generating the CRS when defining zero-knowledge.

We next argue that if there exists a $K$-round fully-quantum BBZK protocol (in the standard model), then it can be converted into a $K$-round BBZK protocol in the NPE model, {\em where all messages sent by the prover are classical}. Roughly, this is because the prover can always use quantum teleportation to transmit the originally quantum prover message. In more detail, whenever the prover needs to send a quantum message $p_k$, it instead does the following: The prover performs a teleportation measurement on the (originally quantum) message $p_k$ and the EPR registers that are meant to be used for this round, to obtain a classical measurement outcome $\tilde{p}_k$ (i.e., the teleportation keys) and sends it to the verifier. Using $\tilde{p}_k$ and the corresponding EPR shares on the verifier's side, the verifier can recover the original message $p_k$. Then, it behaves as in the original protocol --- generating $v_k$, sending $v_k$ to the prover (note that we do not ask the verifier to perform quantum teleportation), and moving to the next round.

The aforementioned transformation clearly preserves the round complexity of the original quantum protocol. Furthermore, it can be shown that completeness, soundness, and zero-knowledge properties are all maintained. This follows from relatively standard reduction techniques, so we will not delve into further detail beyond mentioning the intuition behind the preservation: the EPR pairs are solely used as a means for teleportation, effectively ``de-quantizing'' the prover's messages; they do not alter the inherent properties, such as soundness or the ZK property, of the original protocol. For more details, please refer to \Cref{sec:def:QZK:NPE} and \Cref{sec:model:subsection:QZK-to-NPE}.

With the above claim, it now suffices to establish \Cref{thm:main:intro} in the NPE model and assume that all the prover messages are classical.

Finally, it is worth mentioning that the above de-quantization approach {\em only} provides us with a well-defined random-aborting behavior for the verifier. However, it remains uncertain whether this model truly offers substantial benefits, as both the verifier and the simulator still perform inherently quantum operations. Moreover, the NPE model comes at a price: it introduces pre-shared entanglement between the prover and the verifier, which may potentially complicate establishing impossibility results. It is unclear whether what we pay justifies what we gain! Looking ahead, we find that the techniques we develop in the sequel actually work even in the presence of pre-shared entanglement.



\subsection{Inducing an Order on the Measured Messages}
\label{sec:overview:message-order}

In this section, we introduce new ideas aimed at inducing an order on the measured messages in the MnR game, in order to address the issue mentioned in \hyperlink{benefit3}{\underline{Barrier 3}}. Looking ahead, these ideas also prove to be helpful when we address (in \Cref{sec:overview:sim-rewinding-q}) the issues mentioned in \hyperlink{benefit2}{\underline{Barrier 2}}, as these two barriers are closely related.

We propose that the malicious $\tilde{\Verifier}$ maintain two registers additionally: a {\em global counter} register $\reg{gc}$ and a {\em local counter} register $\reg{lc}$. Both registers are initialized to $0$. Intuitively, the global counter will record the number of times the $\tilde{\Verifier}$ has been invoked, and the local counter will record at which round the current execution is located.

In particular, consider the $k$-th round when $\tilde{\Verifier}$ receives a message $p_{k}$. Assume that the current local counter has a value of $j-1$ (for some $j$) and the current global counter has a value of $k-1$.\footnote{As will become clear shortly, at the beginning of the $k$-th round, the global counter must have a value of $k-1$.} $\tilde{\Verifier}$ first \ul{increases the global counter from $k-1$ to $k$}, and then behaves by comparing these two counters (before the increase of $k-1$ happens):
\begin{enumerate}
\item \label{overview:counter:step:1}
If $(j-1) \ne (k-1)$, then $\tilde{\Verifier}$ does not do anything.

\item  \label{overview:counter:step:2}
If $(j-1) = (k-1)$, then $\tilde{\Verifier}$ behaves as the honest verifier with random-aborting. In particular, she first queries the oracle $H$ to learn $H(p_1, \ldots, p_{j})$.\footnote{We remark that this is not a typo. We mean to invoke $H$ on input $(p_1, \ldots, p_{j})$ when the local counter is $j-1$.} Note that the input to $H$ is determined by the local counter value $j-1$, and in this case of $(j-1) = (k-1)$, the value $H(p_1, \ldots, p_{j})$ is exactly the value $H(p_1, \ldots, p_k)$.

The subsequent movements of $\tilde{\Verifier}$ are controlled by this value:
\begin{enumerate}
\item \label{overview:counter:step:2:1}
If $H(p_1, \ldots, p_{j}) =0$, then $\tilde{\Verifier}$ does not do anything;
\item  \label{overview:counter:step:2:2}
If $H(p_1, \ldots, p_{j}) =1$, then $\tilde{\Verifier}$ generates the response $v_k$ to $p_k$ in the same manner as the honest verifier. After that, $\tilde{\Verifier}$ \ul{increases the local counter from $j-1$ to $j$}.
\end{enumerate}
\end{enumerate}
The most important sentences to notice in the above description are the ones underscored: at each round, the global counter will always be increased; however, {\em the local counter will be increased only if that round is executed successfully, namely when $H(p_1, \ldots, p_{j}) = 1$.}

Let us explain the benefits of the above design. First, consider the ``straight-line'' execution between $\tilde{\Verifier}$ and the honest prover $\Prover$. In this real execution, we claim that $\tilde{\Verifier}$ simply behaves as the random-aborting verifier as the one from \cite{STOC:BarLin02} or \cite{FOCS:CCLY21}. That is, the use of these extra counters at least does not interfere with the ``random-aborting'' behavior in this straight-line execution. This is particularly important to ensure that we can establish the completeness of the $\BQP$ decider, which works in a similar manner as in \cite{STOC:BarLin02,FOCS:CCLY21}. This can be easily seen by tracking the execution: At the beginning, both counters are set to 0. When the first message $p_1$ arrives, the global counter first gets increased to $\ket{1}_{\reg{gc}}$. Next,
\begin{itemize}
\item 
If $H(p_1) = 0$, then nothing will happen;
\item
If $H(p_1) = 1$, then the message $v_1$ is generated, and the local counter is increased to $\ket{1}_{\reg{lc}}$.
\end{itemize}
In summary, if $H(p_1) = 0$ happens, then the counters become $\ket{1}_{\reg{gc}} \ket{0}_{\reg{lc}}$. According to our definition in \Cref{overview:counter:step:1}, the execution is essentially ``dead'' in the sense that nothing will happen in subsequent steps, because the global and local counters are not consistent. This is exactly the same as in \cite{STOC:BarLin02,FOCS:CCLY21}. If $H(p_1) = 1$ happens, then the counters become $\ket{1}_{\reg{gc}} \ket{1}_{\reg{lc}}$ and the execution proceeds to the next round properly. This is also the same as in \cite{STOC:BarLin02,FOCS:CCLY21}. The similar feature holds for every subsequent round, and thus one can see that our counters do not alter the behavior of $\tilde{\Verifier}$ (as in \cite{STOC:BarLin02,FOCS:CCLY21}) in the real, straight-line execution with the honest $\Prover$.

Next, we turn to the real virtue of the above design --- it ensures that the measured messages in the MnR game appear in an increasing order. Let us first show that the first selected (to be measured) query could measure to $p_1$. Recall that the MnR game starts with the null oracle $H_0$. Before the first measurement happens, $H_0$ has not been reprogrammed and thus it outputs 0 on all inputs. In this case, we know that the local counter $j$ must be 0 (recall from \Cref{overview:counter:step:2:2} that the local counter gets increased only in the $H(p_1, \ldots, p_j)=1$ branch). Thus, the first selected query sent to $H_0$ must be a superposition of messages of length $j+1 = 0+1 = 1$. Therefore, the measurement outcome must be some message $p_1$.

Using a similar argument as above, we can indeed show that for the $k$-th selected (to be measured) query to the oracle, the length of the measured outcome could never exceed $k$ ($\forall k \in [K]$). Finally, since there are only $K$ selected measurement opportunities in the MnR game, if the execution of $\msf{SIM}^{H_0}(x)$ finally brings the local counter to value $K$ (which is a necessary condition for the verifier's acceptance), it must be the case that {\em the $k$-th selected query to the oracle measures to exactly a length-$k$ prover's message} (see \Cref{sec:soundness:proof:cleaning:MnR} for a formal treatment). In other words, the measurements will yield prover messages in the desired order.

Note that the above ideas only assist us in achieving the correct order of the measured messages. However, it remains unclear how to construct the malicious prover $\tilde{\Prover}$ to prove the soundness of $\mcal{B}$ because it is uncertain how to manage rewinding queries by $\Sim$. This leads us to \Cref{sec:overview:sim-rewinding-q}.

\subsection{Simulation for Rewinding Queries}
\label{sec:overview:sim-rewinding-q}

We now shift our focus to addressing the challenges mentioned in \hyperlink{benefit2}{\underline{Barrier 2}}, particularly regarding how to manage the rewindings required by $\Sim$ while operating internally within $\tilde{\Prover}$.

Technically, the issue can be accurately described as follows: at a certain round $k$, when $\tilde{\Prover}$ receives the (potentially quantum) message $v_k$ from the external $\Verifier$, we need to devise a mechanism that enables $\tilde{\Prover}$ to ``re-use'' this $v_k$ for future rewinding queries made by $\Sim$ when necessary.

\hypertarget{theIntui}{\para{The Intuition.}} Before delving into the technical intricacies, we aim to provide an intuitive understanding of our approach. Broadly speaking, our goal is to establish an intuition similar to that of \cite{STOC:BarLin02,FOCS:CCLY21}, which helps to extend the ``simulator learns nothing new by rewinding'' concept to our fully quantum setting. This poses a significant challenge due to the fully quantum interaction between $\Sim$ and $\tilde{\Verifier}$, intertwined with intermediate measurements conducted by $\Sim$ and the MnR game.

To tackle this challenge, we will develop a series of analytical tools to characterize key aspects of the interaction between $\Sim$ and $\tilde{\Verifier}$ within the MnR game. These features will serve as crucial threads, enabling us to navigate and maintain a detailed description of the overall state across all registers held by $\Sim$ and $\tilde{\Verifier}$ throughout the execution, at an appropriate resolution. As we will demonstrate shortly, this detailed description will reveal a crucial observation: after each query of $\Sim$, the overall state can always be expressed as a superposition of a ``good'' branch and a ``bad'' branch:
\begin{itemize}
\item 
The good branch will mirror the state of the honest verifier in a real execution, at the appropriate round.

\item 
The bad branch comprises some ``error'' terms that we would ideally like to eliminate, but are unable to do so. Fortunately, we can demonstrate that these error terms possess a structured nature that enables us to assert the following recursive features:
\begin{itemize}
\item 
{\sf Round Slackness:} The bad branch consistently lags behind the good branch. Specifically, if the good branch corresponds to a real execution reaching round $k$, then the bad branch will only contain (in superposition) executions that reached up to round $k-1$.

\item
{\sf Return Simulatability:} In subsequent steps, $\Sim$ may rewind the execution by invoking $\tilde{\Verifier}$'s unitary $\tilde{V}^\dagger$. For these queries, the structure of the bad branch permits us to utilize a ``dummy version'' $\ddot{V}$ to substitute the unitary $\tilde{V}$. This $\ddot{V}$ solely adjusts the global and local counter registers without affecting the overall state. Importantly, $\ddot{V}$ guarantees the following: regardless of how many rounds $\Sim$ rewinds the execution (answered using $\ddot{V}^\dagger$), once the execution (specifically, the good branch) returns to round $k$ (with sufficient $\ddot{V}$ queries by $\Sim$), the state will be essentially identical to the state when it last reached round $k$. Additionally, $\Sim$ cannot discern that $\tilde{V}$ has been replaced with $\ddot{V}$.

\item 
{\sf Error Invariance:} The error terms allow us to make the following recursive claim. After the next move by the simulator $\Sim$, the state can again be partitioned into a good branch and a bad branch. Furthermore, the good branch exhibits the same features as described above, while the bad branch maintains the same {\sf Round Slackness} and {\sf Error Invariance}.
\end{itemize}
\end{itemize}
Next, we demonstrate how these features lead to the desired construction of $\tilde{\Prover}$ for our soundness reduction. The $\tilde{\Prover}$ can be outlined as follows: when $\Sim$ initially requests to observe $v_k$ (and the MnR game measures $H(p_1, \ldots, p_k) = 1$, granting $\Sim$ permission to observe $v_k$), $\tilde{\Prover}$ will transmit the message $p_k$ to the external $\Verifier$, who will subsequently provide $v_k$ in response. $\tilde{\Prover}$ will then relay this $v_k$ to the internal $\Sim$ (corresponding to the good branch at round $k$). Subsequently, all queries from $\Sim$ will be addressed using $\ddot{V}^\dagger$ (or $\ddot{V}$) {\em until the internal execution returns to round $k$ once more}. This same procedure is repeated for each subsequent round. In essence, when $\Sim$ seeks to observe the message $v_{k+1}$, it is provided by the external $\Verifier$ for the first time, and all subsequent queries are answered using $\ddot{V}$ or $\ddot{V}^\dagger$ until $\Sim$ requests the next message $v_{k+2}$, and so forth.

Let us analyze the constructed $\tilde{\Prover}$. Firstly, note that for each $k \in [K]$, $\tilde{\Prover}$ communicates with the external $\Verifier$ precisely once, when $\Sim$ requests to observe $v_k$ for the first time. All subsequent queries are handled internally by $\tilde{\Prover}$ using the dummy unitaries $\ddot{V}$ or $\ddot{V}^\dagger$. Thus, the good branch, {\em at the conclusion of the execution}, mirrors precisely the final state of the external verifier. Therefore, the remaining task is to demonstrate that the good branch at the conclusion of the execution encompasses sufficient ``weight'' on acceptance.

Due to the {\sf Return Simulatability}, the $\Sim$ inside $\tilde{\Prover}$ perceives itself to be operating within the original real MnR game. Hence, it suffices to demonstrate that, at the culmination of the original MnR game, the good branch possesses sufficient weight to warrant acceptance. To this end, the {\sf Error Invariance} property allows us to uphold the clean good-bad state structure throughout the entire execution. Specifically, this structured state persists at the instant $\Sim$ concludes its simulation. Now, we assert that the bad branch does not contribute to acceptance whatsoever. This is because: (1) the bad branch, by virtue of the {\sf Round Slackness}, cannot correspond to the round $K$, and (2) the verifier will not accept an execution that has not reached the final round $K$. On the contrary, recalling our initial assumption (for contradicting soundness), the overall state must contain sufficient weight on acceptance. Therefore, only one possibility remains --- all the weight for acceptance is ``concentrated'' within the good branch at the conclusion of the simulation. This concludes our argument establishing the validity of $\tilde{\Prover}$.

We now transition to a technical discourse, delving into the mathematical instantiation of the above intuition and ideas.

\para{A Simplified Example.} To elucidate our main idea, let us start with an simplified scenario. First, it is crucial to outline certain behaviors of $\Sim$ and $\tilde{\Verifier}$ that bear relevance to the ensuing discussion.

Our construction of $\tilde{\Verifier}$ maintains $K$ registers $\reg{p_1 \ldots p_K}$ intended to store previous messages received from the prover. Additionally, there exists a {\em message transmission} register $\reg{m}$ facilitating the exchange of messages between the prover and verifier. Specifically, to transmit message $p_k$, $\Prover$ loads it into $\reg{m}$ and forwards $\ket{p_k}_{\reg{m}}$ to the verifier. Upon receiving $\ket{p_k}_{\reg{m}}$, the initial step of $\tilde{V}$ --- the unitary operator of our malicious $\tilde{\Verifier}$ --- involves applying a swap operation between register $\reg{p_k}$ and $\reg{m}$. This signifies $\tilde{\Verifier}$ relocating message $p_k$ to its appropriate position (i.e., register $\reg{p_k}$) within her internal space. Subsequently, $\tilde{\Verifier}$ executes requisite computations to generate the response message $v_k$.

We will also utilize a unitary $V_{p_k}$, representing the operation of the {\em honest} verifier that derives $v_k$ from message $p_k$. Notably, in the present model (as per \Cref{sec:overview:NPE}), all messages from the honest prover are classical. Hence, we can presume that at round $k$, the operator of the honest verifier can be represented as $\sum_{p_k} \ketbra{p_k}_{\reg{p_k}}\tensor V_{p_k}$. That is, when the prover's message is $p_k$, the unitary $V_{p_k}$ is applied by the honest verifier.

The simulator $\Sim$ is specified by a ``local'' operator $S$, which operates non-trivially on registers $\reg{m}$ and $\Sim$'s working space $\reg{s}$. Specifically, $\Sim$ operates in two steps: (1) it invokes $S$ to prepare a state over registers $\reg{m}$ and $\reg{s}$, and (2) it makes an oracle call to $\tilde{V}$ (or $\tilde{V}^\dagger$ for rewinding). Crucially, $\Sim$ is not able to observe (let alone modify) the internal registers of $\tilde{\Verifier}$, as we are focusing on {\em black-box} simulation.

Additionally, it is important to note that we are operating within the NPE model as described in \Cref{sec:overview:NPE}. While this restricts us to dealing only with classical prover messages, both the prover and verifier now possess EPR pairs. Thankfully, we can regard these shares as stored in $\Sim$'s internal register $\reg{s}$ and $\tilde{\Verifier}$'s internal registers $\reg{v}$ respectively; the following derivation proceeds without explicitly referencing these shares.

Now, we are ready to describe our simplified example. Let us assume that the execution is currently at round $k$, and the overall state of $\tilde{\Verifier}$ has the following format:
\begin{equation}\label[Expression]{simp-example:initial}
\ket{k}_{\reg{gc}} \ket{k}_{\reg{lc}} \ket{p_1, \ldots, p_k}_{\reg{p_1 \ldots p_k}}  \ket{\rho}_{\reg{v}},
\end{equation}
where $\reg{v}$ represents $\tilde{\Verifier}$'s other registers, and the $(p_1, \ldots, p_k)$ satisfies the current oracle $\hat{H}(p_1, \ldots, p_k) = 1$. (Recall that we are in the MnR game where the initial oracle $H_0$ may get re-programmed during the execution. We use $\hat{H}$ to denote the current oracle.)

\begin{remark}\label{rmk:overview:pure-subnormalized}
Note that in \Cref{simp-example:initial}, we assume for simplicity that the current global counter and local counter are both equal to $k$. Also, note that in the full-fledged game, $\tilde{\Verifier}$'s registers would be in a much more complicated {\em mixed} state. Here, we assume this {\em pure} state format for simplicity. We also make similar simplification assumptions for the following description and derivation regarding the simulator's behavior. These assumptions are meant to help the reader understand our ideas more clearly, without being confused by complex notations of secondary importance. We will present a discussion regarding the full-fledged case later in \Cref{sec:overview:rug}.
\end{remark}
	
Now, assume that $\Sim$ wants to rewind the execution upon receiving input $p_k$. In particular, let us assume that $\Sim$ prepares $\ket{p_k}_{\reg{m}}$ together with some state $\ket{\phi_{p_k}}_{\reg{s}}$ on her internal register $\reg{s}$, and then she rewinds the execution by invoking $\tilde{V}^\dagger$. Next, she applies her local unitary $S$, and finally calls $\tilde{V}$ to bring the execution back. Let us track the overall state during this procedure:
\begin{align*}
& \tilde{V} S \tilde{V}^\dagger \ket{k}_{\reg{gc}} \ket{k}_{\reg{lc}}  \ket{p_1, \ldots, p_k}_{\reg{p_1 \ldots p_k}} \ket{p_k}_{\reg{m}}\ket{\phi_{p_k}}_{\reg{s}} \ket{\rho}_{\reg{v}} 
 \\ 
  = ~ & 
 \tilde{V} S \ket{k-1}_{\reg{gc}} \ket{k-1}_{\reg{lc}}  \ket{p_1, \ldots, p_k}_{\reg{p_1 \ldots p_k}} \ket{p_k}_{\reg{m}} \ket{\phi_{p_k}}_{\reg{s}} V^\dagger_{p_k} \ket{\rho}_{\reg{v}} 
 \numberthis \label{simp-example:real:eq:1} \\ 
 = ~ &
\tilde{V}  \ket{k-1}_{\reg{gc}} \ket{k-1}_{\reg{lc}} \ket{p_1, \ldots, p_k}_{\reg{p_1 \ldots p_k}} \big(S \ket{p_k}_{\reg{m}} \ket{\phi_{p_k}}_{\reg{s}}\big) V^\dagger_{p_k} \ket{\rho}_{\reg{v}} 
\numberthis \label{simp-example:real:eq:2} \\ 
= ~ &
\tilde{V}  \ket{k-1}_{\reg{gc}} \ket{k-1}_{\reg{lc}} \ket{p_1, \ldots, p_k}_{\reg{p_1 \ldots p_k}} \ket{p_k}_{\reg{m}} \ket{\phi'_{p_k}}_{\reg{s}} V^\dagger_{p_k} \ket{\rho}_{\reg{v}} 
~ + ~ \\
& \hspace{8em}
\tilde{V} \ket{k-1}_{\reg{gc}} \ket{k-1}_{\reg{lc}} \ket{p_1, \ldots, p_k}_{\reg{p_1 \ldots p_k}} \sum_{p'_k \ne p_k} \ket{p'_k}_{\reg{m}} \ket{\phi'_{p'_k}}_{\reg{s}} V^\dagger_{p_k} \ket{\rho}_{\reg{v}} 
\numberthis \label{simp-example:real:eq:3} \\ 
= ~ &
\ket{k}_{\reg{gc}} \ket{k}_{\reg{lc}} \ket{p_1, \ldots, p_k}_{\reg{p_1 \ldots p_k}} \ket{p_k}_{\reg{m}} \ket{\phi'_{p_k}}_{\reg{s}} \ket{\rho}_{\reg{v}} 
~ + ~ \\
& \hspace{8em}
\tilde{V} \ket{k-1}_{\reg{gc}} \ket{k-1}_{\reg{lc}} \ket{p_1, \ldots, p_k}_{\reg{p_1 \ldots p_{k-1}}} \sum_{p'_k \ne p_k} \ket{p'_k}_{\reg{p_k}} \ket{p_k}_{\reg{m}} \ket{\phi'_{p'_k}}_{\reg{s}} V^\dagger_{p_k} \ket{\rho}_{\reg{v}} 
\numberthis \label{simp-example:real:eq:4} \\ 
= ~ &
\underbrace{
\ket{k}_{\reg{gc}} \ket{k}_{\reg{lc}}\ket{p_1, \ldots, p_k}_{\reg{p_1 \ldots p_k}} \ket{p_k}_{\reg{m}} \ket{\phi'_{p_k}}_{\reg{s}} \ket{\rho}_{\reg{v}} \vphantom{\sum_{p'_k \ne p_k}} 
}_{good}
~ + ~ 
\underbrace{
\ket{k}_{\reg{gc}} \ket{k-1}_{\reg{lc}} \ket{p_1, \ldots, p_k}_{\reg{p_1 \ldots p_{k-1}}} \sum_{p'_k \ne p_k} \ket{p'_k}_{\reg{p_k}} \ket{p_k}_{\reg{m}} \ket{\phi'_{p'_k}}_{\reg{s}} V^\dagger_{p_k} \ket{\rho}_{\reg{v}} 
}_{bad}
\numberthis \label{simp-example:real:eq:5} 
,\end{align*}
where 
\begin{itemize}
\item 
\Cref{simp-example:real:eq:1} follows from the definition of $\tilde{V}$ --- when $(p_1, \ldots, p_k)$ satisfies the current oracle $\hat{H}$, $\tilde{V}^\dagger$ will simply decrease the global and local counters, and apply the honest verifier's unitary $V^\dagger_{p_k}$.

\item 
\Cref{simp-example:real:eq:2} follows from the fact that $\Sim$'s local unitary $S$ only touches upon registers $\reg{m}$ and $\reg{s}$.

\item 
\Cref{simp-example:real:eq:3} follows from decomposing the state on $\reg{m}$ and $\reg{s}$ in the computational basis on the $\reg{m}$ register, i.e., $S \ket{p_k}_{\reg{m}} \ket{\phi_{p_k}}_{\reg{s}} = \sum_{p'_k} \ket{p'_k}_{\reg{m}} \ket{\phi'_{p'_k}}_{\reg{s}}$. 

\item 
\Cref{simp-example:real:eq:4} again follows from the definition of $\tilde{V}$ --- when $(p_1, \ldots, p_k)$ satisfies the current oracle $\hat{H}$, $\tilde{V}$ will simply decrease the global and local counters, and apply the honest verifier's unitary $V_{p_k}$, which  cancels the $V^\dagger_{p_k}$.

\item 
To see \Cref{simp-example:real:eq:5}, we need to recall that  $\tilde{V}$ indeed first swaps the contents of $\reg{m}$ and $\reg{p_k}$ as we described earlier. After that, the contents in registers $\reg{p_1 \ldots p_k}$ do not satisfy $\hat{H}$ because $p'_k\ne p_k$. Thus, the  $V^\dagger_{p_k}$ and the local counter will be left as they are, and only the global counter gets increased.
\end{itemize}
Next, we define a ``dummy'' operator $\ddot{V}$. It works in the identical manner as $\tilde{V}$, with the only difference that $\ddot{V}$ does not perform the work of $V_{p_k}$. Now, let us track the same execution but with the $\ddot{V}$ in place of $\tilde{V}$.
\begin{align*}
& \ddot{V} S \ddot{V}^\dagger \ket{k}_{\reg{gc}} \ket{k}_{\reg{lc}} \ket{p_1, \ldots, p_k}_{\reg{p_1 \ldots p_k}} \ket{p_k}_{\reg{m}}\ket{\phi_{p_k}}_{\reg{s}} \ket{\rho}_{\reg{v}} 
 \\ 
  = ~ & 
 \ddot{V} S  \ket{k-1}_{\reg{gc}} \ket{k-1}_{\reg{lc}} \ket{p_1, \ldots, p_k}_{\reg{p_1 \ldots p_k}} \ket{p_k}_{\reg{m}} \ket{\phi_{p_k}}_{\reg{s}} \ket{\rho}_{\reg{v}} 
 \\ 
 = ~ &
\ddot{V}  \ket{k-1}_{\reg{gc}} \ket{k-1}_{\reg{lc}} \ket{p_1, \ldots, p_k}_{\reg{p_1 \ldots p_k}} \big(S \ket{p_k}_{\reg{m}} \ket{\phi_{p_k}}_{\reg{s}}\big)  \ket{\rho}_{\reg{v}} 
 \\ 
= ~ &
\ddot{V} \ket{k-1}_{\reg{gc}} \ket{k-1}_{\reg{lc}} \ket{p_1, \ldots, p_k}_{\reg{p_1 \ldots p_k}} \ket{p_k}_{\reg{m}} \ket{\phi'_{p_k}}_{\reg{s}}  \ket{\rho}_{\reg{v}} 
~ + ~ \\
& \hspace{8em}
\ddot{V} \ket{k-1}_{\reg{gc}} \ket{k-1}_{\reg{lc}} \ket{p_1, \ldots, p_k}_{\reg{p_1 \ldots p_k}} \sum_{p'_k \ne p_k} \ket{p'_k}_{\reg{m}} \ket{\phi'_{p'_k}}_{\reg{s}}  \ket{\rho}_{\reg{v}} 
\\ 
= ~ &
\underbrace{
\ket{k}_{\reg{gc}} \ket{k}_{\reg{lc}} \ket{p_1, \ldots, p_k}_{\reg{p_1 \ldots p_k}} \ket{p_k}_{\reg{m}} \ket{\phi'_{p_k}}_{\reg{s}} \ket{\rho}_{\reg{v}} \vphantom{\sum_{p'_k \ne p_k}}
}_{good} 
~ + ~ 
\underbrace{
\ket{k}_{\reg{gc}} \ket{k-1}_{\reg{lc}}  \ket{p_1, \ldots, p_k}_{\reg{p_1 \ldots p_{k-1}}} \sum_{p'_k \ne p_k} \ket{p'_k}_{\reg{p_k}} \ket{p_k}_{\reg{m}} \ket{\phi'_{p'_k}}_{\reg{s}}  \ket{\rho}_{\reg{v}} 
}_{bad}
\numberthis \label{simp-example:dummy:eq:5} 
,\end{align*}
where \Cref{simp-example:dummy:eq:5}  follows from the similar derivation as we did for \Cref{simp-example:real:eq:5}. 
\begin{remark}
As an astute reader may have already noticed, the derivations above rely on an oversimplification that we must now address. Recall that the unitary $V_{p_k}$ generates message $v_k$. Right after that, $v_k$ is stored in some register $\reg{v_k}$ within the internal space of $\tilde{\Verifier}$. To actually deliver this message to the prover, $\tilde{\Verifier}$ needs to apply a swap operator between $\reg{m}$ and $\reg{v_k}$ to load $v_k$ into $\reg{m}$. Thus, $\tilde{\Verifier}$'s operator $\tilde{V}$ indeed also needs to interact with register $\reg{m}$. However, this interaction is not reflected in the above derivations. We note that our actual proof in the main body does account for this case. However, doing so requires demonstrating finer-grained properties of both $\Sim$ and $\tilde{\Verifier}$, which cannot be accommodated within the limited space of this technical overview. Thus, we have chosen to omit these details here. Nonetheless, we assert that this omission does not pose any real problems for the subsequent discussion.
\end{remark}

Now, let us elucidate how the example above illustrates the features outlined in the \hyperlink{theIntui}{The Intuition} part. Firstly, observe that the branches labeled as {\em good} in \Cref{simp-example:real:eq:5,simp-example:dummy:eq:5} are identical, despite being obtained through different procedures $\tilde{V}S\tilde{V}^\dagger$ and $\ddot{V}S\ddot{V}^\dagger$, respectively. This essentially demonstrates what we referred to as {\sf Return Simulatability} earlier. Although \Cref{simp-example:real:eq:5,simp-example:dummy:eq:5} still differ in the bad branches, it is clear that the bad branches in both equations have a local counter value of $k-1$, while the good branch has already progressed to round $k$ (i.e., both the global and local counters are equal to $k$). This aligns with what we call {\sf Round Slackness}. 

Additionally, recall from our design of counters in \Cref{sec:overview:message-order} that when the global counter differs from the local counter, the unitary $\tilde{V}$ does nothing. Therefore, the bad branches in \Cref{simp-example:real:eq:5,simp-example:dummy:eq:5} are essentially ``locked'' there, contributing nothing to the subsequent execution. Moreover, notice that in the good branches of both equations, the state $\ket{\rho}_{\reg{v}}$ over the verifier's internal register $\reg{v}$ turns out to be identical to that in the initial state shown in \Cref{simp-example:initial}. This, along with the uselessness of the bad branches we just explained, essentially means that to handle this ``rewinding then returning back to round $k$'' procedure, one does not need to perform the work $V_{p_k}$ again (as per $\tilde{V}$); the dummy unitary $\ddot{V}$ will suffice.

\para{On the Recursiveness of Error Invariance.} Still, there is an obvious discrepancy between the above simplified example and the {\sf Error Invariance} described in the \hyperlink{theIntui}{The Intuition} part. Recall that {\sf Error Invariance} suggests that the good-bad structure will be {\em recursively} maintained. In our example, we began with an initial state (i.e., \Cref{simp-example:initial}) that only had a good branch (in that it only contains the $(p_1, \dots, p_k)$ satisfying $H$), while to establish  {\sf Error Invariance}, we should have started with a state containing both good and bad branches.

However, the validity of our argument remains intact even if we start with a state containing both good and bad branches. Let us delve into the main intuition. The bad branch in \Cref{simp-example:real:eq:5} differs from that in \Cref{simp-example:dummy:eq:5} only in the presence of $V^\dagger_{p_k}$ in front of $\ket{\rho}_{\reg{v}}$. This difference is not coincidental. If we apply $\tilde{V} S \tilde{V}^\dagger$ (resp.\ $\ddot{V} S \ddot{V}^\dagger$) to the state in \Cref{simp-example:real:eq:5} (resp.\ \Cref{simp-example:dummy:eq:5}), the resulting states will still share an identical good branch, while the bad branches differ by a $V^\dagger_{p_k}$. 

To illustrate, consider applying $\tilde{V} S \tilde{V}^\dagger$ again to \Cref{simp-example:real:eq:5}. The good branch will evolve in exactly the same manner as the above \Cref{simp-example:real:eq:5}, yielding a new good branch $\msf{good}_1$ and a new bad branch $\msf{bad}_1$ (with the same $V^\dagger_{p_k}$ hanging there). 

The evolution of the original bad branch proceeds as follows: Since the global counter $k$ differs from the local counter $k-1$, the first operator $\tilde{V}^\dagger$ only reverts the global counter to $k-1$ and does nothing else. Subsequently, $\Sim$'s local operator $S$ is applied, potentially resulting in a state such as $\sum_{p'_k} \ket{p_k'}_{\reg{m}} \ket{\phi''_{p'_k}}_{\reg{s}}$ as in \Cref{simp-example:real:eq:3}. The subsequent application of $\tilde{V}$ leads to a similar evolution as seen from \Cref{simp-example:real:eq:3} to \Cref{simp-example:real:eq:5}. In
particular, this means that the (original) bad branch will eventually leads to a new good branch $\msf{good}_2$ and a new bad branch $\msf{bad}_2$ (with the same $V^\dagger_{p_k}$ hanging there).

In summary, at the end of the execution, $\msf{good}_1$ and $\msf{good}_2$ merge into a final good branch, while $\msf{bad}_1$ and $\msf{bad}2$ merge into a final bad branch, with a $V^\dagger_{p_k}$ preceding the $\reg{v}$ register. This illustrates that our argument extends even when starting with a superposition of good and bad branches!

This concludes our analysis of the simplified example. In the following \Cref{sec:overview:rug}, we will briefly address the technical challenges that were omitted from the discussion above, but which will necessitate non-trivial effort and novel ideas in the full-fledged setting.

\subsection{On Those under the Rug}
\label{sec:overview:rug}

\para{Branch-Wise Analysis.} The first concern pertains to the assumption of a {\em pure} format in \Cref{simp-example:initial}. In the real execution, the overall state would be {\em mixed} due to two types of measurements. Firstly, at each step, $\Sim$ needs to determine which of the two oracles $\tilde{V}$ and $\tilde{V}^\dagger$ to query. This can be modeled by a special register $\reg{u}$ within $\Sim$'s internal space $\reg{s}$. After $\Sim$ applies the local unitary $S$ over $\reg{m}$ and $\reg{s}$, a superposition such as $\alpha_0\ket{\downarrow} + \alpha_1\ket{\uparrow}$ will be generated on register $\reg{u}$. $\Sim$ then measures this register to decide whether to execute a $\downarrow$-query (i.e., invoking $\tilde{V}$) or an $\uparrow$-query (i.e., invoking $\tilde{V}^\dagger$). Secondly, given that we are operating within the MnR game, certain queries to $H$ will prompt measurements on registers $\reg{p_1 \ldots p_j}$.

We address this issue through the following approach. First, we claim that these measurements occur at predetermined positions. To see that, notice that $\Sim$'s measurements always happen right before each oracle query; these are fixed places we know in advance. Additionally, the MnR measurements are executed at locations sampled at the outset of the game (refer to \Cref{sec:MnR}). Hence, the entire game can be perceived as a series of unitary operators interspersed with measurements at predefined positions. 

To analyze such a procedure,  we can instead examine all possible outcomes of each intermediate measurements. For each fixed ``outcome sequence'' (consisting of the outcomes of all intermediate measurements), we can define a ``sub-normalized'' version of the game, where each intermediate measurement is replaced with a projection that collapses the register to the predetermined outcome specified in the outcome sequence. Any concerned property of the original game (in our case, it is the final decision of the verifier) is essentially the aggregation of that of all possible ``sub-normalized'' games. We formalize this property as a {\em branch-wise equivalence lemma} in \Cref{sec:branchwise}. Leveraging this lemma, we can indeed adopt the pure-state perspective as demonstrated in the aforementioned simplified example.

\para{Structure of the Local Counter.} Next, we address a trickier issue. In the aforementioned simplified example, we assumed that the local counter comprises a classical value $\ket{k}_{\reg{lc}}$. However, in the actual execution, the local counter could carry a superposition of values. In such a scenario, it is not immediately evident whether the state resulting from $\tilde{V} S \tilde{V}^\dagger$ can be expressed in the clear good-bad format illustrated in \Cref{simp-example:real:eq:5}. Specifically, it is uncertain whether the local counter contains a value smaller than the global counter in the bad branch (i.e., the {\sf Round Slackness} property). On the other hand, the earlier discussion about {\sf Round Slackness} w.r.t.\ that simplified example exhibits a recursive nature --- we managed to establish it assuming that {\em either} the initial state lacks a bad branch, {\em or} the bad branch is already round-slack (as explained previously for the case where the initial state already contains a bad branch). This recursive argument now seems to place us in an ``egg-check dilemma.''

We address this issue as follows: Before initiating the derivation depicted in the aforementioned simplified example, we establish a lemma that, in the ``sub-normalized'' game described earlier, offers a full characterization of the structure of global and local counters (refer to \Cref{sec:counter-structure} for details). The principal implication of this lemma (see \Cref{counter-structure:lemma}) can be intuitively summarized as follows: Throughout the (sub-normalized) game, the overall state can be expressed as the sum of pure states in superposition. For the branch where the $\reg{p_1\ldots p_k}$ registers contain precisely the values $p_1, \ldots, p_k$ satisfying the current oracle $H$, both the global counter and the local counter equal $k$. Note that this precisely corresponds to the good branch in the previous discussion. For all other branches (i.e., the bad branch) in the superposition, the local counter is strictly smaller than the global counter; this is exactly what we require for {\sf Round Slackness}! Essentially, this lemma establishes the {\sf Round Slackness} {even before we start the derivation} (as shown in the simplified example), resolving the egg-chicken dilemma.

Moreover, the clear characterization of the counter structures enables us to conduct the derivation without explicitly tracking the local counter. This is because all branches where the local counter is smaller than the global counter (i.e., where {\sf Round Slackness} is satisfied) can be grouped under the name of a single bad branch. This presents a significant advantage for presentation: as previously mentioned (and as demonstrated in \Cref{counter-structure:lemma}), the local counter is in a superposition of many values, resulting in the bad branch being a summation of multiple sub-branches. In such a scenario, explicit enumeration of all the sub-branches in the derivation would be necessary. Although the proof would still hold, it would certainly be overly complex to understand.

\para{Commutativity Lemma and Hybrid Argument.} There is one more issue worth mentioning here. The simplified example discussed above only addresses the scenario where $\Sim$ rewinds the execution by one round and subsequently resumes it. In the full-fledged setting, however, there are $K$ rounds, and at any given point, $\Sim$ can opt to rewind the execution as far back as desired. Moreover, at each move, $\Sim$ may not necessarily provide the input $\ket{p_k}\ket{\phi_{p_k}}_{\reg{s}}$ as illustrated in the example. Instead, it could potentially query the oracle $\tilde{V}^{\dagger}$ with a general state $\rho$ over the $\reg{p_k}$ and $\reg{s}$ registers. Handling this full-fledged setting presents significant challenges in performing the derivation.

We resolve these issues as follows. Firstly, we encapsulate the aforementioned derivation  as a general, information-theoretic lemma concerning the commutativity of certain unitary operators. This lemma, presented in \Cref{sec:EI-Commutativity-Lemma}, is meticulously crafted so that it can be later applied to handle the general case where $\Sim$ queries her oracles with a general state over registers $\reg{m}$ and $\reg{s}$. Essentially, it asserts exactly what we demonstrated in the simplified example: 
\begin{quote}
{\bf Error-Invariant Commutativity (Informal):} If $\Sim$ rewinds the execution by one step and then resumes it, the final state resulting from the real execution (utilizing $\tilde{V}$) and the one from the dummy execution (utilizing $\ddot{V}$) will possess an identical good branch, differing at the bad branch with a {\em structured} error term.
\end{quote}
However, it is important to note that this commutativity lemma is established in the rewinding-one-step-back setting. Handling multiple-round rewinding necessitates new ideas. We address this challenge through a careful design of hybrids. Roughly speaking, we create $K+1$ hybrids, with hybrid $H_0$ representing the real MnR game (utilizing $\tilde{V}$), and the $k$-th hybrid (for $k \in [K]$) structured as follows:

 \subpara{Hybrid $H_{k}$:} This hybrid is identical to $H_{k-1}$, except for the following difference:
\begin{itemize}
\item
The first query (made by $\Sim$) that brings the global counter from $\ket{k-1}_{\reg{gc}}$ to $\ket{k}_{\reg{gc}}$ is answered with $\tilde{V}$ (as in the previous hybrid). However, all subsequent queries that bring the global counter from $\ket{k-1}_{\reg{gc}}$ to $\ket{k}_{\reg{gc}}$ (resp.\ from $\ket{k}_{\reg{gc}}$ to $\ket{k-1}_{\reg{gc}}$) are answered with the ``dummy'' unitary $\ddot{V}$ (resp.\ $\ddot{V}^\dagger$).
\end{itemize}
A helpful approach to understand these hybrids is to examine the  baby case of $K = 2$, which we also include in the main body as a warm-up example (see \Cref{sec:CMP:warm-up}). In this scenario, there are only three hybrids $H_0$, $H_1$, and $H_2$. We illustrate them in \Cref{figure:Overview-3ZK:hybrids}:
\begin{itemize}
 \item
In hybrid $H_0$, it can be seen from \Cref{figure:Overview-3ZK:hybrids:H0} that all the $\downarrow$-queries are answered using $\tilde{V}$ and all the $\uparrow$-queries are answered using $\tilde{V}^\dagger$.  

\item
The hybrid $H_1$ shown in \Cref{figure:Overview-3ZK:hybrids:H1} is identical to hybrid $H_0$ except that the $\downarrow$-queries that bring the global counter from $\ket{0}_{\reg{gc}}$ to $\ket{1}_{\reg{gc}}$ are answered using the dummy-version unitary $\ddot{V}$, and the $\uparrow$-queries that bring the global counter from $\ket{1}_{\reg{gc}}$ to $\ket{0}_{\reg{gc}}$ are answered using the dummy-version unitary $\ddot{V}^\dagger$. This is except for the query labelled as $\msf{sq}(1)$, which represents the first time the global counter successfully reaches value $1$ (because it invokes the measurement on the query to $H$ in the MnR game).

\item
The hybrid $H_2$ shown in \Cref{figure:Overview-3ZK:hybrids:H2} is identical to hybrid $H_1$ except that the $\downarrow$-queries that bring the global counter from $\ket{1}_{\reg{gc}}$ to $\ket{2}_{\reg{gc}}$ are answered using the dummy-version unitary $\ddot{V}$, and the $\uparrow$-queries that bring the global counter from $\ket{2}_{\reg{gc}}$ to $\ket{1}_{\reg{gc}}$ are answered using the dummy-version unitary $\ddot{V}^\dagger$.  This is except for the query labelled as $\msf{sq}(2)$, which represents the first time the global counter successfully reaches value $2$.
 \end{itemize}
\begin{figure}[!tb]
\centering
     \begin{subfigure}[t]{0.47\textwidth}
         \fbox{
         \includegraphics[width=\textwidth,page=1]{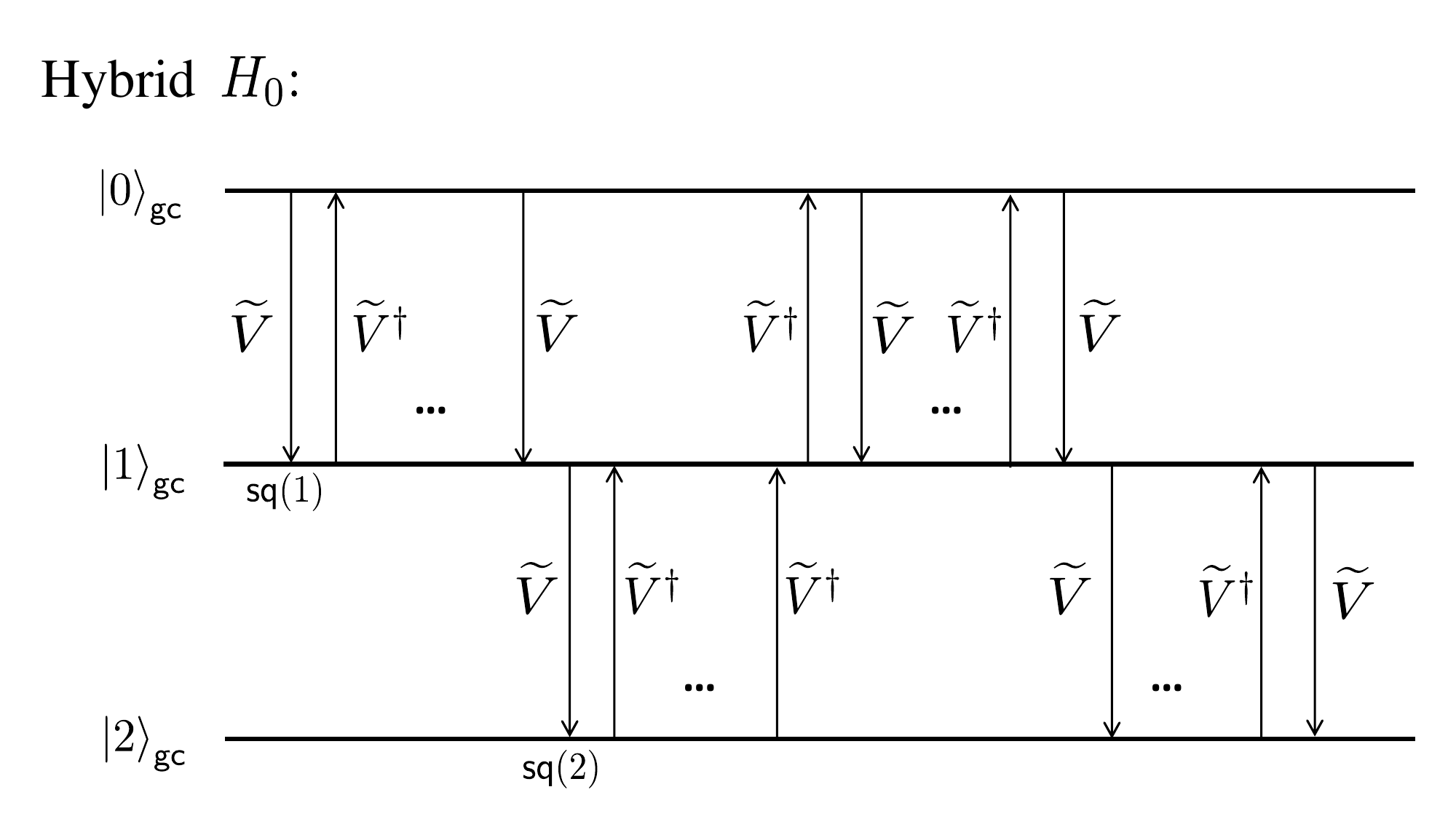}
         }
         \caption{}
         \label{figure:Overview-3ZK:hybrids:H0}
     \end{subfigure}
     \hspace{15pt}
     \begin{subfigure}[t]{0.47\textwidth}
         \centering
         \fbox{
         \includegraphics[width=\textwidth,page=2]{figures/Overview-3ZK-figures.pdf}
         }
         \caption{}
         \label{figure:Overview-3ZK:hybrids:H1}
     \end{subfigure}
     
     \centering\vspace{1em}
     \begin{subfigure}[t]{0.47\textwidth}
         
         \fbox{
         \includegraphics[width=\textwidth,page=3]{figures/Overview-3ZK-figures.pdf}
         }
         \caption{}
         \label{figure:Overview-3ZK:hybrids:H2}
     \end{subfigure}
     \caption{Illustration of Hybrids $H_0$, $H_1$, and $H_2$}
     \label{figure:Overview-3ZK:hybrids}
\end{figure}

It is evident from \Cref{figure:Overview-3ZK:hybrids:H2} that hybrid $H_2$ precisely corresponds to the ``dummy'' MnR game necessary for constructing the malicious prover $\tilde{\Prover}$ in our soundness reduction. Therefore, it is sufficient to demonstrate that the acceptance probability remains unchanged across these three hybrids. To establish this for $H_0$ and $H_1$, we can readily employ the previously mentioned commutativity lemma. The disparity between these two hybrids only arises in the ``first layer'' (i.e., the execution between the line of $\ket{0}_{\reg{gc}}$ and the line of $\ket{1}_{\reg{gc}}$), which corresponds precisely to the one-step-back rewinding scenario addressed by the lemma.

To transition from $H_1$ to $H_2$, we can leverage the same lemma. This is facilitated by the fact that in $H_1$, all the oracles for the first layer have already been substituted with the dummy operator $\ddot{V}$ or $\ddot{V}^\dagger$ (except for the $\msf{sq}(1)$ and $\msf{sq}(2)$ queries). As the dummy $\ddot{V}$ essentially has no impact except for some counter adjustments, the disparity between $H_1$ and $H_2$ can effectively be treated as a one-step-back rewinding in the second layer alone. Consequently, it can be handled by the commutativity lemma once more.

It is conceivable that this argument extends to the scenario with any constant $K$, where we simply invoke the commutativity lemma to argue the indistinguishability between each pair of $H_{k-1}$ and $H_k$ (for all $k \in [K]$) in a similar manner as the above.

This eventually completes our proof.


\subsection{Putting Things Together}
\label{sec:overview:putting-together}

Let us assemble all the techniques discussed thus far, providing a complete picture of how we ultimately establish \Cref{thm:main:intro}.


We first note that our focus will be on establishing a version of \Cref{thm:main:intro} where the simulator runs in {\em strict} QPT, rather than {\em expected} QPT. Once this is accomplished, the extension to expected QPT simulation follows the same technique from \cite{FOCS:CCLY21}. This extension will be covered in \Cref{sec:expected_sim_efficient_verifier}.

For a language $\Lang$, assuming the existence of a constant-round fully-quantum BBZK protocol $\langle \Prover, \Verifier \rangle$, our objective is to construct a $\BQP$ decider $\mcal{B}$ for the language $\Lang$. This construction involves leveraging $\Verifier$ along with the ZK simulator $\Sim$.

Toward that, we first introduce a non-programmable EPR model, where we can assume without loss of generality that all prover messages are classical. Furthermore, we demonstrate that it suffices to establish \Cref{thm:main:intro} in this model. This is covered in \Cref{sec:define-model}.

Subsequently, we define a random-aborting verifier $\tilde{\Verifier}$, akin to the one introduced in \cite{FOCS:CCLY21}. However, we augment it with a global counter and a local counter, as discussed in \Cref{sec:overview:message-order}. These two counters play a crucial role in the subsequent proofs. The precise definition of our $\tilde{\Verifier}$ is elucidated in \Cref{sec:particular-mal-V}.


We then proceed to define the desired $\BQP$ decider $\mcal{B}$ in \Cref{sec:BQP-decider}. Essentially, it executes the measure-and-reprogram (MnR) version of $\Sim^{\tilde{\Verifier}(x)}(x)$, with $\tilde{\Verifier}$'s random function instantiated by $H_0$ and serving as the oracle in the MnR game. Our task then becomes demonstrating that $\mcal{B}$ is both complete and sound.

The proof for completeness is similar to 	that in \cite{FOCS:CCLY21}, and it is presented in \Cref{sec:CPM:decider:lem:completeness:proof}.

The proof for soundness is the most intricate aspect of our work. It entails addressing several challenges discussed earlier in \Cref{sec:overview:message-order,sec:overview:sim-rewinding-q,sec:overview:rug}. In the main body, this is covered in \Cref{sec:CPM:decider:lem:soundness:proof} and \Cref{sec:relate-Dummy-Real} to \Cref{sec:CMP:full}.


\section{Preliminaries}

\para{Notation.} Let $\secpar$ denote the security parameter throughout the paper. For a positive integer $n\in\mathbb{N}$, $[n]$ denotes a set $\{1,2 \ldots, n\}$. For a finite set $\mcal{X}$, $x\pick \mcal{X}$ means that $x$ is uniformly chosen from $\mcal{X}$. For a finite set $\mcal{X}$ and a positive integer $k$, $\mcal{X}^{\leq k}$ is defined to be $\bigcup_{i\in [k]}\mcal{X}^{i}$. 
For finite sets $\mcal{X}$ and $\mcal{Y}$, $\func(\mcal{X},\mcal{Y})$ denotes the set of all functions with domain $\mcal{X}$ and range $\mcal{Y}$.

A function $f:\mathbb{N}\ra [0,1]$ is said to be \emph{negligible} if for all polynomial $p$ and sufficiently large $\secpar \in \mathbb{N}$, we have $f(\secpar)< 1/p(\secpar)$; it is said to be \emph{overwhelming} if $1-f$ is negligible, and said to be \emph{noticeable} if there is a polynomial $p$ such that $f(\secpar)\geq  1/p(\secpar)$ for sufficiently large $\secpar\in \mathbb{N}$. We denote by $\poly$ an unspecified polynomial and by $\negl$ an unspecified negligible function.

We will use the following convention. If we have a state $\ket{\rho}_{\reg{a}\reg{b}}$ over two registers $\reg{a}$ and $\reg{b}$ and a unitary operator $U_{\reg{a}}$ that acts on registers $\reg{a}$ only, we will use $U_{\reg{a}} \ket{\rho}_{\reg{a}\reg{b}}$ as an abbreviation for applying $U_{\reg{a}}\tensor I_{\reg{b}}$ to $\ket{\rho}_{\reg{a}\reg{b}}$ (where $I_{\reg{b}}$ is the identity operator on register $\reg{b}$). Also, if $U_{\reg{b}}$  operates non-trivially only on register $\reg{b}$, we will use $\ket{\psi}_a U_{\reg{b}} \ket{\phi}_b$ as an abbreviation for applying $I_\reg{a}\tensor U_{\reg{b}}$ to $\ket{\psi}_a \ket{\phi}_b$.

\subsection{Technical Lemmas}
The following lemma will be used throughout the paper.
\begin{lemma}[\cite{C:Zhandry12}]\label{lem:ind-hash}
For any sets $\calX$ and $\calY$ of classical strings and $q$-quantum-query algorithm $\A$, we have
\[
\Pr[\A^{H}=1:H\sample \func(\calX,\calY)]= \Pr[\A^{H}=1:H\sample \mathcal{H}_{2q}]
\]
where  $\mathcal{H}_{2q}$ is a family of $2q$-wise independent hash functions from $\calX$ to $\calY$.
\end{lemma}

\para{Differentiating Sparsity Functions as Quantum Oracles.}

\begin{lemma}[{\cite[Lemma 3]{PKC:HulRijSon16}}]\label{lem:ind-sparse-and-zero}
Let $\calX$ be a finite set, $\epsilon\in [0,1]$ be a non-negative real number, and 
$\mathcal{H}_\epsilon$ be a distribution over $H_\epsilon :\calX\rightarrow \bits$ such that we have  $\Pr[H_\epsilon(x)=1]=\epsilon$ independently for each $x\in \calX$.
Let $H_0:\calX\rightarrow \bits$ be the function that returns $0$ for all inputs $x\in \calX$. 
Then for any algorithm $\A$ that makes at most $q$ quantum queries, we have 
\begin{align*}
    \left|\Pr\left[\A^{H_\epsilon}=1:H_\epsilon \sample \mathcal{H}_\epsilon\right]
    -
    \Pr\left[\A^{H_0}=1\right]
    \right|\leq 8q^2\epsilon.
\end{align*}
\end{lemma}

\subsection{(Ordered-Query) Measure-and-Reprogram Lemma}
\label{sec:MnR}
\xiao{I modified this subsection to put the MnR lemma into a clean format that suits our applications.}

We recall the measure-and-reprogram (MnR) lemma developed by \cite{C:DFMS19,C:DonFehMaj20}. In particular, we need the special ``ordered-query'' version of the (MnR) lemma formalized in \cite[Lemma 2.11]{FOCS:CCLY21}. We adapt it with some cosmetic changes to suit our applications.

We first give intuitive explanations for notation, which are taken from \cite{FOCS:CCLY21}.
For a quantumly-accessible classical oracle $\mcal{O}$, we denote by $\mcal{O}\leftarrow \reprogram(\mcal{O},x,y)$ to mean that we reprogram $\oracle$ to output $y$ on input $x$. For a $q$-quantum-query algorithm $\Adv$, function $H:\mcal{X} \rightarrow \mcal{Y}$, and  $\vb{y}=(y_1, \ldots ,y_k)\in \mcal{Y}^k$, we denote by $\Adv[H,\vb{y}]$  to mean an algorithm that runs $\Adv$ w.r.t.\ an oracle that computes $H$ except that randomly chosen $k$ queries are measured and the oracle is reprogrammed to output  $y_i$ on $i$-th measured query. Formal definitions are given below.

\begin{definition}[Reprogramming Oracle \textnormal{\cite[Definition 2.9]{FOCS:CCLY21}}]
\label{MnR:def:reprogram}
Let $\Adv$ be a quantum algorithm with quantumly-accessible oracle $\mcal{O}$ that is initialized to be an oracle that computes some classical function from $\mcal{X}$ to $\mcal{Y}$. At some point in an execution of $\Adv^{\mcal{O}}$, we say that we reprogram $\mcal{O}$ to output $y\in \mcal{Y}$ on $x\in \mcal{X}$ if we update the oracle to compute the function $H_{x,y}$ defined by
\begin{eqnarray*}
    H^{x,y}(x')
    \coloneqq
    \begin{cases} y   &\text{if~}  x'=x \\  
     H(x')  &\text{otherwise}
    \end{cases}
\end{eqnarray*}
where $H$ is the function computed by $\mcal{O}$ before the update. This updated oracle is used in the rest of execution of $\Adv$. We denote by $\mcal{O}\leftarrow \reprogram(\mcal{O},x,y)$ the above reprogramming procedure. 
\end{definition}

\xiao{@All: Indeed, \Cref{MnR:lem} is slightly different from \cite[Lemma 2.11]{FOCS:CCLY21}. Please take a close look to make sure that the current \Cref{MnR:lem} really holds. The difference is: I allow $\Adv$ to additionally output a quantum part $\rho$ as its output, and test a (quantum) predicate in the final inequality relation between the real game and the MnR game. }

\begin{lemma}[Measure-and-Reprogram Lemma \textnormal{\cite[Lemma 2.11]{FOCS:CCLY21}}]
\label{MnR:lem}
Let $\mcal{X}$ and $\mcal{Y}$ be sets of classical strings and $k$ be a positive integer. Let $\Adv$ be a $q$-quantum-query algorithm with quantum oracle access to an oracle that computes a function from $\mcal{X}$ to $\mcal{Y}$ and outputs $\vb{x} \in \mcal{X}^k$ and a (possibly quantum) output $\rho$. For a function $H:\mcal{X}^{\leq k} \ra \mcal{Y}$ and  $\vb{y} =(y_1, \ldots ,y_k) \in \mcal{Y}^k$, we define a MnR game $\Adv[H,\vb{y}]$ in \Cref{MnR:lem:game}.
 
Then, for any positive integer $k$, any $q$-quantum query algorithm $\Adv$, any function $H:\mcal{X}^{\leq k}\rightarrow \mcal{Y}$, any vectors  $\vb{x}^*=(x^*_1, \ldots ,x^*_k) \in \mcal{X}^k$ and $\vb{y}=(y_1, \ldots ,y_k) \in \mcal{Y}^k$, and any (potentially quantum) predicate $\msf{Pred}$, it holds that 
\begin{align*}
& \Pr\left[\big(\vb{x} = \vb{x}^*\big) \wedge \big(\msf{Pred}(\vb{x}, \vb{y}, \rho) = 1\big) ~:~(\vb{x}, \rho) \la \Adv[H,\vb{y}]\right] \\ 
& \hspace{5em} \geq 
\frac{1}{(2q+1)^{2k}} \cdot \Pr\left[\big(\vb{x} = \vb{x}^*\big) \wedge \big(\msf{Pred}(\vb{x}, \vb{y}, \rho) = 1\big) ~:~(\vb{x}, \rho) \la \Adv^{H^{\vb{x}^*,\vb{y}}} \right].
\end{align*}
where $H^{\vb{x}^*,\vb{y}}: \mcal{X}^{\leq k} \rightarrow \mcal{Y}$ is defined as 
$$
    H^{\vb{x}^*,\vb{y}}(\vb{x}')
    \coloneqq
    \begin{cases} y_i   & \text{if~} \exists i\in[k] \text{~s.t.~} \vb{x}'=(x^*_1,...,x^*_i) \\  
     H(\vb{x}')  & \text{otherwise}
    \end{cases}.
$$
\end{lemma}

\begin{remark}
We remark that the original \cite[Lemma 2.11]{FOCS:CCLY21} defines the output of $\Adv$ to be $\vb{x} \in \mcal{X}^k$, without the (possibly quantum) $\rho$ part as in our \Cref{MnR:lem}. Our modification is fine since \cite[Lemma 2.11]{FOCS:CCLY21} is indeed a corollary of the ``un-ordered'' MnR lemma shown in \cite[Lemma 2.10]{FOCS:CCLY21}, which allows $\Adv$ to additionally output a (potentially quantum)\footnote{More accurately, \cite[Lemma 2.10]{FOCS:CCLY21} requires $z$ to be a classical string. But \cite[Lemma 2.10]{FOCS:CCLY21} is indeed an adaption of the original MnR lemma from \cite[Theorem 6]{C:DonFehMaj20}, where $z$ is allowed to be quantum.} part $z$. The derivation of \cite[Lemma 2.11]{FOCS:CCLY21} from \cite[Lemma 2.10]{FOCS:CCLY21} holds even if one takes the quantum $z$ (i.e., the $\rho$ in our notation) into consideration.
\end{remark}

\begin{GameBox}[label={MnR:lem:game}]{Measure-and-Reprogram Game \textnormal{$\Adv[H, \vb{y}]$}}
It  works as follows:
\begin{enumerate}
    \item 
    For each $i\in[k]$, uniformly pick $(j_i,b_i)\in ([q]\times \bits) \cup \{(\bot,\bot)\}$  conditioned on that 
    there exists at most one $i\in [k]$ such that $j_i=j^*$  for all $j^*\in [q]$.
    
    \item \label[Step]{MnR:lem:game:step:relabel}
   Let $s$ denote the number of $j_i$'s in $\Set{j_i}_{i \in [k]}$ that are not $\bot$. We re-label the indices $i$ for pairs $\Set{(j_i, b_i)}_{i \in [k]}$ so that $j_1 < j_2 < \ldots < j_s$ and $j_{s+1} = j_{s+2} =\ldots = j_k = \bot$. (See \Cref{rmk:MnR:game:relabel}.)

    \item 
    Run $\Adv^{\mcal{O}}$ where the oracle $\mcal{O}$ is initialized to be a quantumly-accessible classical oracle that computes $H$, and when $\Adv$ makes its $j$-th query, the oracle is simulated as follows:
    \begin{enumerate}
    \item 
      If $j=j_i$ for some $i\in[s]$, measure $\Adv$'s  query register to obtain $\vb{x}'_i=(x'_{i,1}, \ldots ,x'_{i,k_i})$ where $k_i\in [k]$ is determined by the measurement outcome, and then behaves according to the value $b_i$ as follows:.
        \begin{enumerate}
        \item 
        If $b_i=0$: First reprogram $\mcal{O} \leftarrow \reprogram(\mcal{O},\vb{x}'_i,y_i)$, and then answer the $j_i$-th query using the reprogrammed oracle. (See \Cref{MnR:def:reprogram} for the definition of $\reprogram$.) 

        \item 
        If $b_i=1$: First answer the $j_i$-th query using the oracle before the reprogramming, and then reprogram $\ora\leftarrow \reprogram(\mcal{O},\vb{x}'_i,y_i)$. 
        \end{enumerate}
    
    \item 
    Otherwise (i.e., $j \ne j_i$ $\forall i \in[s]$), answer $\A$'s $j$-th query just using the oracle $\mcal{O}$ without any measurement or reprogramming. 
    \end{enumerate}

    \item 
    Let $(\vb{x}=(x_1, \ldots ,x_k), \rho)$ be $\Adv$'s output.

     \item 
     For all $i \in \Set{s+1, s+2, \ldots, k}$, set $\vb{x}'_i = \vb{x}_i$ where $\vb{x}_i := (x_1, \cdots, x_i)$. 

     \item 
     {\bf Output:} The output of this game is defined as follows
	\begin{itemize}
	\item
	If it holds for all $i \in [k]$ that $\vb{x}'_i$ is a prefix of $\vb{x}'_k$, then output $(\vb{x}'_k, \rho)$;

	\item
	Otherwise, output $(\bot, \bot)$.
	\end{itemize}

\end{enumerate}
\end{GameBox}
\begin{remark}[Regarding Re-labeling]\label{rmk:MnR:game:relabel}
The re-labeling we performed in \Cref{MnR:lem:game:step:relabel} is only cosmetic. It just means to maintain an increasing order for the sequence $\Set{j_i}_{i\in [k]}$ and put all the $j_i$'s which are $\bot$ to the very end in this sequence. It does not affect \Cref{MnR:lem:game} at all because the order of the $j_i$'s is not used anywhere in this game. But we prefer to maintain an order for $\Set{j_i}_{i\in [k]}$ because it will help simplify the presentation of our results and proofs later in this work.
\end{remark}

\subsection{Branch-Wise Equivalence Lemma}
\label{sec:branchwise}
We hereby present a lemma pertaining to quantum computing. Although the proof of this lemma is relatively straightforward, it plays a crucial role in streamlining certain aspects of the proof for our main results.

Let us first provide an intuitive description of this lemma. Consider a quantum game wherein unitaries and measurements are alternately applied to an initial pure state. Subsequently, a binary-outcome POVM is applied, determining the final output of the game. We assert that the probability of outputting 1 in such a quantum game can be expressed equivalently as follows:
\begin{itemize}
 \item
    Firstly, we examine all possible outcomes of each intermediate measurements (except for the final POVM). For each fixed ``outcome sequence'' (consisting of the outcomes of all intermediate measurements), we can define a ``sub-normalized'' version of the game. In this modified version, each intermediate measurement is replaced with a projection that collapses the register to a predetermined outcome (specified in the ``outcome sequence''). Finally, a binary-outcome POVM is applied, determining the final output of this ``subnormalized'' game.
\item
     We then aggregate the probabilities of outputting 1 across all possible ``sub-normalized'' games.
\end{itemize} 

In the following, we present the formal lemma. To enhance clarity, we will first provide a simplified version of this lemma in \Cref{sec:branchwise:babycase}, termed the ``baby-case'' version. In this context, we consider a quantum game involving the alternating application of two unitaries and two measurements. This simplified scenario serves to illustrate the essence of the lemma and the key concepts underlying its proof. Subsequently, we will present the comprehensive, fully elaborated version in \Cref{sec:branchwise:full-fledged}.

\subsubsection{Baby-Case Version}
\label{sec:branchwise:babycase}
 
We first present in \Cref{branchwise:babycase:game} the quantum game of alternating applications of two unitaries and measurements.
\begin{GameBox}[label={branchwise:babycase:game}]{Baby-Case Game $G(\ket{\psi})$ with Alternating Unitary and Measurement}
{\bf Parameters.} Let $\ket{\psi}$ be a pure state over some registers. Let $\reg{m}$ be a $\ell$-qubit subset of these registers. Let $U_1$ and $U_2$ be unitaries over these registers. Let $P = \Set{E_0, E_1}$ be a binary-outcome POVM over all the registers. 

\para{Game $G(\ket{\psi})$:} on input $\ket{\psi}$, $G(\ket{\psi})$ performs the following operations:
\begin{enumerate}
    \item \label[Step]{branchwise:babycase:game:step:1}
    Apply $U_1$;
    \item \label[Step]{branchwise:babycase:game:step:2}
    Measure register $\reg{m}$ in the computational basis;
    \item \label[Step]{branchwise:babycase:game:step:3}
    Apply $U_2$;
    \item \label[Step]{branchwise:babycase:game:step:4}
    Measure register $\reg{m}$ in the computational basis;
    \item \label[Step]{branchwise:babycase:game:step:5}
    Apply the POVM $P$.
\end{enumerate}

\para{Output:} the output is defined to be the POVM measurement outcome. We denote it as $\msf{OUT}\big(G(\ket{\psi})\big)$. 
\end{GameBox}

\begin{lemma}[Branch-Wise Equivalence---Baby Case]\label{branchwise:babycase:lem}
Use the notation in \Cref{branchwise:babycase:game}. For each $m_1, m_2 \in \bits^\ell$, define 
$$\ket{\psi^{(2)}_{m_1, m_2}} \coloneqq \big(\ketbra{m_2}_{\reg{m}} U_2\big) \big(\ketbra{m_1}_{\reg{m}} U_1\big) \ket{\psi}.$$ 
Then, it holds that
$$\Pr[\msf{Out}\big(G(\ket{\psi})\big) = 1] = \sum_{m_1, m_2 \in \bits^\ell} \bra{\psi^{(2)}_{m_1, m_2}}E_1\ket{\psi^{(2)}_{m_1, m_2}}.$$
\end{lemma}
\begin{proof}[Proof of \Cref{branchwise:babycase:lem}]
This lemma can be easily shown by tracking the steps in \Cref{branchwise:babycase:game} with elementary calculation.

\subpara{Tracking \Cref{branchwise:babycase:game}.} Right after \Cref{branchwise:babycase:game:step:1}, the state $U_1 \ket{\psi}$ can be written in the following format
\begin{equation}\label{branchwise:babycase:lem:proof:eq:U1}
U_1 \ket{\psi} = \sum_{m_1\in \bits^\ell} \alpha_{m_1} \ket{m_1}_{\reg{m}}\ket{\dot{\psi}^{(1)}_{m_1}},
\end{equation}
where the $\alpha_{m_1}$'s are amplitudes satisfies $\sum_{m_1} |\alpha_{m_1}|^2 = 1$ and $\ket{\dot{\psi}^{(1)}_{m_1}}$'s are pure states over registers other than $\reg{m}$.
Measuring the $\reg{m}$ register (i.e., \Cref{branchwise:babycase:game:step:2}) results in the following mixture:
$$\text{with probability}~|\alpha_{m_1}|^2,~\text{the overall state collapses to}~\ket{m_1}\ket{\dot{\psi}^{(1)}_{m_1}}.$$

For each possible state $\ket{m_1}\ket{\dot{\psi}^{(1)}_{m_1}}$, applying $U_2$ (i.e., \Cref{branchwise:babycase:game:step:3}) yields
\begin{equation}\label{branchwise:babycase:lem:proof:eq:U2}
U_2 \ket{m_1}\ket{\dot{\psi}^{(1)}_{m_1}} = \sum_{m_2 \in \bits^\ell}\beta_{m_1, m_2} \ket{m_2}\ket{\dot{\psi}^{(2)}_{m_1, m_2}},
\end{equation}
where the $\beta_{m_1,m_2}$'s are amplitudes satisfies $\sum_{m_2} |\beta_{m_1, m_2}|^2 = 1$ and $\ket{\dot{\psi}^{(2)}_{m_1, m_2}}$'s are pure states over registers other than $\reg{m}$. Now, measuring the $\reg{m}$ register (i.e., \Cref{branchwise:babycase:game:step:4}) results in the following mixture:
$$\text{with probability}~|\beta_{m_1,m_2}|^2,~\text{the overall state collapses to}~\ket{m_2}\ket{\dot{\psi}^2_{m_1, m_2}}.$$

In summary, we can say that the state right before the POVM (i.e., \Cref{branchwise:babycase:game:step:5}) is the following mixture:
\begin{equation}\label[Expression]{branchwise:babycase:lem:proof:final:mixture}
\text{with probability}~|\alpha_{m_1}|^2 \cdot |\beta_{m_1,m_2}|^2,~\text{the overall state is}~\ket{m_2}\ket{\dot{\psi}^{(2)}_{m_1, m_2}}.
\end{equation}

\subpara{Analyzing Sub-Normalized States.} To relate \Cref{branchwise:babycase:game} with the state $\ket{\psi^{(2)}_{m_1, m_2}}$ defined in \Cref{branchwise:babycase:lem}, we now define some sub-normalized states. In particular, for each $m_1 \in \bits^\ell$, let 
\begin{align}
\ket{\psi^{(1)}_{m_1}} 
& \coloneqq 
\ketbra{m_1}U_1 \ket{\psi} 
\label{branchwise:babycase:lem:proof:psi1:eq:1}\\ 
&= 
\alpha_{m_1} \ket{m_1}_{\reg{m}}\ket{\dot{\psi}^{(1)}_{m_1}}
\label{branchwise:babycase:lem:proof:psi1:eq:2}
,\end{align}
where \Cref{branchwise:babycase:lem:proof:psi1:eq:2} follows from \Cref{branchwise:babycase:lem:proof:eq:U1}.

Then, it holds for the $\ket{\psi^{(2)}_{m_1, m_2}}$ defined in \Cref{branchwise:babycase:lem} that 
\begin{align}
\ket{\psi^{(2)}_{m_1, m_2}} 
& =  
\ketbra{m_2}_{\reg{m}} U_2 \ket{\psi^{(1)}_{m_1}} 
\label{branchwise:babycase:lem:proof:psi2:eq:1} \\ 
& =
\alpha_{m_1} \ketbra{m_2}_{\reg{m}} U_2  \ket{m_1}_{\reg{m}}\ket{\dot{\psi}^{(1)}_{m_1}} 
\label{branchwise:babycase:lem:proof:psi2:eq:2} \\ 
& = 
\alpha_{m_1} \beta_{m_1, m_2} \ket{m_2} \ket{\dot{\psi}^{(2)}_{m_1, m_2}}
\label{branchwise:babycase:lem:proof:psi2:eq:3}
,\end{align}
where \Cref{branchwise:babycase:lem:proof:psi2:eq:1} follows from  \Cref{branchwise:babycase:lem:proof:psi1:eq:1} and the definition of $\ket{\psi^{(2)}_{m_1, m_2}}$ in \Cref{branchwise:babycase:lem}, \Cref{branchwise:babycase:lem:proof:psi2:eq:2} follows from \Cref{branchwise:babycase:lem:proof:psi1:eq:2}, and \Cref{branchwise:babycase:lem:proof:psi2:eq:3} follows from \Cref{branchwise:babycase:lem:proof:eq:U2}.

\subpara{Establishing Equivalence.} With the above notation, it then follows that 
\begin{align*}
\Pr\big[\msf{Out}\big(G(\ket{\psi})\big) = 1\big] 
& = 
\sum_{m_1, m_2} |\alpha_{m_1}|^2 \cdot |\beta_{m_1,m_2}|^2 \cdot \bra{m_2, \dot{\psi}^{(2)}_{m_1, m_2}}E_1\ket{m_2, \dot{\psi}^{(2)}_{m_1, m_2}} 
\numberthis \label{branchwise:babycase:lem:proof:final:eq:1}\\ 
& = 
\sum_{m_1, m_2} \alpha^*_{m_1}\cdot \alpha_{m_1} \cdot \beta^*_{m_1,m_2} \cdot \beta_{m_1,m_2} \cdot \bra{m_2, \dot{\psi}^{(2)}_{m_1, m_2}}E_1\ket{m_2, \dot{\psi}^{(2)}_{m_1, m_2}} \\ 
& = 
\sum_{m_1, m_2}   \bra{m_2, \dot{\psi}^{(2)}_{m_1, m_2}} \alpha^*_{m_1} \beta^*_{m_1,m_2} E_1 \alpha_{m_1} \beta_{m_1,m_2}\ket{m_2, \dot{\psi}^{(2)}_{m_1, m_2}} \\ 
& = 
\sum_{m_1, m_2} \bra{\psi^{(2)}_{m_1, m_2}}E_1\ket{\psi^{(2)}_{m_1, m_2}}
\numberthis \label{branchwise:babycase:lem:proof:final:eq:2}
,\end{align*}
where \Cref{branchwise:babycase:lem:proof:final:eq:1} follows from \Cref{branchwise:babycase:lem:proof:final:mixture}, \Cref{branchwise:babycase:lem:proof:final:eq:2} follows from \Cref{branchwise:babycase:lem:proof:psi2:eq:3}.

This finishes the proof of \Cref{branchwise:babycase:lem}. 

\end{proof}

\subsubsection{Full-Fledged Version} 
\label{sec:branchwise:full-fledged}

We now extend the baby-case version into the full-fledged scenario. In particular, we generalize the baby case in two aspects: (1) we allow alternating applications of multiple (instead of two) unitaries and measurements, and (2) the intermediate measurements may performed over different registers.

We present the general quantum game in \Cref{branchwise:game} and the lemma in \Cref{branchwise:lem}.
\begin{GameBox}[label={branchwise:game}]{Game $G_k(\ket{\psi})$ with Alternating Unitary and Measurement}
{\bf Parameters.} Let $k$ be an positive integer. Let $\ket{\psi}$ be a pure state over some registers. For each $i\in [k]$, let $\reg{m}_i$ be a $\ell_i$-qubit subset of these registers. Let $\Set{U_i}_{i \in [k]}$ be $k$ unitaries. Let $P = \Set{E_0, E_1}$ be a binary-outcome POVM over all the registers. 

\para{Game $G_k(\ket{\psi})$:} on input $\ket{\psi}$, $G_k(\ket{\psi})$ iterates the following operations for each $i \in [k]$:
\begin{enumerate}
    \item \label[Step]{branchwise:game:step:1}
    Apply $U_i$;
    \item \label[Step]{branchwise:game:step:2}
    Measure register $\reg{m}_i$ in the computational basis;
\end{enumerate}

\para{Output:} finally, apply the POVM $P$ and output the measurement outcome. We denote it as $\msf{OUT}(G_k(\ket{\phi}))$. 
\end{GameBox}

\begin{lemma}[Branch-Wise Equivalence]\label{branchwise:lem}
Use the notation in \Cref{branchwise:game}. 
For each possible vector $\vb{m} = (m_1, \ldots, m_k)$ where $m_i \in \bits^{\ell_i}$ for all $i \in [k]$, define a sub-normalized state
$$
\ket{\psi^{(k)}_{\vb{m}}} \coloneqq  \bigg(\prod^{k}_{i = 1}\ketbra{m_i}_{\reg{m_i}}U_i\bigg) \ket{\psi} = \big(\ketbra{m_k}_{\reg{m_k}}U_k\big)  \big(\ketbra{m_{k-1}}_{\reg{m_{k-1}}} U_{k-1}\big) \cdots \big(\ketbra{m_1}_{\reg{m_1}} U_1\big) \ket{\psi}.
$$ 
Then, it holds that
$$
\Pr[\msf{Out}\big(G_k(\ket{\psi})\big) = 1] 
= 
\sum_{\vb{m}} \bra{\psi^{(k)}_{\vb{m}}}E_1\ket{\psi^{(k)}_{\vb{m}}},
$$
where the summation over $\vb{m}$ means to sum over all $\vb{m} = (m_1, \ldots, m_k) \in \bits^{\ell_1} \times \ldots \times \bits^{\ell_k}$.
\end{lemma}
\begin{proof}[Proof of \Cref{branchwise:lem}]
We prove this lemma via a mathematical induction over the number $k$ of measurements (excluding the final POVM).

\para{Base Case $(k = 1)$.} This corresponds to game $G_1(\ket{\psi})$ which consists of only one unitary $U_1$ and one measurement over register $m_1$ (before the final POVM). In this case, it is obvious that \Cref{branchwise:lem} holds. Indeed, \Cref{branchwise:babycase:lem} can be understood as a special case of $k = 2$, modulo that the two measurement there were performed over the same register $\reg{m}$ (i.e., $\reg{m_1} = \reg{m_2}$ in the \Cref{branchwise:lem} notation).

\para{Induction Step $(k \ge 2)$.} We now assume the lemma holds for $k-1$, and prove it for $k$.

Consider the first iteration of $G_k(\ket{\psi})$. Right after \Cref{branchwise:game:step:1} of the first iteration, the state becomes $U_1 \ket{\psi}$. Such a state can be written in the following format
\begin{equation}\label{branchwise:lem:proof:eq:U1}
U_1 \ket{\psi} = \sum_{m_1\in \bits^\ell} \alpha_{m_1} \ket{m_1}_{\reg{m_1}}\ket{\dot{\psi}^{(1)}_{m_1}},
\end{equation}
where the $\alpha_{m_1}$'s are amplitudes satisfies $\sum_{m_1} |\alpha_{m_1}|^2 = 1$ and the $\ket{\dot{\psi}^{(1)}_{m_1}}$'s are pure states over registers other than $\reg{m_1}$.
Measuring the $\reg{m_1}$ register (i.e., \Cref{branchwise:game:step:2} in the first iteration) results in the following mixture:
$$\text{with probability}~|\alpha_{m_1}|^2,~\text{the overall state collapses to}~\ket{m_1}_{\reg{m_1}}\ket{\dot{\psi}^{(1)}_{m_1}}.$$
Therefore, the output of the game can be described as: 
\begin{equation}\label{branchwise:lem:proof:induction:eq:out}
\Pr[\msf{Out}\big(G_k(\ket{\psi})\big) = 1] = \sum_{m_1 \in \bits^\ell} |\alpha_{m_1}|^2 \cdot \Pr[\msf{Out}\big(G_{2:k}(\ket{m_1}_{\reg{m_1}}\ket{\dot{\psi}^{(1)}_{m_1}})\big) = 1] 
,
\end{equation}
where $G_{2:k}(\ket{m_1}_{\reg{m_1}}\ket{\dot{\psi}^{(1)}_{m_1}})$ denote the remaining party of $G_{k}(\ket{\psi})$ (i.e., right after the the first iteration). For clearness, we present this game in \Cref{branchwise:game:induction}.
\begin{GameBox}[label={branchwise:game:induction}]{Game $G_{2:k}(\ket{m_1}_{\reg{m_1}}\ket{\dot{\psi}^{(1)}_{m_1}})$}
{\bf Parameters.} Same as in \Cref{branchwise:game}. 

\para{Game $G_{2:k}(\ket{m_1}_{\reg{m_1}}\ket{\dot{\psi}^{(1)}_{m_1}})$:} on input $\ket{m_1}_{\reg{m_1}}\ket{\dot{\psi}^{(1)}_{m_1}}$, iterate the following operations for each $i \in \Set{2, 3, \ldots, k}$:
\begin{enumerate}
    \item 
    Apply $U_i$;
    \item 
    Measure register $\reg{m}_i$ in the computational basis;
\end{enumerate}

\para{Output:} apply the POVM $P$ and output the measurement outcome. 
\end{GameBox}
Now, it is important to notice that \Cref{branchwise:game:induction} can be viewed as a version of \Cref{branchwise:game} with parameter $k-1$, because it includes $k-1$ iterations of alternating unitaries and measurements (and then the final POVM). Therefore, we can invoke our induction assumption. In particular, for each possible state $\ket{m_1}_{\reg{m_1}}\ket{\dot{\psi}^{(1)}_{m_1}}$, our induction assumption implies that 
\begin{equation}\label{branchwise:lem:proof:induction:eq:2:k:out}
\Pr[\msf{Out}\big(G_{2:k}(\ket{m_1}_{\reg{m_1}}\ket{\dot{\psi}^{(1)}_{m_1}})\big) = 1]  
= 
\sum_{\vb{m}'} \bra{\dot{\psi}^{(k-1)}_{\vb{m}'}}E_1\ket{\dot{\psi}^{(k-1)}_{\vb{m}'}},
\end{equation}
where the summation over $\vb{m}'$ means to sum over all $\vb{m}' = (m_2, \ldots, m_k) \in \bits^{\ell_2} \times \ldots \times \bits^{\ell_k}$, and the state $\ket{\psi^{(k-1)}_{\vb{m}'}}$ is defined as
\begin{equation}\label{branchwise:lem:proof:induction:eq:def:psi:k-1}
\ket{\dot{\psi}^{(k-1)}_{\vb{m}'}} 
\coloneqq 
\bigg(\prod^{k}_{i = 2}\ketbra{m_i}_{\reg{m_i}}U_i\bigg) \ket{m_1}_{\reg{m_1}}\ket{\dot{\psi}^{(1)}_{m_1}}. 
\end{equation}
On the other hand, we note that
\begin{align*}
\ket{\psi^{(k)}_{\vb{m}}} 
& = 
  \bigg(\prod^{k}_{i = 1}\ketbra{m_i}_{\reg{m_i}}U_i\bigg) \ket{\psi} 
  \numberthis \label{branchwise:lem:proof:induction:psi:k:eq:1} \\ 
& =
\bigg(\prod^{k}_{i = 2} \ketbra{m_i}_{\reg{m_i}}U_i\bigg) \ketbra{m_1}_{\reg{m_1}}U_1 \ket{\psi} 
\\ 
& =
\alpha_{m_1} \bigg(\prod^{k}_{i = 2} \ketbra{m_i}_{\reg{m_i}}U_i\bigg) \ket{m_1}_{\reg{m_1}}
\ket{\dot{\psi}^{(1)}_{m_1}}
\numberthis  \label{branchwise:lem:proof:induction:psi:k:eq:2} \\ 
& = 
\alpha_{m_1} \ket{\dot{\psi}^{(k-1)}_{\vb{m}'}} 
\numberthis  \label{branchwise:lem:proof:induction:psi:k:eq:3}
,\end{align*}
where \Cref{branchwise:lem:proof:induction:psi:k:eq:1} is the definition of $\ket{\psi^{(k)}_{\vb{m}}}$ (see \Cref{branchwise:lem}), \Cref{branchwise:lem:proof:induction:psi:k:eq:2} follows from \Cref{branchwise:lem:proof:eq:U1}, and \Cref{branchwise:lem:proof:induction:psi:k:eq:3} follows from \Cref{branchwise:lem:proof:induction:eq:def:psi:k-1}.

With the above notation, we show the final derivation
\begin{align*}
\Pr[\msf{Out}\big(G_k(\ket{\psi})\big) = 1] 
& = 
\sum_{m_1 \in \bits^\ell} |\alpha_{m_1}|^2 \cdot \Pr[\msf{Out}\big(G_{2:k}(\ket{m_1}_{\reg{m_1}}\ket{\dot{\psi}^{(1)}_{m_1}})\big) = 1] 
\numberthis  \label{branchwise:lem:proof:induction:final:eq:1}\\ 
& = 
\sum_{m_1 \in \bits^\ell} |\alpha_{m_1}|^2 \cdot \sum_{\vb{m}'} \bra{\dot{\psi}^{(k-1)}_{\vb{m}'}}E_1\ket{\dot{\psi}^{(k-1)}_{\vb{m}'}} 
\numberthis  \label{branchwise:lem:proof:induction:final:eq:2} \\ 
& = 
\sum_{\vb{m}', m_1 \in \bits^\ell} \bra{\dot{\psi}^{(k-1)}_{\vb{m}'}}\alpha^*_{m_1} E_1 \alpha_{m_1}\ket{\dot{\psi}^{(k-1)}_{\vb{m}'}} \\ 
& = 
\sum_{\vb{m}} \bra{\psi^{(k)}_{\vb{m}}}E_1\ket{\psi^{(k)}_{\vb{m}}}
\numberthis  \label{branchwise:lem:proof:induction:final:eq:3} 
,\end{align*}
where \Cref{branchwise:lem:proof:induction:final:eq:1} follows from \Cref{branchwise:lem:proof:induction:eq:out}, \Cref{branchwise:lem:proof:induction:final:eq:2} follows from \Cref{branchwise:lem:proof:induction:eq:2:k:out}, and \Cref{branchwise:lem:proof:induction:final:eq:3} follows from \Cref{branchwise:lem:proof:induction:psi:k:eq:3}.

This completes the proof of the induction step.

\vspace{1em}
This completes the proof of \Cref{branchwise:lem}.

\end{proof}

\section{Error-Invariant Commutativity Lemma}
\label{sec:EI-Commutativity-Lemma}
In this section, we present an information-theoretic lemma regarding the commutativity of some unitary operators. As will become evident later, this lemma will serve as a clean abstraction of the behaviors of the ZK simulator and a malicious verifier we designed for the impossibility results. 

\subsection{Statement and Interpretation}

\begin{xiaoenv}{An intuitive explanation of the operators in \Cref{lem:err-inv-com}}
An intuitive explanation of the operators in \Cref{lem:err-inv-com}
\begin{itemize}
\item
$U$ will the verifier's operation to generate the next message
\item
$W$ is the verifiiere's operator to move the message (indeed, its response to $P$) to the correct register.
\item
$\tilde{W}$ is the dummy-up operator 
\item
$S$ is simulator's local unitary.
\end{itemize}
\end{xiaoenv}

\begin{lemma}[Error-Invariant Commutativity Lemma]\label{lem:err-inv-com}
Let $W_0$, $W_1$, and $U_0$ be unitary operators over $\Hil_{\reg{m}}\tensor \Hil_{\reg{t}} \tensor \Hil_{\reg{s}}\tensor  \Hil_{\reg{o}}$ satisfying the following requirements:
\begin{itemize}
\item
$W_1$ acts non-trivially only on $\Hil_{\reg{m}} \tensor \Hil_{\reg{t}} \tensor \Hil_{\reg{o}}$, and is identity on $\Hil_\reg{s}$. 

\item
$W_0$ is the swap operator between registers $\reg{m}$ and $\reg{t}$, and is identity on $\Hil_{\reg{s}}\tensor \Hil_{\reg{o}}$.

\item
$U_1$  acts non-trivially only on $ \Hil_{\reg{m}} \tensor \Hil_{\reg{t}}\tensor \Hil_\reg{o}$, and is identity on $\Hil_{\reg{s}}$. 
\xiao{indeed, this $U_0$ in our application will be identity on $\Hil_{\reg{m}} \tensor \Hil_{\reg{t}}$  as well. But it seems I never used this stronger property when proving the current lemma. So in this lemma, I don't spell it out explicitly.}
\end{itemize}
 
Let $\Hil_{\reg{a}}$ be a Hilbert space of $\ell$ qubits. Let $a \in \bits^{\ell}$ be a fixed classical string. Define the following unitary operators on $\Hil_{\reg{a}} \tensor \Hil_{\reg{m}}\tensor \Hil_{\reg{t}} \tensor \Hil_{\reg{s}}\tensor  \Hil_{\reg{o}}$:
\begin{itemize}
\item
$W \coloneqq \ketbra{a}_{\reg{a}} \tensor W_1 + \sum_{a'\ne a} \ketbra{a'}_{\reg{a}} \tensor W_0$, where $W_0$ and $W_1$ are as specified above.

\item
$\tilde{W} \coloneqq \ketbra{a}_{\reg{a}} \tensor I_{\reg{mtso}}+ \sum_{a'\ne a} \ketbra{a'}_{\reg{a}} \tensor W_0$, where $I_{\reg{mtso}}$ is the identity operator on $\Hil_{\reg{m}}\tensor \Hil_{\reg{t}} \tensor \Hil_{\reg{s}}\tensor  \Hil_{\reg{o}}$ and $W_0$ are as specified above.

\item
$U \coloneqq \ketbra{a}_{\reg{a}} \tensor U_1 + \sum_{a'\ne a} \ketbra{a'}_{\reg{a}} \tensor I_{\reg{mtso}}$, where $U_1$ is as specified above and $I_{\reg{mtso}}$ is the identity operator on $\Hil_{\reg{m}}\tensor \Hil_{\reg{t}} \tensor \Hil_{\reg{s}}\tensor  \Hil_{\reg{o}}$.

\item
$S$ is an operator that acts non-trivially only on $\Hil_\reg{a} \tensor \Hil_\reg{s}$, and is identity on $\Hil_{\reg{m}} \tensor \Hil_{\reg{t}}\tensor  \Hil_{\reg{o}}$. 
\xiao{In our applications, the $S$ may not be a unitary. It will be alternating between unitary and projections. Make sure that the proof can go through.}

\end{itemize}
These operators satisfy the following property:
\begin{itemize}
\item 
Let $\Set{\ket{\rho^{(\msf{in})}_{a'}}_{\reg{mtso}}}_{a' \in \bits^\ell}$ be a sequence of (potentially sub-normalized) pure states over registers $\reg{m}$, $\reg{t}$, $\reg{s}$, and $\reg{o}$. Define states $\ket{\eta^{(\msf{in})}_0}$, $\ket{\eta^{(\msf{in})}_1}$, $\ket{\eta^{(\msf{out})}_0}$, and $\ket{\eta^{(\msf{out})}_1}$ as follows:
\begin{align}
& \ket{\eta^{(\msf{in})}_0} 
\coloneqq 
\ket{a}_{\reg{a}} \ket{\rho^{(\msf{in})}_a}_{\reg{mtso}} + \sum_{a' \in \bits^{\ell} \setminus \Set{a}}  \ket{a'}_{\reg{a}} \red{W_0 U_1^\dagger W_1^\dagger} \ket{\rho^{(\msf{in})}_{a'}}_{\reg{mtso}} 
\label{lem:err-inv-com:def:in-0} \\
& \ket{\eta^{(\msf{in})}_1} 
\coloneqq
\ket{a}_{\reg{a}} \ket{\rho^{(\msf{in})}_a}_{\reg{mtso}} + \sum_{a' \in \bits^{\ell} \setminus \Set{a}}  \ket{a'}_{\reg{a}} \red{W_0}  \ket{\rho^{(\msf{in})}_{a'}}_{\reg{mtso}} 
\label{lem:err-inv-com:def:in-1}\\ 
& \ket{\eta^{(\msf{out})}_0} 
\coloneqq 
\red{W U} S \red{U^\dagger W^\dagger} \cdot \ket{\eta^{(\msf{in})}_0} 
\label{lem:err-inv-com:def:out-0}\\ 
& \ket{\eta^{(\msf{out})}_1} 
\coloneqq 
\red{\tilde{W}} S \red{\tilde{W}^\dagger} \cdot \ket{\eta^{(\msf{in})}_1}
\label{lem:err-inv-com:def:out-1}
\end{align}
Then, there exists a sequence of (potentially sub-normalized) pure states  $\Set{\ket{\rho^{(\msf{out})}_{a'}}_{\reg{mtso}}}_{a' \in \bits^\ell}$ so that the states $\ket{\eta^{(\msf{out})}_0}$ and $\ket{\eta^{(\msf{out})}_1}$ defined above can be written in the following format:
\begin{align}
\ket{\eta^{(\msf{out})}_0}
& = 
\ket{a}_{\reg{a}} \ket{\rho^{(\msf{out})}_a}_{\reg{mtso}} + \sum_{a' \in \bits^{\ell} \setminus \Set{a}}  \ket{a'}_{\reg{a}} \red{W_0 U_1^\dagger W_1^\dagger} \ket{\rho^{(\msf{out})}_{a'}}_{\reg{mtso}} \label{eq:lem:err-inv-com:property-2:real} \\
\ket{\eta^{(\msf{out})}_1}
& = 
 \ket{a}_{\reg{a}} \ket{\rho^{(\msf{out})}_a}_{\reg{mtso}} + \sum_{a' \in \bits^{\ell} \setminus \Set{a}}  \ket{a'}_{\reg{a}} \red{W_0} \ket{\rho^{(\msf{out})}_{a'}}_{\reg{mtso}} \label{eq:lem:err-inv-com:property-2:dummy}
\end{align}
\end{itemize}
\end{lemma}



\subsection{Proof of \Cref{lem:err-inv-com}}

In this proof, we assume for simplicity that each of the registers $\reg{m}$, $\reg{t}$, $\reg{s}$, and $\reg{o}$ consists of $\ell$ qubits. This assumption is without loss of generality because the subsequent derivation works regardless of the length of these registers.

We first note that since $S$ acts non-trivially only on $\Hil_{\reg{a}} \tensor \Hil_{\reg{s}}$, it holds that for any $a'\in\bits^\ell$, $s\in \bits^\ell$, and any pure state $\ket{\rho_{a',s}}_{\reg{mto}}$,
\begin{equation}\label{proof:err-inv-com:property-1:derive:S-format}
S\ket{a'}_{\reg{a}} \ket{s}_{\reg{s}} \ket{\rho_{a',s}}_{\reg{mto}} = \sum_{a^*, s^* \in \bits^{\ell}} \beta^{a', s}_{a^*, s^*} \ket{a^*}_{\reg{a}} \ket{s^*}_{\reg{s}} \ket{\rho_{a', s}}_{\reg{mto}},
\end{equation}
where each $\beta^{a', s}_{a^*, s^*}$ is a complex number that depends on $(a', s, a^*, s^*)$.\footnote{It is worth noting that \Cref{proof:err-inv-com:property-1:derive:S-format} holds for {\em any} $a'\in \bits^\ell$, including the case where $a'$ equals to the $a$ we fixed.} Note that since $S$ may not be unitary, the $\beta$ values may not be normalized. But this does not affect our proof as we work with sub-normalized states anyway.

In the following, we derive \Cref{eq:lem:err-inv-com:property-2:real,eq:lem:err-inv-com:property-2:dummy} one by one.

\subsubsection{Deriving \Cref{eq:lem:err-inv-com:property-2:real}}
We first derive \Cref{eq:lem:err-inv-com:property-2:real}:
\begin{align*}
\ket{\eta^{(\msf{out})}_0}
& = WU SU^\dagger W^\dagger  \cdot \ket{\eta^{(\msf{in})}_0} 
\numberthis \label{proof:err-inv-com:property-2:real:derive:00} \\
& = 
W U SU^\dagger W^\dagger  \cdot 
\Big(
\ket{a}_{\reg{a}} \ket{\rho^{(\msf{in})}_a}_{\reg{mtso}} + \sum_{a' \in \bits^{\ell} \setminus \Set{a}} \ket{a'}_{\reg{a}} W_0 U^\dagger_1 W^\dagger_1 \ket{\rho^{(\msf{in})}_{a'}}_{\reg{mtso}} 
\Big) 
\numberthis \label{proof:err-inv-com:property-2:real:derive:0} \\ 
& = 
W U SU^\dagger W^\dagger  \cdot 
\Big(
\ket{a}_{\reg{a}} \sum_{s \in \bits^\ell}  \ket{s}_{\reg{s}} \ket{\rho^{(\msf{in})}_{a, s}}_{\reg{mto}} ~~+ \\ 
& \hspace{10em}
 \sum_{a' \in \bits^{\ell} \setminus \Set{a}}  \ket{a'}_{\reg{a}} W_0 U^\dagger_1 W^\dagger_1 \sum_{s \in \bits^\ell}  \ket{s}_{\reg{s}} \ket{\rho^{(\msf{in})}_{a', s}}_{\reg{mto}} 
\Big) \\ 
& = 
W U SU^\dagger W^\dagger  \cdot 
\Big(
\ket{a}_{\reg{a}} \sum_{s \in \bits^\ell}  \ket{s}_{\reg{s}} \ket{\rho^{(\msf{in})}_{a, s}}_{\reg{mto}} ~~+ \\ 
& \hspace{10em}
 \sum_{a' \in \bits^{\ell} \setminus \Set{a}} \ket{a'}_{\reg{a}}  \sum_{s \in \bits^\ell}  \ket{s}_{\reg{s}} W_0 U^\dagger_1 W^\dagger_1 \ket{\rho^{(\msf{in})}_{a', s}}_{\reg{mto}} 
\Big) \numberthis \label{proof:err-inv-com:property-2:real:derive:1} \\ 
& =  
W U SU^\dagger \cdot 
\Big(
\ket{a}_{\reg{a}} \sum_{s \in \bits^\ell} \ket{s}_{\reg{s}} W^\dagger_1 \ket{\rho^{(\msf{in})}_{a, s}}_{\reg{mto}} ~~+ \\ 
& \hspace{10em}
 \sum_{a' \in \bits^{\ell} \setminus \Set{a}}  \ket{a'}_{\reg{a}}  \sum_{s \in \bits^\ell}  \ket{s}_{\reg{s}} U^\dagger_1 W^\dagger_1 \ket{\rho^{(\msf{in})}_{a', s}}_{\reg{mto}} 
\Big) \numberthis \label{proof:err-inv-com:property-2:real:derive:2} \\ 
& =  
W U S \cdot 
\Big(
\ket{a}_{\reg{a}} \sum_{s \in \bits^\ell}  \ket{s}_{\reg{s}} U^\dagger_1 W^\dagger_1 \ket{\rho^{(\msf{in})}_{a, s}}_{\reg{mto}} ~~+ \\ 
& \hspace{10em}
 \sum_{a' \in \bits^{\ell} \setminus \Set{a}} \ket{a'}_{\reg{a}}  \sum_{s \in \bits^\ell}  \ket{s}_{\reg{s}} U^\dagger_1 W^\dagger_1 \ket{\rho^{(\msf{in})}_{a', s}}_{\reg{mto}} 
\Big) \numberthis \label{proof:err-inv-com:property-2:real:derive:3} \\ 
& = 
W U S \cdot 
\Big(
\sum_{a' \in \bits^{\ell}} \ket{a'}_{\reg{a}}  \sum_{s \in \bits^\ell} \ket{s}_{\reg{s}} U^\dagger_1 W^\dagger_1 \ket{\rho^{(\msf{in})}_{a', s}}_{\reg{mto}} 
\Big) \\ 
& = 
W U S \cdot 
\Big(
\sum_{a',s \in \bits^\ell} \ket{a'}_{\reg{a}} \ket{s}_{\reg{s}} U^\dagger_1 W^\dagger_1 \ket{\rho^{(\msf{in})}_{a', s}}_{\reg{mto}} 
\Big) \\ 
& = 
W U \cdot 
\Big(
\sum_{a', s \in \bits^\ell} S \ket{a'}_{\reg{a}} \ket{s}_{\reg{s}} U^\dagger_1 W^\dagger_1 \ket{\rho^{(\msf{in})}_{a', s}}_{\reg{mto}} 
\Big)  \\ 
\text{(by \Cref{proof:err-inv-com:property-1:derive:S-format})}~~ & =
W  U \cdot 
\Big(
\sum_{a', s \in \bits^\ell}   \sum_{a^*, s^* \in \bits^{\ell}} \beta^{a', s}_{a^*, s^*} \ket{a^*}_{\reg{a}} \ket{s^*}_{\reg{s}}    U^\dagger_1 W^\dagger_1 \ket{\rho^{(\msf{in})}_{a', s}}_{\reg{mto}} 
\Big) \\ 
& = 
W U \cdot 
\sum_{a', s \in \bits^\ell} 
\Big(
\sum_{ s^* \in \bits^{\ell}} \beta^{a', s}_{a, s^*} \ket{a}_{\reg{a}} \ket{s^*}_{\reg{s}}    U^\dagger_1 W^\dagger_1 \ket{\rho^{(\msf{in})}_{a', s}}_{\reg{mto}} ~~+ \\ 
& \hspace{12em}
\sum_{a^* \in \bits^{\ell} \setminus \Set{a}} \sum_{~s^* \in \bits^{\ell}} \beta^{a', s}_{a^*, s^*} \ket{a^*}_{\reg{a}} \ket{s^*}_{\reg{s}}    U^\dagger_1 W^\dagger_1 \ket{\rho^{(\msf{in})}_{a', s}}_{\reg{mto}}
\Big) \\
& = 
W \cdot 
\sum_{a', s \in \bits^\ell}  
\Big(
\sum_{ s^* \in \bits^{\ell}} \beta^{a', s}_{a, s^*} \ket{a}_{\reg{a}} \ket{s^*}_{\reg{s}}  W^\dagger_1 \ket{\rho^{(\msf{in})}_{a', s}}_{\reg{mto}} ~~+ \\ 
& \hspace{6em}
\sum_{a^* \in \bits^{\ell} \setminus \Set{a}} \sum_{~s^* \in \bits^{\ell}} \beta^{a', s}_{a^*, s^*} \ket{a^*}_{\reg{a}} \ket{s^*}_{\reg{s}}    U^\dagger_1 W^\dagger_1 \ket{\rho^{(\msf{in})}_{a', s}}_{\reg{mto}}
\Big) \numberthis \label{proof:err-inv-com:property-2:real:derive:4} \\ 
& = 
\sum_{a', s \in \bits^\ell}   
\Big(
\sum_{ s^* \in \bits^{\ell}} \beta^{a', s}_{a, s^*} \ket{a}_{\reg{a}} \ket{s^*}_{\reg{s}}  \ket{\rho^{(\msf{in})}_{a', s}}_{\reg{mto}} ~~+ \\ 
& \hspace{6em}
\sum_{a^* \in \bits^{\ell} \setminus \Set{a}} \sum_{~s^* \in \bits^{\ell}} \beta^{a', s}_{a^*, s^*} \ket{a^*}_{\reg{a}} \ket{s^*}_{\reg{s}}    W_0 U^\dagger_1 W^\dagger_1 \ket{\rho^{(\msf{in})}_{a', s}}_{\reg{mto}}
\Big) \numberthis \label{proof:err-inv-com:property-2:real:derive:5} \\ 
& = 
\ket{a}_{\reg{a}}
\sum_{a',s, s^* \in \bits^{\ell}} \beta^{a', s}_{a, s^*}  \ket{s^*}_{\reg{s}}  \ket{\rho^{(\msf{in})}_{a', s}}_{\reg{mto}} ~~+ \\ 
& \hspace{6em}
\sum_{a^* \in \bits^{\ell} \setminus \Set{a}} \ket{a^*}_{\reg{a}}   W_0 U^\dagger_1 W^\dagger_1     \sum_{a', s, s^* \in \bits^{\ell}} \beta^{a', s}_{a^*, s^*}  \ket{s^*}_{\reg{s}}  \ket{\rho^{(\msf{in})}_{a', s}}_{\reg{mto}}
\numberthis \label{proof:err-inv-com:property-2:real:derive:6} \\ 
& = 
\ket{a}_{\reg{a}} \ket{\rho^{(\msf{out})}_{a}}_{\reg{mtso}}
+ 
\sum_{a^* \in \bits^{\ell} \setminus \Set{a}} \ket{a^*}_{\reg{a}}   W_0 U^\dagger_1 W^\dagger_1     \ket{\rho^{(\msf{out})}_{a^*}}_{\reg{mtso}}
\numberthis \label{proof:err-inv-com:property-2:real:derive:7} \\ 
& = 
\ket{a}_{\reg{a}} \ket{\rho^{(\msf{out})}_{a}}_{\reg{mtso}}
+ 
\sum_{a' \in \bits^{\ell} \setminus \Set{a}} \ket{a'}_{\reg{a}}   W_0 U^\dagger_1 W^\dagger_1     \ket{\rho^{(\msf{out})}_{a'}}_{\reg{mtso}}
\numberthis \label{proof:err-inv-com:property-2:real:derive:8}
,\end{align*}
where \Cref{proof:err-inv-com:property-2:real:derive:00} follows from  \Cref{lem:err-inv-com:def:out-0},  \Cref{proof:err-inv-com:property-2:real:derive:0} follows from  \Cref{lem:err-inv-com:def:in-0}, \Cref{proof:err-inv-com:property-2:real:derive:1} follows from the fact that all the unitary  operators $W_0$, $U^\dagger_1$, and $W^\dagger_1$ act as identity on $\Hil_{\reg{s}}$, \Cref{proof:err-inv-com:property-2:real:derive:2} follows from the definition of $W^\dagger$ and the fact that $W^\dagger_0$ and $W^\dagger_1$ act as identity on $\Hil_{\reg{s}}$, \Cref{proof:err-inv-com:property-2:real:derive:3} follow from the definition of $U^\dagger$ and the fact that $U^\dagger_1$ acts as identity on $\Hil_{\reg{s}}$, \Cref{proof:err-inv-com:property-2:real:derive:4} follow from the definition of $U$ and the fact that $U_1$ acts as identity on $\Hil_{\reg{s}}$, \Cref{proof:err-inv-com:property-2:real:derive:5} follow from the definition of $W$ and the fact that $W_0$ and $W_1$ act as identity on $\Hil_{\reg{s}}$, and \Cref{proof:err-inv-com:property-2:real:derive:6} follows from standard algebraic calculation and the fact that all the unitary operators $W_0$, $U^\dagger_1$, and $W^\dagger_1$ act as identity on $\Hil_{\reg{s}}$, \Cref{proof:err-inv-com:property-2:real:derive:7} follows by defining
\begin{equation}\label[Expression]{proof:err-inv-com:property-2:real:derive:def:rho-out}
\forall~a^* \in \bits^\ell,~~~ \ket{\rho^{(\msf{out})}_{a^*}}_{\reg{mtso}} \coloneqq \sum_{a',s, s^* \in \bits^{\ell}} \beta^{a', s}_{a^*, s^*}  \ket{s^*}_{\reg{s}}  \ket{\rho^{(\msf{in})}_{a', s}}_{\reg{mto}},
\end{equation}
and \Cref{proof:err-inv-com:property-2:real:derive:8} follows simply by renaming $a^*$ to $a'$.

This concludes the proof of \Cref{eq:lem:err-inv-com:property-2:real}.

\subsubsection{Deriving \Cref{eq:lem:err-inv-com:property-2:dummy}}

Finally, we derive \Cref{eq:lem:err-inv-com:property-2:dummy}:
\begin{align*}
\ket{\eta^{(\msf{out})}_1}
& = 
\tilde{W} S\tilde{W}^\dagger \cdot \ket{\eta^{(\msf{in})}_1} 
\numberthis \label{proof:err-inv-com:property-2:dummy:derive:00} \\ 
& = 
\tilde{W} S\tilde{W}^\dagger \cdot 
\Big(
 \ket{a}_{\reg{a}} \ket{\rho^{(\msf{in})}_a}_{\reg{mtso}} 
 + \sum_{a' \in \bits^{\ell} \setminus \Set{a}}  \ket{a'}_{\reg{a}} W_0 \ket{\rho^{(\msf{in})}_{a'}}_{\reg{mtso}} 
\Big) 
\numberthis \label{proof:err-inv-com:property-2:dummy:derive:0} \\ 
& = 
\tilde{W} S\tilde{W}^\dagger \cdot 
\Big(
 \ket{a}_{\reg{a}} \sum_{s \in \bits^\ell}  \ket{s}_{\reg{s}} \ket{\rho^{(\msf{in})}_{a, s}}_{\reg{mto}} ~~+ \\ 
& \hspace{10em}
 \sum_{a' \in \bits^{\ell} \setminus \Set{a}}  \ket{a'}_{\reg{a}} W_0 \sum_{s \in \bits^\ell} \ket{s}_{\reg{s}} \ket{\rho^{(\msf{in})}_{a', s}}_{\reg{mto}} 
\Big) \\
& = 
\tilde{W} S \cdot 
\Big(
\ket{a}_{\reg{a}} \sum_{s \in \bits^\ell}  \ket{s}_{\reg{s}} \ket{\rho^{(\msf{in})}_{a, s}}_{\reg{mto}} ~~+ \\ 
& \hspace{10em}
 \sum_{a' \in \bits^{\ell}\setminus \Set{a}}  \ket{a'}_{\reg{a}} \sum_{s \in \bits^\ell}  \ket{s}_{\reg{s}} \ket{\rho^{(\msf{in})}_{a', s}}_{\reg{mto}} 
\Big)  \numberthis \label{proof:err-inv-com:property-2:dummy:derive:1} \\ 
& = 
\tilde{W} S \cdot \Big(
\sum_{a' \in \bits^{\ell}}  \ket{a'}_{\reg{a}} \sum_{s \in \bits^\ell} \ket{s}_{\reg{s}} \ket{\rho^{(\msf{in})}_{a', s}}_{\reg{mto}} 
\Big) \\ 
& = 
\tilde{W}  \cdot 
\Big(
\sum_{a',s \in \bits^{\ell}}  S\ket{a'}_{\reg{a}}\ket{s}_{\reg{s}} \ket{\rho^{(\msf{in})}_{a', s}}_{\reg{mto}} 
\Big) \\ 
\text{(by \Cref{proof:err-inv-com:property-1:derive:S-format})}
~~ & =
\tilde{W} \cdot 
\Big(
\sum_{a', s \in \bits^\ell}  \sum_{a^*, s^* \in \bits^{\ell}} \beta^{a', s}_{a^*, s^*} \ket{a^*}_{\reg{a}} \ket{s^*}_{\reg{s}}    \ket{\rho^{(\msf{in})}_{a', s}}_{\reg{mto}} 
\Big) \\ 
& = 
\tilde{W} \cdot 
\sum_{a', s \in \bits^\ell}  
\Big(
\sum_{ s^* \in \bits^{\ell}} \beta^{a', s}_{a, s^*} \ket{a}_{\reg{a}} \ket{s^*}_{\reg{s}}   \ket{\rho^{(\msf{in})}_{a', s}}_{\reg{mto}} ~~+ \\ 
& \hspace{12em}
\sum_{a^* \in \bits^{\ell} \setminus \Set{a}} \sum_{~s^* \in \bits^{\ell}} \beta^{a', s}_{a^*, s^*} \ket{a^*}_{\reg{a}} \ket{s^*}_{\reg{s}}  \ket{\rho^{(\msf{in})}_{a', s}}_{\reg{mto}}
\Big) \\
& = 
\sum_{a', s \in \bits^\ell}  
\Big(
\sum_{ s^* \in \bits^{\ell}} \beta^{a', s}_{a, s^*} \ket{a}_{\reg{a}} \ket{s^*}_{\reg{s}} \ket{\rho^{(\msf{in})}_{a', s}}_{\reg{mto}} ~~+ \\ 
& \hspace{6em}
\sum_{a^* \in \bits^{\ell} \setminus \Set{a}} \sum_{~s^* \in \bits^{\ell}} \beta^{a', s}_{a^*, s^*} \ket{a^*}_{\reg{a}} \ket{s^*}_{\reg{s}} W_0 \ket{\rho^{(\msf{in})}_{a', s}}_{\reg{mto}}
\Big) \numberthis \label{proof:err-inv-com:property-2:dummy:derive:2} \\ 
& = 
\ket{a}_{\reg{a}}   
\sum_{a', s, s^* \in \bits^{\ell}} \beta^{a', s}_{a, s^*}  \ket{s^*}_{\reg{s}}  \ket{\rho^{(\msf{in})}_{a', s}}_{\reg{mto}} ~~+ \\ 
& \hspace{6em}
\sum_{a^* \in \bits^{\ell} \setminus \Set{a}} \ket{a^*}_{\reg{a}}  W_0  \sum_{a', s, s^* \in \bits^{\ell}}  \beta^{a', s}_{a^*, s^*}  \ket{s^*}_{\reg{s}}  \ket{\rho^{(\msf{in})}_{a', s}}_{\reg{mto}}
\numberthis \label{proof:err-inv-com:property-2:dummy:derive:3} \\ 
& = 
\ket{a}_{\reg{a}}   
 \ket{\rho^{(\msf{out})}_{a}}_{\reg{mtso}} 
+ 
\sum_{a^* \in \bits^{\ell} \setminus \Set{a}} \ket{a^*}_{\reg{a}}  W_0  \ket{\rho^{(\msf{out})}_{a^*}}_{\reg{mtso}} 
\numberthis \label{proof:err-inv-com:property-2:dummy:derive:4} \\ 
& = 
\ket{a}_{\reg{a}}   
 \ket{\rho^{(\msf{out})}_{a}}_{\reg{mtso}} 
+ 
\sum_{a' \in \bits^{\ell} \setminus \Set{a}} \ket{a'}_{\reg{a}}  W_0  \ket{\rho^{(\msf{out})}_{a'}}_{\reg{mtso}} 
\numberthis \label{proof:err-inv-com:property-2:dummy:derive:5}
,\end{align*}
where \Cref{proof:err-inv-com:property-2:dummy:derive:00} follows from  \Cref{lem:err-inv-com:def:out-1},  \Cref{proof:err-inv-com:property-2:dummy:derive:0} follows from  \Cref{lem:err-inv-com:def:in-1}, \Cref{proof:err-inv-com:property-2:dummy:derive:1} follows from the definition of $\tilde{W}^\dagger$, \Cref{proof:err-inv-com:property-2:dummy:derive:2} follows from the definition of $\tilde{W}$ and the fact that $W_0$ acts as identity on $\Hil_{\reg{s}}$, and \Cref{proof:err-inv-com:property-2:dummy:derive:3} follows from standard algebraic calculation and the fact that $W_0$ acts as identity on $\Hil_{\reg{s}}$, \Cref{proof:err-inv-com:property-2:dummy:derive:4} follows from {\em the same definition of} $\Set{\ket{\rho^{(\msf{out})}_{a^*}}}_{a^* \in \bits^\ell}$ as shown in \Cref{proof:err-inv-com:property-2:real:derive:def:rho-out}, and \Cref{proof:err-inv-com:property-2:dummy:derive:5} follows by renaming $a^*$ to $a'$.

This concludes the proof of \Cref{eq:lem:err-inv-com:property-2:dummy}. 

\vspace{1em}

This eventually completes the proof of \Cref{lem:err-inv-com}.

\section{The Models for Quantum Zero-Knowledge}
\label{sec:define-model}

\subsection{Quantum Black-Box Zero-Knowledge Protocols for $\NP$}
\label{sec:def:QZK:standard-model}

We define quantum black-box zero-knowledge proofs (and arguments) for $\NP$. This model is similar to the standard notion of {\em post-quantum} black-box zero-knowledge proofs (and arguments) for $\NP$ (e.g., see \cite{FOCS:CCLY21}) with the following differences:
\begin{itemize}
\item 
The (honest) prover and verifier could be quantum polynomial-time machines. In particular, their communication channel is also quantum. 
\end{itemize}

In the following, we present the formal definitions. We first define quantum interactive proofs (and arguments) for $\NP$ (in \Cref{sec:def:QIP}), and then define zero-knowledge property (in \Cref{sec:def:BBQZK}).

\subsubsection{Quantum Interactive Proofs/Arguments for $\NP$}
\label{sec:def:QIP}

For an $\NP$ language $\Lang$ and $x \in \Lang$, $\Relation_\Lang(x)$ is the set that consists of all (classical) witnesses $w$ such that the verification machine for $\Lang$ accepts $(x,w)$.

A quantum interactive protocol $\langle \Prover, \Verifier \rangle$ is modeled as an interaction between {\em interactive quantum polynomial-time machines}\footnote{We refer to the literature (e.g., \cite{C:HalSmiSon11}) for the standard notion of {\em quantum interactive machines}.} $\Prover$ ( dubbed the prover) and $\Verifier$ (dubbed the verifier). We denote by $\langle \Prover(x_\Prover), \Verifier(x_\Verifier) \rangle (x)$ an execution of the protocol where $x$ is a common input, $x_\Prover$ is the prover's private input, and $x_\Verifier$ is the verifier's private input. We denote by $\rho \la \msf{OUT}_{\Verifier}\langle \Prover(x_\Prover), \Verifier(x_\Verifier) \rangle (x)$ the final output of $\Verifier$, where $\rho$ is a single-qubit state over a special register $\reg{d}$ of $\Verifier$. We also define a quantum predicate $\msf{Acc}(\cdot)$ which, on input a single-qubit state $\rho$, measures $\rho$ in the computational basis and outputs the measurement outcome. Intuitive, the output of $\msf{Acc}$ indicates if the the verifier accepts (outputting 1) or rejects (outputting 0).
\begin{definition}[Quantum Interactive Proofs/Arguments]
\label{def:QIP}
A quantum interactive proof or argument $\langle \Prover, \Verifier \rangle$ for an $\NP$ language $\Lang$ is a quantum interactive protocol between a QPT prover $\Prover$ and a QPT verifier $\Verifier$ that satisfy the following requirements:
\begin{itemize}
\item 
{\bf Completeness.} For each $x \in \Lang$ and each $w \in \Relation_\Lang(x)$, it holds that
$$\Pr\big[\msf{Acc}(\rho) = 1~:~\rho\la\msf{OUT}_{\Verifier}\langle \Prover(w), \Verifier \rangle (x)\big]  \ge 1 -\negl(\secpar).$$

\item 
{\bf Statistical/Computational Soundness.} The protocol is statistically (resp.\ computationally) {\em sound} if for any unbounded (resp.\ non-uniform QPT) cheating prover
$\Prover^*$ and any $x \in \bits^\secpar \setminus \Lang$, it holds that
$$\Pr\big[\msf{Acc}(\rho) = 1~:~\rho\la\msf{OUT}_{\Verifier}\langle \Prover^*, \Verifier \rangle (x)\big]   = \negl(\secpar).$$
The protocol is dubbed an interactive proof (resp.\ argument) if the soundness property is statistical (resp.\ computational).
\end{itemize}
\end{definition}

\begin{remark}
We note that considering the function $\mathsf{Acc}(\cdot)$ as part of the verifier is a valid alternative. However, we prefer to use the above formalism to treat $\mathsf{Acc}(\cdot)$ as an ``external'' measurement. This approach aims to facilitate our subsequent presentation, where we will assume without loss of generality that the verifier $\Verifier$ does not conduct any measurements.
\end{remark}

\begin{remark}
    In a plain standard quantum interactive protocol, $\Verifier$ and $\Prover$ do not pre-share any entanglement at the beginning of the protocol.
\end{remark}

\subsubsection{Black-Box Quantum Zero-Knowledge}
\label{sec:def:BBQZK}
Next, we proceed to define the quantum zero-knowledge property. Toward that, we first need to be more accurate about the malicious verifiers and the way the simulator interacts with the malicious verifier.

\para{Number of Rounds.} We say that a quantum interactive proof/argument has $K$ rounds if the messages exchanged between $\Prover$ and $\Verifier$ are of the following structure
$$(p_1, v_1, p_2, v_2, \ldots, p_{K-1}, v_{K-1}, p_{K}, v_{K}),$$
where $p_1$ is the first message $\Prover$ sends to $\Verifier$, and $v_1$ is $\Verifier$'s response to $p_1$, and $p_2$ is the next message $\Prover$ sends to $\Verifier$ and so one. Some remarks regarding this message structure follow:
\begin{itemize}
\item
All the $p_i$'s and $v_i$'s could be quantum messages, as we allow quantum communication in quantum interactive proofs/arguments.

\item
It is without loss of generality to assume that $\Prover$ sends the first message (i.e., $p_1$). 

\item 
Typically, the last message should be from the prover to the verifier. However, in our model, we ask the verifier to send the last message $v_K$. We explain the reason in \Cref{rmk:msg:vk}.

\item 
We remark that different authors may use the term ``round'' differently. In this work, we use it to refer to {\em a pair} of adjacent messages. For example, $(p_1, v_1)$ constitutes the first round, and $(p_2, v_2)$ constitutes the second round and so on. This is why we say that the protocol, where $2K$ messages are actually exchanged, has $K$ rounds. 
\end{itemize}
\begin{remark}[On the last message $v_K$]\label{rmk:msg:vk}
In this work, we require the honest verifier to send its final decision qubit (i.e., register $\reg{d}$) to the prover, which is exactly the last message $v_K$.  It is important to note that this requirement does not impact completeness or soundness. However, it does influence the definition of zero-knowledge. Specifically, it could potentially make the simulator's task easier, leading to a weaker definition of the zero-knowledge property. Nevertheless, since our goal is to establish impossibility results, this assumption could only make our results stronger.

A related issue arises: If we ask the honest $\Verifier$ to sends its final decision qubit as message $v_K$, how do  we define the final output of $\Verifier$? This can be handled as follows. When the message $v_K$ is generated in register $\reg{m}$, we ask $\Verifier$ to keep a copy of it in a designated register $\reg{d}$ in its working space $\reg{w}$. Technically, $\Verifier$ applies a $\msf{CNOT}$ gate on the register containing the decision qubit $v_K$ and the register $\ket{0}_\reg{d}$ where the former register is the control register, and then $\Verifier$ sends the $\ket{v_K}_\reg{m}$ to the prover as the last message. Note that this effectively collapses the decision qubit $\ket{v_K}_\reg{d}$, but this is fine since the predicate $\msf{Acc}(\cdot)$ anyway performs computational measurement on the decision qubit (see \Cref{def:QIP}).
\end{remark}

\para{Malicious Verifiers.} For a formal definition of black-box quantum zero-knowledge, we give a model of quantum malicious verifiers against quantum interactive protocols. A malicious verifier $\tilde{\Verifier}$ is specified by a sequence of unitary $\tilde{V}_{\secpar}$ over the internal register $\reg{v}$ and the message register $\reg{m}$ (whose details are explained later) and an auxiliary input $\rho_\secpar$ indexed by the security parameter $\secpar \in \Naturals$. We say that  $\tilde{\Verifier}$ is non-uniform QPT if the sizes of $\tilde{V}_{\secpar}$ and $\rho_\secpar$ are polynomial in $\secpar$. In the rest of this paper, $\secpar$ is always is always set to be the length of the statement $x$ to be proven, and thus we omit $\secpar$ for notation simplicity.

Its internal register $\reg{v}$ consists of the instance register $\reg{ins}$, auxiliary input register $\reg{aux}$, and verifier's working register $\reg{w}$. Part of the working space $\reg{w}$ is designated as the output register $\reg{out}$. $\tilde{\Verifier}$ interacts
with an honest prover $\Prover$ on a common input $x$ and $\Prover$'s private input $w \in \Relation_\Lang(x)$ in the following manner:
\begin{enumerate}
\item 
Register $\reg{ins}$ is initialized to $x$, $\reg{aux}$ is initialized to $\rho$, and $\reg{w}$ and $\reg{m}$ are initialized to be $\reg{\vb{0}}$ of sufficient length.

\item
$\Prover$ (with private input $w$) and $\tilde{\Verifier}$ run the $K$-round protocol as follows. On round $k \in [K]$,
\begin{enumerate}
\item 
$\Prover$ sends message $p_k$ by loading it in register $\reg{m}$. (Physically, one can think that register $\reg{m}$ was in the hand of $\Prover$ at the beginning of this round; $\Prover$ loads message $p_k$ in $\reg{m}$ and sends $\reg{m}$ to $\tilde{\Verifier}$.)

\item
$\tilde{\Verifier}$ applies the unitary $\tilde{V}$, which generates $\tilde{\Verifier}$'s response $v_k$ and loads it in register $\reg{m}$. (Physically, one can think that register $\reg{m}$ is then sent to $\Prover$; This is how we model ``$\tilde{\Verifier}$ sends message $v_k$ to $\Prover$''.)
\end{enumerate}

\item 
At the end of the execution, $\tilde{\Verifier}$ outputs the state in register $\reg{out}$ as her final output.
\end{enumerate}
We denote by $\langle \Prover(w), \tilde{\Verifier}(\rho)\rangle(x)$ the above execution and by $\rho_{\msf{out}} \la \msf{OUT}_{\tilde{\Verifier}}\langle \Prover(w), \tilde{\Verifier}(\rho)\rangle(x)$ the final output $\rho_{\msf{out}}$ of $\tilde{\Verifier}$, which is a quantum state over $\reg{out}$.

\para{Black-Box Simulator.} 
We now describe formally how the simulator $\Sim$ interacts with a malicious verifier $\tilde{\Verifier}$ defined above. We note that there are multiple ways to do so and we choose the following one without loss of generality, which is particularly suitable for our proof of impossibility. 

A quantum black-box simulator $\Sim$ is modeled as a quantum oracle Turing machine (e.g., see \cite{bennett1997strengths}). We say that $\Sim$ is expected-QPT (resp.\ strict-QPT) if the expected (resp.\ maximum) number of steps is polynomial in the input length, counting an oracle access as a unit step. 

Note that we focus on black-box simulation. Thus, $\Sim$ can only query $\tilde{\Verifier}$ via her unitary $\tilde{V}$ and its inverse $\tilde{V}^\dagger$, but does not get to see the internal register registers of $\tilde{\Verifier}$. Also, recall that $\reg{m}$ is used to exchange messages as explained above. Therefore, for simulation, $\reg{m}$ is considered as being held by $\Sim$.

In more detail, a black-box simulator $\Sim$ is defined by a ``local'' unitary $S$. It is granted oracle access to $\tilde{V}$ and $\tilde{V}^\dagger$. It makes use of the following registers (in addition to $\tilde{\Verifier}$'s internal registers that she could only access through $\tilde{V}$ or $\tilde{V}^\dagger$ )
\begin{itemize}
\item
Register $\reg{ins}$ initialized to $\ket{x}_{\reg{ins}}$.
\item
Register $\reg{m}$ to exchange messages with the malicious verifier $\tilde{\Verifier}$;
\item
Register $\reg{s}$, which is her ``local'' working space;
\item
A query-type register $\reg{u}$ in a $2$-dimension Hilbert space spanned by the basis $\Set{ \ket{\downarrow}_{\reg{u}}, \ket{\uparrow}_{\reg{u}}}$ (explained shortly).
\end{itemize}
$\Sim$'s behavior is iterations of the following three steps:
\begin{enumerate}
\item
Apply the local operator $S$ on registers $\reg{m}$, $\reg{u}$, and $\reg{s}$;
\item
Measure the $\reg{u}$ register to learn the type of the next query (see the next bullet);
\item
Make the oracle query according to the measurement outcome from the last step---if the measurement outcome is $\ket{\downarrow}_{\reg{u}}$, it query the $\tilde{V}$ oracle on $\reg{m}$; otherwise (i.e., the measurement outcome is $\ket{\uparrow}_{\reg{u}}$), it query the $\tilde{V}^\dagger$ oracle on $\reg{m}$.
\end{enumerate}
$\Sim$ keeps repeating the above three steps until she decides to stop. At that moment, the internal state of $\tilde{\Verifier}$ (i.e., contents of register $\reg{out}$) is defined to be the simulation output. Notation-wise, we denote the above procedure by $\rho_{\msf{out}} \la \Sim^{\tilde{\Verifier}(x;\rho)}(x)$ the final output $\rho_{\msf{out}}$ of $\tilde{\Verifier}$, where $\rho_{\msf{out}}$ the final state $\reg{out}$ (tracing
out all other registers after the execution).

\para{Black-Box Quantum ZK.} With the above notations, we now present the definition of black-box quantum zero-knowledge in \Cref{def:BBQZK}.
\begin{definition}[Black-Box Quantum ZK Proofs/Arguments]
\label{def:BBQZK}
A black-box quantum zero-knowledge proof (resp.\ argument) for an $\NP$ language $\Lang$ is a quantum interactive proof (resp.\ argument) $\langle \Prover, \Verifier \rangle$ for $\Lang$ (as per \Cref{def:QIP}) that additionally satisfies the following property:
\begin{itemize}
\item 
{\bf Black-Box Quantum Zero-Knowledge.} There exists an expected-QPT simulator $\Sim$ such
that for each non-uniform QPT malicious verifier $\tilde{\Verifier}$ with an auxiliary input $\rho$, it holds that
$$
\big\{
	\msf{OUT}_{\tilde{\Verifier}}\langle \Prover(w), \tilde{\Verifier}(\rho)\rangle(x)
\big\}_{\secpar \in \Naturals, x \in \Lang\cap \bits^\secpar, w \in \Relation_\Lang(x)}
~\cind~
\big\{
	\Sim^{\tilde{\Verifier}(x;\rho)}(x)
\big\}_{\secpar \in \Naturals, x \in \Lang\cap \bits^\secpar, w \in \Relation_\Lang(x)}
.$$
\end{itemize}
\end{definition}

\subsection{Quantum ZK in the Non-Programmable EPR Model}
\label{sec:def:QZK:NPE}

In this section, we define a new model for quantum zero-knowledge, which we refer to as the {\em Non-Programmable EPR} (NPE for short) model. Before we dive into the formal definitions, we first provide an intuitive explanation of this model and how it helps us establish our impossibility results for the (standard) QZK (as per \Cref{sec:def:QZK:standard-model}). 

The NPE model is an intermediate model where $\Prover$ and $\Verifier$ pre-share EPR pairs issued by some trusted third party. These EPR pairs can be used to ``de-quantize'' the prover's messages. We will show that given a fully quantum BBZK protocol, we can transform it into a  BBZK protocol in the NPE model with the same round complexity. This allows us to focus on the NPE model when establishing our impossibility result.



We next proceed to the formal definitions.


\para{Non-Programmable EPR Model.} As explained above, QZK in this non-programmable EPR model is almost identical to the standard QZK we defined in \Cref{sec:def:QZK:standard-model}, except for how the prover's message is transmitted. Therefore, instead of presenting all the details (which include repeated descriptions of the same steps as we did for the standard QZK model), we choose to describe the non-programmable EPR model by only highlighting its differences from the standard QZK model.

The non-programmable EPR model (NPE) ZK is identical to the QZK protocol as we defined in \Cref{sec:def:QZK:standard-model}, except for the following differences
\begin{itemize}
\item 
We assume there is a trusted party $\msf{Trust}$ who prepares polynomially many EPR pairs 
$$
(e^{(1)}_1, e^{(1)}_2), \ldots, (e^{(n)}_1, e^{(n)}_2)
,$$
where $n(\secpar)$ is a fixed polynomial specified by the protocol.

\item
At the beginning of the protocol, the first halves of these EPR pairs $(e^{(1)}_1, \ldots, e^{(n)}_1)$ are given to the honest prover $\Prover$ as a part of its input; the second halves of these EPR pairs $(e^{(1)}_2, \ldots, e^{(n)}_2)$ are given to the honest verifier $\Verifier$ as a part of its input.

\item
The computation models for the honest prover $\Prover$ and honest verifier $\Verifier$ are identical to that in the standard QZK defined in \Cref{sec:def:QZK:standard-model}, except that $\Prover$ and $\Verifier$ now can make use of the EPR pairs in their input during the execution, as long as the computation is QPT. 

We note that we can continue to use the same interface between $\Prover$ and $\Verifier$ as described in \Cref{sec:def:BBQZK}. That is, though they additionally hold EPR pairs, the protocol still has the generic form that $\Prover$ and $\Verifier$ exchanges (possibly) quantum messages $(p_1, v_1, \ldots, p_K, v_K)$. The only modifications would be in the specific descriptions of their local unitary, which are anyway left unspecified in the generic description provided in \Cref{sec:def:QZK:standard-model}.

Similarly, we denote by 
$$\langle \Prover\big(w, (e^{(1)}_1, \ldots, e^{(n)}_1) \big), \Verifier(e^{(1)}_2, \ldots, e^{(n)}_2)\rangle(x)
$$ 
the above execution, and by 
$$
\rho \la \msf{OUT}_{\Verifier}\langle \Prover\big(w, (e^{(1)}_1, \ldots, e^{(n)}_1) \big), \Verifier(e^{(1)}_2, \ldots, e^{(n)}_2) \rangle (x)
$$ 
the final output of $\Verifier$, where $\rho$ is a single-qubit state over a special register $\reg{d}$ of $\Verifier$. We re-use the same definition of $\msf{Acc}(\cdot)$ as in \Cref{sec:def:QZK:standard-model}.


\item
The modeling of the malicious verifier $\tilde{\Verifier}$ and the simulator $\Sim$ also remains the same as in \Cref{sec:def:BBQZK}, except that $\Sim$ now gets the $(e^{(1)}_1, \ldots, e^{(n)}_1)$ as additional input (similar as the honest prover). We emphasize the following points regarding the simulator $\Sim$.
\begin{itemize}
\item 
$\Sim$ does not get to choose the EPR pairs. That is, the EPR pairs are always sampled by the trusted party. $\Sim$ gets $(e^{(1)}_1, \ldots, e^{(n)}_1)$ as input, and the other halves $(e^{(1)}_2, \ldots, e^{(n)}_2)$ are given to the (potentially malicious) verifier $\tilde{\Verifier}$. The latter is considered as being stored in $\tilde{\Verifier}$'s internal registers, so $\Sim$ cannot see or modify these shares hold by $\tilde{\Verifier}$ directly (unless through oracle access to $\tilde{\Verifier}$'s unitary $\tilde{V}$ and its inverse $\tilde{V}^\dagger$).	

\item

We note that we can continue to use the same interface between $\Sim$ and $\tilde{\Verifier}$ as described in \Cref{sec:def:BBQZK}. In terms of notation, we can consider $(e^{(1)}_1, \ldots, e^{(n)}_1)$ to be stored in a designated region of $\Sim$'s working space $\reg{s}$, and $(e^{(1)}_2, \ldots, e^{(n)}_2)$ to be stored in a corresponding region of $\tilde{\Verifier}$'s working space $\reg{w}$. Consequently, the generic description of how $\Sim$ interacts with $\tilde{\Verifier}$ remains unchanged---the only modifications would be in the specific descriptions of $\Sim$'s local unitary $S$ and $\tilde{\Verifier}$'s unitary $\tilde{V}$, which are anyway left unspecified in the generic description provided in \Cref{sec:def:BBQZK}.

Similarly, we use $\rho_{\msf{out}} \la \Sim^{\tilde{\Verifier}\big(x, (e^{(1)}_2, \ldots, e^{(n)}_2);\rho\big)}\big( x, (e^{(1)}_1, \ldots, e^{(n)}_1)\big)$ to denote the final output $\rho_{\msf{out}}$ of $\tilde{\Verifier}$, where $\rho_{\msf{out}}$ the final state on $\tilde{\Verifier}$'s internal register $\reg{out}$ (tracing out all other registers after the execution).
\end{itemize}
\end{itemize}
With the above notation, we now present the formal definition for QZK in the NPE model in \Cref{def:NPE-QZK}.

\begin{definition}[Black-Box QZK in the NPE Model]
\label{def:NPE-QZK}
Let $n(\secpar)$ be a fixed polynomial of the security parameter $\secpar$. A black-box quantum zero-knowledge proof (resp.\ argument) for an $\NP$ language $\Lang$, in the $n$-pair non-programmable EPR model, is a quantum interactive protocol between a QPT prover $\Prover$ and a QPT verifier $\Verifier$ that satisfy the following requirements:
\begin{itemize}
\item 
{\bf Completeness.} For any $x \in \Lang \cap \{0,1\}^\lambda$, any $w \in \Relation_\Lang(x)$, and any $n$ EPR pairs $\Set{e^{(i)}_1, e^{(i)}_2}_{i \in [n]}$ generated by $\msf{Trust}$, it holds that
$$\Pr\big[\msf{Acc}(\rho) = 1~:~\rho \la \msf{OUT}_{\Verifier}\langle \Prover\big(w, (e^{(1)}_1, \ldots, e^{(n)}_1) \big), \Verifier(e^{(1)}_2, \ldots, e^{(n)}_2) \rangle (x)\big]  \ge 1 -\negl(\secpar).$$

\item 
{\bf Statistical/Computational Soundness.} The protocol is statistically (resp.\ computationally) {\em sound} if for any unbounded (resp.\ non-uniform QPT) cheating prover
$\Prover^*$ and any $x \in \bits^\secpar \setminus \Lang$, and any  $n$ EPR pairs $\Set{e^{(i)}_1, e^{(i)}_2}_{i \in [n]}$ generated by $\msf{Trust}$, it holds that
$$\Pr\big[\msf{Acc}(\rho) = 1~:~\rho\la\msf{OUT}_{\Verifier}\langle \Prover^*(e^{(1)}_1, \ldots, e^{(n)}_1), \Verifier(e^{(1)}_2, \ldots, e^{(n)}_2) \rangle (x)\big]   = \negl(\secpar).$$
The protocol is dubbed a proof (resp.\ argument) if the soundness property is statistical (resp.\ computational).

\item 
{\bf Black-Box Quantum Zero-Knowledge.} There exists an expected-QPT simulator $\Sim$ such
that for any non-uniform QPT malicious verifier $\tilde{\Verifier}$ with an auxiliary input $\rho$ and any $n$ EPR pairs $\Set{e^{(i)}_1, e^{(i)}_2}_{i \in [n]}$ generated by $\msf{Trust}$, it holds that
\begin{align*}
& \big\{
	\msf{OUT}_{\tilde{\Verifier}}\langle \Prover\big(w, (e^{(1)}_1, \ldots, e^{(n)}_1)\big), \tilde{\Verifier}\big((e^{(1)}_2, \ldots, e^{(n)}_2), \rho\big)\rangle(x)
\big\}_{\secpar \in \Naturals, x \in \Lang\cap \bits^\secpar, w \in \Relation_\Lang(x)} \\
~\cind~ &
\big\{
	\Sim^{\tilde{\Verifier}\big(x, (e^{(1)}_2, \ldots, e^{(n)}_2);\rho\big)}\big(x, (e^{(1)}_1, \ldots, e^{(n)}_1)\big)
\big\}_{\secpar \in \Naturals, x \in \Lang\cap \bits^\secpar, w \in \Relation_\Lang(x)}
.
\end{align*}
\end{itemize}
Henceforth, we use $n$-NPE-BBQZK as the abbreviation for black-box quantum zero-knowledge arguments in the $n$-pair non-programmable EPR model.
\end{definition}

\subsection{Standard BBQZK Implies NPE-BBQZK}
\label{sec:model:subsection:QZK-to-NPE}

In this part, we show in \cref{lem:QZK-to-NPE} that if (standard) BBQZK exists, then BBQZK in the NPE model also exists. Moreover, the prover in the latter protocol only needs to send {\em classical} messages. Looking forward, \cref{lem:QZK-to-NPE} allows us to switch our attention to the NPE model---to prove that there does not exist constant-round black-box quantum zero-knowledge, we only need to show that such protocols do not exist in the NPE model.	
\begin{lemma}\label{lem:QZK-to-NPE}
For a number $K$, assume that there exists a $K$-round black-box QZK proof (resp.\ argument) as per \Cref{def:BBQZK}, where we assume w.l.o.g.\ that each messages exchanged in this protocol are of the same length of $\ell(\secpar)$ qubits. Then, there exists a $K$-round black-box QZK proof (resp.\ argument) in the $(K\cdot \ell)$-pair NPE model as per \Cref{def:NPE-QZK}, where all the messages sent by the prover is classical.
\end{lemma}
\begin{proof}[Proof of \Cref{lem:QZK-to-NPE}]
Consider a $K$-round BBQZK protocol $\langle \Prover, \Verifier \rangle$ as defined in \Cref{def:BBQZK}. We assume without loss of generality that all the $p_k$'s and $v_k$'s message are of the same length $\ell(\secpar)$ qubits, where $\ell(\secpar)$ is some polynomial of the security parameter $\secpar$. To establish \Cref{lem:QZK-to-NPE}, we show a compiler that converts $\langle \Prover, \Verifier \rangle$ to a new protocol $\langle \Prover, \Verifier \rangle_{\textsc{npe}}$ in the $(K\cdot \ell)$-pair NPE model, with all prover's messages being classical. 

At a high level, the compiler simply runs the original  $\langle \Prover, \Verifier \rangle$ with only one difference: all the prover's messages $p_k$'s are sent to the verifier by quantum teleportation. Note that each $p_k$ consists of $\ell$ qubits. Thus, $(K\cdot \ell)$ EPR pairs suffice for teleporting all prover's messages. We present the formal description of the compiler in \Cref{compiler:QZK-to-NPE}, and then prove that the resulting protocol 	$\langle \Prover, \Verifier \rangle_{\textsc{npe}}$ does satisfies the requirements in \Cref{lem:QZK-to-NPE}.

\begin{ProtocolBox}[label={compiler:QZK-to-NPE}]{NPE-BBQZK Protocol \textnormal{$\langle \Prover, \Verifier \rangle_{\textsc{npe}}$}}
Let $n(\secpar) = K \cdot \ell(\secpar)$. Then, a $K$-round QZK in the NPE model is identical to the original QZK protocol $\langle \Prover, \Verifier \rangle$ except for the following differences:
\begin{itemize}
\item
Before the protocol starts, the trusted party prepares $n$ EPR pairs 
$$
(e^{(1)}_1, e^{(1)}_2), \ldots, (e^{(n)}_1, e^{(n)}_2)
.$$
The first halves of these EPR pairs $(e^{(1)}_1, \ldots, e^{(n)}_1)$ are given to the honest prover $\Prover$ as a part of its input (in addition to its original input $x$ and $w$); the second halves of these EPR pairs $(e^{(1)}_2, \ldots, e^{(n)}_2)$ are given to the honest verifier $\Verifier$ as a part of its input (in addition to its original input $x$).

\item
$\Prover$ ane $\Verifier$ then behaves as in the original execution of $\langle \Prover(w), \Verifier \rangle(x)$, except for the following differences:
\begin{itemize}
\item 
When $\Prover$ wants to send message $p_k$ ($\forall k \in [K]$), it does not send $p_k$ directly. Instead, it uses quantum teleportation to transmit it, consuming the EPR shares $(e^{((k-1)\ell + 1)}_1, \ldots, e^{(k\ell)}_1)$. In more detail, $\Prover$ performs the teleportation measurements over the $\ell$-qubit message $p_k$ and the $\ell$-qubit EPR shares $(e^{((k-1)\ell + 1)}_1, \ldots, e^{(k\ell)}_1)$. This leads to $\ell$ pairs of (classical) teleportation keys $\Set{(a^{(i)}_k, b^{(i)}_k)}_{i \in [\ell]}$. The honest prover $\Prover$ then sends $\tilde{p}_k = \Set{(a^{(i)}_k, b^{(i)}_k)}_{i \in [\ell]}$ to the verifier.

\item
When the honest verifier $\Verifier$ receives $\tilde{p}_k = \Set{(a^{(i)}_k, b^{(i)}_k)}_{i \in [\ell]}$, it first recover the quantum message $p_k$ using the teleportation keys contained in $\tilde{p}_k$ and the corresponding halves of the EPR pairs $(e^{((k-1)\ell + 1)}_2, \ldots, e^{(k\ell)}_2)$. Then, $\Verifier$ behaves in the same manner as in the original protocol to generate message $v_k$ and sends it to the prover.	 

\item
At the end of the execution, the verifier output whatever the original $\Verifier$ would output in the original $\langle \Prover(w), \Verifier \rangle(x)$ protocol.
\end{itemize}
\end{itemize}

\end{ProtocolBox}

From the description of \Cref{compiler:QZK-to-NPE}, it is easy to see that the protocol 	$\langle \Prover, \Verifier \rangle_{\textsc{npe}}$ has $K$ rounds, consumes $(K\cdot \ell)$ EPR pairs, and all the prover's messages $\Set{\tilde{p}_k}_{k \in [K]}$ are classical. Also, the completeness follows direct from the completeness of the original $\langle \Prover, \Verifier \rangle$ protocol and the correctness guarantee of quantum teleportation.

To prove soundness, we assume that there is a malicious QPT prover $\tilde{\Prover}_\textsc{npe}$ that breaks the soundness of $\langle \Prover, \Verifier \rangle_{\textsc{npe}}$ shown in  \Cref{compiler:QZK-to-NPE}. Then, we can construct another malicious QPT prover $\tilde{\Prover}$ that breaks the soundness of the original protocol $\langle \Prover, \Verifier \rangle$ as follows. Machine $\tilde{\Prover}$ samples the EPR pairs by herself, and emulates machine $\tilde{\Prover}_{\textsc{npe}}$ internally, on input $(e^{(1)}_1, \ldots, e^{(n)}_1)$. When the internal $\tilde{\Prover}_{\textsc{npe}}$ sends the message $\tilde{p}_k$, $\tilde{\Prover}$ first recover the original quantum message $p_k$ using $(e^{(1)}_2, \ldots, e^{(n)}_2)$ (note that this is possible because $\tilde{\Prover}$ sampled the EPR pairs), and then forward $p_k$ to the external honest verifier $\Verifier$; When the external $\Verifier$ sends message $v_k$, machine $\tilde{\Prover}$ simply forward it to the internal $\tilde{\Prover}_{\textsc{npe}}$. It is straightforward that if $\tilde{\Prover}_\textsc{npe}$ breaks the soundness of $\langle \Prover, \Verifier \rangle_{\textsc{npe}}$, then the constructed  $\tilde{\Prover}$ breaks the soundness of the original protocol $\langle \Prover, \Verifier \rangle$ as well.

The zero-Knowledge property follows from a similar argument as for soundness above. In more detail, the black-box ZK simulator $\Sim_{\textsc{npe}}$ works as follows:
\begin{enumerate}
	\item 	
 Given a malicious verifier $\tilde{\Verifier}_{\textsc{npe}}$ for the protocol \Cref{compiler:QZK-to-NPE}, it first constructs a malicious verifier $\tilde{\Verifier}$ for the original protocol $\langle \Prover, \Verifier \rangle$ as follows:
\begin{itemize}
\item 
Machine $\tilde{\Verifier}$ samples the EPR pairs by herself, and emulates machine $\tilde{\Verifier}_{\textsc{npe}}$ internally, on input $(e^{(1)}_2, \ldots, e^{(n)}_2)$. When the external $\tilde{\Prover}$ sends the message $p_k$ ($\forall k \in [K]$), $\tilde{\Verifier}$ performs teleportation measurements on $p_k$ and the EPR shares
$(e^{((k-1)\ell + 1)}_1, \ldots, e^{(k\ell)}_1)$, which yield the classical keys $\tilde{p}_k = \Set{(a^{(i)}_k, b^{(i)}_k)}_{i \in [\ell]}$. $\tilde{\Verifier}$ forward $\tilde{p}_k$ to the internal $\tilde{\Verifier}_{\textsc{npe}}$. When the internal $\tilde{\Verifier}_{\textsc{npe}}$ sends $v_k$, $\tilde{\Verifier}$ forwards it to the external $\Prover$. At the end of the execution, $\tilde{\Verifier}$ outputs whatever the internal $\tilde{\Verifier}_{\textsc{npe}}$ outputs.
\end{itemize}

\item 
$\Sim_{\textsc{npe}}$ then invokes the simulator $\Sim$ of the original protocol $\tilde{\Verifier}$, providing the $\tilde{\Verifier}$ constructed above as the oracle required by $\Sim$. Finally, $\Sim_{\textsc{npe}}$ outputs whatever $\Sim^{\tilde{\Verifier}}$ outputs.
\end{enumerate} 
We first argue that the above $\Sim_{\textsc{npe}}$ only makes black-box access to $\tilde{\Verifier}_{\textsc{npe}}$. This can be seen by noticing that the construction of $\tilde{\Verifier}$ requires only black-box access to $\tilde{\Verifier}_{\textsc{npe}}$, and that the simulator $\Sim$ of the original protocol makes only black-box to the $\tilde{\Verifier}$ constructed in black-box from $\tilde{\Verifier}_{\textsc{npe}}$.

As for the indistinguishability of the simulation, first notice that in a real execution between the original honest prover $\Prover(x, w)$ and the above constructed $\tilde{\Verifier}$, the final output of $\tilde{\Verifier}$ is identically distributed as the final output of $\tilde{\Verifier}_{\textsc{npe}}$ obtained from an execution with the honest \Cref{compiler:QZK-to-NPE} prover $\Prover_{\textsc{npe}}(x,w)$. This is because $\tilde{\Verifier}$ perfectly emulates the execution for the internal $\tilde{\Verifier}_{\textsc{npe}}$, and eventually outputs whatever the latter outputs. Then, by the ZK property of the original protocol, the output of $\Sim^{\tilde{\Verifier}}$ is computationally indistinguishable with $\tilde{\Verifier}$'s output in the real execution with $\Prover(x,w)$, which, as we just argued, is identical to $\tilde{\Verifier}_{\textsc{npe}}$'s output from the real execution with $\Prover_{\textsc{npe}}(x,w)$. Also notice that $\Sim_{\textsc{npe}}$ is defined to output whatever $\Sim^{\tilde{\Verifier}}$ outputs. Therefore, the final output of $\Sim_{\textsc{npe}}$ is computationally indistinguishable with $\tilde{\Verifier}_{\textsc{npe}}$'s output from the real execution with $\Prover_{\textsc{npe}}(x,w)$.

This completes the proof of \Cref{lem:QZK-to-NPE}.

\end{proof}

\section{Impossibility of Constant-Round Black-Box Quantum Zero-Knowledge}

In this section, we prove the following \Cref{thm:impossibility:QZK}.
\begin{theorem}\label{thm:impossibility:QZK}
For a language $\Lang$, if there exists a constant-round black-box quantum zero-knowledge argument (as per \Cref{def:BBQZK}), then it holds that $\Lang \in \BQP$.
\end{theorem}

It follows from \Cref{lem:QZK-to-NPE} that to prove \Cref{thm:impossibility:QZK}, it suffices to establish the following \Cref{thm:impossibility:NPE}. The rest of this section is devoted to proving \Cref{thm:impossibility:NPE}.
\begin{theorem}\label{thm:impossibility:NPE}
For a language $\Lang$, if there exists a constant-round black-box quantum zero-knowledge argument in the non-programmable EPR model (as per \Cref{def:NPE-QZK}) with prover's messages being classical, then it holds that $\Lang \in \BQP$.
\end{theorem}

\para{Proof Structure.} At a high level, our proof for \Cref{thm:impossibility:NPE} follows the paradigm established in \cite{STOC:BarLin02,FOCS:CCLY21}. Given a fixed language $\Lang$, we begin by assuming the existence of a constant-round BBQZK protocol $\langle \Prover, \Verifier \rangle$ in the non-programmable EPR model, where the prover's messages are classical. We then proceed to construct a malicious verifier $\tilde{\Verifier}$, who behaves identically to the honest $\Verifier$ except for the fact that $\tilde{\Verifier}$ aborts at each round with a carefully chosen probability of $1-\epsilon$. To elaborate further, $\tilde{\Verifier}$ employs a random oracle $H_\epsilon$ that outputs $1$ with probability $\epsilon$; at each round, $\tilde{\Verifier}$ continues the execution only if $H_\epsilon$ outputs $1$ on input the prover's messages $\tilde{\Verifier}$ received thus far.

Since the protocol $\langle \Prover, \Verifier \rangle$ is zero-knowledge, there must exist a black-box simulator $\Sim$ that, when given oracle access to $\tilde{\Verifier}$, is able to simulate the output of $\tilde{\Verifier}$ in the real execution. We then demonstrate that the execution $\Sim^{\tilde{\Verifier}}$ can be converted into a bounded-error quantum polynomial-time decider $\mcal{B}$ for the language $\Lang$, thereby establishing that $\Lang \in \BQP$.

We make two important remarks regarding the above procedure:
\begin{enumerate}
\item \label{rmk:verifier-efficiency}
{\bf On the Efficiency of $\tilde{\Verifier}$.} Note that the final (QPT) decider $\mcal{B}$ is constructed from $\Sim^{\tilde{\Verifier}}$. However, the $\tilde{\Verifier}$ described above is not QPT yet, because the random oracle $H_\epsilon$ does not have a polynomial-size description.

We can address this issue in exactly the same manner as \cite{FOCS:CCLY21}. Specifically, the oracle $H_\epsilon$ can be replaced with a $2q$-wise independent hash function, where $q$ is the number of queries made by the original $\Sim^{\tilde{\Verifier}}$ machine to $H_\epsilon$. It then follows from \Cref{lem:ind-hash} that the constructed $\BQP$ decider $\msf{B}$ will function just as effectively. In the subsequent discussion, we will primarily focus on the scenario where $\tilde{\Verifier}$ uses $H_\epsilon$. For the replacement of $H_\epsilon$ with a $2q$-wise independent hash, we refer to \Cref{sec:expected_sim_efficient_verifier} for more details.

\item \label{rmk:expected-QPT}
{\bf On the Expected QPT Running Time of $\Sim$.} Notice that the simulator in \Cref{def:BBQZK} runs in {\em expected} QPT time. Consequently, our impossibility results shown in \Cref{thm:impossibility:QZK} (and \Cref{thm:impossibility:NPE}) aim to rule out expected QPT simulation. However, the above paradigm does not work if $\Sim$ runs in {\em expected} QPT, as it would result in the decider $\mcal{B}$ running in expected QPT, which does not meet the efficiency requirement (i.e., strictly QPT) for being a valid $\BQP$ decider.

Once again, we resolve this issue in the same manner as \cite{FOCS:CCLY21}. Specifically, we first utilize the above paradigm to demonstrate the impossibility for {\em strictly} QPT simulation only. Then, we extend this impossibility to expected QPT simulation using the identical argument as presented in \cite[Section 3.3]{FOCS:CCLY21}. For a formal treatment of this issue, we refer to \Cref{sec:expected_sim_efficient_verifier}. 
\end{enumerate}
We now proceed to the formal proof of \Cref{thm:impossibility:NPE}. Given the complexity of this proof and its fulfillment across several sections, we provide an overview of the contents of the related sections:
\begin{itemize}
\item 

In \Cref{sec:particular-mal-V}, we define the malicious $\tilde{\Verifier}$; in \Cref{sec:BQP-decider}, we define the $\BQP$ decider $\mcal{B}$. It is important to note that the $\tilde{\Verifier}$ and $\mcal{B}$ defined so far are subject to the efficiency issues mentioned in \Cref{rmk:verifier-efficiency,rmk:expected-QPT}. However, we will proceed with the proof while ignoring these issues, as they will be addressed in \Cref{sec:expected_sim_efficient_verifier} as explained above.

\item 

We next prove that $\mcal{B}$ is indeed a valid $\BQP$ decider. This requires us to demonstrate:

\begin{itemize}
\item {\bf Completeness:} On input $x \in \Lang$, $\mcal{B}(x)$ accepts with $1/\poly(\secpar)$ probability. This is established in \Cref{sec:CPM:decider:lem:completeness:proof}.

\item {\bf Soundness:} On input $x \notin \Lang$, $\mcal{B}(x)$ accepts with $\negl(\secpar)$ probability. This proof is intricate. In \Cref{sec:CPM:decider:lem:soundness:proof}, we complete the proof for soundness assuming the important technical lemma \Cref{CPM:MnR:game:dummy:lem}. Then, we conclude the proof of \Cref{CPM:MnR:game:dummy:lem} in \Cref{sec:relate-Dummy-Real,sec:CMP:warm-up,sec:CMP:full}.
\end{itemize}

\item 
In \Cref{sec:expected_sim_efficient_verifier}, we address the efficiency issues mentioned in \Cref{rmk:verifier-efficiency,rmk:expected-QPT}. This completes the proof of \Cref{thm:impossibility:NPE}.
\end{itemize}

\subsection{A Malicious Verifier}
\label{sec:particular-mal-V}

We start by considering a language $\Lang$. For this $\Lang$, assume that there exists a $K$-round black-box quantum zero-knowledge argument $\langle \Prover, \Verifier \rangle$ in the non-programmable EPR model as per \Cref{def:NPE-QZK}, where all the prover's message are classical. 

Let us recall that structure of the protocol as defined in \Cref{sec:def:QZK:NPE}. The protocol consists $2K$ messages 
$$
(p_1, v_1, p_2, v_2, \ldots, p_K, v_K).
$$
That is, the protocol has $K$ rounds in total. In round $k \in [K]$, $\Prover$ sends message $p_k$ and $\Verifier$ responds with $v_k$. Since we assumed that the prover's messages are classical, all the messages $(p_1, \ldots, p_k)$ are classical strings; but the messages $(v_1, \ldots, v_K)$ could be quantum. W.l.o.g., we assume that all these massages are of the same length $\ell(\secpar)$, which is a polynomial in $\secpar$.

We now proceed to define the particular malicious verifier $\tilde{\Verifier}$.

\subsubsection{$\tilde{\Verifier}$'s registers}
\label{sec:CPM:Vtil:registers} 
$\tilde{\Verifier}$ starts by initializing the following registers:
\begin{equation}\label[Expression]{CMP:V:inital:expr}
\ket{x}_{\reg{ins}} 
\ket{0}_{\reg{gc}}
\ket{0}_{\reg{lc}} 
\ket{\vb{0}}_{\reg{p_1}}
\ldots 
\ket{\vb{0}}_{\reg{p_K}}
\ket{\vb{0}}_{\reg{v_1}}
\ldots 
\ket{\vb{0}}_{\reg{v_K}}
\ket{\vb{0}}_{\reg{m}} 
\ket{\bot}_{\reg{t_1}}
\ldots 
\ket{\bot}_{\reg{t_K}} 
\ket{e^{(1)}_2, \ldots ,e^{(n)}_2, \vb{0}}_\reg{w} 
\ket{H}_{\reg{aux}} 
,\end{equation}
where the meaning of each register is explained below:
\begin{itemize}
\item
Register $\reg{ins}$ stores the statement (or instance) $x$ of language $\Lang$, which is a common input to both the verifier and the prover.

\item 
Register $\reg{gc}$ is called the {\em global counter} register and register $\reg{lc}$ is called the {\em local counter} register. The functionality of these two registers will become clear later when we describe the behavior of $\tilde{\Verifier}$ in \Cref{sec:CPM:Vtil:unitaries}.

\item 
Register $\reg{p_1 \ldots p_K}$ (resp.\ $\reg{v_1 \ldots v_K}$) are to store all the prover's messages (resp.\ verifier's messages). The exact meaning of these registers will become clear later when we describe the behavior of $\tilde{\Verifier}$ in \Cref{sec:CPM:Vtil:unitaries}.

\item 
Register $\reg{m}$ is used to exchange messages between  the prover and the verifier. Again, the exact meaning of $\reg{m}$ will become clear later when we describe the behavior of $\tilde{\Verifier}$ in \Cref{sec:CPM:Vtil:unitaries}.

\item 
Register $\reg{t_1 \ldots t_k}$ will be used when the verifier decides to abort in a particular rounds. See \Cref{sec:CPM:Vtil:unitaries} for how these registers are used.

\item
Register $\reg{w}$ is $\tilde{\Verifier}$'s working space. It contains suffices $0$ states that will be used as ancilla qubits during the computation. Also, recall that we are in the non-programmable EPR model, which means that the verifier gets the second halves of $n$ EPR pairs $(e^{(1)}_2, \ldots, e^{(n)}_2)$. We choose to put then in the $\reg{w}$ register as well. We remark that the subsequent proof does not need to refer to these EPR shares explicitly.

\item 
Register $\reg{aux}$ stores the truth table of a function $H: \mcal{M}^{\le K} \ra \bits$, where $\mcal{M} = \bits^\ell$ (note that each $p_k$ is a classical string of length $\ell$). This $H$ comes from certain distributions that we will specify later in our proof. Here, we remark that this $H$ is the only inefficient part of $\tilde{\Verifier}$, and we will eventually replace it with a $2q$-wise independent hash function as mentioned in \Cref{rmk:verifier-efficiency}. 
\end{itemize}

\subsubsection{$\tilde{V}$'s Unitary.} 
\label{sec:CPM:Vtil:unitaries}

In this part, we define the malicious verifier $\tilde{\Verifier}$ by specifying her unitary $\tilde{V}$. We present $\tilde{V}$ in two equivalent manners. The first description (shown in \Cref{CPM:V:unitary}) is formulated to help the reader understand the behavior of $\tilde{\Verifier}$ at the operational level. However, it is initially unclear whether the description in \Cref{CPM:V:unitary} can be implemented as a unitary circuit. Particularly, the $\tilde{V}$ in \Cref{CPM:V:unitary}, at first glance, appears to utilize controlled gates on the function $H$. This poses a problem because in later parts of the proof, we need to regard $H$ as a quantum oracle. (It is not known whether one can implement controlled gates on $H$ in the model where $H$ is provided as an oracle.)

Therefore, in \Cref{CPM:V:unitary:implementation}, we provide a functionally equivalent description of \Cref{CPM:V:unitary}. It is evident from \Cref{CPM:V:unitary:implementation} that the $\tilde{V}$ defined in \Cref{CPM:V:unitary} can indeed be implemented as a unitary that utilizes $H$ as a quantum oracle, without the need for any controlled gates on it.

In subsequent parts of the proofs, we primarily utilize \Cref{CPM:V:unitary} as it illustrates the operational meaning more effectively. However, we may also reference \Cref{CPM:V:unitary:implementation} in certain instances where it aids in clarifying matters (e.g., in the proof of \Cref{counter-structure:lemma} and the proof of \cref{MnR:game:modify:1:z:non-devreasing:lemma}).

\para{Operationally Clear Description.} We start by defining two unitaries that works on the global counter and local counter registers defined above. 

Note that both $\reg{gc}$ and $\reg{lc}$ registers are initialized to $0$. We set $C = 2^\secpar$,\footnote{We remark that it suffices to set $C$ to any super-polynomial function on $\secpar$. We choose $2^\secpar$ only for concreteness.} and define the unitary $U_{gc}$ as the following mapping
 $$U_{gc} : 	\ket{i}_{\reg{gc}} \mapsto \ket{i+1 \mod C}_{\reg{gc}}.$$
 We also define the unitary to increase the local counter $U_{lc}$ as follows:
 $$U_{lc} : 	\ket{i}_{\reg{lc}} \mapsto \ket{i+1 \mod (K+1)}_{\reg{lc}}.$$
 We remark that modulus $(k+1)$ is good enough for us, because we will show later that the value of the local counter will stay in the set $\Set{0}\cup [K]$.

With the above notation in hand, we now describe $\tilde{\Verifier}$'s unitary $\tilde{V}$ in \Cref{CPM:V:unitary}. 
\begin{AlgorithmBox}[label={CPM:V:unitary}]{Unitary \textnormal{$\tilde{V}$} for the Malicious Verifier \textnormal{$\tilde{\Verifier}$}}

{\bf Helper Unitaries.} Before we described $\tilde{V}$, we first define some helper unitaries.  For each $k \in [K]$:
\begin{itemize}	
\item 
{\bf $A_k$: swap $\reg{p_k}$ and $\reg{m}$.} This unitary swaps the contents of $\reg{p_k}$ and $\reg{m}$.

\item 
{\bf $B_k$: apply $V$'s unitary.}
This unitary perform the following operation in superposition\footnote{Henceforth, when we refer to $(p_1, \ldots, p_k)$, we mean the values stored in registers $\reg{p_1 \ldots p_k}$.}:
\begin{itemize}
\item
If $H(p_1, \ldots, p_k) = 1$, then apply the honest verifier's unitary $V$ to generate message $v_k$. We remark that the generated $v_k$ is now stored in register $\reg{v_k}$.
\item
If $H(p_1, \ldots, p_k) = 0$, then do nothing; 
\end{itemize}

\item 
{\bf $C_k$: deliver $v_k$.} 
This unitary perform the following operation in superposition:
\begin{itemize}
\item
If $H(p_1, \ldots, p_k) = 1$, then swap the contents of registers $\reg{m}$ and $\reg{v_k}$.
\item
If $H(p_1, \ldots, p_k) = 0$, then swap the contents of registers $\reg{m}$ and $\reg{t_k}$.
\end{itemize}

\item 
{\bf $C'_k$: swap $\reg{m}$ and $\reg{t_k}$.} 
This unitary is the swap operator between registers $\reg{m}$ and $\reg{t_k}$.

\item 
{\bf $D_k$: increase local counter.} 
This unitary perform the following operation in superposition:
\begin{itemize}
\item
If $H(p_1, \ldots, p_k) = 1$, then apply the unitary $U_{lc}$ to increase the local counter by 1. 
\item
If $H(p_1, \ldots, p_k) = 0$, then do nothing.
\end{itemize}
\end{itemize}

\para{$\tilde{\Verifier}$'s Unitary $\tilde{V}$.} On each query, it compares the global counter value $k$ with the local counter value $j$ and behaves accordingly. In particular:
\begin{description}
\item[Case $k = j$:] It first applies the unitary $U_{gc}$ to increase the global counter by 1, and then behaves according to the (increased) global counter value $k+1$. In particular:
\begin{itemize}
\item
If $k+1 \in [K]$, then it applies $D_{k+1}C_{k+1}B_{k+1}A_{k+1}$ as defined above.
\item
Otherwise (i.e., $k+1 \notin[K]$), it does nothing (i.e., applies the identity operator).
\end{itemize}

\item[Case $(k \ne j)$:] It first applies the unitary $U_{gc}$ to increase the global counter by 1, and then behaves according to the (increased) global counter value $k+1$. In particular:
\begin{itemize}
\item
If $k+1 \in [K]$, then it applies $C'_{k+1}A_{k+1}$ as defined above.
\item
Otherwise (i.e., $k+1 \notin[K]$), it does nothing (i.e., applies the identity operator).
\end{itemize}
\end{description}

\para{$\tilde{\Verifier}$'s Output:} It output all the registers. Notation-wise, we write the final output as $\big(\vb{p} = (p_1, \ldots, p_K), \rho\big)$, where $p_1, \ldots, p_K$ is the contents in register $\reg{p_1 \ldots p_K}$ and $\rho$ is the state over the remaining registers at halt.	
\end{AlgorithmBox}
\para{Notation.} We will make use of the following notations. We write $\tilde{\Verifier}^{H}$ to refer to the malicious verifier $\tilde{\Verifier}$ when the function in register $\reg{aux}$ is instantiated by $H$ (see \Cref{CMP:V:inital:expr}). We use $\langle \Prover(w), \tilde{\Verifier}^{H}\rangle(x)$ to denote the execution of the protocol between $\tilde{\Verifier}^{H}$ and the honest prover $\Prover$, where the comment input is $x$ and the honest prover holds a witness $w$ as its private input. We use
$
(\vb{p}, \rho) \la \msf{OUT}_{\tilde{\Verifier}}\langle \Prover(w), \tilde{\Verifier}^{H} \rangle (x)
$ 
to denote the final output of $\tilde{\Verifier}$.

\begin{remark}[Omitting EPR Shares]\label{rmk:omit-EPR}
Notice that we are currently in the non-programmable EPR model, and thus $\Prover$ (resp.\ $\tilde{\Verifier}$) also takes the EPR shares $(e^{(1)}_1, \ldots, e^{(n)}_1)$ (resp.\ $(e^{(1)}_2, \ldots, e^{(n)}_2)$) as a part of its input. Henceforth, we omit these EPR pairs in our notation for succinctness, e.g., in $\langle \Prover(w), \tilde{\Verifier}^{H}\rangle(x)$. This is fine because our proof never needs to explicitly refer to these EPR shares.
\end{remark}

\para{On Implementing $\tilde{V}$ as an Unitary.} As discussed at the beginning of \Cref{sec:CPM:Vtil:unitaries} , we now show in \Cref{CPM:V:unitary:implementation} that the $\tilde{V}$ defined in \Cref{CPM:V:unitary} can indeed be implemented as a unitary with quantum-oracle access to $H$. 
\begin{AlgorithmBox}[label={CPM:V:unitary:implementation}]{Unitary Implementation of $\tilde{V}$}
At each round $k \in [K]$, it behaves in the following three steps:

\para{Step 1:} It applies $U_{gc}$ to increase the global counter from $k-1$ to $k$. (Recall that right before round $k$ starts, the global counter is $k-1$.) 

\para{Step 2:} It behaves according to the current (i.e., already increased) global counter value $k$:
\begin{enumerate}
	\item
If $k \in [K]$, then it applies the unitary $A_k$ as defined in \Cref{CPM:V:unitary};
\item
Otherwise (i.e., $k \notin [K]$), it does not do anything (i.e., it applies the identity operator).
\end{enumerate}

\para{Step 3:} It queries the (quantum) oracle $H$ to learn the value $H(p_1, \ldots, p_{j+1})$ and store this value in a temporary register $\reg{tmp}$. We note that $j$ is the value of the current local counter and $(p_1, \ldots, p_{j+1})$ are the values stored in registers $\reg{p_1 \ldots p_{j+1}}$. Also note that in later proofs, the local counter register may contain a superposition of values; that is why we need {\em quantum} oracle access to $H$.

\para{Step 4:} It behaves according to the current (i.e., already increased) global counter value $k$:
\begin{enumerate}
\item
If $k\in [K]$, then it behaves by comparing the local counter value $j$ with $k-1$ (recall again that $k-1$ is the global counter value at the beginning of round $k$, before the $U_{gc}$ at the very beginning is applied): 
\begin{enumerate}
\item 
If $(k-1) = j$, then it applies the $D_kC_kB_k$ defined in \Cref{CPM:V:unitary}. Note that these operators make use of the value $H(p_1, \ldots, p_k)$ as the control (qu)bit. But this is exactly the value stored in register $\reg{tmp}$ defined in {\bf Step 3} (note that in this case $j+1$ is exactly $k$). Thus, $D_kC_kB_k$ can be implemented using $\reg{tmp}$ as the control register.
\item 
Otherwise (i.e., $(k-1) \ne j$), it applies the $C'_k$ defined in \Cref{CPM:V:unitary}.
\end{enumerate}

\item 
Otherwise (i.e., $k \notin [K]$), it does not do anything (i.e., it applies the identity operator)
\end{enumerate}
\end{AlgorithmBox}
The equivalence between \Cref{CPM:V:unitary} and \Cref{CPM:V:unitary:implementation} can be easily seen by comparing their respective descriptions. Moreover, it is evident from the description of \Cref{CPM:V:unitary:implementation} that it can be implemented as a unitary circuit with quantum oracle access to $H$ --- particularly, the query to $H$ is elevated to {\bf Step 3} and is not controlled by any register.

\subsubsection{Understanding $\tilde{\Verifier}$ with Two examples} 
To gain a better understanding of the unitary $\tilde{V}$ defined in \Cref{CPM:V:unitary}, let us consider two examples where we instantiate the function $H$ differently. We recommend that the reader not skip these examples, as we will introduce important notations that will be utilized in the subsequent proofs as well.

\para{The First Example.} In this example, consider a function $H$ that always output $1$ on all input. For this such an $H$, it is not hard to see that the execution $\langle \Prover(w), \tilde{\Verifier}^{H}\rangle(x)$ is effectively identical to the  $\langle \Prover(w), \Verifier\rangle(x)$ (i.e., the real protocol between honest prover and verifier)---in each round $k$, only the branch corresponding to $H(p_1, \ldots, p_k)=1$ would happen; if one track the execution, it is easy to see that only the work performed by $B_k$ actually matters (while $A_k$, $C_k$, and $D_k$ are there to deliver messages or maintain the counters), and $B_k$ is nothing but emulating the honest verifier $\Verifier$ (in the $H(p_1, \ldots, p_k)=1$ branch). Therefore, the execution in this case is indeed a perfect emulation of the real $\langle \Prover(w), \Verifier\rangle(x)$. In particular,  $\tilde{\Verifier}^{H}$'s final output will be accepted by $\msf{Acc}$ with the same probability $1-\negl(\secpar)$ as in the honest execution (see the {\bf Completeness} requirement in  \Cref{def:NPE-QZK}). 

Let us now turn to the local counter register. In this example, it is also straightforward that the local counter register will contain the a classical value $K$ at the end of the execution $\langle \Prover(w), \tilde{\Verifier}^{H}\rangle(x)$. This is simply because the local counter will increase by 1 at the end of each round (due to the unitary $D_k$), and there are $K$ rounds in total.	

Moreover, we also know that the value $(p_1, \ldots, p_K)$ contained in the $\vb{p}$ part of the final output of $\tilde{\Verifier}^{H}$ will be accepted by $H$. In particular, it holds that $\bar{H}(\vb{p}) = \vb{1}$, where $\vb{1}$ is the vector containing $K$ repetition of value 1 and $\bar{H}$ is the vector-valued version of $H$ as defined in the following \Cref{def:vector-function}.
\begin{definition}[Vector-Valued Function]\label{def:vector-function}
Let $\mcal{M} \coloneqq \bits^\ell$. Let $H$ be a function from $\mcal{M}^{\le K}$ to $\bits$. Then, we define the vector-version $\bar{H}$ as follows:
$$\forall \vb{p} = (p_1, \ldots, p_K) \in \mcal{M}^K, ~~\bar{H}(\vb{p}) \coloneqq  \big(H(p_1), H(p_1, p_2), \ldots, H(p_1, \ldots, p_K) \big).$$
\end{definition}
The above observations for this example can be summarized as follows. For any $x \in \Lang$ and any $w \in \Relation_\Lang(x)$, it holds that
\begin{align*}
& \Pr[\msf{Pred}(\rho) = 1 \wedge \bar{H}(\vb{p}) = \vb{1} ~:~ 
(\vb{p}, \rho) \la \msf{OUT}_{\tilde{\Verifier}}\langle \Prover(w), \tilde{\Verifier}^{H} \rangle (x)] \\  	
 = ~&
\Pr[\msf{Acc}(\rho) = 1 ~:~ 
\rho \la \msf{OUT}_{{\Verifier}}\langle \Prover(w), {\Verifier} \rangle (x)
]  \\ 
 = ~& 
 1 - \negl(\secpar),
\end{align*}
where the $\msf{Pred}(\cdot)$ is a quantum predicate defined in the following \Cref{def:predicte:Pred}.
\begin{definition}[Predicate $\msf{Pred}$]\label{def:predicte:Pred}
We define $\msf{Pred}(\cdot)$ as a quantum predicate over the second part (i.e., the $\rho$) in the output of $\tilde{\Verifier}$. In particular, $\msf{Pred}(\rho) = 1$ if and only if the following conditions hold:
\begin{enumerate}
\item
Measure the local counter register $\reg{lc}$ in the computational basis and the measurement outcome is $K$. (Recall that the $\reg{lc}$ register is contained in $\rho$.)
\item
Measure register $\reg{d}$ in the computational basis and the measurement outcome is 1. (Recall that the $\reg{d}$ register is a designated register in $\reg{w}$ containing the verifier's final decision qubit (see also \Cref{rmk:msg:vk}). It is contained in $\rho$ because $\reg{w}$ is so.)
\end{enumerate}
\end{definition}

\para{The Second Example.} A more interesting example is when $H$ is sampled from a family of random functions that output 1 with probability $\epsilon$. Formally, we define the following family of functions.
\begin{definition}[Family of $\epsilon$-Random Functions]\label{def:H-epsilon}
Let $\mcal{M} \coloneqq \bits^\ell$. For a real number $\epsilon \in [0,1]$ and a positive integer $K$, let $\mcal{H}_{\epsilon, K}$ be a distribution over $H_{\epsilon, K}: \mcal{M}^{\le K} \ra \bits$ such that we have 
$$\Pr[H_{\epsilon, K}(p_1, \ldots, p_k) = 1] = \epsilon$$ 
independently for each $(p_1, \ldots, p_k) \in \mcal{M}^{\le K}$. (In the sequel, the value $K$ is always fixed to the number of rounds. Thus, we omit the $K$ from the subscript of $H_{\epsilon, K}$ and simply write is as $H_\epsilon$.)
\end{definition}
Consider a $H_\epsilon$ that sampled from the $\mcal{H}_\epsilon$ defined in \Cref{def:H-epsilon}. Let us compare the the execution $\langle \Prover(w), \tilde{\Verifier}^{H_\epsilon}\rangle(x)$ and the honest execution  $\langle \Prover(w), \Verifier\rangle(x)$. Due to a similar argument as we did in the previous example with the all-accepting function $H$, it is easy to see that the execution $\langle \Prover(w), \tilde{\Verifier}^{H_\epsilon}\rangle(x)$, when conditioned on the fact that $H_\epsilon(p_1, \ldots, p_k) = 1$ for all $k \in [K]$, is identical to the real execution $\langle \Prover(w), \Verifier\rangle(x)$ between the honest parties. In this case (i.e., the conditioned fact happens), all the observations we made in the first example (regarding the accepting probability, the local counter, the predicate $\msf{Pred}$) remain valid.

On the other hand, it follows from \Cref{def:H-epsilon} that the event ``$H_\epsilon(p_1, \ldots, p_k) = 1$ for all $k \in [K]$'' happen with probability exactly $\epsilon^K$ over the randomly sampling of $H_\epsilon$. 

Therefore, the following holds: For any $x \in \Lang$ and any $w \in \Relation_\Lang(x)$, it holds that
\begin{align*}
& \Pr[\msf{Pred}(\rho) = 1 \wedge \bar{H}_\epsilon(\vb{p}) = \vb{1}  ~:~ 
\begin{array}{l}
H_{\epsilon} \la \mcal{H}_{\epsilon}\\
(\vb{p}, \rho) \la \msf{OUT}_{\tilde{\Verifier}}\langle \Prover(w), \tilde{\Verifier}^{H_{\epsilon}} \rangle (x)
\end{array}
]  	
\\
\ge~&  
\epsilon^K \cdot
\Pr[\msf{Acc}(\rho) = 1 ~:~ 
\rho \la \msf{OUT}_{{\Verifier}}\langle \Prover(w), {\Verifier} \rangle (x)
]  \\ 
=~& 
\epsilon^K \cdot
\big(1 - \negl(\secpar)\big) 
\\ 
 = ~&
 \epsilon^K - \negl(\secpar) \numberthis \label{eq:epsilon:real}
,\end{align*}  	
where $\bar{H}_\epsilon$ is the vector-valued version of $H_{\epsilon}$ as defined in \Cref{def:vector-function} and $\msf{Pred}(\cdot)$ is the quantum predicate defined in \Cref{def:predicte:Pred}. 

\subsubsection{Interaction between $\Sim$ and $\tilde{\Verifier}$} 

Recall from \Cref{sec:def:QZK:standard-model,sec:def:QZK:NPE} that the simulator $\Sim$ is granted oracle access to $\tilde{V}$ and $\tilde{V}^\dagger$. She does not get to see the internal registers of $\tilde{\Verifier}$ (except for register $\reg{m}$). She has a local operator $S$, and after each oracle query, her behavior consists of applying $S$ and then measuring a special $\reg{u}$ register to determine the type of the next query.

As for notations, for any function $H$, we use $\msf{SIM}^{H}(x)$ to refer to the execution of $\Sim^{\tilde{\Verifier}(x)}(x)$ where the function in $\tilde{\Verifier}$'s $\reg{aux}$ register  is instantiated by $H$. (Similar as in \Cref{rmk:omit-EPR}, we omit the EPR shares from our notation.) Also, recall from \Cref{sec:def:QZK:standard-model} that the output of the simulator is defined to be the output of the malicious verifier at the end of simulation. For the particular $\tilde{\Verifier}$ defined in \Cref{CPM:V:unitary}, the output of the simulation can be written as $(\vb{p}, \rho) \la \msf{SIM}^{H}(x)$, where $(\vb{p}, \rho)$ has the same meaning as defined toward the end of \Cref{CPM:V:unitary}.

With the above notations, let us know make three simply but useful claims.
\begin{MyClaim}\label{claim:sim-Vtil-prob}
For any $x \in \Lang$, it holds that
$$\Pr[\msf{Pred}(\rho) = 1 \wedge \bar{H}_\epsilon(\vb{p}) = \vb{1}  ~:~ 
\begin{array}{l}
H_{\epsilon} \la \mcal{H}_{\epsilon}\\
(\vb{p}, \rho) \la \msf{SIM}^{H_\epsilon}(x)
\end{array}
]  \ge \frac{\epsilon^K}{4} -\negl(\secpar).$$ 
\end{MyClaim}
\begin{proof}[Proof of \Cref{claim:sim-Vtil-prob}]
This claim simply follows from the ZK property of the protocol and \Cref{eq:epsilon:real}.

In more detail,
\begin{align*}
& \Pr[\msf{Pred}(\rho) = 1 \wedge \bar{H}_\epsilon(\vb{p}) = \vb{1}  ~:~ 
\begin{array}{l}
H_{\epsilon} \la \mcal{H}_{\epsilon}\\
(\vb{p}, \rho) \la \msf{SIM}^{H_\epsilon}(x)
\end{array}
] 
\\
\ge ~ & 
\Pr[\msf{Pred}(\rho) = 1 \wedge \bar{H}_\epsilon(\vb{p}) = \vb{1}  ~:~ 
\begin{array}{l}
H_{\epsilon} \la \mcal{H}_{\epsilon}\\
(\vb{p}, \rho) \la \msf{OUT}_{\tilde{\Verifier}}\langle \Prover(w), \tilde{\Verifier}^{H_{\epsilon}} \rangle (x)
\end{array}
]  	 
- \negl(\secpar)
\numberthis  \label{claim:sim-Vtil-prob:proof:eq:1} \\ 
 \ge ~ & 
\epsilon^K - \negl(\secpar)
\numberthis  \label[Inequality]{claim:sim-Vtil-prob:proof:eq:2}
,\end{align*}
where \Cref{claim:sim-Vtil-prob:proof:eq:1} follows from the ZK property of the $\langle \Prover, \Verifier \rangle$ protocol, and \Cref{claim:sim-Vtil-prob:proof:eq:2} follows form \Cref{eq:epsilon:real}.

Then, \Cref{claim:sim-Vtil-prob} simply follows from \Cref{claim:sim-Vtil-prob:proof:eq:2} and the fact that $\epsilon^K > \frac{\epsilon^K}{4}$.

\end{proof}
\begin{remark}[On Tightness of \Cref{claim:sim-Vtil-prob}]
One may wonder why we state \Cref{claim:sim-Vtil-prob} with a lower bound of $\frac{\epsilon^K}{4} - \negl(\secpar)$, given that the proof of \Cref{claim:sim-Vtil-prob} already achieves a tighter lower bound of $\epsilon^K - \negl(\secpar)$ (i.e., \Cref{claim:sim-Vtil-prob:proof:eq:2}). This is because \Cref{claim:sim-Vtil-prob} serves two purposes in the later part of our proof:
\begin{enumerate}
\item 
First, \Cref{claim:sim-Vtil-prob} will be used when establishing the completeness property of the $\BQP$ decider (in \Cref{CPM:decider:lem:completeness}). For this purpose, we can indeed use \Cref{claim:sim-Vtil-prob} with the tighter lower bound of $\epsilon^K - \negl(\secpar)$.

\item
Second, \Cref{claim:sim-Vtil-prob} will also be used when extending our impossibility results from strictly QPT simulation to {\em expected} QPT simulation in \Cref{sec:expected_sim_efficient_verifier}. 
For this purpose, the tighter lower bound of $\epsilon^K - \negl(\secpar)$ does not suffice, and we have to use the bound of $\frac{\epsilon^K}{4} - \negl(\secpar)$.
\end{enumerate}
Therefore, we choose to state \Cref{claim:sim-Vtil-prob} with the lower bound of $\frac{\epsilon^K}{4} - \negl(\secpar)$.
\end{remark}

\begin{MyClaim}[Classical Global Counter]\label{item:fact:gc}
For any function $H$, throughout the execution of $\msf{SIM}^H(x)$, the global counter register will always contain a classical value.
\end{MyClaim}
\begin{proof}[Proof of \Cref{item:fact:gc}]
Note that $\Sim$ can only modify the content in the global counter register through invoking her oracle $\tilde{V}$ or $\tilde{V}^\dagger$. Then, \Cref{item:fact:gc} can be easily seen from the definition of $\tilde{V}$ in \Cref{CPM:V:unitary}---every invocation of $\tilde{V}$ (resp. $\tilde{V}^\dagger$) will trigger $U_{gc}$ (resp. $U^\dagger_{gc}$) exactly once, increasing (resp. decreasing) the global counter by 1. Thus, the global counter will always contain a classical value, decoherent from other registers. 

This completes the proof of \Cref{item:fact:gc}.

\end{proof}

\begin{MyClaim}\label{number-of-H-queries}
For any function $H$, the execution $\msf{SIM}^{H}(x)$ can be seen as an oracle-aided algorithm where $H$ plays the role of the oracle. If the simulator $\Sim$ makes $q$ quieres to her oracle $\tilde{V}$ and $\tilde{V}^\dagger$ in total, then the oracle $H$ will be queries at most $2Kq$ times.	
\end{MyClaim}
\begin{proof}[Proof of \Cref{number-of-H-queries}]
First, note that $\msf{SIM}^{H}(x)$ can indeed be seen as an oracle-aided algorithm where $H$ plays the role of the oracle. This follows from the description of $\tilde{\Verifier}$, who only makes use of the I/O behavior of $H$.

As for the number of queries, it suffices to show that each query of $\Sim$ (to her oracle $\tilde{V}$ or $\tilde{V}^\dagger$) will invoke at most $2K$ queries to $H$. This is true by definition of $\tilde{V}$ \Cref{CPM:V:unitary}: From the descritpion there, unitaries $B_k$, $C_k$, and $D_k$ all makes $H$ queries. However, we remark that this can be done by query the $H$ once and store the output of $H$ in a seperate register that can be used (as a control register) by $B_k$, $C_k$, and $D_k$. In the manner, each $\tilde{V}$ or $\tilde{V}^\dagger$ query only invoke 2 quries to $H$ (note that we need one more query to uncompute each query to $H$). Then, \Cref{number-of-H-queries} follows because $2q \le 2Kq$ for all positive interger $K$.

This completes the proof of \Cref{number-of-H-queries}.

\end{proof}

\begin{remark}[On Tightness of \Cref{number-of-H-queries}]
The above proof of \Cref{number-of-H-queries} indeed establishes a tighter upper bound on the number of queries --- it shows that $\msf{SIM}^{H}$ can make at most $2q$ queries to $H$. One may wonder why we choose to claim the upper bound as $2Kq$ in the statement of \Cref{number-of-H-queries}. Indeed, our \Cref{number-of-H-queries} can be treated as the analogue of {\bf Observation 3} on Page 20 of \cite{FOCS:CCLY21}, where the upper bound is $2Kq$. We choose to use the same bound because this allows us to reuse the same parameter settings (and certain calculations) in \cite{FOCS:CCLY21}.

One may also wonder why we can save the multiplicative factor $K$ in this bound while \cite{FOCS:CCLY21} cannot in their {\bf Observation 3}. This is due to a difference in the behavior of the malicious verifier at the last round. In our case, the malicious verifier $\tilde{\Verifier}$'s behavior at the last round is syntactically the same as other rounds. In particular, she only queries $H$ once as we argued in the above proof of \Cref{number-of-H-queries} (plus one more query for uncomputing). In contrast, the malicious verifier in \cite{FOCS:CCLY21} behaves differently at the last round. Specifically, at the last round, the \cite{FOCS:CCLY21} verifier will invoke $H$ for $K$ times to learn $H(m_1, \ldots, m_i)$ for all $i \in [K]$ and check if all these values are 1 (see the top of Page 20 of \cite{FOCS:CCLY21}). We remark that our $\tilde{\Verifier}$ does not need to perform such checks at the last round. This is because (looking ahead) our proof will perform a fine-grained analysis of the measure-and-reprogram game with respect to $\msf{SIM}^H$ to establish certain properties of the prover's messages $(p_1, \ldots, p_K)$, which will effectively achieve the same effect as the checks performed by the \cite{FOCS:CCLY21} verifier at the last round.

\end{remark}

\subsection{$\BQP$ Decider}
\label{sec:BQP-decider}

In this section, we define the $\BQP$ decider for the language $\Lang$. 


First, let us take a closer look at the execution of $\msf{SIM}^H$. As we established in \Cref{number-of-H-queries}, $\msf{SIM}^H$ can be treated as an oracle-aided execution, where $H$ plays the role of the oracle. Our first step toward constructing the $\BQP$ decider $\mcal{B}$ is to put $\msf{SIM}^H$ into the format of a measure-and-reprogram (MnR) game, as discussed in \Cref{sec:MnR}.

\para{MnR Game with ZK Simulator.} We first define the MnR game w.r.t.\ the simulator $\Sim$ and $\tilde{\Verifier}(x)$ defined in previous sections. This game is simply an instantiation of MnR game defined in \Cref{MnR:lem:game} with the following notation: 
\begin{enumerate}
\item \label{CPM:MnR:notation:item:1}
We treat the execution of the simulator (with oracle access to $\tilde{V}$ and $\tilde{V}^\dagger$) as an oracle machine $\msf{SIM}^H$ that has quantum oracle access to a random function $H: \mcal{X}^{\le K} \rightarrow \mcal{Y}$, where $\mcal{X}=\bits^\ell$ and $\mcal{Y} = \bits$. Let $\vb{y} = (y_1, \ldots, y_K) \in \mcal{Y}^K$.

\item \label{CPM:MnR:notation:item:2}
Recall that the output of  $\msf{SIM}^H$ 
is defined to be the output $\tilde{\Verifier}$ at halt, which is splitted as $(\vb{p}, \rho)$ (see also \Cref{CPM:V:unitary}).



\item \label{CPM:MnR:notation:item:3}
Recall from \Cref{number-of-H-queries} that $\msf{SIM}$ as an oracle machine makes at most $2Kq$ queries to $H$ in total.
\end{enumerate}

Since this game will be particularly important for the sequel sections, we choose to present it in full detail in \Cref{CPM:MnR:game}.

\begin{GameBox}[label={CPM:MnR:game}]{Measure-and-Reprogram Game \textnormal{$\Sim^{\tilde{\Verifier}}[H, \vb{y}](x)$}}
 It works as follows
\begin{enumerate}
\item \label[Step]{CPM:MnR:game:step:1}
For each $i\in[K]$, uniformly pick $(j_i,b_i)\in ([2Kq]\times \bits) \cup \{(\bot,\bot)\}$  conditioned on that there exists at most one $i\in [K]$ such that $j_i=j^*$  for all $j^*\in [2Kq]$.

\item \label[Step]{CPM:MnR:game:step:2}
    Let $s$ denote the number of $j_i$'s in $\Set{j_i}_{i \in [K]}$ that are not $\bot$.  We re-label the indices $i$ for pairs $\Set{(j_i, b_i)}_{i \in [K]}$ so that $j_1 < j_2 < \ldots < j_s$ and $j_{s+1} = j_{s+2} =\ldots = j_K = \bot$.

\item \label[Step]{CPM:MnR:game:step:3}
Run the oracle machine $\msf{SIM}(x)$ with oracle $\mcal{O}$, which is initialized to be a quantumly-accessible classical oracle that computes $H$, and when $\msf{SIM}(x)$ makes its $j$-th query, the oracle is simulated as follows:
\begin{enumerate}
      \item If $j=j_i$ for some $i\in[K]$, measure this query to obtain $\vb{p}_i=(p_{i,1}, \ldots,p_{i, z_i})$ where $z_i \in [K]$ is determined by the measurement outcome, and then behaves according to the value $b_i$ as follows:
        \begin{enumerate}
        \item 
        If $b_i=0$: First reprogram $\mcal{O} \leftarrow \reprogram(\mcal{O},\vb{p}_i,y_i)$, and then answer the $j_i$-th query using the reprogrammed oracle. 
        \item 
        If $b_i=1$: First answer the $j_i$-th query  using the oracle before the reprogramming, and then reprogram $\mcal{O} \leftarrow \reprogram(\mcal{O},\vb{p}_i,y_i)$. 
        \end{enumerate}
    \item Otherwise (i.e., $j \ne j_i$ $\forall i \in [s]$), answer the $j$-th query just using the oracle $\mcal{O}$ without any measurement or  reprogramming. 
    \end{enumerate}

\item
Let $(\vb{p}^*= (p^*_1, \ldots, p^*_K), \rho_{\msf{out}})$ be the output of $\Sim^{\tilde{\Verifier}}$ at halt (as per \Cref{CPM:MnR:notation:item:2}).

\item
For all $i \in \Set{s+1, s+2, \ldots, K}$, set $\vb{p}_i = \vb{p}^*_i$ where $\vb{p}^*_i \coloneqq (p^*_1, \ldots, p^*_i)$. 

\item
{\bf Output:} The output of this game is defined as follows
\begin{itemize}
\item
If it holds for all $i \in [K]$ that $\vb{p}_i$ is a prefix of $\vb{p}_K$, then output $(\vb{p}_K, \rho_{\msf{out}})$;

\item
Otherwise, output $(\bot, \bot)$.
\end{itemize}
\end{enumerate}
\end{GameBox}

\para{Decider for $\BQP$.} With the above notation in hand, we proceed to define the $\BQP$ decider $\mcal{B}$. Roughly speaking, $\mcal{B}$ on input $x$ simply runs the MnR game $\Sim^{\tilde{\Verifier}}[H, \vb{y}](x)$ as defined in \Cref{CPM:MnR:game} by setting $H$ to an all-zero function and setting $\vb{y}$ to an all-1 vector. It accepts (resp.\ rejects) if the verifier $\tilde{\Verifier}$ accepts (resp.\ rejects) at the end of the MnR game.

The formal description of $\mcal{B}$ is presented in \Cref{CPM:BQP-decider}.
\begin{AlgorithmBox}[label={CPM:BQP-decider}]{\textnormal{$\BQP$} Decider $\mcal{B}(x)$}
On input a classical string $x$, machine $\mcal{B}(x)$ works as follows:
\begin{enumerate}
\item
{\bf (Parameter Setting.)}
Let $\vb{1} \coloneqq (1, \ldots, 1)$ containing $K$ copies of $1$. Let $\mcal{M} \coloneqq \bits^{\ell}$ and $H_0 :\mcal{M}^{\le K} \ra \bits$ be the zero-function, i.e., $H_0(p_1, \ldots, p_i) = 0$ for all $(p_1, \ldots, p_i) \in \mcal{M}^{\le K}$.

\item \label[Step]{CPM:BQP-decider:step:2}
{\bf (MnR Game with $\Sim^{\tilde{V}}$.)} Execute $(\vb{p}, \rho) \la \Sim^{\tilde{\Verifier}}[H_0, \vb{1}](x)$ as per \Cref{CPM:MnR:game}. 
 That is, $\mcal{B}(x)$ executes \Cref{CPM:MnR:game} with $H \coloneqq H_0$ and $\vb{y} \coloneqq (1, \ldots, 1)$ and  denotes the output of this procedure as $(\vb{p}, \rho)$.

\item
{\bf (Output.)} It outputs 1 if and only if $\msf{Pred}(\rho) = 1$, where $\msf{Pred}(\rho)$ is the predicate defined in \Cref{def:predicte:Pred}.

\end{enumerate}
\end{AlgorithmBox}

Next, we argue that $\mcal{B}$ is a valid $\BQP$ decider. Toward that, it suffices to establish the following \Cref{CPM:decider:lem:completeness,CPM:decider:lem:soundness}. We present the proof for \Cref{CPM:decider:lem:completeness} in \Cref{sec:CPM:decider:lem:completeness:proof}. The proof for \Cref{CPM:decider:lem:soundness} is the most technically involved part of this work. It will be covered in \Cref{sec:CPM:decider:lem:soundness:proof} and \Cref{sec:relate-Dummy-Real,sec:CMP:warm-up,sec:CMP:full}.


\begin{lemma}[Completeness]\label{CPM:decider:lem:completeness}
For all $\Lang \in \BQP$ and all $x \in \Lang \cap \bits^{\secpar}$, it holds for the $\mcal{B}$ defined in \Cref{CPM:BQP-decider} that
$$\Pr[\mcal{B}(x) = 1] \ge \frac{1}{8 \cdot (4Kq + 1)^{2K}} -\negl(\secpar),$$
where $q$ is the number of oracle queries made by the $\Sim$ (to her oracle $\tilde{V}$ or $\tilde{V}^\dagger$) during the execution of $\Sim^{\tilde{\Verifier}}[H_0, \vb{1}](x)$
in \Cref{CPM:BQP-decider:step:2} of $\mcal{B}(x)$ (i.e., \Cref{CPM:BQP-decider}).
\end{lemma}

\begin{lemma}[Soundness]\label{CPM:decider:lem:soundness}
For all $\Lang \in \BQP$ and all $x \in \bits^{\secpar} \setminus \Lang$, it holds for the $\mcal{B}$ defined in \Cref{CPM:BQP-decider} that
$$\Pr[\mcal{B}(x) = 1] = \negl(\secpar).$$
\end{lemma}

\subsection{Proof of Completeness}
\label{sec:CPM:decider:lem:completeness:proof}

In this subsection, we prove \Cref{CPM:decider:lem:completeness}.

We first set some parameters that will be used in this proof. Recall that $q$ denotes the number of oracle queries made by the $\Sim$ (to her oracle $\tilde{V}$ or $\tilde{V}^\dagger$) during the execution of $\Sim^{\tilde{\Verifier}}[H_0, \vb{1}](x)$
in \Cref{CPM:BQP-decider:step:2} of $\mcal{B}(x)$ (i.e., \Cref{CPM:BQP-decider}). For this $q$, we define 
\begin{equation}\label[Expression]{define:epsilon-q}
\epsilon_q \coloneqq \frac{1}{256K^2q^2(4Kq+1)^{2K}}.
\end{equation}
In the sequel, let $\epsilon \le \epsilon_q$ be an arbitrary noticeable function in $\secpar$.

\para{Notation.} We first recall some notations that have already been defined previously. They will be used in the sequel derivation:
\begin{itemize}
\item 
We will make use of the family of $\epsilon$-random functions $\mcal{H}_\epsilon$ defined in \Cref{def:H-epsilon}. Note that this $\epsilon$ is the particular noticeable $\epsilon \le \epsilon_q$ we fixed above. When a probability is taken over the random sampling of $H_\epsilon \la \mcal{H}_\epsilon$, we will not explicitly include this expression; instead, we simply write $H_\epsilon$ under the ``$\Pr$'' symbol to indicate the sampling procedure.

\item 
For a function $H_\epsilon$, we will make use of its vector-valued version  $\bar{H}_\epsilon$ as defined in \Cref{def:vector-function}.

\item 
We will make use of $\Sim^{\tilde{V}}[H_\epsilon, \vb{1}](x)$, which is the MnR game \Cref{CPM:MnR:game} instantiated with $H \coloneqq H_\epsilon$ and $\vb{y} \coloneqq \vb{1}$ containing $K$ repetitions of $1$'s.

\item 
When we write $\bs{\beta}$ in under the ``$\Pr$'' symbol, it means we sample $\bs{\beta}$ from a distribution $D_\epsilon$ over $\bits^K$, where each bit of $\bf{\beta}$ takes the value 1 with probability $\epsilon$ independently. 

\item
For a function $H$, we will make use of the symbol $\msf{SIM}^H$, which is the execution of $\Sim$ with oracle access to $\tilde{\Verifier}$ whose random functions is instantiated by $H$. We treat the $\Sim^{\tilde{\Verifier}}$ as an oracle-aided machine with $H$ playing the role of the oracle (as explained in \Cref{number-of-H-queries}).

\item
For $\vb{x} = (x_1, \ldots, x_K) \in \bits^{K\ell}$, $\vb{y} = (y_1, \ldots, y_K) \in \bits^K$, and a function $H: (\bits^\ell)^{\le K} \ra \bits$, recall the function $H^{\vb{x},\vb{y}}$ defined in \Cref{MnR:lem}: 
$$
  \forall i \in [K]~\forall \vb{x}'_i =(x'_1, \ldots, x'_i),~~  
  H^{\vb{x},\vb{y}}(\vb{x}'_i)
    \coloneqq
    \begin{cases} y_i   & \text{if~}  \vb{x}'_i=(x_1,...,x_i) \\  
     H(\vb{x}'_i)  & \text{otherwise}
    \end{cases}.
$$
\end{itemize}

With the above notations in hand, we now establish \Cref{CPM:decider:lem:completeness}.

First, notice that 
\begin{align}
\Pr[\mcal{B}(x) = 1] 
& = 
\Pr[\msf{Pred}(\rho) = 1 ~:~(\vb{p}, \rho) \la \Sim^{\tilde{V}}[H_0, \vb{1}](x)] \label{CPM:decider:lem:completeness:proof:sparsity:1} \\ 
& \ge 
\Pr_{H_\epsilon}[\msf{Pred}(\rho) = 1 ~:~(\vb{p}, \rho) \la \Sim^{\tilde{V}}[H_\epsilon, \vb{1}](x)] - 32K^2q^2\epsilon \label[Inequality]{CPM:decider:lem:completeness:proof:sparsity:2} 
\end{align}
where \Cref{CPM:decider:lem:completeness:proof:sparsity:1} follows from the definition and notation in \Cref{CPM:BQP-decider}, and \Cref{CPM:decider:lem:completeness:proof:sparsity:2} follows from \Cref{lem:ind-sparse-and-zero} and the fact that in the MnR game, there are $2Kq$ queries to the oracle $H_0$ (or $H_\epsilon$) in total (see \Cref{number-of-H-queries}).





Next, we derive a lower bound for the first term in the RHS of \Cref{CPM:decider:lem:completeness:proof:sparsity:2}:
\begin{align*}
& \Pr_{H_\epsilon}[\msf{Pred}(\rho) = 1 ~:~(\vb{p}, \rho) \la \Sim^{\tilde{V}}[H_\epsilon, \vb{1}](x)] \\  
=~&  
\sum_{\vb{p}^* \in \bits^{K\ell} \cup \Set{\bot}}   \Pr_{H_\epsilon}[\vb{p} = \vb{p}^* \wedge \msf{Pred}(\rho) = 1 ~:~(\vb{p}, \rho) \la \Sim^{\tilde{V}}[H_\epsilon, \vb{1}](x)]  
\numberthis \label{CPM:decider:lem:completeness:proof:final:1} \\ 
\ge~& 
\frac{1}{(4Kq+1)^{2K}} \cdot \sum_{\vb{p}^* \in \bits^{K\ell} \cup \Set{\bot}}  \Pr_{H_\epsilon}[\vb{p} = \vb{p}^* \wedge \msf{Pred}(\rho) = 1 ~:~(\vb{p}, \rho) \la \msf{SIM}^{H^{\vb{p}^*, \vb{1}}_\epsilon}(x)] \numberthis \label{CPM:decider:lem:completeness:proof:final:2} \\
=~& 
\frac{\epsilon^{-K}}{(4Kq+1)^{2K}} \cdot \sum_{\vb{p}^* \in \bits^{K\ell}  \cup \Set{\bot}}  \Pr_{H_\epsilon, \bs{\beta}}[\vb{p} = \vb{p}^* \wedge \msf{Pred}(\rho) = 1 \wedge \bs{\beta} = \vb{1} ~:~(\vb{p}, \rho) \la \msf{SIM}^{H^{\vb{p}^*, \bs{\beta}}_\epsilon}(x)] \\
=~& 
\frac{\epsilon^{-K}}{(4Kq+1)^{2K}} \cdot \sum_{\vb{p}^* \in \bits^{K\ell}  \cup \Set{\bot}}  \Pr_{H_\epsilon}[\vb{p} = \vb{p}^* \wedge \msf{Pred}(\rho) = 1 \wedge \bar{H}_\epsilon(\vb{p}) = \vb{1} ~:~(\vb{p}, \rho) \la \msf{SIM}^{H_\epsilon}(x)] \\
=~& 
\frac{\epsilon^{-K}}{(4Kq+1)^{2K}} \cdot \Pr_{H_\epsilon}[ \msf{Pred}(\rho) = 1 \wedge \bar{H}_\epsilon(\vb{p}) = \vb{1} ~:~(\vb{p}, \rho) \la \msf{SIM}^{H_\epsilon}(x)] \numberthis \label{CPM:decider:lem:completeness:proof:final:3} \\ 
\ge ~ & 
\frac{\epsilon^{-K}}{(4Kq+1)^{2K}} \cdot 
\bigg( 
\frac{\epsilon^{K}}{4} 	- \negl(\secpar)
\bigg)
 \numberthis \label[Inequality]{CPM:decider:lem:completeness:proof:final:4} 
,\end{align*}
where \Cref{CPM:decider:lem:completeness:proof:final:1} follows from the Law of Total Probability, \Cref{CPM:decider:lem:completeness:proof:final:2} follows from \Cref{MnR:lem}, \Cref{CPM:decider:lem:completeness:proof:final:3} follows again from the Law of Total Probability, and \Cref{CPM:decider:lem:completeness:proof:final:4} follows from \Cref{claim:sim-Vtil-prob}, 

Finally, it holds that 
\begin{align*}
\Pr[\mcal{B}(x) = 1] 
& \ge 
\frac{\epsilon^{-K}}{(4Kq+1)^{2K}} \cdot \bigg(\frac{\epsilon^K}{4} - \negl(\secpar) \bigg) - 32K^2q^2\epsilon \numberthis \label[Inequality]{CPM:decider:lem:completeness:proof:final:6} \\ 
& \ge 
\frac{1}{4\cdot(4Kq+1)^{2K}} - \negl(\secpar) - 32K^2q^2\epsilon 
\\
& \ge
\frac{1}{8\cdot(4Kq+1)^{2K}} - \negl(\secpar) 
\numberthis \label[Inequality]{CPM:decider:lem:completeness:proof:final:7} 
,\end{align*}
where \Cref{CPM:decider:lem:completeness:proof:final:6} follows from  \Cref{CPM:decider:lem:completeness:proof:sparsity:2,CPM:decider:lem:completeness:proof:final:4}, and \Cref{CPM:decider:lem:completeness:proof:final:7} follows from our parameter setting that $\epsilon \le \epsilon_q$ with the $\epsilon_q$ defined in \Cref{define:epsilon-q}.

This completes the proof of \Cref{CPM:decider:lem:completeness}.

\subsection{Proof of Soundness}
\label{sec:CPM:decider:lem:soundness:proof}

In this section, we prove \Cref{CPM:decider:lem:soundness}.

At a high level, we assume for contradiction that  \Cref{CPM:decider:lem:soundness} is false and show how to construct a malicious prover $\tilde{\Prover}$ breaking the soundness of the original protocol $\langle \Prover, \Verifier \rangle$. Toward that, we first need to make some modifications to the machine $\mcal{B}$ and the MnR game $\Sim^{\tilde{\Verifier}}$, making use of some particular properties of the $\tilde{\Verifier}$ we defined earlier. This is covered in \Cref{sec:soundness:proof:cleaning:MnR,sec:soundness:proof:def:real,sec:soundness:proof:def:dummy}. Next, we will present the malicious prover $\tilde{\Prover}$ and prove that it can indeed break the soundness. This is done in \Cref{sec:soundness:proof:def:malicious-Prover}.

\subsubsection{Cleaning the MnR Game}
\label{sec:soundness:proof:cleaning:MnR}

\subsubsubsection{Structure of $z_i$'s}
In the sequel, we focus on the MnR game $\Sim^{\tilde{\Verifier}}[H_0, \vb{1}](x)$, which is exactly the game \Cref{CPM:MnR:game} with $H$ instantiated by the all-zero function $H_0$ and $\vb{y}$ instantiated by the vector $\vb{1}$ of $K$ repetition of value 1.

Recall that the length $z_i$ of the measurement outcome $(p_{i,1}, \ldots, p_{i,z_i})$ defined in \Cref{CPM:MnR:game:step:3} of the MnR game $\Sim^{\tilde{\Verifier}}[H_0, \vb{1}](x)$ is determined by the measurement. In the following \Cref{MnR:game:modify:1:z:non-devreasing:lemma}, we show that the length values $z_i$'s will indeed exhibit a non-decreasing order. This property will be crucial for the later parts of the proof.


For notational convenience, let us first give a name to $\Sim$'s query to her oracle $\tilde{V}$ (or $\tilde{V}^\dagger$) that invokes the measurement in \Cref{CPM:MnR:game:step:3} of the MnR game $\Sim^{\tilde{\Verifier}}[H_0, \vb{1}](x)$.

\begin{definition}[Special Queries]\label{def:sq-query}
During the execution of $\Sim^{\tilde{\Verifier}}[H_0, \vb{1}](x)$, note that the $j_i$-th query to $\mcal{O}$ must be invoked by $\Sim$'s query to verifier $\tilde{\Verifier}$. We call this query of  $\Sim$ to $\tilde{\Verifier}$ (which in turn invokes the $j_i$-th query to $\mcal{O}$) as the {\em $i$-th special query} and denote it by $\msf{sq}(i)$.
\end{definition}

\begin{lemma}\label{MnR:game:modify:1:z:non-devreasing:lemma}
For the $s$ defined in \Cref{CPM:MnR:game:step:2} and $\Set{z_t}_{t \in [s]}$ defined in defined in \Cref{CPM:MnR:game:step:3} of game $\Sim^{\tilde{\Verifier}}[H_0, \vb{1}](x)$, it holds that\footnote{In \Cref{MnR:game:modify:1:z:non-devreasing:lemma} and its proof, for the corner case $t= 1$, the $\max$ is taken over $\emptyset$. We naturally define $\max(\emptyset) = 0$. }
$$\forall t \in [s],~ z_t \le \max\{z_1, z_2 \ldots, z_{t-1}\} + 1. ~~(\text{In particular, $z_1 = 1$ if $s\ge 1$.})$$
\end{lemma}
\begin{proof}
In the following proof, (arbitrarily) fix a $t \in [s]$.
 
Consider the moment of the execution when $\Sim$ is about to make the $\msf{sq}(t)$ query (i.e., the query that will invoke the measurement of $z_t$). Let us denote the overall state, at this moment, across all the registers as $\rho$. Let us denote $z \coloneqq \max\{z_1, z_2, \ldots, z_{t-1}\}$.

 At this moment, we know by definition of \Cref{CPM:MnR:game} that the oracle $H_0$ has been re-programmed to output 1 for input vectors of length {\em at most $z$}. In particular, if we denote the oracle at this moment as $\hat{H}$, then it outputs $0$ on all vectors $(p_1, \ldots, p_{z+1})$ of length $z+1$.

Next, we claim that 
$$
\forall j \in\Set{z+1, z+2, \ldots, K},~ \Tr[\ketbra{j}_{\reg{lc}} \rho] = 0,
$$ 
where $\ketbra{j}_{\reg{lc}}$ is the projector that projects register $\reg{lc}$ to value $j$. That is, we claim that the $\reg{lc}$ register of state $\rho$ does not have any weights on value $j \ge z+1$ (or equivalently, $j \le z$). This is because:
\begin{enumerate}
 \item
 by the definition of $\tilde{V}$ (see \Cref{CPM:V:unitary}), for $\reg{lc}$ to contain the value $z+1$ or greater, it is necessary that there exists some $(p_1, \ldots, p_{z+1})$ such that $H(p_1, \ldots, p_{z+1}) = 1$;
 \item
 as we just argued, it holds for the current oracle $\hat{H}$ that $\hat{H}(p_1, \ldots, p_{z+1}) = 0$ for all possible $(p_1, \ldots, p_{z+1})$.
 \end{enumerate}
Now, let us example how the $\msf{sq}(t)$ query will be process using the description of $\tilde{\Verifier}$ in \Cref{CPM:V:unitary:implementation}. In particular, according to {\bf Step 3} of  \Cref{CPM:V:unitary:implementation}, the maximum length of each ``branch'' in the super-position query to $H$ is upper bounded by $j+1$. Thus, the value $z_t$ determined by measuring this super-position will not exceed $j+1$ either. It then follows from the above argument that $z_t \le j+1 \le z+1$, satisfying the requirement of \Cref{MnR:game:modify:1:z:non-devreasing:lemma}.



Finally, we remark that the above argument implies $z_1 \le 1$ as a corner case. On the other hand, if $s \ge 1$, then $z_1$ by definition lies in $[K]$. Thus, it follows that $z_1 = 1$. 

This completes the proof of \Cref{MnR:game:modify:1:z:non-devreasing:lemma}.

\end{proof}

We next show in \Cref{MnR:game:modify:1:z:non-devreasing:cor} a powerful corollary of \Cref{MnR:game:modify:1:z:non-devreasing:lemma}. 
\begin{corollary}[of \Cref{MnR:game:modify:1:z:non-devreasing:lemma}]\label{MnR:game:modify:1:z:non-devreasing:cor}
For the game $\Sim^{\tilde{\Verifier}}[H_0, \vb{1}](x)$, define the following event $E_\msf{bad}$
$$E_\msf{Bad} \coloneqq (s \ne K) \vee (z_1 \ne 1) \vee (z_2 \ne 2) \vee \ldots \vee (z_K \ne K).$$
It then follows that
$$\Pr[\msf{Pred}(\rho) = 1 ~\wedge~E_\msf{Bad} ~:~ (\vb{p}, \rho) \la {\Sim}^{\tilde{\Verifier}}[{H_0}, {\vb{1}}](x)] = 0.$$
\end{corollary}
\begin{proof}[Proof of \Cref{MnR:game:modify:1:z:non-devreasing:cor}]
We first define another ``bad'' event $\tilde{E}_{\msf{Bad}}$
$$\tilde{E}_{\msf{Bad}} \coloneqq \big(\max\{z_1, \ldots, z_s\} < K\big).$$
Then, by \Cref{MnR:game:modify:1:z:non-devreasing:lemma}, it is not hard to see that the event that ``$\tilde{E}_{\msf{Bad}}$ happens'' implies the event that ``$E_\msf{Bad}$ happens.''

Therefore, to prove \Cref{MnR:game:modify:1:z:non-devreasing:cor}, it suffices to prove the following inequality:
\begin{equation}\label[Inequality]{MnR:game:modify:1:z:non-devreasing:cor:inter:ineq}
\Pr[\msf{Pred}(\rho) = 1 ~\wedge~\tilde{E}_\msf{Bad} ~:~ (\vb{p}, \rho) \la {\Sim}^{\tilde{\Verifier}}[{H_0}, {\vb{1}}](x)] = 0
\end{equation}
The proof for \Cref{MnR:game:modify:1:z:non-devreasing:cor:inter:ineq} is very similar to that for \Cref{MnR:game:modify:1:z:non-devreasing:lemma}. We include it below for the sake of completeness.   

Let $z \coloneqq \max\{z_1, \ldots, z_s\}$. If event $\tilde{E}_{\msf{Bad}}$ happens, we know that $z < K$. Now, consider the moment when the $\Sim$ halts. At this moment, we know by definition of game ${\Sim}^{\tilde{\Verifier}}[{H_0}, {\vb{1}}](x)$ that the oracle $H_0$ has been re-programmed to output 1 for input vectors of length {\em at most $z < K$}. In particular, if we denote the oracle at this moment as $\hat{H}$, then it outputs $0$ on all vectors $(p_1, \ldots, p_{K})$ of length $K$.

Next, we claim that 
\begin{equation}\label{MnR:game:modify:1:z:non-devreasing:cor:eq:K:weights}
\Tr[\ketbra{K}_{\reg{lc}} \rho] = 0,
\end{equation}
where $\ketbra{K}_{\reg{lc}}$ is the projector that projects register $\reg{lc}$ to value $K$. That is, we claim that the $\reg{lc}$ register of the final state $\rho$ at $\Sim$'s halt does not have any weights on value $K$. This is because:
\begin{enumerate}
 \item
 by the definition of $\tilde{V}$ (see \Cref{CPM:V:unitary}), for $\reg{lc}$ to contain the value $K$, it is necessary that there exists some $(p_1, \ldots, p_K)$ such that $H(p_1, \ldots, p_K) = 1$;
 \item
 as we just argued, it holds for the current oracle $\hat{H}$ that $\hat{H}(p_1, \ldots, p_K) = 0$ for all possible $(p_1, \ldots, p_K)$.
 \end{enumerate}
In this case, \Cref{MnR:game:modify:1:z:non-devreasing:cor:eq:K:weights} implies that $\msf{Pred}(\rho) = 0$ (recall the definition of $\msf{Pred}$ in \Cref{CPM:BQP-decider}). 

This completes the proof of \Cref{MnR:game:modify:1:z:non-devreasing:cor}.

\end{proof}

\Cref{MnR:game:modify:1:z:non-devreasing:cor} is powerful in the sense that it allows us to ``clean'' the game ${\Sim}^{\tilde{V}}[H_0, \vb{1}](x)$ in the following manner: during the execution of ${\Sim}^{\tilde{V}}[H_0, \vb{1}](x)$, once the event $E_\msf{Bad}$ occurs, we can immediately abort the execution (possibly prematurely), without reducing the probability of $\msf{Pred}(\rho) = 1$. We formalize this observation as the following 
\Cref{CPM:MnR:game:modify:2} (where we highlight its difference with ${\Sim}^{\tilde{V}}[H_0, \vb{1}](x)$ in \red{red color}) and \Cref{MnR:game:modify:2:cor}.
\begin{GameBox}[label={CPM:MnR:game:modify:2}]{Measure-and-Reprogram Game \textnormal{$\ddot{\Sim}^{\tilde{\Verifier}}[H_0, \vb{1}](x)$}}
 It works as follows
\begin{enumerate}
\item \label[Step]{CPM:MnR:game:modify:2:step:1}
For each $i\in[K]$, uniformly pick $(j_i,b_i)\in ([2Kq]\times \bits)$ \dlt{$\cup \{(\bot,\bot)\}$}  conditioned on that there exists at most one $i\in [K]$ such that $j_i=j^*$  for all $j^*\in [2Kq]$.

\item \label[Step]{CPM:MnR:game:modify:2:step:2}
 \dlt{Let $s$ denote the number of $j_i$'s in $\Set{j_i}_{i \in [k]}$ that are not $\bot$.  We re-label the indices $i$ for pairs $\Set{(j_i, b_i)}_{i \in [K]}$ so that $j_1 < j_2 < \ldots < j_s$ and $j_{s+1} = j_{s+2} =\ldots = j_K = \bot$.}

\red{We re-label the indices $i$ for pairs $\Set{(j_i, b_i)}_{i \in [K]}$  so that $j_1 < j_2 < \ldots < j_K$.}

\item
Run the oracle machine $\Sim^{\tilde{\Verifier}}$ with oracle $\mcal{O}$, which is initialized to be a quantumly-accessible classical oracle that computes $H_0$, and when $\Sim^{\tilde{\Verifier}}$ makes its $j$-th query, the oracle is simulated as follows:
\begin{enumerate}
      \item \label[Step]{CPM:MnR:game:modify:2:step:4:a} 
      If $j=j_i$ for some $i\in[\red{K}]$, measure this query to obtain $\vb{p}_i=(p_{i,1}, \ldots,p_{i, z_i})$ where $z_i \in [K]$ is determined by the measurement outcome. \red{It halts immediately and outputs $(\bot, \bot)$ if any of the following events happen:}
      \begin{enumerate}
        \item \label[Step]{CPM:MnR:game:modify:2:step:3:a:i} 
        \red{
        	$z_i \ne i$, {\em or}
        }
      \item \label[Step]{CPM:MnR:game:modify:2:step:3:a:ii} 
        \red{ 
        	$\vb{p}_{i-1}$ is not a prefix of $\vb{p}_{i}$. 
        }
      \end{enumerate}
     
      \red{The $\msf{sq}(i)$ queries are defined in the same manner as in \Cref{def:sq-query}.}
       
      It then behaves according to the value $b_i$ as follows:
        \begin{enumerate}
        \item 
        If $b_i=0$: First reprogram $\mcal{O} \leftarrow \reprogram(\mcal{O},\vb{p}_i, {1})$, and then answer the $j_i$-th query using the reprogrammed oracle.
        \item 
        If $b_i=1$: First answer the $j_i$-th query  using the oracle before the reprogramming, and then reprogram $\mcal{O} \leftarrow \reprogram(\mcal{O},\vb{p}_i, {1})$. 
        \end{enumerate}

        
    \item 
    Otherwise (i.e., $j \ne j_i$ $\forall i \in [\red{K}]$), answer the $j$-th query just using the oracle $\mcal{O}$ without any measurement or  reprogramming. 
    \end{enumerate}

\item
Let $(\vb{p}^* = (p^*_1, \ldots, p^*_K), \rho_{\msf{out}})$ be the output of $\Sim^{\tilde{\Verifier}}$ at halt (as per \Cref{CPM:MnR:notation:item:2}).

\item \label[Step]{CPM:MnR:game:modify:2:step:5}
\dlt{For all $i \in \Set{s+1, s+2, \ldots, K}$, set $\vb{p}_i = \vb{p}^*_i$ where $\vb{p}^*_i \coloneqq (p^*_1, \ldots, p^*_i)$.} 

\item \label[Step]{CPM:MnR:game:modify:2:step:6}
{\bf Output:}   \red{directly output $(\vb{p}_K, \rho_{\msf{out}})$} 
\dlt{The output of this game is defined as follows}
\begin{itemize}
\item
\dlt{If it holds for all $i \in [K]$ that $\vb{p}_i$ is a prefix of $\vb{p}_K$, then output $(\vb{p}_K, \rho_{\msf{out}})$;}

\item
\dlt{Otherwise, output $(\bot, \bot)$.}
\end{itemize}

\end{enumerate}
\end{GameBox}

\begin{corollary}[of \Cref{MnR:game:modify:1:z:non-devreasing:cor}]\label{MnR:game:modify:2:cor}
For the $\ddot{\Sim}^{\tilde{\Verifier}}[{H_0}, {\vb{1}}](x)$ defined in \Cref{CPM:MnR:game:modify:2} , it holds that
$$\Pr[\msf{Pred}(\rho) = 1~:~ (\vb{p}, \rho) \la \ddot{\Sim}^{\tilde{\Verifier}}[{H_0}, {\vb{1}}](x)] \ge \Pr[\msf{Pred}(\rho) = 1~:~ (\vb{p}, \rho) \la {\Sim}^{\tilde{\Verifier}}[{H_0}, {\vb{1}}](x)].$$
\end{corollary}
\begin{proof}[Proof of \Cref{MnR:game:modify:2:cor}]
We prove this corollary by considering the \red{red-color} changes in \Cref{CPM:MnR:game:modify:2} one by one.

In \Cref{CPM:MnR:game:modify:2:step:1}, we simply prevent the value $j_i$ from taking the value $\bot$. This follows immediately from \Cref{MnR:game:modify:1:z:non-devreasing:cor}---if there exists some $j_i = \bot$, then it must follow that $s \ne K$, which is a part of the $E_{\msf{Bad}}$ defined in  \Cref{MnR:game:modify:1:z:non-devreasing:cor}. Thus, we can safely ignore this case without reducing the probability of $\msf{Pred}(\rho) = 1$.

The changes made in \Cref{CPM:MnR:game:modify:2:step:2,CPM:MnR:game:modify:2:step:5} are direct consequences of the adjustment made in \Cref{CPM:MnR:game:modify:2:step:1}.

The introduction of \Cref{CPM:MnR:game:modify:2:step:3:a:i} can also be elucidated by \Cref{MnR:game:modify:1:z:non-devreasing:cor}. Specifically, the occurrence of the event $z_i \neq i$ is encompassed within the definition of $E_{\msf{Bad}}$ outlined in \Cref{MnR:game:modify:1:z:non-devreasing:cor}. Consequently, we can safely disregard this scenario without diminishing the probability of $\msf{Pred}(\rho) = 1$.

The inclusion of \Cref{CPM:MnR:game:modify:2:step:3:a:ii} essentially involves relocating the checks conducted in the original \Cref{CPM:MnR:game:modify:2:step:6} to an earlier stage, specifically within the current \Cref{CPM:MnR:game:modify:2:step:3:a:ii}.

This finishes the proof of \Cref{MnR:game:modify:2:cor}.

\end{proof}

\begin{remark}[The Type of $\msf{sq}(i)$ Query]\label{rmk:sq:type} 
We remark that in \Cref{CPM:MnR:game:modify:2}, the $\msf{sq}(i)$ query must be a $\tilde{V}$ query (instead of being  a $\tilde{V}^\dagger$ query). This follows from a similar argument as we did in the proof of \Cref{MnR:game:modify:1:z:non-devreasing:lemma}. In more detail, note that the $\msf{sq}(i)$ query by definition invokes the measurement of $\reg{p_1\ldots p_i}$ in \Cref{CPM:MnR:game:modify:2}, if the game is not aborted in \Cref{CPM:MnR:game:modify:2:step:4:a}. If the $\msf{sq}(i)$ query is invoked by a $\tilde{V}^\dagger$ query, then by the definition of $\tilde{V}$ (see \Cref{CPM:V:unitary}),  both the global counter and the local counter must be of the same value $i$ {\em right before this query}. However, this is simply impossible---using the argument as we did in the proof of \Cref{MnR:game:modify:1:z:non-devreasing:lemma}, the $\reg{lc}$ register has zero weights on value $i$ before the $\msf{sq}(i)$ query is processed. Therefore, the $\msf{sq}(i)$ query must be a $\tilde{V}$ query. 
\end{remark}

\subsubsubsection{Simplification Assumptions} 
Henceforth, we make three assumptions regarding the behavior of the simulator $\Sim$  that would greatly simplify the remaining proof. We first present two assumptions which are easy to state and validate.
\begin{itemize}
\item 
\hypertarget{CPM:simp-assumption:1}{{\bf Assumption 1:}} When the global counter is $\ket{0}_{\reg{gc}}$,\footnote{Note that due to \Cref{item:fact:gc}, we do not need to consider the case where the global counter is a superposition.} $\Sim$ will not make a $\tilde{V}^\dagger$ query.

\item 
\hypertarget{CPM:simp-assumption:2}{{\bf Assumption 2:}} When the global counter is $\ket{K}_{\reg{gc}}$, $\Sim$ will not make a $\tilde{V}$ query.
\end{itemize}
We argue that these assumptions will not decrease the probability of $\msf{Pred}(\rho) = 1$ for \Cref{CPM:MnR:game:modify:2}. To see that, consider that scenario where the $\Sim$ makes a $\tilde{V}^\dagger$ (resp.\ $\tilde{V}$) query when global counter is $\ket{0}_{\reg{gc}}$ (resp.\ $\ket{K}_{\reg{gc}}$). By the definition in \Cref{CPM:V:unitary}, the effect of this $\tilde{V}^\dagger$ (resp.\ $\tilde{V}$) query is to apply the identity operator (i.e., do nothing). Thus, such a query has no effects on the overall states at all and can be ignored. More accurately, one can define another machine $\Sim^*$ that behaves identically to $\Sim$ except that when $\Sim$ tries to make a $\tilde{V}^\dagger$ (resp.\ $\tilde{V}$) query at $\ket{0}_{\reg{gc}}$ (resp.\ $\ket{K}_{\reg{gc}}$), $\Sim^*$ does not do anything and pretend that this query has been answered. The probability of $\msf{Pred}(\rho) = 1$ for \Cref{CPM:MnR:game:modify:2} w.r.t.\ this new $\Sim^*$ is then identical to that of the original \Cref{CPM:MnR:game:modify:2}. 

Next, we state and validate the third assumption.
\begin{itemize}
\item 
\hypertarget{CPM:simp-assumption:3}{{\bf Assumption 3:}} For each $k \in \Set{0, 1, \ldots, K-1}$, {\em before the $\msf{sq}(k+1)$ query is made}, we allow $\Sim$ to operate directly on registers $\reg{t_z}$ for all $z \in \Set{k+1, k+2, \ldots, K}$. Then, for each $k \in [K]$, before the $\msf{sq}(k+1)$ query is made, $\Sim$ will not try to make a $\tilde{V}$-query when the global counter is $\ket{k}_{\reg{gc}}$.
\end{itemize}
We now argue that \hyperlink{CPM:simp-assumption:3}{{\bf Assumption 3}} will not decrease the probability of $\msf{Pred}(\rho) = 1$ for \Cref{CPM:MnR:game:modify:2}.
Consider the scenario where before the $\msf{sq}(k+1)$ query is made, $\Sim$ makes a $\tilde{V}$ when the global counter is $\ket{k}_{\reg{gc}}$. At this moment, via the same argument as in the proof of \Cref{MnR:game:modify:1:z:non-devreasing:lemma} and \Cref{MnR:game:modify:1:z:non-devreasing:cor}, we know that the local counter register does not have any weights on value $k+1$. Therefore, the only effect of this application of $\tilde{V}$, according to \Cref{CPM:V:unitary}, it to increase the global counter to $k+1$, apply $A_{k+1}$, and then apply $C_{k+1, 0}$. From $\Sim$'s view point, this is nothing but swapping the registers $\reg{t_{k+1}}$ and $\reg{m}$.   
Moreover, note that after this query, the global counter has been increase to the value $k+1$ while the local counter does not increase at all. Therefore, future $\tilde{V}$ queries (before $\msf{sq}(k+1)$) will always been answered by $C'_{k+1}A_{k+1}$ according to the global counter value $k+1$, {\em until $\Sim$ makes enough $\tilde{V}^\dagger$ queries to bring the global counter back to value $k$}. Therefore, instead of allowing $\Sim$ to perform these queries, one can define another machine $\Sim^*$ that behaves identically to $\Sim$ except that when $\Sim$ tries to make a $\tilde{V}$ query at $\ket{g}_{\reg{gc}}$ ($\forall g \in \Set{k, k+1, \ldots, K-1}$) before the $\msf{sq}(k+1)$ query, $\Sim^*$ dose not make this query; instead it operates directly on register $\reg{t_{g+1}}$. The probability of $\msf{Pred}(\rho) = 1$ for \Cref{CPM:MnR:game:modify:2} w.r.t.\ this new $\Sim^*$ is then identical to that of the original \Cref{CPM:MnR:game:modify:2}. This is exactly what we allowed in \hyperlink{CPM:simp-assumption:3}{{\bf Assumption 3}}.

\subsubsection{Defining the Real Game} 
\label{sec:soundness:proof:def:real}

\para{Looking into $\Sim^{\tilde{\Verifier}}$.} Next, we re-state the game $\ddot{\Sim}^{\tilde{\Verifier}}[H_0, \vb{1}](x)$ defined in \Cref{CPM:MnR:game:modify:2}. This time, we will look into the execution of $\Sim^{\tilde{\Verifier}}$ and their registers. This is different from previous games where we always treat $\Sim^{\tilde{\Verifier}}$ as a ``black-box'' and only focus on how it accesses the oracle $H_0$. We denote this game as $\msf{Real}(\Sim, \tilde{\Verifier})$ and present it in \Cref{CPM:MnR:game:real}.       

It is straightforward that \Cref{CPM:MnR:game:real} is identical to \Cref{CPM:MnR:game:modify:2} with only cosmetic changes. Therefore, it holds that
\begin{equation}\label{CPM:MnR:to:real:ind-eq}
\Pr[\msf{Pred}(\rho) = 1~:~ (\vb{p}, \rho) \la \ddot{\Sim}^{\tilde{\Verifier}}[{H_0}, {\vb{1}}](x)] = \Pr[\msf{Pred}(\rho) = 1~:~ (\vb{p}, \rho) \la \msf{Real}(\Sim, \tilde{\Verifier})].
\end{equation}

\begin{GameBox}[label={CPM:MnR:game:real}]{Game \textnormal{$\msf{Real}(\Sim, \tilde{\Verifier})$}}
For each $i\in[K]$, uniformly pick $(j_i,b_i)\in [2Kq]\times \bits$ conditioned on that there exists at most one $i\in [K]$ such that $j_i=j^*$  for all $j^*\in [2Kq]$.  We re-label the indices $i$ for $\Set{(j_i, b_i)}_{i \in [K]}$  so that $j_1 < j_2 < \ldots < j_K$. Initialize $H^{(0)}_0 \coloneqq H_0$ and $\vb{p}_0 = \emptyset$ and run the oracle machine $\Sim^{\tilde{\Verifier}}(x)$ in the following manner: 

\vspace{0.5em}
For each $i \in [K]$: 
\begin{itemize}
\item
{\bf For \text{$\Sim$}'s Queries between \text{$\msf{sq}(i-1)$} and \text{$\msf{sq}(i)$}:} Answer such queries with $\tilde{V}$ and $\tilde{V}^\dagger$ accordingly, where $\tilde{\Verifier}$ uses $H^{(i-1)}_0$ as the oracle. (see below for the definitions of $H^{(i-1)}_0$ for $i\ge 2$.)

\item
{\bf For Query $\msf{sq}(i)$:} Note that by definition (and \Cref{rmk:sq:type}), the $\msf{sq}(i)$ query is a $\tilde{V}$-query and it invokes a measurement over registers $\reg{p_1}\ldots \reg{p_i}$, resulting in the measurement outcome $\vb{p}_i = (p_1, \ldots, p_i)$. If $\vb{p}_{i-1}$ is not a prefix of $\vb{p}_{i}$, the execution halts immediately with output $(\bot, \bot)$.   Otherwise,  it  defines $H^{(i)}_0 \coloneqq H^{\vb{p}_i, \vb{1}}_0$ and then behaves according to the value $b_i$:
\begin{itemize}
\item 
If $b_i = 0$, applies $\tilde{V}$ using $H^{(i)}_0$ as the oracle.

\item 
If $b_i = 1$, applies $\tilde{V}$ using $H^{(i-1)}_0$ as the oracle.
\end{itemize}
\end{itemize}

\para{Output:} Let $(\vb{p}^*, \rho_{\msf{out}})$ be the output of $\Sim^{\tilde{\Verifier}}$ at halt. This game $\msf{Real}(\Sim, \tilde{\Verifier})$ outputs $(\vb{p}_K, \rho_{\msf{out}})$.

\end{GameBox}

\subsubsection{Defining the Dummy Game} 
\label{sec:soundness:proof:def:dummy}

\para{A Dummy Version of $\tilde{\Verifier}$.} We first define a unitary $\ddot{V}$, which should be treated as the ``dummy version'' of $\tilde{\Verifier}$'s unitary $\tilde{V}$. We present it in \Cref{CPM:V-dummy}, highlighting its difference with $\tilde{V}$ (defined in \Cref{CPM:V:unitary}) in \red{red color}.
\begin{AlgorithmBox}[label={CPM:V-dummy}]{Unitary \textnormal{$\ddot{V}$} for a Dummy-Version of \textnormal{$\tilde{\Verifier}$}}
{\bf Helper Unitaries.} We first define additional helper unitaries. For each $k \in [K]$:
\begin{itemize}
 \item
{\bf $\ddot{C}_k$: swap $\reg{m}$ and $\reg{t_k}$.} 
This unitary perform the following operation in superposition:
\begin{itemize}
\item
If $H(p_1, \ldots, p_k) = 1$, \red{then do nothing}. 
\item
If $H(p_1, \ldots, p_k) = 0$, then swap the contents of registers $\reg{t_k}$ and $\reg{m}$.
\end{itemize}
 \end{itemize} 

\para{Unitary $\ddot{V}$.} On each query, it compares the global counter value $k$ with the local counter value $j$ and behaves accordingly. In particular:
\begin{description}
\item[Case $(k = j)$:] It first applies the unitary $U_{gc}$ to increase the global counter by 1, and then behaves according to the (increased) global counter value $k+1$. In particular:
\begin{itemize}
\item
If $k+1 \in [K]$, then it applies $D_{k+1}\red{\ddot{C}_{k+1}}A_{k+1}$, where $A_{k+1}$ and $D_{k+1}$ is  defined in \Cref{CPM:V:unitary} and $\ddot{C}_{k+1}$ is defined above.
\item
Otherwise (i.e., $k+1 \notin[K]$), it does nothing (i.e., applies the identity operator).
\end{itemize}

\item[Case $(k \ne j)$:] It first applies the unitary $U_{gc}$ to increase the global counter by 1, and then behaves according to the (increased) global counter value $k+1$. In particular:
\begin{itemize}
\item
If $k+1 \in [K]$, then it applies $C'_{k+1}A_{k+1}$, where $A_{k+1}$ and $C_{k+1}$ is  defined in \Cref{CPM:V:unitary}
\item
Otherwise (i.e., $k+1 \notin[K]$), it does nothing (i.e., applies the identity operator).
\end{itemize}
\end{description}
\end{AlgorithmBox}

\para{A Dummy MnR Game.} We now define another game $\msf{Dummy}(\Sim, \tilde{\Verifier})$. This game will essentially serve as a malicious prover $\tilde{\Prover}$ that could break the soundness of the original ZK protocol, leading to our desired contradiction. Intuitively, $\msf{Dummy}(\Sim, \tilde{\Verifier})$ behaves identical to $\msf{Real}(\Sim, \tilde{\Verifier})$ except for that it answers $\Sim$'s queries between $\msf{sq}(i-1)$ and $\msf{sq}(i)$ using the dummy version of $\tilde{\Verifier}$ as we defined in \Cref{CPM:V-dummy}. We present the game in \Cref{CPM:MnR:game:dummy} and highlight its difference $\msf{Real}(\Sim, \tilde{\Verifier})$ (in \Cref{CPM:MnR:game:real}) in \red{red color}. 
\begin{GameBox}[label={CPM:MnR:game:dummy}]{Game \textnormal{$\msf{Dummy}(\Sim, \tilde{\Verifier})$}}
For each $i\in[K]$, uniformly pick $(j_i,b_i)\in [2Kq]\times \bits$ conditioned on that there exists at most one $i\in [K]$ such that $j_i=j^*$  for all $j^*\in [2Kq]$.  We re-label the indices $i$ for $\Set{(j_i, b_i)}_{i \in [K]}$  so that $j_1 < j_2 < \ldots < j_K$. Initialize $H^{(0)}_0 \coloneqq H_0$ and $\vb{p}_0 = \emptyset$ and run the oracle machine $\Sim^{\tilde{\Verifier}}(x)$ in the following manner: 

\vspace{0.5em}
For each $i \in [K]$: 
\begin{itemize}
\item
{\bf For $\Sim$'s Queries between $\msf{sq}(i-1)$ and $\msf{sq}(i)$:} Answer such queries with \red{$\ddot{V}$} and \red{$\ddot{V}^\dagger$} accordingly, using $H^{(i-1)}_0$ as the oracle. (Note that \red{$\ddot{V}$} is defined in \Cref{CPM:V-dummy}.)

\item
{\bf For Query $\msf{sq}(i)$:} Answered in the same manner as in \Cref{CPM:MnR:game:real}.
\end{itemize}

\para{Output:} Let $(\vb{p}^*, \rho_{\msf{out}})$ be the output of $\Sim^{\tilde{\Verifier}}$ at halt. This game $\msf{Dummy}(\Sim, \tilde{\Verifier})$ outputs $(\vb{p}_K, \rho_{\msf{out}})$.

\end{GameBox}
The following is the most involved lemma. We defer its proof to \Cref{sec:relate-Dummy-Real}. In the following, we show (in \Cref{sec:soundness:proof:def:malicious-Prover}) how to finish the current proof of \Cref{CPM:decider:lem:soundness} assuming that \Cref{CPM:MnR:game:dummy:lem} holds.

\begin{lemma}\label{CPM:MnR:game:dummy:lem}
For the $\msf{Dummy}(\Sim, \tilde{\Verifier})$ defined in \Cref{CPM:MnR:game:dummy} and the $\msf{Real}(\Sim, \tilde{\Verifier})$ defined in \Cref{CPM:MnR:game:real}, it holds that
$$\Pr[\msf{Pred}(\rho) = 1~:~ (\vb{p}, \rho) \la \msf{Dummy}(\Sim, \tilde{\Verifier})] = \Pr[\msf{Pred}(\rho) = 1~:~ (\vb{p}, \rho) \la \msf{Real}(\Sim, \tilde{\Verifier})].$$
\end{lemma}

\subsubsection{Finishing the Proof of \Cref{CPM:decider:lem:soundness}}
\label{sec:soundness:proof:def:malicious-Prover}
In this part, we finish the proof of \Cref{CPM:decider:lem:soundness} assuming that \Cref{CPM:MnR:game:dummy:lem} holds. 

We start by describing the malicious prover $\tilde{\Prover}$ that we will use to break the soundness of the original protocol $\langle \Prover, \Verifier \rangle$.

\begin{AlgorithmBox}[label={malicious-Prover}]{Malicious Prover $\tilde{\Prover}(x)$}
On input $x$, $\tilde{\Prover}(x)$ internally emulates \Cref{CPM:MnR:game:dummy} with the help of an external honest verifier $\Verifier$ in the following manner.

\para{Registers:} Note that to emulate \Cref{CPM:MnR:game:dummy}, $\tilde{\Prover}$ needs to prepare registers as in \Cref{CMP:V:inital:expr}. To do that, $\tilde{\Prover}$ prepares the following registers
\begin{equation}\label[Expression]{malicious-P:inital:expr}
\ket{x}_{\reg{ins}} 
\ket{0}_{\reg{gc}}
\ket{0}_{\reg{lc}} 
\ket{\vb{0}}_{\reg{p_1}}
\ldots
\ket{\vb{0}}_{\reg{p_K}}
\ket{\vb{0}}_{\reg{m}} 
\ket{\bot}_{\reg{t_1}}
\ldots
\ket{\bot}_{\reg{t_K}}  
\ket{H_0}_{\reg{aux}} 
\end{equation}
Comparing \Cref{malicious-P:inital:expr} with \Cref{CMP:V:inital:expr}, one can see that registers $\reg{v_1\ldots v_K}$ and $\reg{w}$ are missing. As we will explain shortly, these registers can be think of being held by the external honest verifier $\Verifier$.

\para{Execution:} $\tilde{\Prover}$ emulates \Cref{CPM:MnR:game:dummy} in the following manner. For each $i \in [K]$,
\begin{itemize}
\item
{\bf For $\Sim$'s Queries between $\msf{sq}(i-1)$ and $\msf{sq}(i)$:} $\tilde{\Prover}$ answers such queries with in exactly the same manner as \Cref{CPM:MnR:game:dummy}. We emphasize that such queries will be answered using the $\ddot{V}$ or $\ddot{V}^\dagger$. Note that $\ddot{V}$ by definition (see \Cref{CPM:V-dummy}) will only make use of the registers shown in \Cref{malicious-P:inital:expr}. Therefore, $\tilde{\Prover}$ could emulate them internally, without touching upon the external honest $\Verifier$'s registers.

\item
{\bf For Query $\msf{sq}(i)$:} $\tilde{\Prover}$ first behaves in the same manner as \Cref{CPM:MnR:game:dummy} until she obtains the measurement outcome $\vb{p}_i = (p_{1}, \ldots, p_i)$. Also, same as \Cref{CPM:MnR:game:dummy}, if the value  $\vb{p}_{i-1}$ obtained earlier is not a prefix of the current $\vb{p}_i$, $\tilde{\Prover}$ aborts the execution directly.

If the execution has not been aborted so far, the next step in \Cref{CPM:MnR:game:dummy} is to apply the unitary $\tilde{V}$. We emphasize that this is the (only) step that $\tilde{\Prover}$ cannot finish by herself internally, because the $\tilde{V}$ operator defined in \Cref{CPM:V:unitary} (in particular, the operator $B_i$) needs to work on the external honest $\Verifier$'s registers $\reg{v_i}$ and $\reg{w}$. To do that, $\tilde{\Prover}$ simply puts $p_i$ in to the $\reg{m}$ register and sends it to the external $\Verifier$ (note that $p_i$ is a classical message). Then, by definition, the honest $\Verifier$ will compute the (possibly quantum) response $v_2$, put it in the $\reg{m}$ register, and send $\reg{m}$ back to $\tilde{\Prover}$. Using this returned $\reg{m}$ register, $\tilde{\Prover}$ can easily finish the remaining steps in exactly the same manner as \Cref{CPM:MnR:game:dummy}.

\end{itemize}

\para{Output:} We do not define any output for $\tilde{\Prover}$. (In \Cref{CPM:MnR:game:dummy}, the $\rho_{\msf{out}}$ part in the output is on the register $\reg{w}$. But this register is held by the external honest $\Verifier$ in the current game.)
\end{AlgorithmBox}

From the above description, it is easy to see that the view of $\Sim$ in \Cref{malicious-Prover} is identical to that in \Cref{CPM:MnR:game:dummy}. Thus, during the execution $\langle \tilde{\Prover}, \Verifier\rangle(x)$, the overall state across the joint registers held by both the malicious $\tilde{\Prover}$ (as defined in \Cref{malicious-Prover}) and the honest $\Verifier$ evolves in exactly the same manner as in \Cref{CPM:MnR:game:dummy}. Therefore, it follows that
\begin{equation}\label[Inequality]{eq:malicious-P:dummy}
\Pr[\msf{Acc}(\rho) = 1~:~ \rho \la \msf{OUT}_{\Verifier}\langle \tilde{\Prover}, \Verifier\rangle(x)] 
\ge \Pr[\msf{Pred}(\rho) = 1~:~ (\vb{p}, \rho) \la \msf{Dummy}(\Sim, \tilde{\Verifier})]
,\end{equation}
where the `$>$' part in the ``$\ge$'' sign in \Cref{eq:malicious-P:dummy} is due to the following reason: Note that the predicate $\msf{Pred}$ checks if the $\reg{lc}$ register contains $K$ {\em in addition to the check that the verifier's decision bit is 1} (see \Cref{def:predicte:Pred}); however, the LHS in \Cref{eq:malicious-P:dummy} only checks if the  verifier's decision bit is 1.

\para{Deriving the Final Contradiction.} Now, we are ready to finish the proof of \Cref{CPM:decider:lem:soundness}. Toward that, we assume for contradiction that \Cref{CPM:decider:lem:soundness} does not hold. Formally, we assume that there exist a $\Lang \in \BQP$, a $x \in \bits^{\secpar} \setminus \Lang$, and a polynomial $\delta(\cdot)$ so that for infinitely many $\secpar$ it holds that  
\begin{equation}\label[Inequality]{decider:lem:soundness:proof:eq:contradiction}
\Pr[\mcal{B}(x) = 1] \ge  \frac{1}{\delta(\secpar)}.
\end{equation}

It then follows that
\begin{align*}
& \Pr[\msf{Pred}(\rho) = 1~:~ (\vb{p}, \rho)  \la \msf{Out}_{\Verifier}\big(\langle \tilde{\Prover}, \Verifier\rangle(x)\big)] \\
\ge ~ & 
\Pr[\msf{Pred}(\rho) = 1~:~ (\vb{p}, \rho) \la \msf{Dummy}(\Sim, \tilde{\Verifier})]
\numberthis \label[Inequality]{decider:lem:soundness:proof:eq:final:1} \\ 
 = ~ &
\Pr[\msf{Pred}(\rho) = 1~:~ (\vb{p}, \rho) \la \msf{Real}(\Sim, \tilde{\Verifier})]
\numberthis \label{decider:lem:soundness:proof:eq:final:2} \\ 
 = ~ &
\Pr[\msf{Pred}(\rho) = 1~:~ (\vb{p}, \rho) \la \ddot{\Sim}^{\tilde{\Verifier}}[{H_0}, {\vb{1}}]]
\numberthis \label{decider:lem:soundness:proof:eq:final:3} \\ 
 \ge ~ &
\Pr[\msf{Pred}(\rho) = 1~:~ (\vb{p}, \rho) \la {\Sim}^{\tilde{\Verifier}}[{H_0}, {\vb{1}}]]
\numberthis \label[Inequality]{decider:lem:soundness:proof:eq:final:4} \\ 
 = ~ &
 \Pr[\mcal{B}(x) = 1] 
\numberthis \label[Inequality]{decider:lem:soundness:proof:eq:final:5} \\ 
 \ge ~ &
\frac{1}{\delta(\secpar)}
\numberthis \label[Inequality]{decider:lem:soundness:proof:eq:final:6} 
,\end{align*}
where \Cref{decider:lem:soundness:proof:eq:final:1} follows form \Cref{eq:malicious-P:dummy}, \Cref{decider:lem:soundness:proof:eq:final:2} follows from \Cref{CPM:MnR:game:dummy:lem}, \Cref{decider:lem:soundness:proof:eq:final:3} follows from \Cref{CPM:MnR:to:real:ind-eq}, \Cref{decider:lem:soundness:proof:eq:final:4} follows from \Cref{MnR:game:modify:2:cor}, \Cref{decider:lem:soundness:proof:eq:final:5} follows from the definition of $\mcal{B}(x)$ (see \Cref{CPM:BQP-decider}), and \Cref{decider:lem:soundness:proof:eq:final:6} follows from \Cref{decider:lem:soundness:proof:eq:contradiction}.

Note that \Cref{decider:lem:soundness:proof:eq:final:6} contradicts the soundness of the protocol $\langle \Prover, \Verifier \rangle$.

\vspace{1em}
This eventually completes the proof of \Cref{CPM:decider:lem:soundness}.

\section{Relating the Dummy and Real Games}
\label{sec:relate-Dummy-Real}

In this section (and the next two sections), we establish \Cref{CPM:MnR:game:dummy:lem}.

\subsection{Branch-Wise Equivalence Suffices}
We first notice that \Cref{CPM:MnR:game:real} can be treated as a sequence of alternating unitaries and measurements. In particular, there are two types of measurements (and $(q+K)$ measurements are made in total):
\begin{enumerate}
\item
\label[Type]{CPM:branchwise:measure:type:1}
{\bf Type-1:} Recall that the simulator $\Sim$ simply alternates between her local unitary $S$ and the measurement on register $\reg{u}$, which determines the type of her next query (i.e., either $\tilde{V}$ or $\tilde{V}^\dagger$). Note that there are $q$ such measurements on $\reg{u}$ since $\Sim$ makes $q$ queries in total.
\item
\label[Type]{CPM:branchwise:measure:type:2}
{\bf Type-2:}
When $\Sim$ makes the $\msf{sq}(i)$-th query, it triggers a measurement on registers $\reg{p_1} \ldots \reg{p_i}$. Note that there are $K$ such measurements (if the game is not aborted prematurely).
\end{enumerate}
Also, notice that in \Cref{CPM:MnR:game:real}, the values $\Set{(j_i, b_i)}_{i \in [K]}$ are sampled at the very beginning and after that the timing of all the measurements are fixed. It then follows from \Cref{branchwise:lem} that the accepting probability $\Pr[\msf{Pred}(\rho)=1]$ w.r.t.\  \Cref{CPM:MnR:game:real} can be equivalently expressed as the summation of all accepting probabilities w.r.t.\ a ``sub-normalized'' version of \Cref{CPM:MnR:game:real} where all the $(q+ K)$ measurements are replaced with projections. Formally, we defined the ``sub-normalized'' \Cref{CPM:MnR:game:real} (and similarly for \Cref{CPM:MnR:game:dummy}) in \Cref{CPM:MnR:game:subnormalized}.

\begin{GameBox}[label={CPM:MnR:game:subnormalized}]{Sub-Normalized Games \textnormal{$\msf{Real}(\Sim, \tilde{\Verifier}, J, \msf{pat})$} and \textnormal{$\msf{Real}(\Sim, \tilde{\Verifier}, J, \msf{pat})$}}
{\bf Parameters.} Fix $J = \Set{(j_i, b_i)}_{i \in [K]}$ where each $j_i \in [2Kq]$, each $b_i \in \bits$,  and $j_1 < j_2 < \ldots < j_K$. Also fix a classical string of $(q+K)$ symbols $\msf{pat} = (u_1, u_2, \ldots, u_q, p_1, p_2,  \ldots, p_K)$ where each $u_i \in \Set{\uparrow, \downarrow}$, and each $p_i \in \Set{0, 1}^{\ell}$. The following games are parameterized by the fixed $(J, \msf{pat})$.

\para{Game $\msf{Real}(\Sim, \tilde{\Verifier}, J, \msf{pat})$.} It behaves identically to \Cref{CPM:MnR:game:real} except for the following differences: 
\begin{enumerate}
	\item
	It does not sample the $(j_i, b_i)$ pairs; instead, it uses the $\Set{(j_i, b_i)}_{i \in [K]}$  contained in $J$.

\item
When $\Sim$ needs to make the $i$-th measurement on $\reg{u}$ (corresponding to \Cref{CPM:branchwise:measure:type:1} above), it instead applies the projector $\ketbra{u_i}$ to $\reg{u}$; (Note that the value $u_i$ is contained in $\msf{pat}$.)
\item
When the measurement on registers $\reg{p_1\ldots p_i}$ is triggered (corresponding to \Cref{CPM:branchwise:measure:type:2} above), it instead applies the projector $\ketbra{p_1, \ldots, p_i}$ registers $\reg{p_1\ldots p_i}$; (Note that the values $p_1, \ldots, p_i$ are contained in $\msf{pat}$.)
\end{enumerate}

\para{Game $\msf{Dummy}(\Sim, \tilde{\Verifier}, J, \msf{pat})$.} This is defined in a similar way as the above. Namely, It behaves identically to \Cref{CPM:MnR:game:dummy} except for the following differences: 
\begin{enumerate}
\item
	It does not sample the $(j_i, b_i)$ pairs; instead, it uses the $\Set{(j_i, b_i)}_{i \in [K]}$  contained in $J$.

\item
When $\Sim$ needs to make the $i$-th measurement on $\reg{u}$ (corresponding to \Cref{CPM:branchwise:measure:type:1} above), it instead applies the projector $\ketbra{u_i}$ to $\reg{u}$; (Note that the value $u_i$ is contained in $\msf{pat}$.)
\item
When the measurement on registers $\reg{p_1\ldots p_i}$ is triggered (corresponding to \Cref{CPM:branchwise:measure:type:2} above), it instead applies the projector $\ketbra{p_1, \ldots, p_i}$ registers $\reg{p_1\ldots p_i}$; (Note that the values $p_1, \ldots, p_i$ are contained in $\msf{pat}$.)
\end{enumerate}
\end{GameBox}

It then follows from \Cref{branchwise:lem} that
\begin{align*}
& \Pr[\msf{Pred}(\rho) = 1~:~ (\vb{p}, \rho) \la \msf{Real}(\Sim, \tilde{\Verifier})] \\
= ~&  
\frac{1}{\binom{2Kq}{K} \cdot 2^K} \cdot \sum_{J, \msf{pat}}\Pr[\msf{Pred}(\rho) = 1~:~ (\vb{p}, \rho) \la \msf{Real}(\Sim, \tilde{\Verifier}, J, \msf{pat})], 
\end{align*}
and that
\begin{align*}
& \Pr[\msf{Pred}(\rho) = 1~:~ (\vb{p}, \rho) \la \msf{Dummy}(\Sim, \tilde{\Verifier})] \\
= ~& 
\frac{1}{\binom{2Kq}{K} \cdot 2^K} \cdot \sum_{J, \msf{pat}}\Pr[\msf{Pred}(\rho) = 1~:~ (\vb{p}, \rho) \la \msf{Dummy}(\Sim, \tilde{\Verifier}, J, \msf{pat})],
\end{align*}
where the summations are taken over all possible $J$ and $\msf{pat}$ satisfying the requirements in \Cref{CPM:MnR:game:subnormalized}. (Note that the multiplicative factor $1/\big(\binom{2Kq}{K} \cdot 2^K\big)$ is to compensate for the fact that the $J$ is fixed in \Cref{CPM:MnR:game:subnormalized}, instead of being sampled randomly as in the original real  \Cref{CPM:MnR:game:real} and dummy \Cref{CPM:MnR:game:dummy}.)

Therefore, to prove \Cref{CPM:MnR:game:dummy:lem}, it suffices to prove the following \Cref{CPM:branch-wise:suffices}.
\begin{lemma}[Branch-Wise Equivalence Suffices]\label{CPM:branch-wise:suffices}
For all $(J, \msf{pat})$ satisfying the requirements in \Cref{CPM:MnR:game:subnormalized}, it holds that 
$$\Pr[\msf{Pred}(\rho) = 1~:~ (\vb{p}, \rho) \la \msf{Real}(\Sim, \tilde{\Verifier}, J, \msf{pat})] = \Pr[\msf{Pred}(\rho) = 1~:~ (\vb{p}, \rho) \la \msf{Dummy}(\Sim, \tilde{\Verifier}, J, \msf{pat})].$$
\end{lemma}
The proof of \Cref{CPM:branch-wise:suffices} is involved. In the following \Cref{sec:orignal:hybrids}, we prove \Cref{CPM:branch-wise:suffices} making use of a technical lemma (i.e., \Cref{lem:H:IND}). Then, the later proofs (i.e., \Cref{sec:counter-structure}, \Cref{sec:CPM:hybrids:restate}, \Cref{sec:CMP:warm-up},  and \Cref{sec:CMP:full}) are devoted to establishing \Cref{lem:H:IND}.

\subsection{Proving \Cref{CPM:branch-wise:suffices}: Define Hybrids}
\label{sec:orignal:hybrids}
Henceforth, we consider a fixed $(J, \msf{pat})$ pair satisfying the requirements in \Cref{CPM:MnR:game:subnormalized}. We first define a sequence of hybrids $H^{(J, \msf{pat})}_0, H^{(J, \msf{pat})}_1, \ldots, H^{(J, \msf{pat})}_K$. 

\para{Hybrid $H_0^{(J, \msf{pat})}$.} This is exactly the execution of $\msf{Real}(\Sim, \tilde{\Verifier}, J, \msf{pat})$. Therefore, it holds that 
\begin{equation}\label{CMP:soundness:proof:hyb:first-real}
\Pr\big[\msf{Pred}(\rho) = 1~:~(\vb{p}, \rho)\la H_0 \big] = \Pr\big[\msf{Pred}(\rho) = 1~:~(\vb{p}, \rho)\la\msf{Real}(\Sim, \tilde{\Verifier}, J, \msf{pat})\big].
\end{equation}

\para{Hybrid $H_k^{(J, \msf{pat})}$ $(k \in [K])$.} This hybrid is identical to $H^{(J, \msf{pat})}_{k-1}$, except for the following difference:
\begin{itemize}
\item
The first query (made by $\Sim$) that brings the global counter from $\ket{k-1}_{\reg{gc}}$ to $\ket{k}_{\reg{gc}}$ (i.e., query $\msf{sq}(k)$) is answered with $\tilde{V}$ (as in the previous hybrid). However, all subsequent queries that brings the global counter from $\ket{k-1}_{\reg{gc}}$ to $\ket{k}_{\reg{gc}}$ (resp.\ from $\ket{k}_{\reg{gc}}$ to $\ket{k-1}_{\reg{gc}}$) are answered with the ``dummy'' unitary $\ddot{V}$ (resp.\ $\ddot{V}^\dagger$).
\end{itemize}
This completes the description of the hybrids.

\para{Intuition.} Intuitively, we use these hybrids to replace the $\tilde{V}$ (and $\tilde{V}^\dagger$) to its dummy version $\ddot{V}$ (and $\ddot{V}^\dagger$) ``one layer by one layer,'' where by ``layer'' we mean the execution that brings the global counter from some value $k$ to $k+1$. Note that such replacement does not affect the special $\msf{sq}(k)$ queries, which are always answered by $\tilde{V}$ across all the hybrids.

To further aid in interpreting these hybrids, we recommend referring to \Cref{figure:CPM-3ZK:hybrids} on \Cpageref{figure:CPM-3ZK:hybrids}, where we illustrate these hybrids for the simplified case of $K = 2$. In more detail, 
\begin{itemize}
\item 
\Cref{figure:CPM-3ZK:hybrids:H0} is the real game $\msf{Real}(\Sim, \tilde{\Verifier}, J, \msf{pat})$ where all the $\downarrow$ queries (resp.\ $\uparrow$ queries) are answered by $\tilde{V}$ (resp.\ $\tilde{V}^\dagger$); 

\item 
In \Cref{figure:CPM-3ZK:hybrids:H1} (corresponding to $H_1^{(J, \msf{pat})}$), all the $\downarrow$ queries (resp.\ $\uparrow$ queries) in the first ``layer'' (i.e., the area between the horizontal line for $\ket{0}_{\reg{gc}}$ and the horizontal line for $\ket{1}_{\reg{gc}}$) are replaced with $\ddot{V}$ (resp.\ $\ddot{V}^\dagger$), except for the $\msf{sq}(1)$ query, which is still answered by $\tilde{V}$. 

\item 
In \Cref{figure:CPM-3ZK:hybrids:H2} (corresponding to $H_2^{(J, \msf{pat})}$), all the $\downarrow$ queries (resp.\ $\uparrow$ queries) in the first two layers are replaced with $\ddot{V}$ (resp.\ $\ddot{V}^\dagger$), except for the $\msf{sq}(1)$ and $\msf{sq}(2)$ query, which is still answered by $\tilde{V}$. 
\end{itemize}
Note that in this baby case of $K = 2$, the hybrid $H_2^{(J, \msf{pat})}$ illustrated by \Cref{figure:CPM-3ZK:hybrids:H2} is the last hybrid, and obviously, it is exactly the dummy game $\msf{Dummy}(\Sim, \tilde{\Verifier}, J, \msf{pat})$.

From the definition of hybrids (and the intuitive explanation above), it is easy to see that hybrid $H^{(J, \msf{pat})}_K$ is exactly the execution of $\msf{Dummy}(\Sim, \tilde{\Verifier}, J, \msf{pat})$. Thus, it holds that
\begin{equation}\label{CMP:soundness:proof:hyb:last-dummy}
\Pr\big[\msf{Pred}(\rho) = 1~:~(\vb{p}, \rho)\la H^{(J, \msf{pat})}_K \big] = \Pr\big[\msf{Pred}(\rho) = 1~:~(\vb{p}, \rho)\la\msf{Dummy}(\Sim, \tilde{\Verifier}, J, \msf{pat})\big].\end{equation}

It then following from \Cref{CMP:soundness:proof:hyb:first-real} and \Cref{CMP:soundness:proof:hyb:last-dummy} that to prove \Cref{CPM:branch-wise:suffices}, it suffices to establish the following \Cref{lem:H:IND}. 
\begin{lemma}\label{lem:H:IND}
For all $(J, \msf{pat})$ satisfying the requirements in \Cref{CPM:MnR:game:subnormalized} and all $k \in [K]$, it holds that 
$$\Pr\big[\msf{Pred}(\rho) = 1~:~(\vb{p}, \rho)\la H^{(J, \msf{pat})}_{k-1} \big] = \Pr\big[\msf{Pred}(\rho) = 1~:~(\vb{p}, \rho)\la H^{(J, \msf{pat})}_k \big].$$
\end{lemma}
\para{Proof Structure of \Cref{lem:H:IND}.} The remaining part of the proof focuses on establishing \Cref{lem:H:IND}. Its proof is quite intricate and spans several sections: \Cref{sec:counter-structure}, \Cref{sec:CPM:hybrids:restate}, \Cref{sec:CMP:warm-up}, and \Cref{sec:CMP:full}. Here is an overview of what each section covers:
\begin{itemize}
\item 
First, we remark that the hybrids described above is stated in a manner for ease of understanding; however, this presentation is not conducive to the mathematical derivations needed to establish \Cref{lem:H:IND}. Therefore, we need to restate these hybrids in a more mathematically friendly form. To accomplish this, we introduce a lemma characterizing the structure of the local counter and global counter registers in \Cref{sec:counter-structure}. Armed with this lemma, we proceed to provide an alternative description of the hybrids in \Cref{sec:CPM:hybrids:restate}. This new description is essentially equivalent but offers a more suitable framework for the subsequent mathematical analysis.

\item 
Having established the alternative description of the hybrids in \Cref{sec:CPM:hybrids:restate}, we are now prepared to demonstrate the indistinguishability between adjacent hybrids, as mandated by \Cref{lem:H:IND}. However, given the complexity of this proof, we opt to initially focus on the simplified case of $K=2$, aiming to elucidate the core concept in a more streamlined context. This treatment is presented in \Cref{sec:CMP:warm-up}.

\item 
Ultimately, we provide the full proof for the general case (for an arbitrary constant $K$). This is elucidated in \Cref{sec:CMP:full}.
\end{itemize}

\subsection{Prove \Cref{lem:H:IND}: Structure of Counters}
\label{sec:counter-structure}

\para{Intuition.} Let us first provide a high-level overview of this part. We assert that within any hybrid $H^{(J, \msf{pat})}_{k}$, throughout its execution, the global counter and local counter will exhibit a nice structure as defined in the following \Cref{counter-structure:lemma}. Intuitively, \Cref{counter-structure:lemma} can be interpreted as follows: during the execution of $H^{(J, \msf{pat})}_{k}$, the overall state can be expressed as the sum of pure states in superposition, and within each superposition, the local counter value does not exceed the global counter value. Furthermore, for the superposition where the $\reg{p_1\ldots p_i}$ registers contain precisely the values $p_1, \ldots, p_i$ (specified in $\msf{pat}$), both the global counter and the local counter equal $i$.

Looking ahead, \Cref{counter-structure:lemma} bears significance for the subsequent proof for the following reason. As per \Cref{CPM:V:unitary}, the behavior of $\tilde{\Verifier}$ is (partially) dictated by the global and local counters. In other words, it necessitates comparing these two counters to determine the appropriate unitaries to employ in \Cref{CPM:V:unitary}. Hence, if we aim to monitor the evolution of the overall state throughout the execution, we must monitor these two counters. It is not immediately evident whether this is feasible, as the global and local counter registers might be in a complex superposition that defies a neat mathematical expression.

Now, armed with \Cref{counter-structure:lemma}, we possess a complete and clear understanding of the structure of these two counters. This structured framework enables us to derive a ``finer-grained'' description of the unitaries $\tilde{V}$ and $\ddot{V}$ simply by examining the global counter, which, as noted in \Cref{item:fact:gc}, maintains a classical value throughout the execution. As will become clear later in \Cref{sec:CMP:warm-up,sec:CMP:full}, such a ``finer-grained'' characterization serves as the linchpin for tracking the overall states throughout the hybrid executions. This capability, in turn, facilitates the establishment of the indistinguishability conditions stipulated in \Cref{lem:H:IND}.

We next proceed to the formal treatment.
\begin{lemma}[Counter Structure.]\label{counter-structure:lemma}
For each  pair $(J, \msf{pat})$ satisfying the requirements in \Cref{CPM:MnR:game:subnormalized}, each $k \in [K] \cup \Set{0}$, each hybrid $H^{(J, \msf{pat})}_{k}$, let $\ket{\phi^{(t,i)}}$ denote the overall state right after the $\Sim$'s query that leads to the global counter's $t$-th arrival at value $i$. Then, the following holds:
\begin{enumerate}
\item \label[Case]{counter-structure:lemma:state:case:1}
If this arrival at value $i$ is due to the $\msf{sq}(i)$ query {\em and} the value $b_i = 1$, then 
the state $\ket{\phi^{(t,i)}}$ can be written in the following format:
\begin{equation}\label{counter-strucutre:psi:format:case:1}
\ket{\phi^{(t,i)}} 
 =  
\ket{i}_{\reg{gc}} \ket{i-1}_{\reg{lc}} \ket{p_1, \ldots, p_i}_{\reg{p_1, \ldots, p_i}}\ket{\rho^{(t,i)}}. 
\end{equation}

\item \label[Case]{counter-structure:lemma:state:case:2}
In all the other cases (i.e., either this arrival at value $i$ is not caused by the $\msf{sq}(i)$ query or it is but $b_i = 0$), the state $\ket{\phi^{(t,i)}}$ can be written in the following format:
\begin{align*}
\ket{\phi^{(t,i)}} 
& =  
\ket{i}_{\reg{gc}} \ket{i}_{\reg{lc}} \ket{p_1, \ldots, p_i}_{\reg{p_1, \ldots, p_i}}\ket{\rho^{(t,i)}} ~+  \\
& \hspace{8em} 
\ket{i}_{\reg{gc}} \sum^{i-1}_{j = 0} \ket{j}_{\reg{lc}} \ket{p_1, \ldots, p_j}_{\reg{p_1, \ldots, p_j}} \sum_{p'_{j+1} \ne p_{j+1}}\ket{p'_{j+1}}_{\reg{p_{j+1}}} \ket{\rho^{(t,i)}_{p'_{j+1}}}
\numberthis \label{counter-strucutre:psi:format:case:2}
,\end{align*}
where the summation over $p'_{j+1} \ne p_{j+1}$ means to take the summation over $p'_{j+1} \in \bits^\ell \setminus \Set{p_{j+1}}$.
\end{enumerate}
For ease of understanding, we expand the above succinct expression in \Cref{counter-strucutre:psi:format:case:2} as follows:
\begin{align*}
\ket{\phi^{(t,i)}}
= & 
\ket{i}_{\reg{gc}} \ket{i}_{\reg{lc}} \ket{p_1, \ldots, p_i}_{\reg{p_1, \ldots, p_i}}\ket{\rho^{(t,i)}} ~+  \\ 
&
\ket{i}_{\reg{gc}} \ket{i-1}_{\reg{lc}} \ket{p_1, \ldots, p_{i-1}}_{\reg{p_1, \ldots, p_{i-1}}} \sum_{p'_i \ne p_i} \ket{p'_i}_{\reg{p_i}} \ket{\rho^{(t,i)}_{ p'_i}} ~+  \\ 
&
\ket{i}_{\reg{gc}} \ket{i-2}_{\reg{lc}} \ket{p_1, \ldots, p_{i-2}}_{\reg{p_1, \ldots, p_{i-2}}} \sum_{p'_{i-1} \ne p_{i-1}} \ket{p'_{i-1}}_{\reg{p_{i-1}}} \ket{\rho^{(t,i)}_{p'_{i-1}}} ~+  \\ 
& \cdots \\ 
&
\ket{i}_{\reg{gc}} \ket{1}_{\reg{lc}} \ket{p_1}_{\reg{p_1}} \sum_{p'_{2} \ne p_2} \ket{p'_2}_{\reg{p_2}} \ket{\rho^{(t,i)}_{p'_2}} ~+  \\ 
&
\ket{i}_{\reg{gc}} \ket{0}_{\reg{lc}} \sum_{p'_{1}} \ket{p'_1}_{\reg{p_1}} \ket{\rho^{(t,i)}_{p'_1 \ne p_1}} 
.\end{align*}
\end{lemma}

\begin{proof}[Proof of \Cref{counter-structure:lemma}]
We prove this lemma via mathematical induction over the operations performed by the simulator $\Sim$. That is, we first prove that the initial state at the very beginning of the MnR game is of the format shown in \Cref{counter-strucutre:psi:format:case:2} (i.e., the {\bf Base Case}), and then show that each movement of $\Sim$ will lead to a new state of the same format shown in \Cref{counter-strucutre:psi:format:case:1} or \Cref{counter-strucutre:psi:format:case:1}, depending on which case between \Cref{counter-structure:lemma:state:case:1,counter-structure:lemma:state:case:2} will happen  (i.e., the {\bf Induction Step}).

In the sequel, we consider a fixed pair $(J, \msf{pat})$ and $H^{(J, \msf{pat})}_k$. We emphasize that the lower-case ``$k$'' denotes the hybrid index. Do not confuse it with the capitalized ``$K$'' which denotes the round complexity of the original $\langle \Prover, \Verifier\rangle$. 

\para{Base Case.} This corresponds to the state at the very beginning of the MnR game. By definition, the initial sate is of the format $\ket{0}_{\reg{gc}}\ket{0}_{\reg{lc}}\ket{\rho}$. It satisfies the format shown in \Cref{counter-strucutre:psi:format:case:2} (by setting $\ket{\rho^{(0,0)}} \coloneqq \ket{\rho}$).

\para{Induction Step.} We assume that the lemma is true for the global counter's $t$-th arrival at value $i$, namely, the current overall state $\ket{\phi^{(t,i)}}$ satisfies the format shown in \Cref{counter-structure:lemma}. We prove that the simulator's next query, no matter what this query is, will lead to a new state that satisfies the format shown in \Cref{counter-structure:lemma} as well.

Toward that, let us first classify the possible movements of $\Sim$ when the state is $\ket{\phi^{(t,i)}}$.
\begin{itemize}
 \item
 \underline{Type 1:} after applying the local operator $S$ and measuring register $\reg{u}$, $\Sim$ makes the $\msf{sq}(i+1)$ query for some $i \in [K]$. (Recall that when the global counter is $i$ and $\Sim$ is about to make a $\msf{sq}(j)$ query, then it must be the $\msf{sq}(i+1)$ query.) 

 \item
 \underline{Type 2:} after applying the local operator $S$ and measuring register $\reg{u}$, $\Sim$ makes a $\tilde{V}$ query that is not the $\msf{sq}(i+1)$ query. (This could happen only if the global counter $i \ge k$)

\item
 \underline{Type 3:} after applying the local operator $S$ and measuring register $\reg{u}$, $\Sim$ makes a $\tilde{V}^\dagger$ query. (This could happen only if the global counter $i > k$)

 \item
 \underline{Type 4:} after applying the local operator $S$ and measuring register $\reg{u}$, $\Sim$ makes a $\ddot{V}$ query. (This could happen if the global counter $i < k$.)

 \item
 \underline{Type 5:} after applying the local operator $S$ and measuring register $\reg{u}$, $\Sim$ makes a $\ddot{V}^\dagger$ query. (This could happen if the global counter $i \le k$.)
 \end{itemize} 
The proof for the above five cases are very similar. In the following, we only show the proofs for \underline{Type 1}, \underline{Type 2}, and \underline{Type 3}. They are the most representative cases because  \underline{Type 1} illustrates why the statement of \Cref{counter-structure:lemma} contains two separate cases (i.e., \Cref{counter-structure:lemma:state:case:1,counter-structure:lemma:state:case:2}), and \underline{Type 2} and \underline{Type 3} cover the both the ``going-up'' $\uparrow$-query and the ``going-down'' $\downarrow$-query.

\subpara{Proof of Type 1:} From our induction assumption, we know that right before the $\msf{sq}(i+1)$ query, the state $\ket{\phi^{(t,i)}}$ is of the format shown in either \Cref{counter-strucutre:psi:format:case:1} or \Cref{counter-strucutre:psi:format:case:2}.

Next, we further claim that the state $\ket{\phi^{(t,i)}}$ must be of the format shown in \Cref{counter-strucutre:psi:format:case:2} (i.e., we know for sure that we are in \Cref{counter-structure:lemma:state:case:2}). Let us prove this: Assume for contradiction that this is not the case, then $\ket{\phi^{(t,i)}}$ is of the format shown in \Cref{counter-strucutre:psi:format:case:1} and in particular, the local counter register contains the (classical) value $i - 1$. Now, since the next query is the $\msf{sq}(i+1)$ query, it will invoke the a measurement on the query made to the oracle $H$ oracle query. Then, from {\bf Step 3} in \Cref{CPM:V:unitary:implementation}, we know that $H$ will be queried on registers $\reg{p_1 \ldots p_i}$ and thus the measurement outcome $\vb{p}_{i+1}$ is of length exactly $z_{i+1} = i$. However, this contradicts the requirement at \Cref{CPM:MnR:game:modify:2:step:4:a} (or otherwise the game is aborted prematurely at this step) of \Cref{CPM:MnR:game:modify:2}. Thus, the state $\ket{\phi^{(t,i)}}$ must be of the format shown in \Cref{counter-strucutre:psi:format:case:2}.

Therefore, the $\msf{sq}(i+1)$ will be handled as follows:
\begin{enumerate}
\item 
$U_{gc}$ is first applied to increase the global counter from $\ket{i}_{\reg{gc}}$ to $\ket{i+1}_{\reg{gc}}$. 

\item 
The $A_{i+1}$ is applied;
\item 
The query to $H$ will be made, which invokes the projection on the registers $\reg{p_1 \ldots p_i}$ to value $\ket{p_1, \ldots, p_{i+1}}_{\reg{p_1 \ldots p_{i+1}}}$. (Recall that we are currently in the ``sub-normalized'' game $H^{(J, \msf{pat})}_k$. Thus, the measurement has been replaced with this projection.) Also, as we just argued, the initial state $\ket{\phi^{(t,i)}}$ was of the format shown in \Cref{counter-strucutre:psi:format:case:2}. Thus, all the branches in \Cref{counter-strucutre:psi:format:case:2}, except for the first one,  will be ``killed'' by the projector $\ket{p_1, \ldots, p_{i+1}}_{\reg{p_1 \ldots p_{i+1}}}$. Therefore, the resulting state would be of the following format:
\begin{equation} \label[Expression]{counter-structure:proof:type-1:eq:1} 
\ket{i+1}_{\reg{gc}} \ket{i}_{\reg{lc}} \ket{p_1, \ldots, p_{i}}_{\reg{p_1, \ldots, p_{i}}}\ket{p_{i+1}}_{\reg{p_{i+1}}} \ket{\rho}.
\end{equation}

\item
Next, the behavior depends on the the local counter value (in superposition) and the value $b_{i+1}$ specified in $\msf{pat}$. In particular:
\begin{enumerate}
\item 
If $b_{i+1} = 0$: the operator $D_{i+1}C_{i+1}B_{i+1}$ (defined in \Cref{CPM:V:unitary}) will be applied to \Cref{counter-structure:proof:type-1:eq:1}, resulting in the following state:
\begin{align*}
 & D_{i+1}C_{i+1}B_{i+1} \ket{i+1}_{\reg{gc}} \ket{i}_{\reg{lc}} \ket{p_1, \ldots, p_{i}}_{\reg{p_1, \ldots, p_{i}}}\ket{p_{i+1}}_{\reg{p_{i+1}}} \ket{\rho} \\ 
= ~ &
\ket{i+1}_{\reg{gc}} \ket{i+1}_{\reg{lc}} \ket{p_1, \ldots, p_{i}}_{\reg{p_1, \ldots, p_{i}}}\ket{p_{i+1}}_{\reg{p_{i+1}}} C_{i+1,0}B_{i+1,0}\ket{\rho} 
\numberthis \label{counter-structure:proof:type-1:derivation:0:1} \\ 
= ~ &
\ket{i+1}_{\reg{gc}} \ket{i+1}_{\reg{lc}} \ket{p_1, \ldots, p_{i+1}}_{\reg{p_1, \ldots, p_{i+1}}} \ket{\rho^{(z, i+1)}}
\numberthis \label{counter-structure:proof:type-1:derivation:0:2} 
 \end{align*} 
 where \Cref{counter-structure:proof:type-1:derivation:0:1} follows from the definition of $D_{i+1}$, $C_{i+1}$, and $B_{i+1}$ (see \Cref{CPM:V:unitary}), and \Cref{counter-structure:proof:type-1:derivation:0:2} follows by defining $\ket{\rho^{(z, i+1)}} \coloneqq C_{i+1,0}B_{i+1,0}\ket{\rho}$. 

 Obviously, \Cref{counter-structure:proof:type-1:derivation:0:2} satisfies the format of \Cref{counter-strucutre:psi:format:case:2} with $t$ updated to $z$ and $i$ updated to $i+1$ (i.e., we assume this is the global counter's $z$-th arrival at value $i+1$).

\item 
If $b_{i+1} = 1$: In this case, first the query will be answered and then the oracle is programmed. Note that before the oracle is programmed, it must hold that $H(p_1, \ldots, p_{i+1}) = 0$. Therefore,  by definition of $\tilde{V}$, only the operator $C_{i+1,0}$ will be effectively applied, resulting in the following state:
\begin{align*}
 &  \ket{i+1}_{\reg{gc}} \ket{i}_{\reg{lc}} \ket{p_1, \ldots, p_{i}}_{\reg{p_1, \ldots, p_{i}}}\ket{p_{i+1}}_{\reg{p_{i+1}}} C_{i+1, 0}\ket{\rho} \\ 
= ~&
\ket{i+1}_{\reg{gc}} \ket{i}_{\reg{lc}} \ket{p_1, \ldots, p_{i+1}}_{\reg{p_1, \ldots, p_{i+1}}} \ket{\rho^{(z, i+1)}}
\numberthis \label{counter-structure:proof:type-1:derivation:1:1} 
, \end{align*} 
where \Cref{counter-structure:proof:type-1:derivation:1:1} follows by defining $\ket{\rho^{(z, i+1)}} \coloneqq C_{i+1,0}\ket{\rho}$. 

 Obviously, \Cref{counter-structure:proof:type-1:derivation:1:1} satisfies the format of \Cref{counter-strucutre:psi:format:case:1} with $t$ updated to $z$ and $i$ updated to $i+1$ (i.e., we assume this is the global counter's $z$-th arrival at value $i+1$).
\end{enumerate}
\end{enumerate}

This finishes the proof of \underline{Type 1}.

\subpara{Proof of Type 2:} From our induction assumption, we know that right before the $\msf{sq}(i+1)$ query, the state $\ket{\phi^{(t,i)}}$ is of the format shown in either \Cref{counter-strucutre:psi:format:case:1} or \Cref{counter-strucutre:psi:format:case:2}.

Next, we further claim that the state $\ket{\phi^{(t,i)}}$ must be of the format shown in \Cref{counter-strucutre:psi:format:case:2} (i.e., we know for sure that we are in \Cref{counter-structure:lemma:state:case:2}). Let us prove this: Assume for contradiction that this is not the case, then the last query is exactly the $\msf{sq}(i)$ query. At this moment, the $\msf{sq}(i+1)$ query has not been made. Then, it follows from \hyperlink{CPM:simp-assumption:3}{{\bf Assumption 3}} that the next query is not a $\tilde{V}$ query.
However, this contradicts the fact that this \underline{Type 2} is a $\tilde{V}$ query.

Therefore, in the following, we only need to show the derivation starting from the state $\ket{\phi^{(t,i)}}$ shown in \Cref{counter-strucutre:psi:format:case:2}.


In this type, $\Sim$ first applies the local operator $S$. Then, she measures the $\reg{u}$ register. Recall that we are in the ``sub-normalized'' game $H^{(J, \msf{pat})}_{k}$, and thus this measurement is replaced with a projector $\ketbra{\downarrow}_{\reg{u}}$ (it must be a $\downarrow$ because in this case the coming query is $\tilde{V}$). Finally, $\Sim$ makes a $\tilde{V}$ query. Notation-wise, we assume this leads to the global counter's $z$-th arrival at value $i+1$.

In summary, the state $\ket{\phi^{(t,i)}}$ evolves as follows 
\begin{align*}
 \ket{\phi^{(z,i+1)}} 
 &= 
 \tilde{V} \ketbra{\downarrow}_{\reg{u}} S \ket{\phi^{(t,i)}} 
\\
& = 
\tilde{V} \ketbra{\downarrow}_{\reg{u}} S 
\bigg(
\ket{i}_{\reg{gc}} \ket{i}_{\reg{lc}} \ket{p_1, \ldots, p_i}_{\reg{p_1, \ldots, p_i}}\ket{\rho^{(t,i)}} ~+  \\
& \hspace{8em} 
\ket{i}_{\reg{gc}} \sum^{i-1}_{j = 0} \ket{j}_{\reg{lc}} \ket{p_1, \ldots, p_j}_{\reg{p_1, \ldots, p_j}} \sum_{p'_{j+1} \ne p_{j+1}}\ket{p'_{j+1}}_{\reg{p_{j+1}}} \ket{\rho^{(t,i)}_{p'_{j+1}}}
\bigg) 
\numberthis \label{counter-strucutre:proof:case2:eq:1} \\ 
& =
\tilde{V}
\bigg(\ket{i}_{\reg{gc}} \ket{i}_{\reg{lc}} \ket{p_1, \ldots, p_i}_{\reg{p_1, \ldots, p_i}} \ketbra{\downarrow}_{\reg{u}}S \ket{\rho^{(t,i)}} ~+  \\
& \hspace{6em} 
\ket{i}_{\reg{gc}} \sum^{i-1}_{j = 0} \ket{j}_{\reg{lc}} \ket{p_1, \ldots, p_j}_{\reg{p_1, \ldots, p_j}} \sum_{p'_{j+1} \ne p_{j+1}}\ket{p'_{j+1}}_{\reg{p_{j+1}}} \ketbra{\downarrow}_{\reg{u}}S \ket{\rho^{(t,i)}_{p'_{j+1}}}
\bigg) 
\numberthis \label{counter-strucutre:proof:case2:eq:2} \\ 
& =
D_{i+1}C_{i+1}B_{i+1}A_{i+1}\ket{i+1}_{\reg{gc}} \ket{i}_{\reg{lc}} \ket{p_1, \ldots, p_i}_{\reg{p_1, \ldots, p_i}} \ketbra{\downarrow}_{\reg{u}}S \ket{\rho^{(t,i)}}
~+ \\ 
& \hspace{2em}
C'_{i+1}A_{i+1}\ket{i+1}_{\reg{gc}} \sum^{i-1}_{j = 0} \ket{j}_{\reg{lc}} \ket{p_1, \ldots, p_j}_{\reg{p_1, \ldots, p_j}} \sum_{p'_{j+1} \ne p_{j+1}}\ket{p'_{j+1}}_{\reg{p_{j+1}}} \ketbra{\downarrow}_{\reg{u}}S \ket{\rho^{(t,i)}_{p'_{j+1}}} 
\numberthis \label{counter-strucutre:proof:case2:eq:3} \\ 
& =
D_{i+1}C_{i+1}B_{i+1}\ket{i+1}_{\reg{gc}} \ket{i}_{\reg{lc}} \ket{p_1, \ldots, p_i}_{\reg{p_1, \ldots, p_i}} A_{i+1} \ketbra{\downarrow}_{\reg{u}}S \ket{\rho^{(t,i)}}
~+ \\ 
& \hspace{2em}
\ket{i+1}_{\reg{gc}} \sum^{i-1}_{j = 0} \ket{j}_{\reg{lc}} \ket{p_1, \ldots, p_j}_{\reg{p_1, \ldots, p_j}} \sum_{p'_{j+1} \ne p_{j+1}}\ket{p'_{j+1}}_{\reg{p_{j+1}}} C'_{i+1}A_{i+1} \ketbra{\downarrow}_{\reg{u}}S \ket{\rho^{(t,i)}_{p'_{j+1}}} 
\numberthis \label{counter-strucutre:proof:case2:eq:4} \\
& =
D_{i+1}C_{i+1}B_{i+1}\ket{i+1}_{\reg{gc}} \ket{i}_{\reg{lc}} \ket{p_1, \ldots, p_i}_{\reg{p_1, \ldots, p_i}}  \sum_{p'_{i+1} \in \bits^{\ell}} \ket{p'_{i+1}}_{\reg{p_{i+1}}} \ket{\rho^{(t,i)}_{p'_{i+1}}}
~+ \\ 
& \hspace{2em}
\ket{i+1}_{\reg{gc}} \sum^{i-1}_{j = 0} \ket{j}_{\reg{lc}} \ket{p_1, \ldots, p_j}_{\reg{p_1, \ldots, p_j}} \sum_{p'_{j+1} \ne p_{j+1}}\ket{p'_{j+1}}_{\reg{p_{j+1}}} C'_{i+1}A_{i+1} \ketbra{\downarrow}_{\reg{u}}S \ket{\rho^{(t,i)}_{p'_{j+1}}} 
\numberthis \label{counter-strucutre:proof:case2:eq:5} \\  
& =
D_{i+1}C_{i+1}B_{i+1}\ket{i+1}_{\reg{gc}} \ket{i}_{\reg{lc}} \ket{p_1, \ldots, p_i}_{\reg{p_1, \ldots, p_i}}  
\bigg( 
\ket{p_{i+1}}_{\reg{p_{i+1}}} \ket{\rho^{(t,i)}_{p_{i+1}}} 
~+ \\ 
& \hspace{22em}
\sum_{p'_{i+1} \ne p_{i+1}} \ket{p'_{i+1}}_{\reg{p_{i+1}}} \ket{\rho^{(t,i)}_{p'_{i+1}}}
\bigg)
~+ \\ 
& \hspace{4em}
\ket{i+1}_{\reg{gc}} \sum^{i-1}_{j = 0} \ket{j}_{\reg{lc}} \ket{p_1, \ldots, p_j}_{\reg{p_1, \ldots, p_j}} \sum_{p'_{j+1} \ne p_{j+1}}\ket{p'_{j+1}}_{\reg{p_{j+1}}} C'_{i+1}A_{i+1} \ketbra{\downarrow}_{\reg{u}}S \ket{\rho^{(t,i)}_{p'_{j+1}}} 
\\  
& =
D_{i+1}\ket{i+1}_{\reg{gc}} \ket{i}_{\reg{lc}} \ket{p_1, \ldots, p_i}_{\reg{p_1, \ldots, p_i}}  
\bigg( 
\ket{p_{i+1}}_{\reg{p_{i+1}}} C_{i+1,1}B_{i+1,1} \ket{\rho^{(t,i)}_{p_{i+1}}} 
~+ \\ 
& \hspace{16em}
\sum_{p'_{i+1} \ne p_{i+1}} \ket{p'_{i+1}}_{\reg{p_{i+1}}} C_{i+1,0}B_{i+1,0}\ket{\rho^{(t,i)}_{p'_{i+1}}}
\bigg)
~+ \\ 
& \hspace{2em}
\ket{i+1}_{\reg{gc}} \sum^{i-1}_{j = 0} \ket{j}_{\reg{lc}} \ket{p_1, \ldots, p_j}_{\reg{p_1, \ldots, p_j}} \sum_{p'_{j+1} \ne p_{j+1}}\ket{p'_{j+1}}_{\reg{p_{j+1}}} C'_{i+1}A_{i+1} \ketbra{\downarrow}_{\reg{u}}S \ket{\rho^{(t,i)}_{p'_{j+1}}} 
\numberthis \label{counter-strucutre:proof:case2:eq:6} \\ 
& =
D_{i+1}\ket{i+1}_{\reg{gc}} \ket{i}_{\reg{lc}} \ket{p_1, \ldots, p_i}_{\reg{p_1, \ldots, p_i}}  
\bigg( 
\ket{p_{i+1}}_{\reg{p_{i+1}}}  \ket{\rho^{(z,i+1)}} 
~+\\ 
& \hspace{16em}
\sum_{p'_{i+1} \ne p_{i+1}} \ket{p'_{i+1}}_{\reg{p_{i+1}}} \ket{\rho^{(z,i+1)}_{p'_{i+1}}}
\bigg)
~+ \\ 
& \hspace{8em}
\ket{i+1}_{\reg{gc}} \sum^{i-1}_{j = 0} \ket{j}_{\reg{lc}} \ket{p_1, \ldots, p_j}_{\reg{p_1, \ldots, p_j}} \sum_{p'_{j+1} \ne p_{j+1}}\ket{p'_{j+1}}_{\reg{p_{j+1}}} \ket{\rho^{(z,i+1)}_{p'_{j+1}}} 
\numberthis \label{counter-strucutre:proof:case2:eq:7} \\ 
& =
\ket{i+1}_{\reg{gc}} \ket{i+1}_{\reg{lc}} \ket{p_1, \ldots, p_i}_{\reg{p_1, \ldots, p_i}}  
\ket{p_{i+1}}_{\reg{p_{i+1}}}  \ket{\rho^{(z,i+1)}} 
~+~ \\ 
& \hspace{6em}
\ket{i+1}_{\reg{gc}} \ket{i}_{\reg{lc}} \ket{p_1, \ldots, p_i}_{\reg{p_1, \ldots, p_i}} \sum_{p'_{i+1} \ne p_{i+1}} \ket{p'_{i+1}}_{\reg{p_{i+1}}} \ket{\rho^{(z,i+1)}_{p'_{i+1}}}
~+ \\ 
& \hspace{8em}
\ket{i+1}_{\reg{gc}} \sum^{i-1}_{j = 0} \ket{j}_{\reg{lc}} \ket{p_1, \ldots, p_j}_{\reg{p_1, \ldots, p_j}} \sum_{p'_{j+1} \ne p_{j+1}}\ket{p'_{j+1}}_{\reg{p_{j+1}}} \ket{\rho^{(z,i+1)}_{p'_{j+1}}} 
\numberthis \label{counter-strucutre:proof:case2:eq:8} \\ 
& =
\ket{i+1}_{\reg{gc}} \ket{i+1}_{\reg{lc}} \ket{p_1, \ldots, p_{i+1}}_{\reg{p_1, \ldots, p_{i+1}}}\ket{\rho^{(z,i+1)}} 
~+~ \\ 
& \hspace{8em}
\ket{i+1}_{\reg{gc}} \sum^{i}_{j = 0} \ket{j}_{\reg{lc}} \ket{p_1, \ldots, p_j}_{\reg{p_1, \ldots, p_j}} \sum_{p'_{j+1} \ne p_{j+1}}\ket{p'_{j+1}}_{\reg{p_{j+1}}} \ket{\rho^{(z,i+1)}_{p'_{j+1}}} 
\numberthis \label{counter-strucutre:proof:case2:eq:9} 
,\end{align*}
where
\begin{itemize}
\item
\Cref{counter-strucutre:proof:case2:eq:1} follows from our induction assumption that the starting state $\ket{\phi^{(t,i)}}$ is as shown in \Cref{counter-strucutre:psi:format:case:2}.

\item
\Cref{counter-strucutre:proof:case2:eq:2} follows from the fact that $\ketbra{\downarrow}_{\reg{u}}S$ does not operate on the $\reg{gc}$, $\reg{lc}$, and $\reg{p_1 \ldots p_i}$ registers. 

\item
\Cref{counter-strucutre:proof:case2:eq:3} follows from the definition of $\tilde{V}$ (see \Cref{CPM:V:unitary}).

\item
\Cref{counter-strucutre:proof:case2:eq:4} follows from the fact that $C'_{i+1}$ and $A_{i+1}$ do not act on  registers $\reg{p_1 \ldots p_i}$.

\item
\Cref{counter-strucutre:proof:case2:eq:5} follows by expanding the $\reg{p_{i+1}}$ register under the computational basis for the state:
$$A_{i+1} \ketbra{\downarrow}_{\reg{u}}S \ket{\rho^{(t,i)}} = \sum_{p'_{i+1} \in \bits^{\ell}} \ket{p'_{i+1}}_{\reg{p_{i+1}}} \ket{\rho^{(t,i)}_{p'_{i+1}}}.$$

\item
\Cref{counter-strucutre:proof:case2:eq:6} follows from the following definitions: 
\begin{itemize}
\item
Define $C_{i+1, 1}$ and $B_{i+1, 1}$ as the parts of $C_{i+1}$ and $B_{i+1}$ (see \Cref{CPM:V:unitary}) w.r.t.\ to the $H(p_1, \ldots, p_{i+1}) = 1$ branch, and define $C_{i+1, 0}$ and $B_{i+1, 0}$ as the parts of $C_{i+1}$ and $B_{i+1}$ (see \Cref{CPM:V:unitary}) w.r.t.\ the $H(p_1, \ldots, p_{i+1}) = 0$ branch. (Therefore, none of $C_{i+1, 1}$, $C_{i+1, 0}$, $B_{i+1, 1}$, or $B_{i+1, 0}$ act on registers $\reg{p_1 \ldots p_{i+1}}$.) 
\end{itemize}  

\item
\Cref{counter-strucutre:proof:case2:eq:7} follows by defining
\begin{align*}
 \ket{\rho^{(z,i+1)}} 
 & \coloneqq  
 C_{i+1,1}B_{i+1,1} \ket{\rho^{(t,i)}_{p_{i+1}}}   
 \\
  \ket{\rho^{(z,i+1)}_{p'_{i+1}}} 
 & \coloneqq  
C_{i+1,0}B_{i+1,0}\ket{\rho^{(t,i)}_{p'_{i+1}}} ~~\big(\forall p'_{i+1} \ne p_{i+1}\big)
 \\
\ket{\rho^{(z,i+1)}_{p'_{j+1}}} 
  & \coloneqq 
  C'_{i+1}A_{i+1} \ketbra{\downarrow}_{\reg{u}}S \ket{\rho^{(t,i)}_{p'_{j+1}}}  ~~\big(\forall j \in \Set{0,1, \ldots, i-1},~\forall p'_{j+1} \ne p_{j+1}\big)
.\end{align*}  

\item
\Cref{counter-strucutre:proof:case2:eq:8} follows by the definition of $D_{i+1}$ (see \Cref{CPM:V:unitary}).
\end{itemize}
Obviously, \Cref{counter-strucutre:proof:case2:eq:9} satisfies the format of \Cref{counter-strucutre:psi:format:case:2} with $t$ updated to $z$ and $i$ updated to $i+1$. This finishes the proof of \underline{Type 2}.

\subpara{Proof of Type 3:} From our induction assumption, we know that right before the $\msf{sq}(i+1)$ query, the state $\ket{\phi^{(t,i)}}$ is of the format shown in either \Cref{counter-strucutre:psi:format:case:1} or \Cref{counter-strucutre:psi:format:case:2}.

In the following, we show the derivation assuming the starting state $\ket{\phi^{(t,i)}}$ is of the \Cref{counter-strucutre:psi:format:case:2} format, because the other case (i.e., $\ket{\phi^{(t,i)}}$ is of the \Cref{counter-strucutre:psi:format:case:1} format) can be established using the same derivation.

In this type, $\Sim$ first applies the local operator $S$. Then, she measures the $\reg{u}$ register. Recall that we are in the ``sub-normalized'' game $H^{(J, \msf{pat})}_{k}$, and thus this measurement is replaced with a projector $\ketbra{\uparrow}_{\reg{u}}$ (it must be a $\uparrow$ because in this case the coming query is $\tilde{V}^\dagger$). Finally, $\Sim$ makes a $\tilde{V}^\dagger$ query. Notation-wise, we assume this leads to the global counter's $z$-th arrival at value $i-1$.

In summary, the state $\ket{\phi^{(t,i)}}$ evolves as follows 
\begin{align*}
 \ket{\phi^{(z,i-1)}} 
 &= 
 \tilde{V}^\dagger \ketbra{\uparrow}_{\reg{u}} S \ket{\phi^{(t,i)}} 
\\
& = 
\tilde{V}^\dagger \ketbra{\uparrow}_{\reg{u}} S 
\bigg(
\ket{i}_{\reg{gc}} \ket{i}_{\reg{lc}} \ket{p_1, \ldots, p_i}_{\reg{p_1, \ldots, p_i}}\ket{\rho^{(t,i)}} ~+  \\
& \hspace{8em} 
\ket{i}_{\reg{gc}} \sum^{i-1}_{j = 0} \ket{j}_{\reg{lc}} \ket{p_1, \ldots, p_j}_{\reg{p_1, \ldots, p_j}} \sum_{p'_{j+1} \ne p_{j+1}}\ket{p'_{j+1}}_{\reg{p_{j+1}}} \ket{\rho^{(t,i)}_{p'_{j+1}}}
\bigg) 
\numberthis \label{counter-strucutre:proof:case3:eq:1} \\ 
& = 
\tilde{V}^\dagger  
\bigg(
\ket{i}_{\reg{gc}} \ket{i}_{\reg{lc}} \ket{p_1, \ldots, p_i}_{\reg{p_1, \ldots, p_i}} \ketbra{\uparrow}_{\reg{u}} S \ket{\rho^{(t,i)}} ~+  \\
& \hspace{6em} 
\ket{i}_{\reg{gc}} \sum^{i-1}_{j = 0} \ket{j}_{\reg{lc}} \ket{p_1, \ldots, p_j}_{\reg{p_1, \ldots, p_j}} \sum_{p'_{j+1} \ne p_{j+1}}\ket{p'_{j+1}}_{\reg{p_{j+1}}} \ketbra{\uparrow}_{\reg{u}} S \ket{\rho^{(t,i)}_{p'_{j+1}}}
\bigg) 
\numberthis \label{counter-strucutre:proof:case3:eq:2} \\ 
& =
U^\dagger_{gc} A^\dagger_i B^\dagger_i C^\dagger_i D^\dagger_i  \ket{i}_{\reg{gc}} \ket{i}_{\reg{lc}} \ket{p_1, \ldots, p_i}_{\reg{p_1, \ldots, p_i}} \ketbra{\uparrow}_{\reg{u}}S \ket{\rho^{(t,i)}}
~+ \\ 
& \hspace{3em}
U^\dagger_{gc} A^\dagger_{i} C'^\dagger_{i} \ket{i}_{\reg{gc}} \sum^{i-1}_{j = 0} \ket{j}_{\reg{lc}} \ket{p_1, \ldots, p_j}_{\reg{p_1, \ldots, p_j}} \sum_{p'_{j+1} \ne p_{j+1}}\ket{p'_{j+1}}_{\reg{p_{j+1}}} \ketbra{\uparrow}_{\reg{u}}S \ket{\rho^{(t,i)}_{p'_{j+1}}} 
\numberthis \label{counter-strucutre:proof:case3:eq:3} \\ 
& =
\ket{i-1}_{\reg{gc}} \ket{i-1}_{\reg{lc}} \ket{p_1, \ldots, p_{i-1}}_{\reg{p_1, \ldots, p_{i-1}}} A^\dagger_i \ket{p_i}_{\reg{p_i}} B^\dagger_{i,1} C^\dagger_{i,1}  \ketbra{\uparrow}_{\reg{u}}S \ket{\rho^{(t,i)}}
~+ \\ 
& \hspace{2em}
 \ket{i-1}_{\reg{gc}} \sum^{i-1}_{j = 0} \ket{j}_{\reg{lc}} \ket{p_1, \ldots, p_j}_{\reg{p_1, \ldots, p_j}} \sum_{p'_{j+1} \ne p_{j+1}} A^\dagger_{i}  \ket{p'_{j+1}}_{\reg{p_{j+1}}} C'^\dagger_{i} \ketbra{\uparrow}_{\reg{u}}S \ket{\rho^{(t,i)}_{p'_{j+1}}} 
\numberthis \label{counter-strucutre:proof:case3:eq:4} \\ 
& =
\ket{i-1}_{\reg{gc}} \ket{i-1}_{\reg{lc}} \ket{p_1, \ldots, p_{i-1}}_{\reg{p_1, \ldots, p_{i-1}}} A^\dagger_i \ket{p_i}_{\reg{p_i}} B^\dagger_{i,1} C^\dagger_{i,1}  \ketbra{\uparrow}_{\reg{u}}S \ket{\rho^{(t,i)}}
~+ \\ 
& \hspace{2em}
 \ket{i-1}_{\reg{gc}} \ket{i-1}_{\reg{lc}} \ket{p_1, \ldots, p_{i-1}}_{\reg{p_1, \ldots, p_{i-1}}} \sum_{p'_{i} \ne p_{i}} A^\dagger_{i}  \ket{p'_{i}}_{\reg{p_{i}}} C'^\dagger_{i} \ketbra{\uparrow}_{\reg{u}}S \ket{\rho^{(t,i)}_{p'_{i}}}   
 \\
& \hspace{3em}
 \ket{i-1}_{\reg{gc}} \sum^{i-2}_{j = 0} \ket{j}_{\reg{lc}} \ket{p_1, \ldots, p_j}_{\reg{p_1, \ldots, p_j}} \sum_{p'_{j+1} \ne p_{j+1}}  \ket{p'_{j+1}}_{\reg{p_{j+1}}} A^\dagger_{i}  C'^\dagger_{i} \ketbra{\uparrow}_{\reg{u}}S \ket{\rho^{(t,i)}_{p'_{j+1}}} 
\numberthis \label{counter-strucutre:proof:case3:eq:5} \\ 
& =
\ket{i-1}_{\reg{gc}} \ket{i-1}_{\reg{lc}} \ket{p_1, \ldots, p_{i-1}}_{\reg{p_1, \ldots, p_{i-1}}} \ket{\rho^{(z,i-1)}}
~+  \\
& \hspace{6em}
 \ket{i-1}_{\reg{gc}} \sum^{i-2}_{j = 0} \ket{j}_{\reg{lc}} \ket{p_1, \ldots, p_j}_{\reg{p_1, \ldots, p_j}} \sum_{p'_{j+1} \ne p_{j+1}}  \ket{p'_{j+1}}_{\reg{p_{j+1}}} \ket{\rho^{(z,i-1)}_{p'_{j+1}}} 
\numberthis \label{counter-strucutre:proof:case3:eq:6} 
,\end{align*}
where 
\begin{itemize}
\item
\Cref{counter-strucutre:proof:case3:eq:1} follows from our induction assumption that the starting state $\ket{\phi^{(t,i)}}$ is as shown in \Cref{counter-strucutre:psi:format:case:2}.

\item
\Cref{counter-strucutre:proof:case3:eq:2} follows from the fact that $\ketbra{\uparrow}_{\reg{u}}S$ does not operate on the $\reg{gc}$, $\reg{lc}$, and $\reg{p_1 \ldots p_i}$ registers. 

\item
\Cref{counter-strucutre:proof:case3:eq:3} follows from the definition of $\tilde{V}$ (see \Cref{CPM:V:unitary}).

\item
\Cref{counter-strucutre:proof:case3:eq:4} follows from the definition of $D_i$ and $U_{gc}$ (see \Cref{CPM:V:unitary}),   the fact that $A_{i}$ does not act on registers $\reg{p_1\ldots p_{i-1}}$, the fact that $C'_{i}$ does not act on registers $\reg{p_1\ldots p_{i}}$, and the following definitions. 
\begin{itemize}
\item
Define $C_{i, 1}$ and $B_{i, 1}$ as the part of $C_{i}$ and $B_{i}$ (see \Cref{CPM:V:unitary}) corresponding to the $H(p_1, \ldots, p_i) = 1$ branch (and thus $C_{i, 1}$ and $B_{i, 1}$ do not act on registers $\reg{p_1 \ldots p_i}$).
\end{itemize} 

\item
\Cref{counter-strucutre:proof:case3:eq:5}  follows by isolating the $(j = i-1)$ term and that $A_{i}$ does not act on registers $\reg{p_1\ldots p_{i-1}}$.

\item
\Cref{counter-strucutre:proof:case3:eq:6} follows by defining
\begin{align*}
 \ket{\rho^{(z,i-1)}} 
 & \coloneqq  
 A^\dagger_i \ket{p_i}_{\reg{p_i}} B^\dagger_{i,1} C^\dagger_{i,1}  \ketbra{\uparrow}_{\reg{u}}S \ket{\rho^{(t,i)}} 
 + 
 \sum_{p'_{i} \ne p_{i}} A^\dagger_{i}  \ket{p'_{i}}_{\reg{p_{i}}} C'^\dagger_{i} \ketbra{\uparrow}_{\reg{u}}S \ket{\rho^{(t,i)}_{p'_{i}}}  
 \\
\ket{\rho^{(z,i-1)}_{p'_{j+1}}} 
  & \coloneqq 
  A^\dagger_{i}  C'^\dagger_{i} \ketbra{\uparrow}_{\reg{u}}S \ket{\rho^{(t,i)}_{p'_{j+1}}} ~~\big(\forall j \in \Set{0,1, \ldots, i-2},~\forall p'_{j+1} \ne p_{j+1}\big)
.\end{align*}  
\end{itemize}
Obviously, \Cref{counter-strucutre:proof:case3:eq:6} satisfies the format of \Cref{counter-strucutre:psi:format:case:2} with $t$ updated to $z$ and $i$ updated to $i-1$. This finishes the proof of \underline{Type 3}.

\vspace{1em}
This finishes the proof of \Cref{counter-structure:lemma}.

\end{proof}

\subsection{Prove \Cref{lem:H:IND}: Re-Stating the Hybrids}
\label{sec:CPM:hybrids:restate}

As explained at the beginning of \Cref{sec:counter-structure}, we now introduce an alternative characterization of the hybrids. This perspective will prove valuable for the forthcoming mathematical derivations in \Cref{sec:CMP:warm-up,sec:CMP:full}.

Formally, we first define some auxiliary unitaries in \Cref{re-define:unitaries}. Subsequently, we present the new description for the hybrids in \Cref{MnR:game:redefine:hybrids}. Finally, in \Cref{MnR:game:redefine:hybrids:cor}, we demonstrate the equivalence of this new description to the original one introduced in \Cref{sec:orignal:hybrids}.

For a fixed $(J, \msf{pat})$, recall that $\msf{pat}$ contain a fix sequence $\vb{p} = (p_1, \ldots, p_K)$. For such a fixed $\vb{p}$, we re-define (in \Cref{re-define:unitaries}) the unitaries $\Set{(A_k, B_k, C_k, D_k)}_{k \in [K]}$ {\em that depend on this $\vb{p}$}. We emphasize that these unitaries were originally defined in \Cref{CPM:V:unitary}. Now, we re-load them w.r.t.\ a fix $\vb{p}$. (Technically, we should have include $\vb{p}$ in the superscript, such as $A^{\vb{p}}_k$, to indicate the dependence on $\vb{p}$. But we choose to omit it for succinct notation.)

\begin{AlgorithmBox}[label={re-define:unitaries}]{Re-Define Verifier's Unitaries for Fixed \textnormal{$(p_1, \ldots, p_K)$}}
For the $\vb{p} = (p_1, \ldots, p_K)$ contained in $(J, \msf{pat})$, we keep the $U_{gc}$, $U_{lc}$, and $A_k$ the same as in \Cref{CPM:V:unitary}, but we re-define the unitaries $B_k$, $C_k$, $\ddot{C}_k$, and $D_k$ ($\forall k \in [K]$) as follows:
\begin{enumerate} 
\item \label{CPM:renaming:Bi}
Let $B_k = \ketbra{p_1, \ldots, p_k}_{\reg{p_1}\ldots\reg{p_k}} \tensor B_{k,1} + \sum_{(p'_1, \ldots, p'_k) \ne (p_1, \ldots, p_k)}  \ketbra{p'_1, \ldots, p'_k}_{\reg{p_1}\ldots\reg{p_k}} \tensor I$. Note that $B_{k,1}$ is exactly the honest verifier's unitary $V$ to generate message $v_k$ (see \Cref{CPM:V:unitary}), which acts non-trivially only on $\reg{v_k}$ and $\reg{w}$, and works as identity on other registers.

\item \label{CPM:renaming:Ci}
Let $C_k = \ketbra{p_1, \ldots, p_k}_{\reg{p_1}\ldots\reg{p_k}} \tensor C_{k,1} + \sum_{(p'_1, \ldots, p'_k) \ne (p_1, \ldots, p_k)}  \ketbra{p'_1, \ldots, p'_k}_{\reg{p_1}\ldots\reg{p_k}} \tensor C_{k,0}$. Note that $C_{k,1}$ is the swap operator between registers $\reg{m}$ and $\reg{v_k}$, and $C_{k,0}$ is the swap operator between registers $\reg{m}$ and $\reg{t_k}$.

\item \label{CPM:renaming:ddotCi}
Let $\ddot{C}_k = \ketbra{p_1, \ldots, p_k}_{\reg{p_1}\ldots\reg{p_k}} \tensor I + \sum_{(p'_1, \ldots, p'_k) \ne (p_1, \ldots, p_k)}  \ketbra{p'_1, \ldots, p'_k}_{\reg{p_1}\ldots\reg{p_k}}  \tensor C_{k,0}$. (Recall the definition of $C_{k,0}$ from \Cref{CPM:renaming:Ci}.)

\item \label{CPM:renaming:Di}
Let $D_k = \ketbra{p_1, \ldots, p_k}_{\reg{p_1}\ldots\reg{p_k}}  \tensor D_{k,1} + \sum_{(p'_1, \ldots, p'_k) \ne (p_1, \ldots, p_k)}  \ketbra{p'_1, \ldots, p'_k}_{\reg{p_1}\ldots\reg{p_k}}  \tensor I$. Note that $D_{k,1}$ is the local-counter increasing unitary $U_{lc}$, which acts non-trivially only on $\reg{lc}$. 
\end{enumerate}
\end{AlgorithmBox}

It will be beneficial to monitor the registers on which the unitaries (in \Cref{re-define:unitaries}) have non-trivial effects. For the reader's ease, we summarize this in \Cref{CPM:non-trivial-registers:table}. For a unitary listed in the table, it operates as the identity on registers not indicated in its row.
\xiao{Note that {\bf Assumptions 3} does not have any effect.}
\begin{table}[!htb]
\centering
\caption{Registers on which the unitaries acts non-trivially}
\label{CPM:non-trivial-registers:table}
\vspace{0.5em}
\begin{tabular}{C{50pt} C{70pt} c C{50pt} C{70pt}}
\toprule
Unitary operators & Non-trivial registers & \phantom{abcde}  & Unitary operators & Non-trivial registers \\ 
\cmidrule{1-2} \cmidrule{4-5}
\addlinespace[0.3em]
$A_k$    	& $\reg{p_k}, \reg{m}$ & & 
$\ddot{C}_k$    	&  $\reg{p_1 \ldots p_k}$,  $\reg{t_k}$, $\reg{m}$   \\ 
\addlinespace[0.3em]
$B_k$ 		& $\reg{p_1 \ldots p_k}$,  $\reg{v_k}$, $\reg{w}$   & & 
$D_k$		&  $\reg{p_1 \ldots p_k}$, $\reg{lc}$ \\
\addlinespace[0.3em]
$B_{k,1}$ 	& $\reg{v_k}$, $\reg{w}$  & & 
$D_{k,1}$		& $\reg{lc}$   \\ 
\addlinespace[0.3em]
$C_k$    	&  $\reg{p_1 \ldots p_k}$,  $\reg{v_k}$, $\reg{t_k}$, $\reg{m}$ & & 
$U_{gc}$ & $\reg{gc}$  \\ 
\addlinespace[0.3em]
$C_{k,1}$   & $\reg{v_k}$, $\reg{m}$  & & 
$S$ 		& $\reg{m}$, $\reg{u}$, $\reg{s}$  \\ 
\addlinespace[0.3em]
$C_{k,0}$  	& $\reg{t_k}$, $\reg{m}$ & & 
&  \\   
\bottomrule
\end{tabular}
\end{table}

Next, we re-state the hybrids $\Set{H^{(J, \msf{pat})}_k}_{k \in [K] \cup \Set{0}}$ in \Cref{MnR:game:redefine:hybrids} and prove in \Cref{MnR:game:redefine:hybrids:cor} that they are indeed equivalent to the original version defined in \Cref{sec:orignal:hybrids}.

\begin{GameBox}[label={MnR:game:redefine:hybrids}]{Re-Define the Hybrids \textnormal{$\Set{H^{(J, \msf{pat})}_k}_{k \in [K] \cup \Set{0}}$} }
{\bf Parameters.} Use the same $(J, \msf{pat})$ as defined \Cref{CPM:MnR:game:subnormalized}. Use the unitaries defined in \Cref{re-define:unitaries}. We emphasize that this game inherits the ``sub-normalized'' nature of \Cref{CPM:MnR:game:subnormalized} for the fixed $(J, \msf{pat})$ (this was already there in its original version defined in \Cref{sec:orignal:hybrids}), where all the measurements are replaced with projectors to the corresponding value contained in $\msf{pat}$.

\para{Hybrid $H^{(J, \msf{pat})}_k$ $(k \in [K] \cup \Set{0})$.} The hybrid behaves as follows for all $i \in [K]$:
\begin{enumerate}
	\item 
	{\bf When $\Sim$ makes the $\msf{sq}(i)$ Query:} Note that by definition, this is the first $\downarrow$-query that brings the global counter from $\ket{i-1}_{\reg{gc}}$ to $\ket{i}_{\reg{gc}}$. The hybrid behaves according to the value $b_i$ contained in $J$:
	\begin{itemize}
	 \item 
	 If $b_i = 0$: answers this query by applying the following operator the overall state:
	$$D_{i}C_{i}B_{i} \ketbra{p_1, \ldots, p_i}_{\reg{p_1\ldots p_i}} A_{i} U_{gc},$$ 
	where recall that the values $p_1, \ldots, p_i$ are specified in $\msf{pat}$.

	 \item
	 If $b_i = 1$: answers this query by applying the following operator the overall state:
	$$C_{i,0} \ketbra{p_1, \ldots, p_i}_{\reg{p_1\ldots p_i}} A_{i} U_{gc},$$
	where recall that the values $p_1, \ldots, p_i$ are specified in $\msf{pat}$.
	 \end{itemize}

	\item
	{\bf When $\Sim$ makes a qeury between the $\msf{sq}(j-1)$ and $\msf{sq}(j)$ queries:} it behaves based on the query type and the value of the global counter value $z \in [K]\cup\Set{0}$:
	\begin{enumerate}
	\item \label[Case]{MnR:game:redefine:hybrids:case:a}
	{\bf For $\downarrow$-query and $z < k$:} This corresponds to a $\ddot{V}$-query bringing the global counter from $\ket{z}_{\reg{gc}}$ to $\ket{z+1}_{\reg{gc}}$. The hybrid answers it by applying $D_{z+1}\ddot{C}_{z+1} A_{z+1} U_{gc}$ to the overall state.

	\item
	{\bf For $\downarrow$-query and $z \ge k$:} This corresponds to a $\tilde{V}$-query bringing the global counter from $\ket{z}_{\reg{gc}}$ to $\ket{z+1}_{\reg{gc}}$. The hybrid answers it by applying $D_{z+1} C_{z+1} B_{z+1} A_{z+1} U_{gc}$ to the overall state.

	\item
	{\bf For $\uparrow$-query and $z \le k$:} This corresponds to a $\ddot{V}^\dagger$-query bringing the global counter from $\ket{z}_{\reg{gc}}$ to $\ket{z-1}_{\reg{gc}}$. The hybrid answers it by applying $U^\dagger_{gc} A^\dagger_z \ddot{C}^\dagger_z D^\dagger_z$ to the overall state.

	\item \label[Case]{MnR:game:redefine:hybrids:case:d}
	{\bf For $\uparrow$-query and $z > k$:} This corresponds to a ${V}^\dagger$-query bringing the global counter from $\ket{z}_{\reg{gc}}$ to $\ket{z-1}_{\reg{gc}}$. The hybrid answers it by applying $U^\dagger_{gc} A^\dagger_z {B}^\dagger_z {C}^\dagger_z {D}^\dagger_z$ to the overall state.
	\end{enumerate}
\end{enumerate}

\end{GameBox}

\begin{corollary}[of \Cref{counter-structure:lemma}]\label{MnR:game:redefine:hybrids:cor}
The hybrids $\Set{H^{(J, \msf{pat})}_k}_{k \in [K] \cup \Set{0}}$ defined in \Cref{MnR:game:redefine:hybrids} are equivalent to those defined in \Cref{sec:orignal:hybrids}.
\end{corollary}
\begin{proof}[Proof of \Cref{MnR:game:redefine:hybrids:cor}] 
In the following, consider a fixed pair $(J, \msf{pat})$ and hybrid $H^{(J, \msf{pat})}_k$.

The main idea of this proof as follows. First, note that the original $H^{(J, \msf{pat})}_k$ (in \Cref{sec:orignal:hybrids}) makes use of the $\tilde{V}$ and $\ddot{V}$, which by definition determine what to do by comparing the global counter and the local counter (see \Cref{CPM:V:unitary,CPM:V-dummy}). On the other hand, \Cref{counter-structure:lemma} provides a full characterization of the relation among the global counter, the local counters, and the $\reg{p_1 \ldots p_K}$ registers throughout the execution, which essentially allows the hybrid to determine what to do by checking the (classical) value contained in the global counter only.	This can be easily seen by comparing the original  $H^{(J, \msf{pat})}_k$ (in \Cref{sec:orignal:hybrids}) and the one defined in \Cref{MnR:game:redefine:hybrids} (together with the re-named unitaries in \Cref{re-define:unitaries}). In the following, we show why they are equivalent for a representative type of query as an example. Other types of queries can be argued in the same manner.

\para{A Representative Example.} Consider the case when $\Sim$ makes the $\msf{sq}(i)$ query for some $i \in [K]$. By definition (and \Cref{rmk:sq:type}), this is a $\tilde{V}$ query that brings the global counter from $\ket{i-1}_{\reg{gc}}$ to $\ket{i}_{\reg{gc}}$. This query is handled as follows in the original $H^{(J, \msf{pat})}_k$ (in \Cref{sec:orignal:hybrids}) according to the definition of $\tilde{\Verifier}$ in \Cref{CPM:V:unitary:implementation}:
\begin{enumerate}
\item \label[Step]{MnR:game:redefine:hybrids:cor:proof:example:1:step:1} 
$U_{gc}$ is first applied to increase the global counter from $\ket{i-1}_{\reg{gc}}$ to $\ket{i}_{\reg{gc}}$. 

\item 
The $A_i$ is applied;
\item 
Then the query to $H$ will be made, which invokes the projection on the registers $\reg{p_1 \ldots p_i}$ to value $\ket{p_1, \ldots, p_i}_{\reg{p_1 \ldots p_i}}$. (Recall that we are currently in the ``sub-normalized'' game $H^{(J, \msf{pat})}_k$. Thus, the measurement on register $\reg{p_1 \ldots p_i}$ has been replaced with this projection.) It then follows from \Cref{counter-structure:lemma} that this projection will retain only the ``branch'' corresponding to $\ket{i-1}_{\reg{lc}}$ and ``kill'' all other branches in the superposition. Thus, 
the current overall state must be of the following format
\begin{equation} \label[Expression]{MnR:game:redefine:hybrids:cor:proof:eq:example:1} 
\ket{i}_{\reg{gc}} \ket{i-1}_{\reg{lc}} \ket{p_1, \ldots, p_{i-1}}_{\reg{p_1, \ldots, p_{i-1}}}\ket{p_i}_{\reg{p_i}} \ket{\rho}.
\end{equation}

\item
Next, the behavior depends on the the local counter value (in superposition) and the value $b_i$ specified in $\msf{pat}$. In particular:
\begin{enumerate}
\item 
If $b_i = 0$: In this case, first the oracle will be programmed and then the query will be answered using the programmed oracle. By definition of $\tilde{V}$, the operator $D_{i}C_{i}B_{i}$ (defined in \Cref{CPM:V:unitary}) will be applied to \Cref{MnR:game:redefine:hybrids:cor:proof:eq:example:1}. 

\item 
If $b_i = 1$: In this case, first the query will be answered and then the oracle is programmed. Note that before the oracle is programmed, it must hold that $H(p_1, \ldots, p_i) = 0$. Therefore,  by definition of $\tilde{V}$, only the operator $C_{i,0}$ will be effectively applied.
\end{enumerate}
\end{enumerate}
In summary, the above four steps can be unified as 
\begin{itemize}
	 \item 
	 If $b_i = 0$: apply
	$D_{i}C_{i}B_{i} \ketbra{p_1, \ldots, p_i}_{\reg{p_1\ldots p_i}} A_{i} U_{gc}$.

	 \item
	 If $b_i = 1$: apply
	$C_{i,0} \ketbra{p_1, \ldots, p_i}_{\reg{p_1\ldots p_i}} A_{i} U_{gc}$.
	 \end{itemize} 
This is exactly what happens in \Cref{MnR:game:redefine:hybrids}.

\vspace{1em}
This completes the proof of \Cref{MnR:game:redefine:hybrids:cor}.

\end{proof}

\section{Proving \Cref{lem:H:IND} (Warm-Up Case)}
\label{sec:CMP:warm-up}

In this section, we prove \Cref{lem:H:IND} for the special case $K = 2$.

In this setting, there are only three hybrids to consider, namely $H^{(J, \msf{pat})}_0$, $H^{(J, \msf{pat})}_1$, and $H^{(J, \msf{pat})}_2$. To provide better intuition, we illustrate them in \Cref{figure:CPM-3ZK:hybrids}, with difference between adjacent hybrids highlighted in \red{red color}. In more detail, \Cref{figure:CPM-3ZK:hybrids} illustrates these three hybrids for some particular $(J, \msf{pat})$:
\begin{itemize}
 \item
In hybrid $H^{(J, \msf{pat})}_0$, it can be seen from \Cref{figure:CPM-3ZK:hybrids:H0} that all the $\downarrow$-queries are answered using $\tilde{V}$ and all the $\uparrow$-queries are answered using $\tilde{V}^\dagger$.  

\item
The hybrid $H^{(J, \msf{pat})}_1$ shown in \Cref{figure:CPM-3ZK:hybrids:H1} is identical to hybrid $H^{(J, \msf{pat})}_0$ except that the $\downarrow$-queries that bring the global counter from $\ket{0}_{\reg{gc}}$ to $\ket{1}_{\reg{gc}}$ (except for  the  $\msf{sq}(1)$ query) are answered using the ``dummy-version'' unitary$\ddot{V}$, and the $\uparrow$-queries that bring the global counter from $\ket{1}_{\reg{gc}}$ to $\ket{0}_{\reg{gc}}$ are answered using the ``dummy-version'' unitary $\ddot{V}^\dagger$.

\item
The hybrid $H^{(J, \msf{pat})}_2$ shown in \Cref{figure:CPM-3ZK:hybrids:H2} is identical to hybrid $H^{(J, \msf{pat})}_1$ except that the $\downarrow$-queries that bring the global counter from $\ket{1}_{\reg{gc}}$ to $\ket{2}_{\reg{gc}}$ (except for  the  $\msf{sq}(2)$ query) are answered using the ``dummy-version'' unitary$\ddot{V}$, and the $\uparrow$-queries that bring the global counter from $\ket{2}_{\reg{gc}}$ to $\ket{1}_{\reg{gc}}$ are answered using the ``dummy-version'' unitary $\ddot{V}^\dagger$.
 \end{itemize}

\begin{figure}[!tb]
\centering
     \begin{subfigure}[t]{0.47\textwidth}
         \fbox{
         \includegraphics[width=\textwidth,page=1]{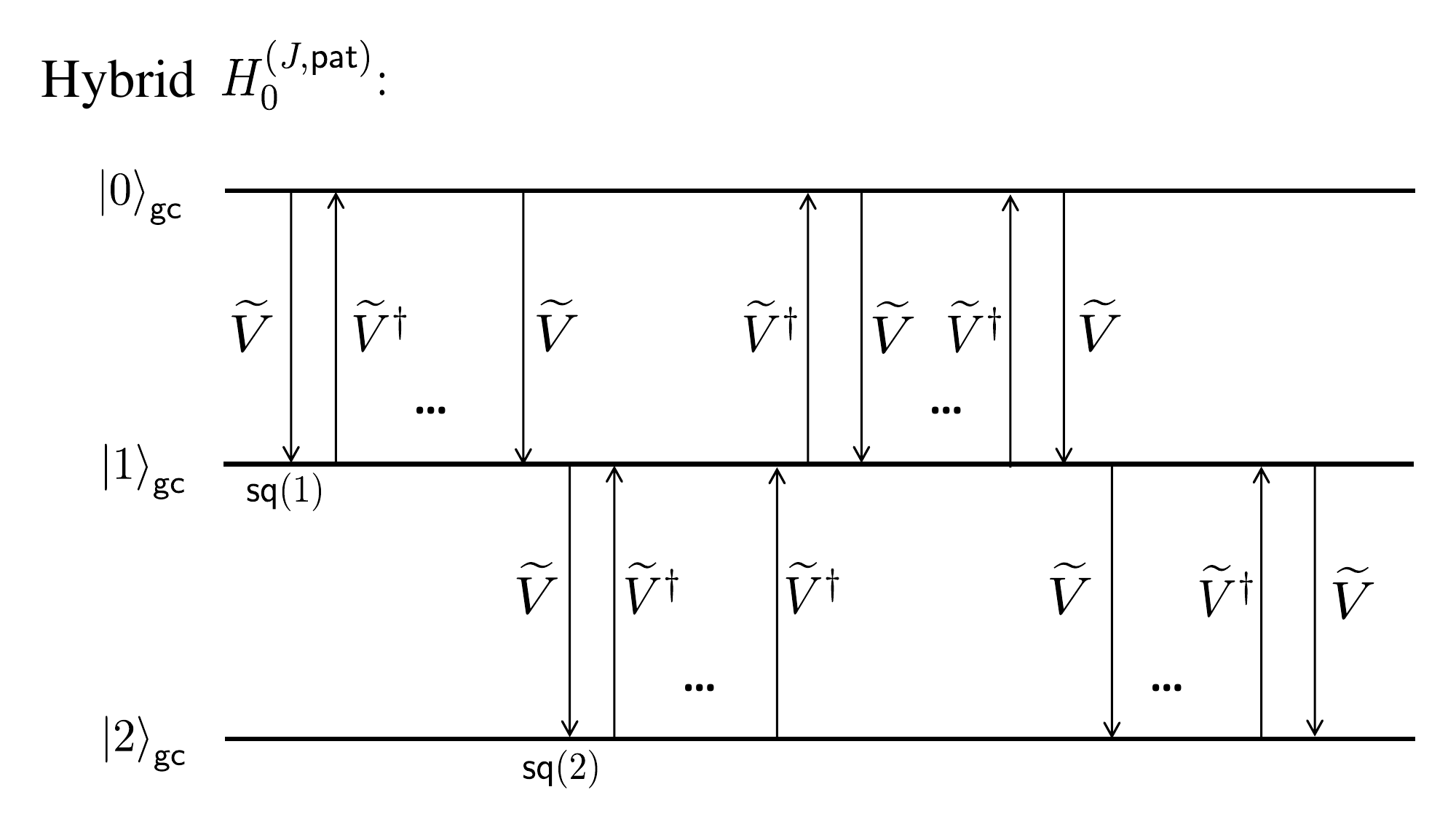}
         }
         \caption{}
         \label{figure:CPM-3ZK:hybrids:H0}
     \end{subfigure}
     \hspace{15pt}
     \begin{subfigure}[t]{0.47\textwidth}
         \centering
         \fbox{
         \includegraphics[width=\textwidth,page=2]{figures/CPM-3ZK-figures.pdf}
         }
         \caption{}
         \label{figure:CPM-3ZK:hybrids:H1}
     \end{subfigure}
     
     \centering\vspace{1em}
     \begin{subfigure}[t]{0.47\textwidth}
         
         \fbox{
         \includegraphics[width=\textwidth,page=3]{figures/CPM-3ZK-figures.pdf}
         }
         \caption{}
         \label{figure:CPM-3ZK:hybrids:H2}
     \end{subfigure}
     \caption{Illustration of Hybrids $H^{(J, \msf{pat})}_0$, $H^{(J, \msf{pat})}_1$, and $H^{(J, \msf{pat})}_2$}
     \label{figure:CPM-3ZK:hybrids}
\end{figure}

For these hybrids, to prove \Cref{lem:H:IND}, it now suffices to establish the following \Cref{lem:CPM-3ZK:H01,lem:CPM-3ZK:H12}. 
\begin{lemma}\label{lem:CPM-3ZK:H01}
For all $(J, \msf{pat})$ satisfying the requirements in \Cref{CPM:MnR:game:subnormalized}, it holds that 
$$\Pr\big[\msf{Pred}(\rho) = 1~:~(\vb{p}, \rho)\la H^{(J, \msf{pat})}_0 \big] = \Pr\big[\msf{Pred}(\rho) = 1~:~(\vb{p}, \rho)\la H^{(J, \msf{pat})}_1 \big].$$
\end{lemma}

\begin{lemma}\label{lem:CPM-3ZK:H12}
For all $(J, \msf{pat})$ satisfying the requirements in \Cref{CPM:MnR:game:subnormalized}, it holds that 
$$\Pr\big[\msf{Pred}(\rho) = 1~:~(\vb{p}, \rho)\la H^{(J, \msf{pat})}_1 \big] = \Pr\big[\msf{Pred}(\rho) = 1~:~(\vb{p}, \rho)\la H^{(J, \msf{pat})}_2 \big].$$
\end{lemma}

In the sequel, we prove these two lemmas in \Cref{sec:lem:CPM-3ZK:H01:proof,sec:lem:CPM-3ZK:H12:proof} respectively. We emphasize that in the following, we use the notation established in \Cref{sec:CPM:hybrids:restate}.

\subsection{Proving \Cref{lem:CPM-3ZK:H01}}
\label{sec:lem:CPM-3ZK:H01:proof}

We present a lemma (\Cref{lem:CPM-3ZK-hyb01:proof:invariant}) that characterizes how the overall state evolves in $H^{(J, \msf{pat})}_0$ and $H^{(J, \msf{pat})}_1$. This lemma is the major workhorse that implies \Cref{lem:CPM-3ZK:H01}. In the sequel, we first show how to establish \Cref{lem:CPM-3ZK:H01} assuming that \Cref{lem:CPM-3ZK-hyb01:proof:invariant} holds. After that, we will present the proof of \Cref{lem:CPM-3ZK-hyb01:proof:invariant}.

\begin{lemma}[Invariance in $H^{(J, \msf{pat})}_0$ and $H^{(J, \msf{pat})}_1$]\label{lem:CPM-3ZK-hyb01:proof:invariant}
Assume that during the execution of $H^{(J, \msf{pat})}_0$ (and $H^{(J, \msf{pat})}_1$), the global counter reaches value $1$ for $T$ times in total. For each $t \in [T]$, there exist (possibly sub-normalized) pure states $\Set{\ket{\rho^{(t)}_{p'_1}}}_{p'_1 \in \bits^\ell}$ so that the following holds: in hybrid $H^{(J, \msf{pat})}_0$ (resp.\ $H^{(J, \msf{pat})}_1$), when the global counter reaches value $1$ for the $t$-th time, the overall state can be written as $\ket{\phi^{(t)}}$ (resp.\ $\ket{\psi^{(t)}}$) defined as follows
\begin{align}
\ket{\phi^{(t)}} 
& =  \ket{p_1}_{\reg{p_1}}\ket{\rho^{(t)}_{p_1}}  ~+~
   \sum_{p'_1 \in \bits^\ell \setminus \Set{p_1}} \ket{p'_1}_{\reg{p_1}} C_{1,0} B^\dagger_{1,1} C^\dagger_{1,1} \ket{\rho^{(t)}_{p'_1}}, \label{lem:CPM-3ZK-hyb01:proof:target-format:H0}\\ 
\ket{\psi^{(t)}} 
& =  \ket{p_1}_{\reg{p_1}}\ket{\rho^{(t)}_{p_1}}  ~+~
 \sum_{p'_1 \in \bits^\ell \setminus \Set{p_1}} \ket{p'_1}_{\reg{p_1}} \red{C_{1,0}} \ket{\rho^{(t)}_{p'_1}}. \label{lem:CPM-3ZK-hyb01:proof:target-format:H1}
\end{align}
where the $p_1$ is defined in $\msf{pat}$ and  the unitaries $B_{1,1}$, $C_{1,0}$, and $C_{1,1}$  are as defined in \Cref{re-define:unitaries}.
\end{lemma}

\para{Finishing the Proof of \Cref{lem:CPM-3ZK:H01}}. Let us consider the last time when the global counter reaches value $1$. By \Cref{lem:CPM-3ZK-hyb01:proof:invariant}, the overall states in $H^{(J, \msf{pat})}_0$ and $H^{(J, \msf{pat})}_1$ would be of the following format respectively
\begin{align}
\ket{\phi^{(T)}} 
& = 
\overbrace{
\ket{p_1}_{\reg{p_1}}\ket{\rho^{(T)}_{p_1}}
}^{\msf{Good}}  ~+~
\overbrace{
\sum_{p'_1 \in \bits^\ell \setminus \Set{p_1}} \ket{p'_1}_{\reg{p_1}} C_{1,0} B^\dagger_{1,1} C^\dagger_{1,1} \ket{\rho^{(T)}_{p'_1}}
}^{\msf{Bad}} ,\\ 
\ket{\psi^{(T)}}
& = 
\overbrace{
\ket{p_1}_{\reg{p_1}}\ket{\rho^{(T)}_{p_1}}
}^{\msf{Good}}  ~+~
\overbrace{
 \sum_{p'_1 \in \bits^\ell \setminus \Set{p_1}} \ket{p'_1}_{\reg{p_1}} C_{1,0} \ket{\rho^{(T)}_{p'_1}}
}^{\msf{Bad}}. 
\end{align}
Since this is the last time the global counter reaches value $1$, the two hybrids $H^{(J, \msf{pat})}_0$ and $H^{(J, \msf{pat})}_1$ by definition behave identically after that (see \Cref{figure:CPM-3ZK:hybrids}). That is, the states $\ket{\phi^{(T)}}$ and $\ket{\psi^{(T)}}$ will go through the same quantum procedure (let us denote it by $\mcal{F}$) and finally measured by the predicate $\msf{Pred}$.

Next, note that the states $\ket{\phi^{(T)}}$ and $\ket{\psi^{(T)}}$ share the identical ``branch'' corresponding to $\ket{p_1}_{\reg{p_1}}$ (labeled as $\msf{Good}$ in the above equations). Also, the $\msf{Bad}$ branch will not contribute any shares to the accepting probability of the predicate $\msf{Pred}$---this is because the quantum procedure $\mcal{F}$ will not change the fact that $p'_1 \ne p_1$ for the $\msf{Bad}$ branch, and recall that, as a consequence of \Cref{counter-structure:lemma}, $\msf{Pred}$ will output $0$ if a branch in the superposition does not have any non-zero amplitude for $\ket{p_1}_{\reg{p_1}}$.	

In summary, only the $\msf{Good}$ branch matters for the remaining execution (i.e., that after the last time when the global counter reaches value 1), contributing to the event $\msf{Pred}(\rho) = 1$. Therefore, \Cref{lem:CPM-3ZK:H01} holds simply because $\ket{\phi^{(T)}}$ and $\ket{\psi^{(T)}}$ share the same $\msf{Good}$ branch.

\para{Proving \Cref{lem:CPM-3ZK-hyb01:proof:invariant}.} Now, the only task remaining is to prove \Cref{lem:CPM-3ZK-hyb01:proof:invariant}. We will demonstrate this lemma through mathematical induction on the number $t \in [T]$, indicating the time at which the global counter reaches the value $1$. Throughout this proof, we will monitor the evolution of the overall states in both $H^{(J, \msf{pat})}_0$ and $H^{(J, \msf{pat})}_1$ simultaneously. This is done in \Cref{sec:lem:CPM-3ZK-hyb01:proof:invariant:proof:base:sec,sec:lem:CPM-3ZK-hyb01:proof:invariant:proof:induction:sec} respectively.

\subsubsection{Base Case ($t = 1$)}
\label{sec:lem:CPM-3ZK-hyb01:proof:invariant:proof:base:sec} 

We first derive how the overall state evolves in $H^{(J, \msf{pat})}_0$. 

This case corresponds to the very first time the global counter reaches $1$. By definition, this is due to the $\msf{sq}(1)$ query. In more detail,  $\Sim$ will first apply her local unitary $S$, followed by the projector $\ketbra{\downarrow}_{\reg{u}}$. Next, according to the notation in \Cref{MnR:game:redefine:hybrids},
\begin{itemize}
\item
If $b_1 = 0$, then the operator $D_1C_1B_1 \ketbra{p_1}_{\reg{p_1}} A_1 U_{gc}$ will be applied;
\item 
If $b_1 = 1$, then the operator $C_{1,0} \ketbra{p_1}_{\reg{p_1}} A_1 U_{gc}$ will be applied;
\end{itemize}
In the following, we show the proof for $b_1 = 0$ only; the other case (i.e., $b_1 = 1$) can be established using the same argument.

If we assume that the initial state across all the registers are $\ket{\rho}$, then the state $\ket{\phi^{(1)}}$ will be as follows:
\begin{align*}
\ket{\phi^{(1)}} 
& =  
D_1C_1B_1 \ketbra{p_1}_{\reg{p_1}} A_1 U_{gc} \ketbra{\downarrow}_{\reg{u}} S  
\ket{\rho}
\\
& =
D_1C_1B_1  
 \ket{p_1}_{\reg{p_1}} \ket{\rho_{p_1}} 
\numberthis \label{lem:CPM-3ZK-hyb01:proof:invariant:proof:base:eq:4} \\ 
& =  
 \ket{p_1}_{\reg{p_1}}  D_{1,1}  C_{1,1}B_{1,1} \ket{\rho_{p_1}}
\numberthis \label{lem:CPM-3ZK-hyb01:proof:invariant:proof:base:eq:5} \\ 
& = 
\ket{p_1}_{\reg{p_1}} \ket{\rho^{(1)}_{p_1}} 
\numberthis \label{lem:CPM-3ZK-hyb01:proof:invariant:proof:base:eq:7} 
,\end{align*}
where 
\begin{itemize}
\item
\Cref{lem:CPM-3ZK-hyb01:proof:invariant:proof:base:eq:4} follows by defining $\ket{\rho_{p_1}} \coloneqq \bra{p_1}_{\reg{p_1}} A_1 U_{gc} \ketbra{\downarrow}_{\reg{u}} S  
\ket{\rho}$;
\item
\Cref{lem:CPM-3ZK-hyb01:proof:invariant:proof:base:eq:5} follows from the definition of $B_1$  $C_1$, and $D_1$ (see \Cref{re-define:unitaries});
\item
\Cref{lem:CPM-3ZK-hyb01:proof:invariant:proof:base:eq:7} follows by defining $\ket{\rho^{(1)}_{p_1}} \coloneqq D_{1,1} C_{1,1}B_{1,1} \ket{\rho_{p_1}}$.
\end{itemize}
It is straightforward that the $\ket{\phi^{(1)}}$ shown in \Cref{lem:CPM-3ZK-hyb01:proof:invariant:proof:base:eq:7} satisfies the format shown in \Cref{lem:CPM-3ZK-hyb01:proof:target-format:H0} when $t = 1$.

Also, note that the hybrids $H^{(J, \msf{pat})}_0$ and $H^{(J, \msf{pat})}_1$ are identical so far and thus $\ket{\phi^{(1)}} = \ket{\psi^{(1)}}$. Therefore, it follows from \Cref{lem:CPM-3ZK-hyb01:proof:invariant:proof:base:eq:7} that 
$$\ket{\psi^{(1)}} =  \ket{p_1}_{\reg{p_1}} \ket{\rho^{(1)}_{p_1}} .$$ Such a $\ket{\psi^{(1)}}$ satisfies the format shown in \Cref{lem:CPM-3ZK-hyb01:proof:target-format:H1} with $t = 1$ as well.

This finishes the proof for the base case $t =1$.

\subsubsection{Induction Step ($t \ge 2$)} 
\label{sec:lem:CPM-3ZK-hyb01:proof:invariant:proof:induction:sec}

We assume that $\ket{\phi^{(t-1)}}$ and $\ket{\psi^{(t-1)}}$ satisfy \Cref{lem:CPM-3ZK-hyb01:proof:invariant}, and show in the following that \Cref{lem:CPM-3ZK-hyb01:proof:invariant} holds when the global counter reaches $1$ for the $t$-th time.

We establish this claim by considering the following MECE (mutually exclusive and collectively exhaustive) cases:
\begin{enumerate}
\item \label[Case]{CPM-3ZK-hyb01:proof:invariant:proof:induction:case-1}
{\bf Case 1:} The $t$-th arrival at value $1$ is due to an immediate $\uparrow \downarrow$ after the $(t-1)$-th arrival. That is, after the $(t-1)$-th arrival, $\Sim$ first makes a $\uparrow$ query, bringing the global counter to $0$, and then makes an $\downarrow$ query, bringing the global counter back to $1$. 
\item \label[Case]{CPM-3ZK-hyb01:proof:invariant:proof:induction:case-2}
{\bf Case 2:} The $t$-th arrival at value $1$  is due to an immediate $\downarrow \uparrow $ after the $(t-1)$-th arrival. That is, after the $(t-1)$-th arrival, $\Sim$ first makes a $\downarrow$ query, bringing the global counter to $2$, and then makes an $\uparrow$ query, bringing the global counter back to $1$. 
\end{enumerate}

\subsubsubsection{Proof for \Cref{CPM-3ZK-hyb01:proof:invariant:proof:induction:case-1}}

We first describe formally how $\ket{\phi^{(t-1)}}$ (resp.\ $\ket{\psi^{(t-1)}}$) evolves into $\ket{\phi^{(t)}}$ (resp.\ $\ket{\psi^{(t)}}$) in \Cref{CPM-3ZK-hyb01:proof:invariant:proof:induction:case-1}:
\begin{enumerate}
 \item
$\Sim$'s local unitary $S$ is applied, followed by the projector $\ketbra{\uparrow}_{\reg{u}}$. Note that this step is identical for both $H^{(J, \msf{pat})}_0$ and $H^{(J, \msf{pat})}_1$. 

\begin{remark}[Hiding the Projector on Register $\reg{u}$]\label{rmk:CPM-3ZK:hid-proj-u}
In the sequel, we overload the notation $S$ to think of it as already including the projection $\ketbra{\uparrow}_{\reg{u}}$, and thus do not spell the $\ketbra{\uparrow}_{\reg{u}}$ out explicitly. We remark that this will not affect our proof, because what matters for $S$ in the following proof is the registers on which it operates, and the original $S$ already operates non-trivially on register $\reg{u}$ (see \Cref{CPM:non-trivial-registers:table}). 
\end{remark}

\item
An $\uparrow$-query is made, bringing the global counter from 1 to 0. According to the notation in \Cref{MnR:game:redefine:hybrids}: 
\begin{itemize}
    \item
    In $H^{(J, \msf{pat})}_0$, this corresponds to applying $U^{\dagger}_{gc} A^\dagger_1 B^\dagger_1 C^\dagger_1 D^\dagger_1$; 

    \item
    In $H^{(J, \msf{pat})}_1$, this corresponds to applying $U^{\dagger}_{gc} A^\dagger_1 \ddot{C}^\dagger_1 D^\dagger_1$. 
\end{itemize}
\item
$\Sim$ will apply her local operation $S$ again (followed by $\ketbra{\downarrow}_{\reg{u}}$ which we hide as per \Cref{rmk:CPM-3ZK:hid-proj-u}).  Note that this step is again identical for both $H^{(J, \msf{pat})}_0$ and $H^{(J, \msf{pat})}_1$.
\item
An $\downarrow$-query is made, bringing the global counter from 0 back to 1, which is the global counter's $t$-th arrival at value 1. According to the notation in \Cref{MnR:game:redefine:hybrids}: 
\begin{itemize}
    \item
    In $H^{(J, \msf{pat})}_0$, this corresponds to applying $D_1C_1B_1A_1U_{gc}$; 
    \item
    In $H^{(J, \msf{pat})}_1$, this corresponds to applying $D_1\ddot{C}_1A_1U_{gc}$. 
\end{itemize} 
\end{enumerate}
It follows from the above description (and \Cref{rmk:CPM-3ZK:hid-proj-u}) that the states $\ket{\phi^{(t)}}$ and $\ket{\psi^{(t)}}$ can be written as:
\begin{align}
\ket{\phi^{(t)}}
& = D_1C_1B_1A_1U_{gc} S U^{\dagger}_{gc} A^\dagger_1 B^\dagger_1 C^\dagger_1 D^\dagger_1 S  \ket{\phi^{(t-1)}}, \label{lem:CPM-3ZK-hyb01:proof:invariant:proof:induction:eq:k:H0}\\ 
\ket{\psi^{(t)}}
& = D_1 \ddot{C}_1A_1U_{gc} S  U^{\dagger}_{gc} A^\dagger_1 \ddot{C}^\dagger_1 D^\dagger_1 S \ket{\psi^{(t-1)}}. \label{lem:CPM-3ZK-hyb01:proof:invariant:proof:induction:eq:k:H1}
\end{align}


\para{High-Level Idea for the Sequel.}
Recall that our eventual goal is to prove that the states $\ket{\phi^{(t)}}$ and $\ket{\psi^{(t)}}$ are of the format shown in \Cref{lem:CPM-3ZK-hyb01:proof:target-format:H0,lem:CPM-3ZK-hyb01:proof:target-format:H1} in \Cref{lem:CPM-3ZK-hyb01:proof:invariant}. At a high level, we prove it by applying \Cref{lem:err-inv-com} to \Cref{lem:CPM-3ZK-hyb01:proof:invariant:proof:induction:eq:k:H0,lem:CPM-3ZK-hyb01:proof:invariant:proof:induction:eq:k:H1}. But we first need to perform some preparation work, putting \Cref{lem:CPM-3ZK-hyb01:proof:invariant:proof:induction:eq:k:H0,lem:CPM-3ZK-hyb01:proof:invariant:proof:induction:eq:k:H1} into a format that is more ``compatible'' with \Cref{lem:err-inv-com}. In the sequel, we first perform the preparation work in \Cref{lem:CPM-3ZK-hyb01:proof:invariant:proof:induction:case1:preparation-claim,lem:CPM-3ZK-hyb01:proof:invariant:proof:induction:case1:preparation-claim:2}. Then, we show on \Cpageref{lem:CPM-3ZK-hyb01:proof:invariant:proof:induction:case1:finish} how to use \Cref{lem:CPM-3ZK-hyb01:proof:invariant:proof:induction:case1:preparation-claim,lem:CPM-3ZK-hyb01:proof:invariant:proof:induction:case1:preparation-claim:2} to complete the proof for \Cref{CPM-3ZK-hyb01:proof:invariant:proof:induction:case-1}.

\begin{MyClaim}\label{lem:CPM-3ZK-hyb01:proof:invariant:proof:induction:case1:preparation-claim}
There exist (possibly sub-normalized) pure states $\Set{\dot{\rho}^{(t-1)}_{p'_1}}_{p'_1 \in \bits^\ell}$ so that the following holds
\begin{align}
\ket{\phi^{(t)}}
& =  
D_1C_1 B_1 U_{gc} S^{\reg{p_1}/\reg{m}} U^\dagger_{gc} B^\dagger_1 C^\dagger_1
\bigg( 
\ket{p_1}_{\reg{p_1}} \ket{\dot{\rho}^{(t-1)}_{p_1}}  
+  
\sum_{p'_1 \in \bits^\ell \setminus \Set{p_1}}  \ket{p'_1}_{\reg{p_1}}  C_{1,0} B^\dagger_{1,1} C^\dagger_{1,1}  \ket{\dot{\rho}^{(t-1)}_{p'_1}} 
\bigg) \label{lem:CPM-3ZK-hyb01:proof:invariant:proof:induction:case1:preparation-claim:eq:target:phi} \\ 
\ket{\psi^{(t)}}
& = 
D_1\ddot{C}_1 U_{gc} S^{\reg{p_1}/\reg{m}} U^\dagger_{gc}   \ddot{C}^\dagger_1 
\bigg( 
\ket{p_1}_{\reg{p_1}} \ket{\dot{\rho}^{(t-1)}_{p_1}}  +  \sum_{p'_1 \in \bits^\ell \setminus \Set{p_1}}  \ket{p'_1}_{\reg{p_1}}  C_{1,0} \ket{\dot{\rho}^{(t-1)}_{p'_1}} 
\bigg) 
\label{lem:CPM-3ZK-hyb01:proof:invariant:proof:induction:case1:preparation-claim:eq:target:psi} 
,\end{align}
where $S^{\reg{p_1}/\reg{m}}$ is identical to $S$ except that it treats register $\reg{p_1}$ as register $\reg{m}$.
\end{MyClaim}
\begin{proof}[Proof of \Cref{lem:CPM-3ZK-hyb01:proof:invariant:proof:induction:case1:preparation-claim}]

First, notice that  
\begin{align*}
 A_1 U_{gc} S U^\dagger_{gc} A^\dagger_1 
 & = 
 A_1 U_{gc} S A^\dagger_1 U^\dagger_{gc}
 \numberthis \label{lem:CPM-3ZK-hyb01:proof:invariant:proof:induction:case1:preparation-claim:proof:eq:swap-S:1} \\ 
 & = 
 A_1 U_{gc} A^\dagger_1 S^{\reg{p_1}/\reg{m}}  U^\dagger_{gc}
 \numberthis \label{lem:CPM-3ZK-hyb01:proof:invariant:proof:induction:case1:preparation-claim:proof:eq:swap-S:2} \\
  & = 
 A_1 A^\dagger_1 U_{gc} S^{\reg{p_1}/\reg{m}}  U^\dagger_{gc}
 \numberthis \label{lem:CPM-3ZK-hyb01:proof:invariant:proof:induction:case1:preparation-claim:proof:eq:swap-S:3} \\
 & = 
U_{gc} S^{\reg{p_1}/\reg{m}}  U^\dagger_{gc}
 \numberthis \label{lem:CPM-3ZK-hyb01:proof:invariant:proof:induction:case1:preparation-claim:proof:eq:swap-S} 
,\end{align*}
where 
\begin{itemize}
\item
\Cref{lem:CPM-3ZK-hyb01:proof:invariant:proof:induction:case1:preparation-claim:proof:eq:swap-S:1,lem:CPM-3ZK-hyb01:proof:invariant:proof:induction:case1:preparation-claim:proof:eq:swap-S:3} follows from the fact that $U_{gc}$ acts on different registers from $A_1$ (see \Cref{CPM:non-trivial-registers:table});

\item
\Cref{lem:CPM-3ZK-hyb01:proof:invariant:proof:induction:case1:preparation-claim:proof:eq:swap-S:2} from the fact that $S$ acts as the identity operator on $\reg{p_1}$ (see \Cref{CPM:non-trivial-registers:table}) and that $A_1$ is nothing but a swap operator between $\reg{m}$ and $\reg{p_1}$ (see \Cref{CPM:V:unitary}). 
\end{itemize}

\subpara{Proving \Cref{lem:CPM-3ZK-hyb01:proof:invariant:proof:induction:case1:preparation-claim:eq:target:phi}.} We now show the derivation for $\ket{\phi^{(t)}}$:
\begin{align*}
& \ket{\phi^{(t)}} \\
=~&
D_1C_1B_1A_1U_{gc} S U^{\dagger}_{gc} A^\dagger_1 B^\dagger_1 C^\dagger_1 D^\dagger_1  S \ket{\phi^{(t-1)}}
\numberthis \label{lem:CPM-3ZK-hyb01:proof:invariant:proof:induction:case1:preparation-claim:proof:phi-derivation:0} \\ 
=~&
D_1C_1B_1  U_{gc} S^{\reg{p_1}/\reg{m}}  U^\dagger_{gc} B^\dagger_1 C^\dagger_1 D^\dagger_1  S \ket{\phi^{(t-1)}}
\numberthis \label{lem:CPM-3ZK-hyb01:proof:invariant:proof:induction:case1:preparation-claim:proof:phi-derivation:1} \\ 
=~&
D_1C_1B_1  U_{gc} S^{\reg{p_1}/\reg{m}}  U^\dagger_{gc} B^\dagger_1 C^\dagger_1 D^\dagger_1 S 
\bigg(
\ket{p_1}_{\reg{p_1}}\ket{\rho^{(t-1)}_{p_1}}  +
 \sum_{p'_1 \in \bits^\ell \setminus \Set{p_1}} \ket{p'_1}_{\reg{p_1}} C_{1,0} B^\dagger_{1,1} C^\dagger_{1,1} \ket{\rho^{(t-1)}_{p'_1}}
\bigg)
\numberthis \label{lem:CPM-3ZK-hyb01:proof:invariant:proof:induction:case1:preparation-claim:proof:phi-derivation:2} \\ 
=~&
D_1C_1B_1  U_{gc} S^{\reg{p_1}/\reg{m}}  U^\dagger_{gc} B^\dagger_1 C^\dagger_1 D^\dagger_1  
\bigg(
\ket{p_1}_{\reg{p_1}} S \ket{\rho^{(t-1)}_{p_1}}  +
 \sum_{p'_1 \in \bits^\ell \setminus \Set{p_1}} \ket{p'_1}_{\reg{p_1}} S C_{1,0} B^\dagger_{1,1} C^\dagger_{1,1} \ket{\rho^{(t-1)}_{p'_1}}
\bigg)
\numberthis \label{lem:CPM-3ZK-hyb01:proof:invariant:proof:induction:case1:preparation-claim:proof:phi-derivation:3} \\ 
=~&
D_1C_1B_1  U_{gc} S^{\reg{p_1}/\reg{m}}  U^\dagger_{gc} B^\dagger_1 C^\dagger_1  
\bigg(
\ket{p_1}_{\reg{p_1}} D^\dagger_{1,1} S \ket{\rho^{(t-1)}_{p_1}}  + \\ 
& \hspace{18em} 
 \sum_{p'_1 \in \bits^\ell \setminus \Set{p_1}} \ket{p'_1}_{\reg{p_1}} S C_{1,0} B^\dagger_{1,1} C^\dagger_{1,1} \ket{\rho^{(t-1)}_{p'_1}}
\bigg)
\numberthis \label{lem:CPM-3ZK-hyb01:proof:invariant:proof:induction:case1:preparation-claim:proof:phi-derivation:4} \\
=~&
D_1C_1B_1  U_{gc} S^{\reg{p_1}/\reg{m}}  U^\dagger_{gc}  B^\dagger_1 C^\dagger_1
\bigg(
\ket{p_1}_{\reg{p_1}} D^\dagger_{1,1} S \ket{\rho^{(t-1)}_{p_1}}  + \\ 
& \hspace{18em} 
 \sum_{p'_1 \in \bits^\ell \setminus \Set{p_1}} \ket{p'_1}_{\reg{p_1}} C_{1,0} S^{\reg{t_1}/\reg{m}} B^\dagger_{1,1} C^\dagger_{1,1} \ket{\rho^{(t-1)}_{p'_1}}
\bigg)
\numberthis \label{lem:CPM-3ZK-hyb01:proof:invariant:proof:induction:case1:preparation-claim:proof:phi-derivation:5} \\ 
=~&
D_1C_1B_1  U_{gc} S^{\reg{p_1}/\reg{m}}  U^\dagger_{gc}  B^\dagger_1 C^\dagger_1
\bigg(
\ket{p_1}_{\reg{p_1}} D^\dagger_{1,1} S \ket{\rho^{(t-1)}_{p_1}}  + \\ 
& \hspace{18em} 
 \sum_{p'_1 \in \bits^\ell \setminus \Set{p_1}} \ket{p'_1}_{\reg{p_1}} C_{1,0}  B^\dagger_{1,1} C^\dagger_{1,1} S^{\reg{t_1}/\reg{m}} \ket{\rho^{(t-1)}_{p'_1}}
\bigg)
\numberthis \label{lem:CPM-3ZK-hyb01:proof:invariant:proof:induction:case1:preparation-claim:proof:phi-derivation:6} \\ 
=~&
D_1C_1B_1  U_{gc} S^{\reg{p_1}/\reg{m}}  U^\dagger_{gc}  B^\dagger_1 C^\dagger_1
\bigg(
\ket{p_1}_{\reg{p_1}} \ket{\dot{\rho}^{(t-1)}_{p_1}}  ~+~ 
 \sum_{p'_1 \in \bits^\ell \setminus \Set{p_1}} \ket{p'_1}_{\reg{p_1}} C_{1,0} B^\dagger_{1,1} C^\dagger_{1,1} \ket{\dot{\rho}^{(t-1)}_{p'_1}}
\bigg)
\numberthis \label{lem:CPM-3ZK-hyb01:proof:invariant:proof:induction:case1:preparation-claim:proof:phi-derivation:7} 
,\end{align*}
where 
\begin{itemize}
\item
\Cref{lem:CPM-3ZK-hyb01:proof:invariant:proof:induction:case1:preparation-claim:proof:phi-derivation:0} follows from \Cref{lem:CPM-3ZK-hyb01:proof:invariant:proof:induction:eq:k:H0};

\item
\Cref{lem:CPM-3ZK-hyb01:proof:invariant:proof:induction:case1:preparation-claim:proof:phi-derivation:1} follows from \Cref{lem:CPM-3ZK-hyb01:proof:invariant:proof:induction:case1:preparation-claim:proof:eq:swap-S};

\item
\Cref{lem:CPM-3ZK-hyb01:proof:invariant:proof:induction:case1:preparation-claim:proof:phi-derivation:2} follows from our induction assumption;

\item
\Cref{lem:CPM-3ZK-hyb01:proof:invariant:proof:induction:case1:preparation-claim:proof:phi-derivation:3} from  from the fact that $S$ acts as the identity operator on $\reg{p_1}$ (see \Cref{CPM:non-trivial-registers:table});

\item
\Cref{lem:CPM-3ZK-hyb01:proof:invariant:proof:induction:case1:preparation-claim:proof:phi-derivation:4} from  from the definition of $D_1$ (see \Cref{re-define:unitaries});

\item
\Cref{lem:CPM-3ZK-hyb01:proof:invariant:proof:induction:case1:preparation-claim:proof:phi-derivation:5} from the fact that $S$ acts as the identity operator on $\reg{t_1}$ (see \Cref{CPM:non-trivial-registers:table}) and $C_{1,0}$ is nothing but a swap operator between $\reg{t_1}$ and $\reg{m}$ (see \Cref{re-define:unitaries}); (Note that $S^{\reg{t_1}/\reg{m}}$ is defined to be an operator that is identical to $S$ except that it treats $\reg{t_1}$ as $\reg{m}$.)

\item
\Cref{lem:CPM-3ZK-hyb01:proof:invariant:proof:induction:case1:preparation-claim:proof:phi-derivation:6} follows from the fact that $S^{\reg{t_1}/\reg{m}}$ acts non-trivially on different registers from $B_{1,1}$ and $C_{1,1}$ (see \Cref{CPM:non-trivial-registers:table});

\item
\Cref{lem:CPM-3ZK-hyb01:proof:invariant:proof:induction:case1:preparation-claim:proof:phi-derivation:7} follows by defining 
\begin{equation}
\label[Expression]{lem:CPM-3ZK-hyb01:proof:invariant:proof:induction:case1:preparation-claim:proof:def:rho-dot}
\ket{\dot{\rho}^{(t-1)}_{p_1}}   \coloneqq D^\dagger_{1,1} S \ket{\rho^{(t-1)}_{p_1}} ~~~\text{and}~~~
\ket{\dot{\rho}^{(t-1)}_{p'_1}}  \coloneqq S^{\reg{t_1}/\reg{m}} \ket{\rho^{(t-1)}_{p'_1}} ~~(\forall p'_1 \in \bits^\ell \setminus \Set{p_1}).
\end{equation}
\end{itemize}
\Cref{lem:CPM-3ZK-hyb01:proof:invariant:proof:induction:case1:preparation-claim:proof:phi-derivation:7} finishes the proof of \Cref{lem:CPM-3ZK-hyb01:proof:invariant:proof:induction:case1:preparation-claim:eq:target:phi} in \Cref{lem:CPM-3ZK-hyb01:proof:invariant:proof:induction:case1:preparation-claim}.

\subpara{Proving \Cref{lem:CPM-3ZK-hyb01:proof:invariant:proof:induction:case1:preparation-claim:eq:target:psi}.} We now show the derivation for $\ket{\psi^{(t)}}$. This is almost identical to the above proof for \Cref{lem:CPM-3ZK-hyb01:proof:invariant:proof:induction:case1:preparation-claim:eq:target:phi}. Nevertheless, we present it for the sake of completeness.
\begin{align*}
\ket{\psi^{(t)}}
& =
D_1 \ddot{C}_1 A_1U_{gc} S U^{\dagger}_{gc} A^\dagger_1 \ddot{C}^\dagger_1 D^\dagger_1  S \ket{\psi^{(t-1)}}
\numberthis \label{lem:CPM-3ZK-hyb01:proof:invariant:proof:induction:case1:preparation-claim:proof:psi-derivation:0} \\ 
& =
D_1 \ddot{C}_1  U_{gc} S^{\reg{p_1}/\reg{m}}  U^\dagger_{gc}  \ddot{C}^\dagger_1 D^\dagger_1  S \ket{\psi^{(t-1)}}
\numberthis \label{lem:CPM-3ZK-hyb01:proof:invariant:proof:induction:case1:preparation-claim:proof:psi-derivation:1} \\ 
& =
D_1 \ddot{C}_1  U_{gc} S^{\reg{p_1}/\reg{m}}  U^\dagger_{gc}  \ddot{C}^\dagger_1 D^\dagger_1  S  
\bigg(
\ket{p_1}_{\reg{p_1}}\ket{\rho^{(t-1)}_{p_1}}  ~+~
 \sum_{p'_1 \in \bits^\ell \setminus \Set{p_1}} \ket{p'_1}_{\reg{p_1}} C_{1,0} \ket{\rho^{(t-1)}_{p'_1}}
\bigg)
\numberthis \label{lem:CPM-3ZK-hyb01:proof:invariant:proof:induction:case1:preparation-claim:proof:psi-derivation:2} \\ 
& =
D_1\ddot{C}_1  U_{gc} S^{\reg{p_1}/\reg{m}}  U^\dagger_{gc}  \ddot{C}^\dagger_1
\bigg(
\ket{p_1}_{\reg{p_1}} D^\dagger_{1,1} S \ket{\rho^{(t-1)}_{p_1}}  ~+~ 
 \sum_{p'_1 \in \bits^\ell \setminus \Set{p_1}} \ket{p'_1}_{\reg{p_1}} C_{1,0}  S^{\reg{t_1}/\reg{m}} \ket{\rho^{(t-1)}_{p'_1}}
\bigg)
\numberthis \label{lem:CPM-3ZK-hyb01:proof:invariant:proof:induction:case1:preparation-claim:proof:psi-derivation:6} \\ 
& =
D_1\ddot{C}_1  U_{gc} S^{\reg{p_1}/\reg{m}}  U^\dagger_{gc}  \ddot{C}^\dagger_1
\bigg(
\ket{p_1}_{\reg{p_1}} \ket{\dot{\rho}^{(t-1)}_{p_1}}  ~+~ 
 \sum_{p'_1 \in \bits^\ell \setminus \Set{p_1}} \ket{p'_1}_{\reg{p_1}} C_{1,0} \ket{\dot{\rho}^{(t-1)}_{p'_1}}
\bigg)
\numberthis \label{lem:CPM-3ZK-hyb01:proof:invariant:proof:induction:case1:preparation-claim:proof:psi-derivation:7} 
,\end{align*}
where 
\begin{itemize}
\item
\Cref{lem:CPM-3ZK-hyb01:proof:invariant:proof:induction:case1:preparation-claim:proof:psi-derivation:0} follows from \Cref{lem:CPM-3ZK-hyb01:proof:invariant:proof:induction:eq:k:H1};

\item
\Cref{lem:CPM-3ZK-hyb01:proof:invariant:proof:induction:case1:preparation-claim:proof:psi-derivation:1} follows from \Cref{lem:CPM-3ZK-hyb01:proof:invariant:proof:induction:case1:preparation-claim:proof:eq:swap-S};

\item
\Cref{lem:CPM-3ZK-hyb01:proof:invariant:proof:induction:case1:preparation-claim:proof:psi-derivation:2}  follows from our induction assumption;

\item
\Cref{lem:CPM-3ZK-hyb01:proof:invariant:proof:induction:case1:preparation-claim:proof:psi-derivation:6} follows from a similar argument as we did to derive \Cref{lem:CPM-3ZK-hyb01:proof:invariant:proof:induction:case1:preparation-claim:proof:phi-derivation:6} from \Cref{lem:CPM-3ZK-hyb01:proof:invariant:proof:induction:case1:preparation-claim:proof:phi-derivation:2};

\item
\Cref{lem:CPM-3ZK-hyb01:proof:invariant:proof:induction:case1:preparation-claim:proof:psi-derivation:7} follows from {\em the same definitions} of $\ket{\dot{\rho}^{(t-1)}_{p_1}}$ and $\ket{\dot{\rho}^{(t-1)}_{p'_1}}$ as shown in \Cref{lem:CPM-3ZK-hyb01:proof:invariant:proof:induction:case1:preparation-claim:proof:def:rho-dot}.
\end{itemize}
\Cref{lem:CPM-3ZK-hyb01:proof:invariant:proof:induction:case1:preparation-claim:proof:psi-derivation:7} finishes the proof of \Cref{lem:CPM-3ZK-hyb01:proof:invariant:proof:induction:case1:preparation-claim:eq:target:psi} in \Cref{lem:CPM-3ZK-hyb01:proof:invariant:proof:induction:case1:preparation-claim}.

\vspace{1em}
This completes the proof of \Cref{lem:CPM-3ZK-hyb01:proof:invariant:proof:induction:case1:preparation-claim}.

\end{proof}

\vspace{1em}

\begin{MyClaim}\label{lem:CPM-3ZK-hyb01:proof:invariant:proof:induction:case1:preparation-claim:2}
Let $S^{\reg{p_1}/\reg{m}}$ and $\Set{\dot{\rho}^{(t-1)}_{p'_1}}_{p'_1 \in \bits^\ell}$ be as defined in \Cref{lem:CPM-3ZK-hyb01:proof:invariant:proof:induction:case1:preparation-claim}. Let
\begin{align}
\ket{\gamma^{(t-1)}_0} 
& \coloneqq 
\ket{p_1}_{\reg{p_1}} \ket{\dot{\rho}^{(t-1)}_{p_1}}  
+  
\sum_{p'_1 \in \bits^\ell \setminus \Set{p_1}}  \ket{p'_1}_{\reg{p_1}}  C_{1,0} B^\dagger_{1,1} C^\dagger_{1,1}  \ket{\dot{\rho}^{(t-1)}_{p'_1}} 
\label{lem:CPM-3ZK-hyb01:proof:invariant:proof:induction:case1:preparation-claim:2:eq:in:0} \\ 
\ket{\gamma^{(t-1)}_1} 
& \coloneqq 
\ket{p_1}_{\reg{p_1}} \ket{\dot{\rho}^{(t-1)}_{p_1}}  +  \sum_{p'_1 \in \bits^\ell \setminus \Set{p_1}}  \ket{p'_1}_{\reg{p_1}}  C_{1,0} \ket{\dot{\rho}^{(t-1)}_{p'_1}} 
\label{lem:CPM-3ZK-hyb01:proof:invariant:proof:induction:case1:preparation-claim:2:eq:in:1} \\ 
\ket{\gamma^{(t)}_0} 
& \coloneqq 
C_1 B_1 U_{gc} S^{\reg{p_1}/\reg{m}}  U^\dagger_{gc}    B^\dagger_1 C^\dagger_1 \ket{\gamma^{(t-1)}_0} 
\label{lem:CPM-3ZK-hyb01:proof:invariant:proof:induction:case1:preparation-claim:2:eq:out:0} \\ 
\ket{\gamma^{(t)}_1}
& \coloneqq 
\ddot{C}_1 U_{gc} S^{\reg{p_1}/\reg{m}}  U^\dagger_{gc}   \ddot{C}^\dagger_1  \ket{\gamma^{(t-1)}_1}
\label{lem:CPM-3ZK-hyb01:proof:invariant:proof:induction:case1:preparation-claim:2:eq:out:1} 
.\end{align}
Then, there exist (possibly sub-normalized) pure states $\Set{\ket{\dot{\rho}^{(t)}_{p'_1}}}_{p'_1 \in \bits^\ell}$ so that the following holds:
\begin{align}
\ket{\gamma^{(t)}_0} 
& =
\ket{p_1}_{\reg{p_1}} \ket{\dot{\rho}^{(t)}_{p_1}}  
+  
\sum_{p'_1 \in \bits^\ell \setminus \Set{p_1}}  \ket{p'_1}_{\reg{p_1}}  C_{1,0} B^\dagger_{1,1} C^\dagger_{1,1}  \ket{\dot{\rho}^{(t)}_{p'_1}} 
\label{lem:CPM-3ZK-hyb01:proof:invariant:proof:induction:case1:preparation-claim:2:eq:target:0} \\ 
\ket{\gamma^{(t)}_1} 
& = 
\ket{p_1}_{\reg{p_1}} \ket{\dot{\rho}^{(t)}_{p_1}}  +  \sum_{p'_1 \in \bits^\ell \setminus \Set{p_1}}  \ket{p'_1}_{\reg{p_1}}  C_{1,0} \ket{\dot{\rho}^{(t)}_{p'_1}} 
\label{lem:CPM-3ZK-hyb01:proof:invariant:proof:induction:case1:preparation-claim:2:eq:target:1} 
.\end{align}
\end{MyClaim}
\begin{proof}[Proof of \Cref{lem:CPM-3ZK-hyb01:proof:invariant:proof:induction:case1:preparation-claim:2}]
This claim follows from an application of \Cref{lem:err-inv-com}, with the notation correspondence listed in \Cref{lem:CPM-3ZK-hyb01:proof:case-1:cor-table}. We provide a detailed explanation below.
\begin{table}[!htb]
\centering
\caption{Notation Correspondence between \Cref{lem:err-inv-com} and \Cref{lem:CPM-3ZK-hyb01:proof:invariant:proof:induction:case1:preparation-claim:2}}
\label{lem:CPM-3ZK-hyb01:proof:case-1:cor-table}
\vspace{0.5em}
\begin{tabular}{ C{50pt} C{60pt} c C{50pt} C{60pt} c C{50pt} C{60pt} }
\toprule
 \multicolumn{2}{c}{Registers}   & \phantom{abc}   & \multicolumn{2}{c}{Operators}   &
\phantom{abc}   & \multicolumn{2}{c}{Random Variables}  \\
\cmidrule{1-2} \cmidrule{4-5} \cmidrule{7-8}
In \Cref{lem:err-inv-com} & In \Cref{lem:CPM-3ZK-hyb01:proof:invariant:proof:induction:case1:preparation-claim:2} & & In \Cref{lem:err-inv-com} & In \Cref{lem:CPM-3ZK-hyb01:proof:invariant:proof:induction:case1:preparation-claim:2} & & In \Cref{lem:err-inv-com} & In \Cref{lem:CPM-3ZK-hyb01:proof:invariant:proof:induction:case1:preparation-claim:2}  \\ 
\midrule
\addlinespace[0.3em]
$\reg{a}$   & $\reg{p_1}$ & & $W_1$ & $C_{1,1}$ & & $\ket{a}_{\reg{a}}$    & $\ket{p_1}_{\reg{p_1}}$  \\ 
\addlinespace[0.3em]
$\reg{m}$    & $\reg{m}$ & & $W_0$ & $C_{1,0}$ & & $\ket{a'}_{\reg{a}}$    & $\ket{p'_1}_{\reg{p_1}}$  \\
\addlinespace[0.3em]
$\reg{t}$    & $\reg{t_1}$ & & $W$   & $C_{1}$ & & $\ket{\rho^{(\msf{in})}_a}_{\reg{mtso}}$    & $\ket{\dot{\rho}^{(t-1)}_{p_1}}$  \\ 
\addlinespace[0.3em]
$\reg{s}$    & $\reg{u}$, $\reg{s}$, $\reg{gc}$ & & $\tilde{W}$ & $\ddot{C}_1$ & & $\ket{\rho^{(\msf{in})}_{a'}}_{\reg{mtso}}$    & $\ket{\dot{\rho}^{(t-1)}_{p'_1}}$ \\ 
\addlinespace[0.3em]
$\reg{o}$    & other registers & & $U_1$ & $B_{1,1}$ & &   $\ket{\eta^{(\msf{in})}_0}$ & $\ket{\gamma^{(t-1)}_0}$ \\ 
\addlinespace[0.3em]
   & & & $U$   & $B_1$ & & $\ket{\eta^{(\msf{in})}_1}$ & $\ket{\gamma^{(t-1)}_1}$  \\ 
\addlinespace[0.3em]
 &  & &   $S$   & $U_{gc} S^{\reg{p_1}/\reg{m}}  U^\dagger_{gc}$ & &  $\ket{\rho^{(\msf{out})}_a}_\reg{mtso}$    & $\ket{\dot{\rho}^{(t)}_{p_1}}$  \\ 
\addlinespace[0.3em]
  &  & &   &  & &  $\ket{\rho^{(\msf{out})}_{a'}}_\reg{mtso}$     & $\ket{\dot{\rho}^{(t)}_{p'_1}}$  \\  
  \addlinespace[0.3em]
 &  & & & & &  $\ket{\eta^{(\msf{out})}_0}$    & $\ket{\gamma^{(t)}_0}$  \\
\addlinespace[0.3em] 
  &  & &   &  & &  $\ket{\eta^{(\msf{out})}_1}$    & $\ket{\gamma^{(t)}_1}$  \\   
\bottomrule
\end{tabular}
\end{table}

First, we argue that the premises in \Cref{lem:err-inv-com} are satisfied with the notation listed in \Cref{lem:CPM-3ZK-hyb01:proof:case-1:cor-table}:
\begin{itemize}
\item
\Cref{lem:err-inv-com} requires that $W_1$ should work as the identity operator on register $\reg{s}$. In terms of the \Cref{lem:CPM-3ZK-hyb01:proof:invariant:proof:induction:case1:preparation-claim:2} notation, this is satisfied by $C_{1,1}$ (playing the role of $W_1$) who works as identity on registers $\reg{u}$, $\reg{s}$, and $\reg{gc}$ (playing the role of registers $\reg{s}$). (Recall $C_{1,1}$ from \Cref{CPM:non-trivial-registers:table}.)

\item
\Cref{lem:err-inv-com} requires that $W_0$ should be a swap operator between $\reg{m}$ and $\reg{t}$. In terms of the \Cref{lem:CPM-3ZK-hyb01:proof:invariant:proof:induction:case1:preparation-claim:2} notation, this is satisfied by $C_{1,0}$ (playing the role of $W_0$), who is a swap operator between registers $\reg{m}$ and $\reg{t_1}$ (playing the role of $\reg{m}$ and $\reg{t}$ respectively). (Recall $C_{1,0}$ from \Cref{re-define:unitaries}.)

\item
\Cref{lem:err-inv-com} requires that $\tilde{W}$ is the identity operator on branch $\ket{a}_{\reg{a}}$ and is identical to $W_0$ on branches $\ket{a'}_{\reg{a}}$ with $a' \ne a$. In terms of the \Cref{lem:CPM-3ZK-hyb01:proof:invariant:proof:induction:case1:preparation-claim:2} notation, this is satisfied by $\ddot{C}_1$ (playing the role of $\tilde{W}$), who is the identity operator on branch $\ket{p_1}_{\reg{p_1}}$ (playing the role of $\ket{a}_{\reg{a}}$) and is identical to $C_{1,0}$ (playing the role of $W_0$) on branches $\ket{p'_1}_{\reg{p_1}}$ (playing the role of $\ket{a'}_{\reg{a}}$) with $p'_1 \ne p_1$. (Recall $\ddot{C}_1$ from \Cref{re-define:unitaries})

\item
 \Cref{lem:err-inv-com} requires that $U_1$ should work as identity on register $\reg{s}$. In terms of the \Cref{lem:CPM-3ZK-hyb01:proof:invariant:proof:induction:case1:preparation-claim:2} notation, this is satisfied by $B_{1,1}$ (playing the role of $U_1$), who works as identity on registers $\reg{u}$, $\reg{s}$, and $\reg{gc}$ (playing the role of register $\reg{s}$). (Recall $B_{1,1}$ from \Cref{CPM:non-trivial-registers:table}.)

\item
 \Cref{lem:err-inv-com} requires that $S$ should act non-trivially {\em only} on registers $\reg{a}$ and $\reg{s}$. In terms of the \Cref{lem:CPM-3ZK-hyb01:proof:invariant:proof:induction:case1:preparation-claim:2} notation, this is satisfied by $U_{gc}S^{\reg{p_1}/\reg{m}}U^\dagger_{gc}$ (playing the role of $S$), who does not touch registers beyond $\reg{p_1}$, $\reg{u}$, $\reg{s}$, and $\reg{gc}$ (playing the role of registers $\reg{a}$ and $\reg{s}$). (Recall $S$ from \Cref{CPM:non-trivial-registers:table} and the fact that $S^{\reg{p_1}/\reg{m}}$ is identical to $S$ except that it treats $\reg{p_1}$ as $\reg{m}$.)
\end{itemize}

Finally, we apply \Cref{lem:err-inv-com} (with the notation in \Cref{lem:CPM-3ZK-hyb01:proof:case-1:cor-table}) to the $\ket{\gamma^{(t-1)}_0}$ and $\ket{\gamma^{(t-1)}_1}$ defined in \Cref{lem:CPM-3ZK-hyb01:proof:invariant:proof:induction:case1:preparation-claim:2:eq:in:0,lem:CPM-3ZK-hyb01:proof:invariant:proof:induction:case1:preparation-claim:2:eq:in:1}  (playing the role of $\ket{\eta^{(\msf{in})}_0}$ and $\ket{\eta^{(\msf{in})}_1}$ in \Cref{lem:err-inv-com}). This implies the existence of (possibly sub-normalized) pure states $\Set{\ket{\dot{\rho}^{(t)}_{p'_1}}}_{p'_1\in\bits^\ell }$  (playing the role of $\Set{\ket{\rho^{(\msf{out})}_{a'}}_{\reg{mtso}}}_{a'\in\bits^\ell }$ in \Cref{lem:err-inv-com}) such that the following holds
\begin{align*}
\ket{\gamma^{(t)}_0} 
& = \ket{p_1}_{\reg{p_1}}\ket{\dot{\rho}^{(t)}_{p_1}}  +  \sum_{p'_1 \in \bits^\ell \setminus \Set{p_1}} \ket{p'_1}_{\reg{p_1}} C_{1,0} B^\dagger_{1,1} C^\dagger_{1,1} \ket{\dot{\rho}^{(t)}_{p'_1}} \\ 
\ket{\gamma^{(t)}_1} 
& = \ket{p_1}_{\reg{p_1}}\ket{\dot{\rho}^{(t)}_{p_1}}  +  \sum_{p'_1 \in \bits^\ell \setminus \Set{p_1}} \ket{p'_1}_{\reg{p_1}} C_{1,0} \ket{\dot{\rho}^{(t)}_{p'_1}} 
,\end{align*}
which are exactly \Cref{lem:CPM-3ZK-hyb01:proof:invariant:proof:induction:case1:preparation-claim:2:eq:target:0,lem:CPM-3ZK-hyb01:proof:invariant:proof:induction:case1:preparation-claim:2:eq:target:1} in  \Cref{lem:CPM-3ZK-hyb01:proof:invariant:proof:induction:case1:preparation-claim:2}.

This completes the proof of \Cref{lem:CPM-3ZK-hyb01:proof:invariant:proof:induction:case1:preparation-claim:2}.

\end{proof}

\para{Finishing the Proof for \Cref{CPM-3ZK-hyb01:proof:invariant:proof:induction:case-1}.}\label{lem:CPM-3ZK-hyb01:proof:invariant:proof:induction:case1:finish} With \Cref{lem:CPM-3ZK-hyb01:proof:invariant:proof:induction:case1:preparation-claim,lem:CPM-3ZK-hyb01:proof:invariant:proof:induction:case1:preparation-claim:2} at hand, we now finish the proof for \Cref{CPM-3ZK-hyb01:proof:invariant:proof:induction:case-1}.

\subpara{Proof for \Cref{lem:CPM-3ZK-hyb01:proof:target-format:H0}.} We first establish \Cref{lem:CPM-3ZK-hyb01:proof:target-format:H0}:
\begin{align}
\ket{\phi^{(t)}} 
& = D_1 \ket{\gamma^{(t)}_0} 
\label{CPM-3ZK-hyb01:proof:invariant:proof:induction:case-1:final:phi:1} \\ 
& = D_1 
\bigg(
\ket{p_1}_{\reg{p_1}} \ket{\dot{\rho}^{(t)}_{p_1}}  
+  
\sum_{p'_1 \in \bits^\ell \setminus \Set{p_1}}  \ket{p'_1}_{\reg{p_1}}  C_{1,0} B^\dagger_{1,1} C^\dagger_{1,1}  \ket{\dot{\rho}^{(t)}_{p'_1}} 
\bigg) 
\label{CPM-3ZK-hyb01:proof:invariant:proof:induction:case-1:final:phi:2} \\ 
& = 
\ket{p_1}_{\reg{p_1}} D_{1,1}\ket{\dot{\rho}^{(t)}_{p_1}}  
+  
\sum_{p'_1 \in \bits^\ell \setminus \Set{p_1}}  \ket{p'_1}_{\reg{p_1}}  C_{1,0} B^\dagger_{1,1} C^\dagger_{1,1}  \ket{\dot{\rho}^{(t)}_{p'_1}} 
\label{CPM-3ZK-hyb01:proof:invariant:proof:induction:case-1:final:phi:3} \\ 
& = 
\ket{p_1}_{\reg{p_1}}\ket{\rho^{(t)}_{p_1}}  
+  
\sum_{p'_1 \in \bits^\ell \setminus \Set{p_1}}  \ket{p'_1}_{\reg{p_1}}  C_{1,0} B^\dagger_{1,1} C^\dagger_{1,1}  \ket{\rho^{(t)}_{p'_1}} 
\label{CPM-3ZK-hyb01:proof:invariant:proof:induction:case-1:final:phi:4} 
,\end{align}
where 
\begin{itemize}
\item
\Cref{CPM-3ZK-hyb01:proof:invariant:proof:induction:case-1:final:phi:1} follows from \Cref{lem:CPM-3ZK-hyb01:proof:invariant:proof:induction:case1:preparation-claim:eq:target:phi} in \Cref{lem:CPM-3ZK-hyb01:proof:invariant:proof:induction:case1:preparation-claim} and \Cref{lem:CPM-3ZK-hyb01:proof:invariant:proof:induction:case1:preparation-claim:2:eq:out:0} in \Cref{lem:CPM-3ZK-hyb01:proof:invariant:proof:induction:case1:preparation-claim:2}; 

\item
\Cref{CPM-3ZK-hyb01:proof:invariant:proof:induction:case-1:final:phi:2} follows from \Cref{lem:CPM-3ZK-hyb01:proof:invariant:proof:induction:case1:preparation-claim:2:eq:target:0} in \Cref{lem:CPM-3ZK-hyb01:proof:invariant:proof:induction:case1:preparation-claim:2};

\item
\Cref{CPM-3ZK-hyb01:proof:invariant:proof:induction:case-1:final:phi:3} follows from the definition of $D_1$ (see \Cref{re-define:unitaries});

\item
\Cref{CPM-3ZK-hyb01:proof:invariant:proof:induction:case-1:final:phi:4} follows by defining
\begin{equation}
\label[Expression]{lem:CPM-3ZK-hyb01:proof:invariant:proof:induction:case1:preparation-claim:proof:def:rho}
\ket{\rho^{(t)}_{p_1}}   \coloneqq D_{1,1} \ket{\dot{\rho}^{(t)}_{p_1}} ~~~\text{and}~~~
\ket{\rho^{(t)}_{p'_1}}  \coloneqq \ket{\dot{\rho}^{(t)}_{p'_1}} ~~(\forall p'_1 \in \bits^\ell \setminus \Set{p_1}).
\end{equation}
\end{itemize}
Note that \Cref{CPM-3ZK-hyb01:proof:invariant:proof:induction:case-1:final:phi:4} is exactly 
\Cref{lem:CPM-3ZK-hyb01:proof:target-format:H0} in \Cref{lem:CPM-3ZK-hyb01:proof:invariant}.

\subpara{Proof for \Cref{lem:CPM-3ZK-hyb01:proof:target-format:H1}.}  Next, we establish \Cref{lem:CPM-3ZK-hyb01:proof:target-format:H1}:

\begin{align}
\ket{\psi^{(t)}} 
& = D_1 \ket{\gamma^{(t)}_1} 
\label{CPM-3ZK-hyb01:proof:invariant:proof:induction:case-1:final:psi:1} \\ 
& = D_1 
\bigg(
\ket{p_1}_{\reg{p_1}} \ket{\dot{\rho}^{(t)}_{p_1}}  
+  
\sum_{p'_1 \in \bits^\ell \setminus \Set{p_1}}  \ket{p'_1}_{\reg{p_1}}  C_{1,0} \ket{\dot{\rho}^{(t)}_{p'_1}} 
\bigg) 
\label{CPM-3ZK-hyb01:proof:invariant:proof:induction:case-1:final:psi:2} \\ 
& = 
\ket{p_1}_{\reg{p_1}} D_{1,1}\ket{\dot{\rho}^{(t)}_{p_1}}  
+  
\sum_{p'_1 \in \bits^\ell \setminus \Set{p_1}}  \ket{p'_1}_{\reg{p_1}}  C_{1,0}  \ket{\dot{\rho}^{(t)}_{p'_1}} 
\label{CPM-3ZK-hyb01:proof:invariant:proof:induction:case-1:final:psi:3} \\ 
& = 
\ket{p_1}_{\reg{p_1}}\ket{\rho^{(t)}_{p_1}}  
+  
\sum_{p'_1 \in \bits^\ell \setminus \Set{p_1}}  \ket{p'_1}_{\reg{p_1}}  C_{1,0} \ket{\rho^{(t)}_{p'_1}} 
\label{CPM-3ZK-hyb01:proof:invariant:proof:induction:case-1:final:psi:4} 
,\end{align}
where 
\begin{itemize}
\item
\Cref{CPM-3ZK-hyb01:proof:invariant:proof:induction:case-1:final:psi:1} follows from \Cref{lem:CPM-3ZK-hyb01:proof:invariant:proof:induction:case1:preparation-claim:eq:target:psi} in \Cref{lem:CPM-3ZK-hyb01:proof:invariant:proof:induction:case1:preparation-claim} and \Cref{lem:CPM-3ZK-hyb01:proof:invariant:proof:induction:case1:preparation-claim:2:eq:out:1} in \Cref{lem:CPM-3ZK-hyb01:proof:invariant:proof:induction:case1:preparation-claim:2}; 

\item
\Cref{CPM-3ZK-hyb01:proof:invariant:proof:induction:case-1:final:psi:2} follows from \Cref{lem:CPM-3ZK-hyb01:proof:invariant:proof:induction:case1:preparation-claim:2:eq:target:1} in \Cref{lem:CPM-3ZK-hyb01:proof:invariant:proof:induction:case1:preparation-claim:2};

\item
\Cref{CPM-3ZK-hyb01:proof:invariant:proof:induction:case-1:final:psi:3} follows from the definition of $D_1$ (see \Cref{re-define:unitaries});

\item
\Cref{CPM-3ZK-hyb01:proof:invariant:proof:induction:case-1:final:psi:4} follows from {\em the same definitions} of $\ket{\rho^{(t)}_{p_1}}$ and $\ket{\rho^{(t)}_{p'_1}}$ in \Cref{lem:CPM-3ZK-hyb01:proof:invariant:proof:induction:case1:preparation-claim:proof:def:rho}.
\end{itemize}
Note that \Cref{CPM-3ZK-hyb01:proof:invariant:proof:induction:case-1:final:psi:4} is exactly 
\Cref{lem:CPM-3ZK-hyb01:proof:target-format:H1} in \Cref{lem:CPM-3ZK-hyb01:proof:invariant}. 

~\\ 
This eventually completes the proof for \Cref{CPM-3ZK-hyb01:proof:invariant:proof:induction:case-1}.

\subsubsubsection{Proof for \Cref{CPM-3ZK-hyb01:proof:invariant:proof:induction:case-2}} 
We first remark that in \Cref{CPM-3ZK-hyb01:proof:invariant:proof:induction:case-2}, the computation that brings $\ket{\phi^{(t-1)}}$ to $\ket{\phi^{(t)}}$ is identical to that brings $\ket{\psi^{(t-1)}}$ to $\ket{\psi^{(t)}}$, because $H^{(J, \msf{pat})}_0$ and $H^{(J, \msf{pat})}_1$ are identical when the global counter ``jumps'' between $1$ and $2$. (This can be also seen pictorially by comparing \Cref{figure:CPM-3ZK:hybrids:H0} and \Cref{figure:CPM-3ZK:hybrids:H1}.) In the following, we refer to this computation as $\Lambda$. That is, we have 
\begin{align}
\ket{\phi^{(t)}} 
& = 
\Lambda \ket{\phi^{(t-1)}} 
\label{lem:CPM-3ZK-hyb01:proof:invariant:proof:induction:case-2:initial:phi} \\ 
\ket{\psi^{(t)}} 
& = 
\Lambda \ket{\psi^{(t-1)}}
\label{lem:CPM-3ZK-hyb01:proof:invariant:proof:induction:case-2:initial:psi}
.\end{align} 

\para{Structure of $\Lambda$.} While the exact format of $\Lambda$ will not be substantial, our proof of \Cref{CPM-3ZK-hyb01:proof:invariant:proof:induction:case-2} will rely on certain properties of $\Lambda$, which we formalize in the following \Cref{lem:CPM-3ZK-hyb01:proof:case-2-1:claim:operator-commute}.

\begin{MyClaim}\label{lem:CPM-3ZK-hyb01:proof:case-2-1:claim:operator-commute}
For the operator $\Lambda$ defined above, there exist two operators $\Lambda_0$ and $\Lambda_1$ so that
\begin{itemize}
\item
both  $\Lambda_1$ and $\Lambda_0$ act as the identity operator on $\reg{p_1}$;
\item
$\Lambda_0$ acts as the identity operator on $\reg{t_1}$;
\end{itemize}
and the following holds
\begin{align}
& 
\Lambda = \ketbra{p_1}_{\reg{p_1}} \tensor \Lambda_1 + \sum_{p'_1 \in \bits^\ell \setminus \Set{p_1}}\ketbra{p'_1}_{\reg{p_1}} \tensor \Lambda_0, \label{lem:CPM-3ZK-hyb01:proof:case-2-1:claim:operator-commute:eq:1} \\ 
& 
\Lambda_0 C_{1,0}  = C_{1,0} \Lambda^{\reg{t_1}/\reg{m}}_0, \label{lem:CPM-3ZK-hyb01:proof:case-2-1:claim:operator-commute:eq:2} \\ 
& 
\Lambda_0 C_{1,0}  B^\dagger_{1,1} C^\dagger_{1,1} = C_{1,0}  B^\dagger_{1,1} C^\dagger_{1,1} \Lambda^{\reg{t_1}/\reg{m}}_0 \label{lem:CPM-3ZK-hyb01:proof:case-2-1:claim:operator-commute:eq:3} 
,\end{align}
where $\Lambda^{\reg{t_1}/\reg{m}}_0$ is identical to $\Lambda_0$ except that it treats $\reg{t_1}$ as $\reg{m}$.  
\end{MyClaim}
\begin{proof}[Proof of \Cref{lem:CPM-3ZK-hyb01:proof:case-2-1:claim:operator-commute}]

First, note that $\Lambda$ can be written as $\Lambda = \Gamma_z \Gamma_{z-1} \cdots \Gamma_1$ (for some integer $z$), where each $\Gamma_i$ ($i \in [z]$) comes from the set of operators $\Set{S, A_2, B_2, C_2, D_2, U_{gc}, \ketbra{p_1, p_2}_{\reg{p_1p_2}}}$. This is because $\Lambda$ only corresponds to the operations that happen when the global counter is no less than value $1$ (see  \Cref{MnR:game:redefine:hybrids} and \Cref{figure:CPM-3ZK:hybrids:H1}). We remark that $\Sim$ may also apply projectors on $\reg{u}$, but we consider it as a part of $S$ as per \Cref{rmk:CPM-3ZK:hid-proj-u}.

Therefore, to prove \Cref{lem:CPM-3ZK-hyb01:proof:case-2-1:claim:operator-commute}, it suffices to show that for each operator 
$$\Gamma \in \Set{\ketbra{p_1, p_2}_{\reg{p_1p_2}}, U_{gc}, S, A_2, B_2, C_2, D_2},$$ 
there exist two operators $\Gamma_0$ and $\Gamma_1$ so that
\begin{itemize}
\item
both  $\Gamma_1$ and $\Gamma_0$ act as the identity operator on $\reg{p_1}$;
\item
$\Gamma_0$ acts as the identity operator on $\reg{t_1}$;
\end{itemize}
and the following holds
\begin{align*}
\Gamma 
& = 
\ketbra{p_1}_{\reg{p_1}} \tensor \Gamma_1 + \sum_{p'_1 \in \bits^\ell \setminus \Set{p_1}}\ketbra{p'_1}_{\reg{p_1}} \tensor \Gamma_0,\\ 
\Gamma_0 C_{1,0}  
& = 
C_{1,0} \Gamma^{\reg{t_1}/\reg{m}}_0, \\ 
\Gamma_0 C_{1,0}  B^\dagger_{1,1} C^\dagger_{1,1} 
& = 
C_{1,0}  B^\dagger_{1,1} C^\dagger_{1,1} \Gamma^{\reg{t_1}/\reg{m}}_0 
.\end{align*}
In the following, we prove it for each possible $\Gamma$.

First, notice that the above is true for $\Gamma =\ketbra{p_1,p_2}_{\reg{p_1}\reg{p_2}}$, simply because such a $\Gamma$ can be written in the following format:
$$\Gamma =\ketbra{p_1,p_2}_{\reg{p_1}\reg{p_2}} = \ketbra{p_1}_{\reg{p_1}} \tensor \Gamma_1  + \sum_{p'_1 \in \bits^\ell \setminus \Set{p_1}}\ketbra{p'_1}_{\reg{p_1}} \tensor \Gamma_0,$$
with $\Gamma_1  \coloneqq \ketbra{p_2}_{\reg{p_2}}$ and $\Gamma_0  \coloneqq I$, and such a $\Gamma_0$ vacuously satisfies the requires $\Gamma_0 C_{1,0}  = C_{1,0} \Gamma^{\reg{t_1}/\reg{m}}_0$ and $\Gamma_0 C_{1,0}  B^\dagger_{1,1} C^\dagger_{1,1} = C_{1,0}  B^\dagger_{1,1} C^\dagger_{1,1} \Gamma^{\reg{t_1}/\reg{m}}_0$.

The above is true for $\Gamma = U_{gc}$ as well, because $U_{gc}$ acts non-trivially only on register $\reg{gc}$.

The above is also true for $\Gamma = S$ because (1) $S$ does not act on $\reg{p_1}$, (2) $S$ does not work on $\reg{t_1}$ and $C_{1,0}$ is nothing but a swap operator between $\reg{m}$ and $\reg{t_1}$, and (3) $S^{\reg{t_1}/\reg{m}}$ acts non-trivially on different registers from $B_{1,1}$ and $C_{1,1}$ (see \Cref{CPM:non-trivial-registers:table}).

The above is true for $\Gamma = A_2$ as well, because $A_2$ is nothing but the swap operator between registers $\reg{p_2}$ and $\reg{m}$ (see \Cref{CPM:V:unitary}).

The only  cases left are $\Gamma \in \Set{B_2, C_2, D_2}$. The proof for these cases are (almost) identical. In the following, we only prove it for $\Gamma = C_2$.
\begin{align*}
\Gamma = C_2 
& = 
\ketbra{p_1, p_2}_{\reg{p_1}\reg{p_2}} \tensor C_{2,1} + \sum_{(p'_1, p'_2) \in \bits^{2\ell} \setminus \Set{(p_1, p_2)}}  \ketbra{p'_1, p'_2}_{\reg{p_1}\reg{p_2}} \tensor C_{2,0} \numberthis \label{lem:CPM-3ZK-hyb01:proof:case-2-1:claim:operator-commute:proof:C2:1} \\ 
& =
\ketbra{p_1}_{\reg{p_1}} \tensor \ketbra{p_2}_{\reg{p_2}} \tensor C_{2,1}
+ 
\sum_{(p'_1, p'_2) \in \bits^{2\ell} \setminus \Set{(p_1, p_2)}}  \ketbra{p'_1}_{\reg{p_1}} \tensor \ketbra{p'_2}_{\reg{p_2}} \tensor C_{2,0} \\ 
& =
\ketbra{p_1}_{\reg{p_1}} \tensor \ketbra{p_2}_{\reg{p_2}} \tensor C_{2,1} 
+ \ketbra{p_1}_{\reg{p_1}} \tensor \big(\sum_{p'_2 \in \bits^{\ell} \setminus \Set{p_2}}  \ketbra{p'_2}_{\reg{p_2}} \big) \tensor C_{2,0} \\ 
& \hspace{15em} + \sum_{p'_1 \in \bits^\ell \setminus \Set{p_1}} \ketbra{p'_1}_{\reg{p_1}} \tensor \big(\sum_{p'_2 \in \bits^{\ell}}  \ketbra{p'_2}_{\reg{p_2}} \big) \tensor C_{2,0} \\ 
& = \ketbra{p_1}_{\reg{p_1}} \tensor \Gamma_1 + \sum_{p'_1 \in \bits^\ell \setminus \Set{p_1}} \ketbra{p'_1}_{\reg{p_1}} \tensor \Gamma_0 \numberthis \label{lem:CPM-3ZK-hyb01:proof:case-2-1:claim:operator-commute:proof:C2:2}
,\end{align*}
where \Cref{lem:CPM-3ZK-hyb01:proof:case-2-1:claim:operator-commute:proof:C2:1} follows from the definition of $C_2$ (see \Cref{re-define:unitaries}), and \Cref{lem:CPM-3ZK-hyb01:proof:case-2-1:claim:operator-commute:proof:C2:2} follows by defining $\Gamma_0$ and $\Gamma_1$ as follows 
\begin{align*}
 \Gamma_1 & \coloneqq \ketbra{p_2}_{\reg{p_2}} \tensor C_{2,1} 
+ \big(\sum_{p'_2 \in \bits^{\ell} \setminus \Set{p_2}}  \ketbra{p'_2}_{\reg{p_2}} \big) \tensor C_{2,0} \\ 
 \Gamma_0 & \coloneqq \big(\sum_{p'_2 \in \bits^{\ell}}  \ketbra{p'_2}_{\reg{p_2}} \big) \tensor C_{2,0}.
\end{align*} 
Clearly, such a $\Gamma_0$ satisfies the requires $\Gamma_0 C_{1,0}  = C_{1,0} \Gamma^{\reg{t_1}/\reg{m}}_0$ and $\Gamma_0 C_{1,0}  B^\dagger_{1,1} C^\dagger_{1,1} = C_{1,0}  B^\dagger_{1,1} C^\dagger_{1,1} \Gamma^{\reg{t_1}/\reg{m}}_0$, because (1) $C_{2,0}$ is nothing but a swap operator between $\reg{m}$ and $\reg{t_2}$, and (2) $C_{2,0}^{\reg{t_1}/\reg{m}}$ acts non-trivially on different registers from $B_{1,1}$ and $ C_{1,1}$ (see \Cref{CPM:non-trivial-registers:table}).

\vspace{1em}
This finishes the proof of \Cref{lem:CPM-3ZK-hyb01:proof:case-2-1:claim:operator-commute}.	

\end{proof}

\para{Finishing the Proof for \Cref{CPM-3ZK-hyb01:proof:invariant:proof:induction:case-2}.} With \Cref{lem:CPM-3ZK-hyb01:proof:case-2-1:claim:operator-commute} in hand, we now show how to finish the proof for \Cref{CPM-3ZK-hyb01:proof:invariant:proof:induction:case-2}.

\subpara{Proof of \Cref{lem:CPM-3ZK-hyb01:proof:target-format:H0}.} First, we establish \Cref{lem:CPM-3ZK-hyb01:proof:target-format:H0}:
\begin{align}
 \ket{\phi^{(t)}} 
 & = 
 \Lambda  \ket{\phi^{(t-1)}}  
 \label{lem:CPM-3ZK-hyb01:proof:case-2:final:derivation:eq:0}\\ 
& = 
\Lambda  
  \bigg(
  \ket{p_1}_{\reg{p_1}}\ket{\rho^{(t-1)}_{p_1}}  ~+~
   \sum_{p'_1 \in \bits^\ell \setminus \Set{p_1}} \ket{p'_1}_{\reg{p_1}} C_{1,0} B^\dagger_{1,1} C^\dagger_{1,1} \ket{\rho^{(t-1)}_{p'_1}}
  \bigg)  
  \label{lem:CPM-3ZK-hyb01:proof:case-2:final:derivation:eq:1}\\ 
  & =  
  \ket{p_1}_{\reg{p_1}} \Lambda_1 \ket{\rho^{(t-1)}_{p_1}}  ~+~
   \sum_{p'_1 \in \bits^\ell \setminus \Set{p_1}} \ket{p'_1}_{\reg{p_1}} \Lambda_0 C_{1,0} B^\dagger_{1,1} C^\dagger_{1,1} \ket{\rho^{(t-1)}_{p'_1}} 
  \label{lem:CPM-3ZK-hyb01:proof:case-2:final:derivation:eq:2}\\ 
  & =  
  \ket{p_1}_{\reg{p_1}} \Lambda_1 \ket{\rho^{(t-1)}_{p_1}}  ~+~
   \sum_{p'_1 \in \bits^\ell \setminus \Set{p_1}} \ket{p'_1}_{\reg{p_1}} C_{1,0} B^\dagger_{1,1} C^\dagger_{1,1} \Lambda^{\reg{t_1}/\reg{m}}_0 \ket{\rho^{(t-1)}_{p'_1}} 
  \label{lem:CPM-3ZK-hyb01:proof:case-2:final:derivation:eq:3}\\ 
    & =  
  \ket{p_1}_{\reg{p_1}}\ket{\rho^{(t)}_{p_1}}  ~+~
   \sum_{p'_1 \in \bits^\ell \setminus \Set{p_1}} \ket{p'_1}_{\reg{p_1}} C_{1,0} B^\dagger_{1,1} C^\dagger_{1,1} \ket{\rho^{(t)}_{p'_1}} 
  \label{lem:CPM-3ZK-hyb01:proof:case-2:final:derivation:eq:4}
,\end{align}
where  
\begin{itemize}
\item

\Cref{lem:CPM-3ZK-hyb01:proof:case-2:final:derivation:eq:0} follows from \Cref{lem:CPM-3ZK-hyb01:proof:invariant:proof:induction:case-2:initial:phi}; 

\item
\Cref{lem:CPM-3ZK-hyb01:proof:case-2:final:derivation:eq:1} follows from our induction assumption; 

\item
\Cref{lem:CPM-3ZK-hyb01:proof:case-2:final:derivation:eq:2}  follows from \Cref{lem:CPM-3ZK-hyb01:proof:case-2-1:claim:operator-commute:eq:1} in \Cref{lem:CPM-3ZK-hyb01:proof:case-2-1:claim:operator-commute}; 

\item
\Cref{lem:CPM-3ZK-hyb01:proof:case-2:final:derivation:eq:3} follows from \Cref{lem:CPM-3ZK-hyb01:proof:case-2-1:claim:operator-commute:eq:3} in \Cref{lem:CPM-3ZK-hyb01:proof:case-2-1:claim:operator-commute}; 

\item 
\Cref{lem:CPM-3ZK-hyb01:proof:case-2:final:derivation:eq:4} follows by defining
\begin{equation}\label[Expression]{lem:CPM-3ZK-hyb01:proof:case-2:final:derivation:def:rho-t}
\ket{\rho^{(t)}_{p_1}} 
\coloneqq 
 \Lambda_1 \ket{\rho^{(t-1)}_{p_1}} ~~~\text{and}~~~ 
\ket{\rho^{(t)}_{p'_1}} 
\coloneqq 
\Lambda^{\reg{t_1}/\reg{m}}_0 \ket{\rho^{(t-1)}_{p'_1}} ~~(\forall p'_1 \in \bits^\ell \setminus \Set{p_1}). 
\end{equation}
\end{itemize}
Clearly, \Cref{lem:CPM-3ZK-hyb01:proof:case-2:final:derivation:eq:4} is of the same format as \Cref{lem:CPM-3ZK-hyb01:proof:target-format:H0} in \Cref{lem:CPM-3ZK-hyb01:proof:invariant}.

\subpara{Proof of \Cref{lem:CPM-3ZK-hyb01:proof:target-format:H1}.} Next, we present the derivation for  \Cref{lem:CPM-3ZK-hyb01:proof:target-format:H1}:
\begin{align}
 \ket{\psi^{(t)}} 
 & = 
 \Lambda  \ket{\psi^{(t-1)}}  
 \label{lem:CPM-3ZK-hyb01:proof:case-2:final:derivation:psi:eq:0}\\ 
& = 
\Lambda  
  \bigg(
  \ket{p_1}_{\reg{p_1}}\ket{\rho^{(t-1)}_{p_1}}  ~+~
   \sum_{p'_1 \in \bits^\ell \setminus \Set{p_1}} \ket{p'_1}_{\reg{p_1}} C_{1,0}  \ket{\rho^{(t-1)}_{p'_1}}
  \bigg)  
  \label{lem:CPM-3ZK-hyb01:proof:case-2:final:derivation:psi:eq:1}\\ 
  & =  
  \ket{p_1}_{\reg{p_1}} \Lambda_1 \ket{\rho^{(t-1)}_{p_1}}  ~+~
   \sum_{p'_1 \in \bits^\ell \setminus \Set{p_1}} \ket{p'_1}_{\reg{p_1}} \Lambda_0 C_{1,0}  \ket{\rho^{(t-1)}_{p'_1}} 
  \label{lem:CPM-3ZK-hyb01:proof:case-2:final:derivation:psi:eq:2}\\ 
  & =  
  \ket{p_1}_{\reg{p_1}} \Lambda_1 \ket{\rho^{(t-1)}_{p_1}}  ~+~
   \sum_{p'_1 \in \bits^\ell \setminus \Set{p_1}} \ket{p'_1}_{\reg{p_1}} C_{1,0} \Lambda^{\reg{t_1}/\reg{m}}_0 \ket{\rho^{(t-1)}_{p'_1}} 
  \label{lem:CPM-3ZK-hyb01:proof:case-2:final:derivation:psi:eq:3}\\ 
    & =  
  \ket{p_1}_{\reg{p_1}}\ket{\rho^{(t)}_{p_1}}  ~+~
   \sum_{p'_1 \in \bits^\ell \setminus \Set{p_1}} \ket{p'_1}_{\reg{p_1}} C_{1,0} \ket{\rho^{(t)}_{p'_1}} 
  \label{lem:CPM-3ZK-hyb01:proof:case-2:final:derivation:psi:eq:4}
,\end{align}
where  
\begin{itemize}
\item
\Cref{lem:CPM-3ZK-hyb01:proof:case-2:final:derivation:psi:eq:0} follows from \Cref{lem:CPM-3ZK-hyb01:proof:invariant:proof:induction:case-2:initial:psi}, 

\item
\Cref{lem:CPM-3ZK-hyb01:proof:case-2:final:derivation:psi:eq:1} follows from our induction assumption, 

\item
\Cref{lem:CPM-3ZK-hyb01:proof:case-2:final:derivation:psi:eq:2}  follows from \Cref{lem:CPM-3ZK-hyb01:proof:case-2-1:claim:operator-commute:eq:1} in \Cref{lem:CPM-3ZK-hyb01:proof:case-2-1:claim:operator-commute}, 

\item
\Cref{lem:CPM-3ZK-hyb01:proof:case-2:final:derivation:psi:eq:3} follows from \Cref{lem:CPM-3ZK-hyb01:proof:case-2-1:claim:operator-commute:eq:2} in \Cref{lem:CPM-3ZK-hyb01:proof:case-2-1:claim:operator-commute}, 

\item 
\Cref{lem:CPM-3ZK-hyb01:proof:case-2:final:derivation:psi:eq:4} follows from {\em the same definitions} of $\ket{\rho^{(t)}_{p_1}}$ and $\ket{\rho^{(t)}_{p'_1}}$ as shown in \Cref{lem:CPM-3ZK-hyb01:proof:case-2:final:derivation:def:rho-t}.
\end{itemize}
Clearly, \Cref{lem:CPM-3ZK-hyb01:proof:case-2:final:derivation:psi:eq:4} is of the same format as \Cref{lem:CPM-3ZK-hyb01:proof:target-format:H1} in \Cref{lem:CPM-3ZK-hyb01:proof:invariant}.

This finishes the proof for \Cref{CPM-3ZK-hyb01:proof:invariant:proof:induction:case-2}.

\vspace{1em}
Finally, we remark that our proof for the base case in \Cref{sec:lem:CPM-3ZK-hyb01:proof:invariant:proof:base:sec} and the proof for the induction step in \Cref{sec:lem:CPM-3ZK-hyb01:proof:invariant:proof:induction:sec} together finish the proof of \Cref{lem:CPM-3ZK-hyb01:proof:invariant}, which in turn finishes the proof of \Cref{lem:CPM-3ZK:H01} eventually.

\subsection{Proof of \Cref{lem:CPM-3ZK:H12}}
\label{sec:lem:CPM-3ZK:H12:proof}

Due to a similar argument as we did at the beginning of \Cref{sec:lem:CPM-3ZK:H01:proof}, we claim that: to prove \Cref{lem:CPM-3ZK:H12}, it suffices to establish the following \Cref{lem:CPM-3ZK-hyb12:proof:invariant}.

\begin{lemma}[Invariance in $H^{(J, \msf{pat})}_1$ and $H^{(J, \msf{pat})}_2$]\label{lem:CPM-3ZK-hyb12:proof:invariant}
Assume that during the execution of $H^{(J, \msf{pat})}_1$ (and $H^{(J, \msf{pat})}_2$), the global counter reaches value $2$ for $T$ times in total. For each $t \in [T]$, there exist (possibly sub-normalized) pure states $\Set{\ket{\rho^{(t)}_{p'_1, p'_2}}}_{p'_1, p'_2 \in \bits^\ell \times \bits^\ell}$ so that the following holds: in hybrid $H^{(J, \msf{pat})}_1$ (resp.\ $H^{(J, \msf{pat})}_2$), when the global counter reaches value $2$ for the $t$-th time, the overall state can be written as $\ket{\phi^{(t)}}$ (resp.\ $\ket{\psi^{(t)}}$) defined as follows
\begin{align}
\ket{\phi^{(t)}} 
& =  \ket{p_1}_{\reg{p_1}} \ket{p_2}_{\reg{p_2}} \ket{\rho^{(t)}_{p_1, p_2}}  ~+~
   \sum_{(p'_1, p'_2)\ne (p_1, p_2)} \ket{p'_1}_{\reg{p_1}} \ket{p'_2}_{\reg{p_2}} C_{2,0} B^\dagger_{2,1} C^\dagger_{2,1} \ket{\rho^{(t)}_{p'_1, p'_2}}, \label{lem:CPM-3ZK-hyb12:proof:target-format:H1}\\ 
\ket{\psi^{(t)}} 
& =  \ket{p_1}_{\reg{p_1}} \ket{p_2}_{\reg{p_2}} \ket{\rho^{(t)}_{p_1, p_2}}  ~+~
 \sum_{(p'_1, p'_2)\ne (p_1, p_2)} \ket{p'_1}_{\reg{p_1}} \ket{p'_2}_{\reg{p_2}} \red{C_{2,0}} \ket{\rho^{(t)}_{p'_1, p'_2}} \label{lem:CPM-3ZK-hyb12:proof:target-format:H2}
,\end{align}
where the summations are taken over all $(p'_1, p'_2) \in \bits^{\ell} \times  \bits^{\ell} \setminus \Set{(p_1, p_2)}$ (abbreviated as $(p'_1, p'_2) \ne (p_1, p_2)$ in the above), the $(p_1,p_2)$ are defined in $\msf{pat}$, and  the unitaries $B_{2,1}$, $C_{2,0}$, and $C_{2,1}$  are as defined in \Cref{re-define:unitaries}. 
\end{lemma}

\para{Proving \Cref{lem:CPM-3ZK-hyb12:proof:invariant}.} Similar as the proof for \Cref{lem:CPM-3ZK-hyb01:proof:invariant}, 
we establish \Cref{lem:CPM-3ZK-hyb12:proof:invariant} through mathematical induction on the number $t \in [T]$, indicating the time at which the global counter reaches the value $2$. Throughout this proof, we will monitor the evolution of the overall states in both $H^{(J, \msf{pat})}_1$ and $H^{(J, \msf{pat})}_2$ simultaneously. This is done in \Cref{sec:lem:CPM-3ZK-hyb12:proof:invariant:proof:base:sec,sec:lem:CPM-3ZK-hyb12:proof:invariant:proof:induction:sec} respectively.

\subsubsection{Base Case ($t = 1$)}
\label{sec:lem:CPM-3ZK-hyb12:proof:invariant:proof:base:sec}

We first derive how the overall state evolves in $H^{(J, \msf{pat})}_1$. 

This case corresponds to the very first time the global counter reaches $2$. By definition, this is due to the $\msf{sq}(2)$ query. We assume w.l.o.g.\ that the overall state right before this query is some pure state $\ket{\rho}$.

\begin{remark}
Actually, we know the exact format of this $\ket{\rho}$ from the already-established \Cref{lem:CPM-3ZK-hyb01:proof:invariant}. That is, this state must be of the format  $\ket{\rho} = \ket{p_1}_{\reg{p_1}}\ket{\rho_{p_1}} + \sum_{p'_1 \in \bits^\ell \setminus \Set{p_1}} \ket{p'_1}_{\reg{p_1}} C_{1,0} \ket{\rho_{p'_1}}$. However, we remark that the exact format of $\ket{\rho}$ is not useful anywhere in this proof, and thus we do not need to refer to this long expression in the following derivation.
\end{remark}

By definition, the $\msf{sq}(2)$ query corresponds to the following procedure:  $\Sim$ will first apply her local unitary $S$ (followed by the projector $\ketbra{\downarrow}_{\reg{u}}$ which we hide as per \Cref{rmk:CPM-3ZK:hid-proj-u}). Next, according to the notation in \Cref{MnR:game:redefine:hybrids}, 
\begin{itemize}
\item
If $b_2 = 0$, then the operator $D_2C_2B_2 \ketbra{p_1,p_2}_{\reg{p_1p_2}} A_2 U_{gc}$ will be applied;
\item 
If $b_2 = 1$, then the operator $C_{2,0} \ketbra{p_1,p_2}_{\reg{p_1p_2}} A_2 U_{gc}$ will be applied;
\end{itemize}
In the following, we show the proof for $b_2 = 0$ only; the other case (i.e., $b_2 = 1$) can be established using the same argument.

It follows from the above discussion that the state $\ket{\phi^{(1)}}$ in $H^{(J, \msf{pat})}_1$ can be written as follows:
\begin{align*}
\ket{\phi^{(1)}} 
& =  
D_2C_2B_2\ketbra{p_1, p_2}_{\reg{p_1p_2}}A_2U_{gc} S   
\ket{\rho}
\\
& =  
D_2C_2B_2\ket{p_1, p_2}_{\reg{p_1p_2}} \ket{\rho_{p_1, p_2}}  
\numberthis \label{lem:CPM-3ZK-hyb12:proof:invariant:proof:base:eq:1} \\
& =  
\ket{p_1, p_2}_{\reg{p_1p_2}} 
D_{2,1} C_{2,1} B_{2,1}\ket{\rho_{p_1, p_2}}  
\numberthis \label{lem:CPM-3ZK-hyb12:proof:invariant:proof:base:eq:2} \\ 
& =  
\ket{p_1, p_2}_{\reg{p_1p_2}}\ket{\rho^{(1)}_{p_1, p_2}} 
\numberthis \label{lem:CPM-3ZK-hyb12:proof:invariant:proof:base:eq:3}
,\end{align*}
where
\begin{itemize}
\item 
\Cref{lem:CPM-3ZK-hyb12:proof:invariant:proof:base:eq:1} follows by defining $\ket{\rho_{p_1, p_2}}  \coloneqq \bra{p_1, p_2}_{\reg{p_1p_2}}A_2U_{gc} S \ket{\rho}$;

\item 
\Cref{lem:CPM-3ZK-hyb12:proof:invariant:proof:base:eq:2} follows from the definitions of $B_2$, $C_2$, and $D_2$ (see \Cref{re-define:unitaries});

\item 
\Cref{lem:CPM-3ZK-hyb12:proof:invariant:proof:base:eq:3} follows by defining $\ket{\rho^{(1)}_{p_1, p_2}} \coloneqq D_{2,1} C_{2,1} B_{2,1}\ket{\rho_{p_1, p_2}}$.
\end{itemize}
It is straightforward that the $\ket{\phi^{(1)}}$ shown in \Cref{lem:CPM-3ZK-hyb12:proof:invariant:proof:base:eq:3} satisfies the format shown in \Cref{lem:CPM-3ZK-hyb12:proof:target-format:H1} when $t = 1$.

Also, note that the game $H^{(J, \msf{pat})}_1$ and $H^{(J, \msf{pat})}_2$ are identical so far and thus $\ket{\phi^{(1)}} = \ket{\psi^{(1)}}$. Therefore, it follows from \Cref{lem:CPM-3ZK-hyb12:proof:invariant:proof:base:eq:3} that 
$$\ket{\psi^{(1)}} =  \ket{p_1, p_2}_{\reg{p_1p_2}} \ket{\rho^{(1)}_{p_1,p_2}}.$$ Such a $\ket{\psi^{(1)}}$ satisfies the format shown in \Cref{lem:CPM-3ZK-hyb12:proof:target-format:H2} with $t = 1$ as well.

This finishes the proof for the base case $t =1$.

\subsubsection{Induction Step ($t \ge 2$)} 
\label{sec:lem:CPM-3ZK-hyb12:proof:invariant:proof:induction:sec} 

We assume $\ket{\phi^{(t-1)}}$ and $\ket{\psi^{(t-1)}}$ satisfy \Cref{lem:CPM-3ZK-hyb12:proof:invariant}, and show in the following that \Cref{lem:CPM-3ZK-hyb12:proof:invariant} holds when the global counter reaches $2$ for the $t$-th time.

We first describe formally how $\ket{\phi^{(t-1)}}$ (resp.\ $\ket{\psi^{(t-1)}}$) evolves into $\ket{\phi^{(t)}}$ (resp.\ $\ket{\psi^{(t)}}$):
\begin{enumerate}
 \item
$\Sim$'s local unitary $S$ is applied (followed by the projector $\ketbra{\uparrow}_{\reg{u}}$ which we hide as per \Cref{rmk:CPM-3ZK:hid-proj-u}). Note that this step is identical for both $H^{(J, \msf{pat})}_1$ and $H^{(J, \msf{pat})}_2$. 

\item
An $\uparrow$-query is made, bringing the global counter from 2 to 1. According to the notation in \Cref{MnR:game:redefine:hybrids}: 
\begin{itemize}
    \item
    In $H^{(J, \msf{pat})}_1$, this corresponds to applying $U^{\dagger}_{gc} A^\dagger_2 B^\dagger_2 C^\dagger_2 D^\dagger_2$; 
    \item
    In $H^{(J, \msf{pat})}_2$, this corresponds to applying $U^{\dagger}_{gc} A^\dagger_2 \ddot{C}^\dagger_2 D^\dagger_2$. 
\end{itemize}
\item \label[Step]{lem:CPM-3ZK-hyb12:proof:invariant:proof:induction:description:def:lambda}
At this point, there are two cases to consider: 
\begin{enumerate}
\item \label[Case]{lem:CPM-3ZK-hyb12:proof:invariant:proof:case-2-1}
 $\Sim$ can make a $\downarrow$-query immediately, bringing the global counter back to $2$; {\em or } 
\item \label[Case]{lem:CPM-3ZK-hyb12:proof:invariant:proof:case-2-2}
 $\Sim$ can rewind the verifier $\tilde{\Verifier}$ further, bringing the global counter to $0$.  
\end{enumerate}
{\bf Defining Operator $\Lambda$:} Looking ahead, our proof can handle these two cases at one stroke. To do that, we denote the execution from this point until (exclusively) {\em the next $\downarrow$ query that brings the global counter to value $2$} (i.e., the exact query that yields the global counter's $t$-th arrival at value $2$) as an operator $\Lambda$. $\Lambda$ might be $\Sim$ local operator $S$ {\em only}, corresponding to \Cref{lem:CPM-3ZK-hyb12:proof:invariant:proof:case-2-1}; $\Lambda$ might also be the combonation of some other operators, representing the operations performed in \Cref{lem:CPM-3ZK-hyb12:proof:invariant:proof:case-2-2} before the global counter's next arrival at value 2. 

\item
After the application of the operator $\Lambda$ defined above, an $\downarrow$-query is made, bringing the global counter from 1 back to 2, which is the global counter's $t$-th arrival at value 2. According to the notation in \Cref{MnR:game:redefine:hybrids}:  
\begin{itemize}
    \item
    In $H^{(J, \msf{pat})}_0$, this corresponds to applying $D_2C_2B_2A_2U_{gc}$; 
    \item
    In $H^{(J, \msf{pat})}_1$, this corresponds to applying $D_2\ddot{C}_2A_2U_{gc}$. 
\end{itemize} 
\end{enumerate}
It follows from the above description that the states $\ket{\phi^{(t)}}$ and $\ket{\psi^{(t)}}$ can be written as:
\begin{align}
\ket{\phi^{(t)}}
& = D_2C_2B_2A_2U_{gc} \Lambda U^{\dagger}_{gc} A^\dagger_2 B^\dagger_2 C^\dagger_2 D^\dagger_2 S  \ket{\phi^{(t-1)}}, \label{lem:CPM-3ZK-hyb12:proof:invariant:proof:induction:eq:k:H1}\\ 
\ket{\psi^{(t)}}
& = D_2 \ddot{C}_2A_2U_{gc} \Lambda U^{\dagger}_{gc} A^\dagger_2 \ddot{C}^\dagger_2 D^\dagger_2 S \ket{\psi^{(t-1)}}. \label{lem:CPM-3ZK-hyb12:proof:invariant:proof:induction:eq:k:H2}
\end{align}

We first make a useful claim that specifies the registers on which the $\Lambda$ defined above acts non-trivially. 
\begin{MyClaim}[Non-Trivial Registers for $\Lambda$]\label{lem:CPM-3ZK-hyb12:proof:invariant:proof:induction:Lambda-register:claim}
The $\Lambda$ defined in \Cref{lem:CPM-3ZK-hyb12:proof:invariant:proof:induction:description:def:lambda} acts non-trivially only on registers $\reg{m}$, $\reg{u}$, $\reg{s}$, $\reg{p_1}$, $\reg{t_1}$, $\reg{lc}$, and $\reg{gc}$ (i.e., it works as the identity operator on the other registers). 
\end{MyClaim}
\begin{proof}[Proof of \Cref{lem:CPM-3ZK-hyb12:proof:invariant:proof:induction:Lambda-register:claim}]

By definition, $\Lambda$ only contains operations that happen when the global counter is $1$ or smaller. This is best illustrated by \Cref{figure:CPM-3ZK:hybrids:H1}---$\Lambda$ only contains operations that happen in between the line corresponding to $\ket{0}_{\reg{gc}}$ and the line corresponding to $\ket{1}_{\reg{gc}}$. That is, all the $\downarrow$ and $\uparrow$ queries are answered by the ``dummy'' operators $\ddot{V}$ and $\ddot{V}^\dagger$. (One exception is the $\tilde{V}$ corresponding $\msf{sq}(1)$. But this does not matter in the current proof---Before the application of operator $\Lambda$, the global counter has already been $2$ (for $t-1$ times), which implies that the $\msf{sq}(1)$ query has already happened and thus cannot be a part of the operator $\Lambda$.)

Therefore, according to our notation in \Cref{MnR:game:redefine:hybrids}, operator $\Lambda$ can be written as $\Lambda = \Gamma_z \Gamma_{z-1} \cdots \Gamma_1$ (for some integer $z$) where each $\Gamma_i$ ($i \in [z]$) comes from the set of operators $\Set{S, A_1,  \ddot{C}_1, D_1, U_{gc}}$. It then follows from \Cref{CPM:non-trivial-registers:table} that $\Lambda$ acts non-trivially only on registers $\reg{m}$, $\reg{u}$, $\reg{s}$, $\reg{p_1}$, $\reg{t_1}$, $\reg{lc}$, and $\reg{gc}$.

This completes the proof of \Cref{lem:CPM-3ZK-hyb12:proof:invariant:proof:induction:Lambda-register:claim}.

\end{proof}

\para{High-Level Idea for the Sequel.} Recall that our eventual goal is to prove that the states $\ket{\phi^{(t)}}$ and $\ket{\psi^{(t)}}$ are of the format shown in \Cref{lem:CPM-3ZK-hyb12:proof:target-format:H1,lem:CPM-3ZK-hyb12:proof:target-format:H2} in \Cref{lem:CPM-3ZK-hyb12:proof:invariant}. At a high level, we prove it by applying \Cref{lem:err-inv-com} to \Cref{lem:CPM-3ZK-hyb12:proof:invariant:proof:induction:eq:k:H1,lem:CPM-3ZK-hyb12:proof:invariant:proof:induction:eq:k:H2}. But we first need to perform some preparation work, putting \Cref{lem:CPM-3ZK-hyb12:proof:invariant:proof:induction:eq:k:H1,lem:CPM-3ZK-hyb12:proof:invariant:proof:induction:eq:k:H2} into a format that is more ``compatible'' with \Cref{lem:err-inv-com}. In the sequel, we first perform the preparation work in \Cref{lem:CPM-3ZK-hyb12:proof:invariant:proof:induction:preparation-claim,lem:CPM-3ZK-hyb12:proof:invariant:proof:induction:preparation-claim:2}. Then, we show on \Cpageref{lem:CPM-3ZK-hyb12:proof:invariant:proof:induction:finish} how to use \Cref{lem:CPM-3ZK-hyb12:proof:invariant:proof:induction:preparation-claim,lem:CPM-3ZK-hyb12:proof:invariant:proof:induction:preparation-claim:2} to complete the proof for the induction step.

The following \Cref{lem:CPM-3ZK-hyb12:proof:invariant:proof:induction:preparation-claim} can be viewed as an analogue of \Cref{lem:CPM-3ZK-hyb01:proof:invariant:proof:induction:case1:preparation-claim}.
\begin{MyClaim}\label{lem:CPM-3ZK-hyb12:proof:invariant:proof:induction:preparation-claim}
There exist (possibly sub-normalized) pure states $\Set{\dot{\rho}^{(t-1)}_{p'_1, p'_2}}_{(p'_1,p'_2 ) \in \bits^\ell\times\bits^\ell}$ so that the following holds
\begin{align*}
\ket{\phi^{(t)}}
& =  
D_2C_2 B_2 U_{gc} \Lambda^{\reg{p_2}/\reg{m}} U_{gc}^\dagger   B^\dagger_2 C^\dagger_2
\bigg( 
\ket{p_1}_{\reg{p_1}} \ket{p_2}_{\reg{p_2}} \ket{\dot{\rho}^{(t-1)}_{p_1, p_2}}  
+ \\ 
& \hspace{16em} 
\sum_{(p'_1, p'_2)\ne (p_1, p_2)}  \ket{p'_1}_{\reg{p_1}} \ket{p'_2}_{\reg{p_2}}   C_{2,0} B^\dagger_{2,1} C^\dagger_{2,1}  \ket{\dot{\rho}^{(t-1)}_{p'_1, p'_2}} 
\bigg) \numberthis \label{lem:CPM-3ZK-hyb12:proof:invariant:proof:induction:preparation-claim:eq:target:phi} \\ 
\ket{\psi^{(t)}}
& = 
D_2\ddot{C}_2 U_{gc}\Lambda^{\reg{p_2}/\reg{m}}  U_{gc}^\dagger \ddot{C}^\dagger_2 
\bigg( 
\ket{p_1}_{\reg{p_1}} \ket{p_2}_{\reg{p_2}} \ket{\dot{\rho}^{(t-1)}_{p_1, p_2}}   +  
\sum_{(p'_1, p'_2)\ne (p_1, p_2)}  \ket{p'_1}_{\reg{p_1}} \ket{p'_2}_{\reg{p_2}}   C_{2,0}  \ket{\dot{\rho}^{(t-1)}_{p'_1, p'_2}} 
\bigg) 
\numberthis \label{lem:CPM-3ZK-hyb12:proof:invariant:proof:induction:preparation-claim:eq:target:psi} 
\end{align*}
where $\Lambda^{\reg{p_2}/\reg{m}}$ is identical to the $\Lambda$ defined in \Cref{lem:CPM-3ZK-hyb12:proof:invariant:proof:induction:description:def:lambda}, except that it treats register $\reg{p_2}$ as $\reg{m}$, and the summations are taken over all $(p'_1, p'_2) \in \bits^{\ell} \times  \bits^{\ell} \setminus \Set{(p_1, p_2)}$ (abbreviated as $(p'_1, p'_2) \ne (p_1, p_2)$ in the above).
\end{MyClaim}
\begin{proof}[Proof of \Cref{lem:CPM-3ZK-hyb12:proof:invariant:proof:induction:preparation-claim}]
First, notice that  
\begin{align*}
 A_2 U_{gc} \Lambda U^\dagger_{gc} A^\dagger_2 
 & = 
 A_2 U_{gc} \Lambda A^\dagger_2 U^\dagger_{gc}
 \numberthis \label{lem:CPM-3ZK-hyb12:proof:invariant:proof:induction:preparation-claim:proof:eq:swap-S:1} \\ 
 & = 
 A_2 U_{gc} A^\dagger_2 \Lambda^{\reg{p_2}/\reg{m}}  U^\dagger_{gc}
 \numberthis \label{lem:CPM-3ZK-hyb12:proof:invariant:proof:induction:preparation-claim:proof:eq:swap-S:2} \\
  & = 
 A_2 A^\dagger_2 U_{gc} \Lambda^{\reg{p_2}/\reg{m}}  U^\dagger_{gc}
 \numberthis \label{lem:CPM-3ZK-hyb12:proof:invariant:proof:induction:preparation-claim:proof:eq:swap-S:3} \\
 & = 
U_{gc} \Lambda^{\reg{p_2}/\reg{m}}  U^\dagger_{gc}
 \numberthis \label{lem:CPM-3ZK-hyb12:proof:invariant:proof:induction:preparation-claim:proof:eq:swap-S} 
,\end{align*}
where 
\begin{itemize}
\item
\Cref{lem:CPM-3ZK-hyb12:proof:invariant:proof:induction:preparation-claim:proof:eq:swap-S:1,lem:CPM-3ZK-hyb12:proof:invariant:proof:induction:preparation-claim:proof:eq:swap-S:3} follows from the fact that $U_{gc}$ acts non-trivially on different registers from $A_2$;

\item
\Cref{lem:CPM-3ZK-hyb12:proof:invariant:proof:induction:preparation-claim:proof:eq:swap-S:2} from the fact that $\Lambda$ acts as the identity operator on $\reg{p_2}$ (see \Cref{lem:CPM-3ZK-hyb12:proof:invariant:proof:induction:Lambda-register:claim}) and that $A_2$ is nothing but a swap operator between $\reg{m}$ and $\reg{p_2}$ (see \Cref{CPM:V:unitary}). 
\end{itemize}

\subpara{Proving \Cref{lem:CPM-3ZK-hyb12:proof:invariant:proof:induction:preparation-claim:eq:target:phi}.} We now show the derivation for $\ket{\phi^{(t)}}$:
\begin{align*}
\ket{\phi^{(t)}}
& =
D_2C_2B_2A_2U_{gc} \Lambda U^{\dagger}_{gc} A^\dagger_2 B^\dagger_2 C^\dagger_2 D^\dagger_2  S \ket{\phi^{(t-1)}}
\numberthis \label{lem:CPM-3ZK-hyb12:proof:invariant:proof:induction:preparation-claim:proof:phi-derivation:0} \\ 
& =
D_2C_2B_2  U_{gc} \Lambda^{\reg{p_2}/\reg{m}}  U^\dagger_{gc} B^\dagger_2 C^\dagger_2 D^\dagger_2  S \ket{\phi^{(t-1)}}
\numberthis \label{lem:CPM-3ZK-hyb12:proof:invariant:proof:induction:preparation-claim:proof:phi-derivation:1} \\ 
& =
D_2C_2B_2  U_{gc} \Lambda^{\reg{p_2}/\reg{m}}  U^\dagger_{gc} B^\dagger_2 C^\dagger_2 D^\dagger_2 S 
\bigg(
\ket{p_1}_{\reg{p_1}}\ket{p_2}_{\reg{p_2}}\ket{\rho^{(t-1)}_{p_1,p_2}}  + \\ 
& \hspace{16em}
 \sum_{(p'_1, p'_2) \ne (p_1, p_2)} \ket{p'_1}_{\reg{p_1}} \ket{p'_2}_{\reg{p_2}} C_{2,0} B^\dagger_{2,1} C^\dagger_{2,1} \ket{\rho^{(t-1)}_{p'_1,p'_2}}
\bigg)
\numberthis \label{lem:CPM-3ZK-hyb12:proof:invariant:proof:induction:preparation-claim:proof:phi-derivation:2} \\ 
& =
D_2C_2B_2  U_{gc} \Lambda^{\reg{p_2}/\reg{m}}  U^\dagger_{gc} B^\dagger_2 C^\dagger_2 D^\dagger_2  
\bigg(
\ket{p_1}_{\reg{p_1}}\ket{p_2}_{\reg{p_2}}S\ket{\rho^{(t-1)}_{p_1,p_2}}  + \\ 
& \hspace{16em}
 \sum_{(p'_1, p'_2) \ne (p_1, p_2)} \ket{p'_1}_{\reg{p_1}} \ket{p'_2}_{\reg{p_2}} SC_{2,0} B^\dagger_{2,1} C^\dagger_{2,1} \ket{\rho^{(t-1)}_{p'_1,p'_2}}
\bigg)
\numberthis \label{lem:CPM-3ZK-hyb12:proof:invariant:proof:induction:preparation-claim:proof:phi-derivation:3} \\ 
& =
D_2C_2B_2  U_{gc} \Lambda^{\reg{p_2}/\reg{m}}  U^\dagger_{gc} B^\dagger_2 C^\dagger_2  
\bigg(
\ket{p_1}_{\reg{p_1}}\ket{p_2}_{\reg{p_2}} D^\dagger_{2,1} S\ket{\rho^{(t-1)}_{p_1,p_2}}  + \\ 
& \hspace{16em}
 \sum_{(p'_1, p'_2) \ne (p_1, p_2)} \ket{p'_1}_{\reg{p_1}} \ket{p'_2}_{\reg{p_2}} S C_{2,0}  B^\dagger_{2,1} C^\dagger_{2,1} \ket{\rho^{(t-1)}_{p'_1,p'_2}}
\bigg)
\numberthis \label{lem:CPM-3ZK-hyb12:proof:invariant:proof:induction:preparation-claim:proof:phi-derivation:4} \\ 
& =
D_2C_2B_2  U_{gc} \Lambda^{\reg{p_2}/\reg{m}}  U^\dagger_{gc} B^\dagger_2 C^\dagger_2  
\bigg(
\ket{p_1}_{\reg{p_1}}\ket{p_2}_{\reg{p_2}} D^\dagger_{2,1} S\ket{\rho^{(t-1)}_{p_1,p_2}}  + \\ 
& \hspace{14em}
 \sum_{(p'_1, p'_2) \ne (p_1, p_2)} \ket{p'_1}_{\reg{p_1}} \ket{p'_2}_{\reg{p_2}}  C_{2,0} S^{\reg{t_2}/\reg{m}}  B^\dagger_{2,1} C^\dagger_{2,1} \ket{\rho^{(t-1)}_{p'_1,p'_2}}
\bigg)
\numberthis \label{lem:CPM-3ZK-hyb12:proof:invariant:proof:induction:preparation-claim:proof:phi-derivation:5} \\ 
& =
D_2C_2B_2  U_{gc} \Lambda^{\reg{p_2}/\reg{m}}  U^\dagger_{gc} B^\dagger_2 C^\dagger_2  
\bigg(
\ket{p_1}_{\reg{p_1}}\ket{p_2}_{\reg{p_2}} D^\dagger_{2,1} S\ket{\rho^{(t-1)}_{p_1,p_2}}  + \\ 
& \hspace{14em}
 \sum_{(p'_1, p'_2) \ne (p_1, p_2)} \ket{p'_1}_{\reg{p_1}} \ket{p'_2}_{\reg{p_2}}  C_{2,0} B^\dagger_{2,1} C^\dagger_{2,1}  S^{\reg{t_2}/\reg{m}}  \ket{\rho^{(t-1)}_{p'_1,p'_2}}
\bigg)
\numberthis \label{lem:CPM-3ZK-hyb12:proof:invariant:proof:induction:preparation-claim:proof:phi-derivation:6} \\ 
& =
D_2C_2B_2  U_{gc} \Lambda^{\reg{p_2}/\reg{m}}  U^\dagger_{gc} B^\dagger_2 C^\dagger_2  
\bigg(
\ket{p_1}_{\reg{p_1}}\ket{p_2}_{\reg{p_2}} \ket{\dot{\rho}^{(t-1)}_{p_1,p_2}}  + \\ 
& \hspace{16em}
 \sum_{(p'_1, p'_2) \ne (p_1, p_2)} \ket{p'_1}_{\reg{p_1}} \ket{p'_2}_{\reg{p_2}}  C_{2,0} B^\dagger_{2,1} C^\dagger_{2,1}     \ket{\dot{\rho}^{(t-1)}_{p'_1,p'_2}}
\bigg)
\numberthis \label{lem:CPM-3ZK-hyb12:proof:invariant:proof:induction:preparation-claim:proof:phi-derivation:7} 
,\end{align*}
where 
\begin{itemize}
\item
\Cref{lem:CPM-3ZK-hyb12:proof:invariant:proof:induction:preparation-claim:proof:phi-derivation:0} follows from \Cref{lem:CPM-3ZK-hyb12:proof:invariant:proof:induction:eq:k:H1};

\item
\Cref{lem:CPM-3ZK-hyb12:proof:invariant:proof:induction:preparation-claim:proof:phi-derivation:1} follows from \Cref{lem:CPM-3ZK-hyb12:proof:invariant:proof:induction:preparation-claim:proof:eq:swap-S};

\item
\Cref{lem:CPM-3ZK-hyb12:proof:invariant:proof:induction:preparation-claim:proof:phi-derivation:2} follows from our induction assumption;

\item
\Cref{lem:CPM-3ZK-hyb12:proof:invariant:proof:induction:preparation-claim:proof:phi-derivation:3} from  from the fact that $S$ acts as the identity operator on $\reg{p_1}$ and $\reg{p_2}$ (see \Cref{CPM:non-trivial-registers:table});

\item
\Cref{lem:CPM-3ZK-hyb12:proof:invariant:proof:induction:preparation-claim:proof:phi-derivation:4} from the definition of $D_2$ (see \Cref{re-define:unitaries});

\item
\Cref{lem:CPM-3ZK-hyb12:proof:invariant:proof:induction:preparation-claim:proof:phi-derivation:5} from the fact that $S$ acts as the identity operator on $\reg{t_2}$ (see \Cref{CPM:non-trivial-registers:table}) and $C_{2,0}$ is nothing but a swap operator between $\reg{t_2}$ and $\reg{m}$ (see \Cref{re-define:unitaries}); (Note that $S^{\reg{t_2}/\reg{m}}$ is defined to be an operator that is identical to $S$ except that it treats $\reg{t_2}$ as $\reg{m}$.)

\item
\Cref{lem:CPM-3ZK-hyb12:proof:invariant:proof:induction:preparation-claim:proof:phi-derivation:6} follows from the fact that $S^{\reg{t_2}/\reg{m}}$ acts non-trivially on different registers from $B_{2,1}$ and $C_{2,1}$ (see \Cref{CPM:non-trivial-registers:table});

\item
\Cref{lem:CPM-3ZK-hyb12:proof:invariant:proof:induction:preparation-claim:proof:phi-derivation:7} follows by defining 
\begin{equation}
\label[Expression]{lem:CPM-3ZK-hyb12:proof:invariant:proof:induction:preparation-claim:proof:def:rho-dot}
\ket{\dot{\rho}^{(t-1)}_{p_1,p_2}}   \coloneqq D^\dagger_{2,1} S \ket{\rho^{(t-1)}_{p_1,p_2}} ~~~\text{and}~~~
\ket{\dot{\rho}^{(t-1)}_{p'_1,p'_2}}  \coloneqq S^{\reg{t_2}/\reg{m}} \ket{\rho^{(t-1)}_{p'_1,p'_2}} ~~\big(\forall (p'_1,p'_2) \in \big(\bits^\ell \times \bits^\ell\big) \setminus \Set{(p_1,p_2)}\big).
\end{equation}
\end{itemize}
\Cref{lem:CPM-3ZK-hyb12:proof:invariant:proof:induction:preparation-claim:proof:phi-derivation:7} finishes the proof of \Cref{lem:CPM-3ZK-hyb12:proof:invariant:proof:induction:preparation-claim:eq:target:phi} in \Cref{lem:CPM-3ZK-hyb12:proof:invariant:proof:induction:preparation-claim}.

\subpara{Proving \Cref{lem:CPM-3ZK-hyb12:proof:invariant:proof:induction:preparation-claim:eq:target:psi}.} We now show the derivation for $\ket{\psi^{(t)}}$. This is almost identical to the above proof for \Cref{lem:CPM-3ZK-hyb12:proof:invariant:proof:induction:preparation-claim:eq:target:phi}. Nevertheless, we present it for the sake of completeness.

\begin{align*}
\ket{\psi^{(t)}}
& =
D_2\ddot{C}_2A_2U_{gc} \Lambda U^{\dagger}_{gc} A^\dagger_2 \ddot{C}^\dagger_2 D^\dagger_2  S \ket{\psi^{(t-1)}}
\numberthis \label{lem:CPM-3ZK-hyb12:proof:invariant:proof:induction:preparation-claim:proof:psi-derivation:0} \\ 
& =
D_2 \ddot{C}_2  U_{gc} \Lambda^{\reg{p_2}/\reg{m}}  U^\dagger_{gc} \ddot{C}^\dagger_2 D^\dagger_2  S \ket{\psi^{(t-1)}}
\numberthis \label{lem:CPM-3ZK-hyb12:proof:invariant:proof:induction:preparation-claim:proof:psi-derivation:1} \\ 
& =
D_2 \ddot{C}_2  U_{gc} \Lambda^{\reg{p_2}/\reg{m}}  U^\dagger_{gc} \ddot{C}^\dagger_2 D^\dagger_2 S 
\bigg(
\ket{p_1}_{\reg{p_1}}\ket{p_2}_{\reg{p_2}}\ket{\rho^{(t-1)}_{p_1,p_2}} 
+
 \sum_{(p'_1, p'_2) \ne (p_1, p_2)} \ket{p'_1}_{\reg{p_1}} \ket{p'_2}_{\reg{p_2}} C_{2,0} \ket{\rho^{(t-1)}_{p'_1,p'_2}}
\bigg)
\numberthis \label{lem:CPM-3ZK-hyb12:proof:invariant:proof:induction:preparation-claim:proof:psi-derivation:2} \\  
& =
D_2 \ddot{C}_2  U_{gc} \Lambda^{\reg{p_2}/\reg{m}}  U^\dagger_{gc} \ddot{C}^\dagger_2  
\bigg(
\ket{p_1}_{\reg{p_1}}\ket{p_2}_{\reg{p_2}} D^\dagger_{2,1} S\ket{\rho^{(t-1)}_{p_1,p_2}}  
+ 
 \sum_{(p'_1, p'_2) \ne (p_1, p_2)} \ket{p'_1}_{\reg{p_1}} \ket{p'_2}_{\reg{p_2}}  C_{2,0}  S^{\reg{t_2}/\reg{m}}  \ket{\rho^{(t-1)}_{p'_1,p'_2}}
\bigg)
\numberthis \label{lem:CPM-3ZK-hyb12:proof:invariant:proof:induction:preparation-claim:proof:psi-derivation:6} \\ 
& =
D_2 \ddot{C}_2  U_{gc} \Lambda^{\reg{p_2}/\reg{m}}  U^\dagger_{gc}  \ddot{C}^\dagger_2  
\bigg(
\ket{p_1}_{\reg{p_1}}\ket{p_2}_{\reg{p_2}} \ket{\dot{\rho}^{(t-1)}_{p_1,p_2}} 
+
 \sum_{(p'_1, p'_2) \ne (p_1, p_2)} \ket{p'_1}_{\reg{p_1}} \ket{p'_2}_{\reg{p_2}}  C_{2,0}  \ket{\dot{\rho}^{(t-1)}_{p'_1,p'_2}}
\bigg)
\numberthis \label{lem:CPM-3ZK-hyb12:proof:invariant:proof:induction:preparation-claim:proof:psi-derivation:7} 
,\end{align*}
where 
\begin{itemize}
\item
\Cref{lem:CPM-3ZK-hyb12:proof:invariant:proof:induction:preparation-claim:proof:psi-derivation:0} follows from \Cref{lem:CPM-3ZK-hyb12:proof:invariant:proof:induction:eq:k:H2};

\item
\Cref{lem:CPM-3ZK-hyb12:proof:invariant:proof:induction:preparation-claim:proof:psi-derivation:1} follows from \Cref{lem:CPM-3ZK-hyb12:proof:invariant:proof:induction:preparation-claim:proof:eq:swap-S};

\item
\Cref{lem:CPM-3ZK-hyb12:proof:invariant:proof:induction:preparation-claim:proof:psi-derivation:2}  follows from our induction assumption;

\item
\Cref{lem:CPM-3ZK-hyb12:proof:invariant:proof:induction:preparation-claim:proof:psi-derivation:6} follows from a similar argument as we did to derive \Cref{lem:CPM-3ZK-hyb12:proof:invariant:proof:induction:preparation-claim:proof:phi-derivation:6} from \Cref{lem:CPM-3ZK-hyb12:proof:invariant:proof:induction:preparation-claim:proof:phi-derivation:2};

\item
\Cref{lem:CPM-3ZK-hyb12:proof:invariant:proof:induction:preparation-claim:proof:psi-derivation:7} follows from {\em the same definitions} of $\ket{\dot{\rho}^{(t-1)}_{p_1,p_2}}$ and $\ket{\dot{\rho}^{(t-1)}_{p'_1,p'_2}}$ as shown in \Cref{lem:CPM-3ZK-hyb12:proof:invariant:proof:induction:preparation-claim:proof:def:rho-dot}.
\end{itemize}
\Cref{lem:CPM-3ZK-hyb12:proof:invariant:proof:induction:preparation-claim:proof:psi-derivation:7} finishes the proof of \Cref{lem:CPM-3ZK-hyb12:proof:invariant:proof:induction:preparation-claim:eq:target:psi} in \Cref{lem:CPM-3ZK-hyb12:proof:invariant:proof:induction:preparation-claim}.

\vspace{1em}
This finishes the proof of \Cref{lem:CPM-3ZK-hyb12:proof:invariant:proof:induction:preparation-claim}.

\end{proof}

The following \Cref{lem:CPM-3ZK-hyb12:proof:invariant:proof:induction:preparation-claim:2} can be treated as an analogue of \Cref{lem:CPM-3ZK-hyb01:proof:invariant:proof:induction:case1:preparation-claim:2}. 
\begin{MyClaim}\label{lem:CPM-3ZK-hyb12:proof:invariant:proof:induction:preparation-claim:2}
Let $\Lambda^{\reg{p_2}/\reg{m}}$ and $\Set{\dot{\rho}^{(t-1)}_{p'_1, p'_2}}_{(p'_1,p'_2 ) \in \bits^\ell\times\bits^\ell}$ be as defined in \Cref{lem:CPM-3ZK-hyb12:proof:invariant:proof:induction:preparation-claim}. Let
\begin{align}
\ket{\gamma^{(t-1)}_0} 
& \coloneqq 
\ket{p_1}_{\reg{p_1}} \ket{p_2}_{\reg{p_2}} \ket{\dot{\rho}^{(t-1)}_{p_1, p_2}}  
+  
\sum_{(p'_1, p'_2)\ne (p_1, p_2)}  \ket{p'_1}_{\reg{p_1}} \ket{p'_2}_{\reg{p_2}}   C_{2,0} B^\dagger_{2,1} C^\dagger_{2,1}  \ket{\dot{\rho}^{(t-1)}_{p'_1, p'_2}} 
\label{lem:CPM-3ZK-hyb12:proof:invariant:proof:induction:preparation-claim:2:eq:in:0} \\ 
\ket{\gamma^{(t-1)}_1} 
& \coloneqq 
\ket{p_1}_{\reg{p_1}} \ket{p_2}_{\reg{p_2}} \ket{\dot{\rho}^{(t-1)}_{p_1, p_2}}  
+  
\sum_{(p'_1, p'_2)\ne (p_1, p_2)}  \ket{p'_1}_{\reg{p_1}} \ket{p'_2}_{\reg{p_2}}   C_{2,0}   \ket{\dot{\rho}^{(t-1)}_{p'_1, p'_2}} 
\label{lem:CPM-3ZK-hyb12:proof:invariant:proof:induction:preparation-claim:2:eq:in:1} \\ 
\ket{\gamma^{(t)}_0} 
& \coloneqq 
C_2 B_2 U_{gc} \Lambda^{\reg{p_2}/\reg{m}} U_{gc}^\dagger   B^\dagger_2 C^\dagger_2 \ket{\gamma^{(t-1)}_0} 
\label{lem:CPM-3ZK-hyb12:proof:invariant:proof:induction:preparation-claim:2:eq:out:0} \\ 
\ket{\gamma^{(t)}_1}
& \coloneqq 
\ddot{C}_2 U_{gc}\Lambda^{\reg{p_2}/\reg{m}}  U_{gc}^\dagger \ddot{C}^\dagger_2 
 \ket{\gamma^{(t-1)}_1}
\label{lem:CPM-3ZK-hyb12:proof:invariant:proof:induction:preparation-claim:2:eq:out:1} 
.\end{align}
Then, there exist (possibly sub-normalized) pure states $\Set{\dot{\rho}^{(t)}_{p'_1, p'_2}}_{(p'_1,p'_2 ) \in \bits^\ell\times\bits^\ell}$ so that the following holds:
\begin{align}
\ket{\gamma^{(t)}_0} 
& =
\ket{p_1}_{\reg{p_1}} \ket{p_2}_{\reg{p_2}} \ket{\dot{\rho}^{(t)}_{p_1, p_2}}  
+  
\sum_{(p'_1, p'_2)\ne (p_1, p_2)}  \ket{p'_1}_{\reg{p_1}} \ket{p'_2}_{\reg{p_2}}   C_{2,0} B^\dagger_{2,1} C^\dagger_{2,1}  \ket{\dot{\rho}^{(t)}_{p'_1, p'_2}} 
\label{lem:CPM-3ZK-hyb12:proof:invariant:proof:induction:preparation-claim:2:eq:target:0} \\ 
\ket{\gamma^{(t)}_1} 
& = 
\ket{p_1}_{\reg{p_1}} \ket{p_2}_{\reg{p_2}} \ket{\dot{\rho}^{(t)}_{p_1, p_2}}  
+  
\sum_{(p'_1, p'_2)\ne (p_1, p_2)}  \ket{p'_1}_{\reg{p_1}} \ket{p'_2}_{\reg{p_2}}   C_{2,0}   \ket{\dot{\rho}^{(t)}_{p'_1, p'_2}} 
\label{lem:CPM-3ZK-hyb12:proof:invariant:proof:induction:preparation-claim:2:eq:target:1} 
.\end{align}
\end{MyClaim}
\begin{proof}[Proof of \Cref{lem:CPM-3ZK-hyb12:proof:invariant:proof:induction:preparation-claim:2}]
This claim follows from an application of \Cref{lem:err-inv-com}, with the notation correspondence listed in \Cref{lem:CPM-3ZK-hyb12:proof:cor-table}. We provide a detailed explanation below.
\begin{table}[!htb]
\centering
\caption{Notation Correspondence between \Cref{lem:err-inv-com} and \Cref{lem:CPM-3ZK-hyb12:proof:invariant:proof:induction:preparation-claim:2}}
\label{lem:CPM-3ZK-hyb12:proof:cor-table}
\vspace{0.5em}
\begin{tabular}{ C{50pt} C{60pt} c C{50pt} C{60pt} c C{50pt} C{60pt} }
\toprule
 \multicolumn{2}{c}{Registers}   & \phantom{abc}   & \multicolumn{2}{c}{Operators}   &
\phantom{abc}   & \multicolumn{2}{c}{Random Variables}  \\
\cmidrule{1-2} \cmidrule{4-5} \cmidrule{7-8}
In \Cref{lem:err-inv-com} & In \Cref{lem:CPM-3ZK-hyb12:proof:invariant:proof:induction:preparation-claim:2} & & In \Cref{lem:err-inv-com} & In \Cref{lem:CPM-3ZK-hyb12:proof:invariant:proof:induction:preparation-claim:2} & & In \Cref{lem:err-inv-com} & In \Cref{lem:CPM-3ZK-hyb12:proof:invariant:proof:induction:preparation-claim:2}  \\ 
\midrule
\addlinespace[0.3em]
$\reg{a}$    & $\reg{p_1}$, $\reg{p_2}$ & & $W_1$ & $C_{2,1}$ & & $\ket{a}_{\reg{a}}$    & $\ket{p_1,p_2}_{\reg{p_1p_2}}$  \\ 
\addlinespace[0.3em]
$\reg{m}$    & $\reg{m}$ & & $W_0$ & $C_{2,0}$ & & $\ket{a'}_{\reg{a}}$    & $\ket{p'_1,p'_2}_{\reg{p_1p_2}}$  \\
\addlinespace[0.3em]
$\reg{t}$    & $\reg{t_2}$ & & $W$   & $C_{2}$ & & $\ket{\rho^{(\msf{in})}_a}_{\reg{mtso}}$    & $\ket{\dot{\rho}^{(t-1)}_{p_1,p_2}}$  \\ 
\addlinespace[0.3em]
$\reg{s}$    & $\reg{u}$, $\reg{s}$, $\reg{t_1}$, $\reg{gc}$, $\reg{lc}$ & & $\tilde{W}$ & $\ddot{C}_2$ & & $\ket{\rho^{(\msf{in})}_{a'}}_{\reg{mtso}}$    & $\ket{\dot{\rho}^{(t-1)}_{p'_1,p'_2}}$ \\ 
\addlinespace[0.3em]
$\reg{o}$    & other registers & & $U_1$ & $B_{2,1}$ & &   $\ket{\eta^{(\msf{in})}_0}$ & $\ket{\gamma^{(t-1)}_0}$ \\ 
\addlinespace[0.3em]
   & & & $U$   & $B_2$ & & $\ket{\eta^{(\msf{in})}_1}$ & $\ket{\gamma^{(t-1)}_1}$  \\ 
\addlinespace[0.3em]
 &  & &   $S$   & $U_{gc}\Lambda^{\reg{p_2}/\reg{m}}U^\dagger_{gc}$ & &  $\ket{\rho^{(\msf{out})}_a}_\reg{mtso}$    & $\ket{\dot{\rho}^{(t)}_{p_1,p_2}}$  \\ 
\addlinespace[0.3em]
  &  & &   &  & &  $\ket{\rho^{(\msf{out})}_{a'}}_\reg{mtso}$     & $\ket{\dot{\rho}^{(t)}_{p'_1,p'_2}}$  \\  
  \addlinespace[0.3em]
 &  & & & & &  $\ket{\eta^{(\msf{out})}_0}$    & $\ket{\gamma^{(t)}_0}$  \\
\addlinespace[0.3em] 
  &  & &   &  & &  $\ket{\eta^{(\msf{out})}_1}$    & $\ket{\gamma^{(t)}_1}$  \\   
\bottomrule
\end{tabular}
\end{table}

First, we argue that the premises in \Cref{lem:err-inv-com} are satisfied with the notation listed in \Cref{lem:CPM-3ZK-hyb12:proof:cor-table}:
\begin{itemize}
\item
\Cref{lem:err-inv-com} requires that $W_1$ should work as the identity operator on register $\reg{s}$. In terms of the \Cref{lem:CPM-3ZK-hyb12:proof:invariant:proof:induction:preparation-claim:2} notation, this is satisfied by $C_{2,1}$ (playing the role of $W_1$) who works as identity on registers $\reg{u}$, $\reg{s}$, $\reg{t_1}$, $\reg{gc}$, and $\reg{lc}$ (playing the role of registers $\reg{s}$). (Recall $C_{2,1}$ from \Cref{CPM:non-trivial-registers:table}.)

\item
\Cref{lem:err-inv-com} requires that $W_0$ should be the swap operator between $\reg{m}$ and $\reg{t}$. In terms of the \Cref{lem:CPM-3ZK-hyb12:proof:invariant:proof:induction:preparation-claim:2} notation, this is satisfied by $C_{2,0}$ (playing the role of $W_0$), who is the swap operator between registers $\reg{m}$ and $\reg{t_2}$ (playing the role of $\reg{m}$ and $\reg{t}$ respectively). (Recall $C_{2,0}$ from \Cref{re-define:unitaries}.)

\item
\Cref{lem:err-inv-com} requires that $\tilde{W}$ is the identity operator on branch $\ket{a}_{\reg{a}}$ and is identical to $W_0$ on branches $\ket{a'}_{\reg{a}}$ with $a' \ne a$. In terms of the \Cref{lem:CPM-3ZK-hyb12:proof:invariant:proof:induction:preparation-claim:2} notation, this is satisfied by $\ddot{C}_2$ (playing the role of $\tilde{W}$), who is the identity operator on branch $\ket{p_1, p_2}_{\reg{p_1p_2}}$ (playing the role of $\ket{a}_{\reg{a}}$) and is identical to $C_{2,0}$ (playing the role of $W_0$) on branches $\ket{p'_1, p'_2}_{\reg{p_1p_2}}$ (playing the role of $\ket{a'}_{\reg{a}}$) with $(p'_1, p'_2) \ne (p_1, p_2)$. (Recall $\ddot{C}_2$ from \Cref{re-define:unitaries})

\item
 \Cref{lem:err-inv-com} requires that $U_1$ should work as identity on register $\reg{s}$. In terms of the \Cref{lem:CPM-3ZK-hyb12:proof:invariant:proof:induction:preparation-claim:2} notation, this is satisfied by $B_{2,1}$ (playing the role of $U_1$), who works as identity on registers $\reg{u}$, $\reg{s}$, $\reg{t_1}$, $\reg{gc}$, and $\reg{lc}$ (playing the role of register $\reg{s}$). (Recall $B_{2,1}$ from \Cref{CPM:non-trivial-registers:table}.)

\item
 \Cref{lem:err-inv-com} requires that $S$ should act non-trivially {\em only} on registers $\reg{a}$ and $\reg{s}$. In terms of the \Cref{lem:CPM-3ZK-hyb12:proof:invariant:proof:induction:preparation-claim:2} notation, this is satisfied by $U_{gc}\Lambda^{\reg{p_2}/\reg{m}}U^\dagger_{gc}$ (playing the role of $S$)---recall from \Cref{lem:CPM-3ZK-hyb12:proof:invariant:proof:induction:Lambda-register:claim} that $\Lambda$ acts non-trivially on registers $\reg{m}$, $\reg{u}$, $\reg{s}$, $\reg{p_1}$, $\reg{t_1}$, $\reg{lc}$, and $\reg{gc}$. Thus, $U_{gc}\Lambda^{\reg{p_2}/\reg{m}}U^\dagger_{gc}$ acts non-trivially on registers $\reg{p_2}$, $\reg{u}$, $\reg{s}$, $\reg{p_1}$, $\reg{t_1}$, $\reg{lc}$, and $\reg{gc}$, which constitute registers $\reg{a}$ and $\reg{s}$ (see \Cref{lem:CPM-3ZK-hyb12:proof:cor-table}). 
\end{itemize}

Finally, we apply \Cref{lem:err-inv-com} (with the notation in \Cref{lem:CPM-3ZK-hyb12:proof:cor-table}) to the $\ket{\gamma^{(t-1)}_0}$ and $\ket{\gamma^{(t-1)}_1}$ defined in \Cref{lem:CPM-3ZK-hyb12:proof:invariant:proof:induction:preparation-claim:2:eq:in:0,lem:CPM-3ZK-hyb12:proof:invariant:proof:induction:preparation-claim:2:eq:in:1}  (playing the role of $\ket{\eta^{(\msf{in})}_0}$ and $\ket{\eta^{(\msf{in})}_1}$ in \Cref{lem:err-inv-com}). This implies the existence of (possibly sub-normalized) pure states $\Set{\ket{\dot{\rho}^{(t)}_{p'_1,p'_2}}}_{(p'_1, p'_2)\in\bits^\ell \times \bits^\ell}$  (playing the role of $\Set{\ket{\rho^{(\msf{out})}_{a'}}_{\reg{mtso}}}_{a'\in\bits^\ell }$ in \Cref{lem:err-inv-com}) such that the following holds
\begin{align*}
\ket{\gamma^{(t)}_0} 
& =
\ket{p_1}_{\reg{p_1}} \ket{p_2}_{\reg{p_2}} \ket{\dot{\rho}^{(t)}_{p_1, p_2}}  
+  
\sum_{(p'_1, p'_2)\ne (p_1, p_2)}  \ket{p'_1}_{\reg{p_1}} \ket{p'_2}_{\reg{p_2}}   C_{2,0} B^\dagger_{2,1} C^\dagger_{2,1}   \ket{\dot{\rho}^{(t)}_{p'_1, p'_2}} \\ 
\ket{\gamma^{(t)}_1} 
& = 
\ket{p_1}_{\reg{p_1}} \ket{p_2}_{\reg{p_2}} \ket{\dot{\rho}^{(t)}_{p_1, p_2}}  
+  
\sum_{(p'_1, p'_2)\ne (p_1, p_2)}  \ket{p'_1}_{\reg{p_1}} \ket{p'_2}_{\reg{p_2}}   C_{2,0}   \ket{\dot{\rho}^{(t)}_{p'_1, p'_2}}  
,\end{align*}
which are exactly \Cref{lem:CPM-3ZK-hyb12:proof:invariant:proof:induction:preparation-claim:2:eq:target:0,lem:CPM-3ZK-hyb12:proof:invariant:proof:induction:preparation-claim:2:eq:target:1} in  \Cref{lem:CPM-3ZK-hyb12:proof:invariant:proof:induction:preparation-claim:2}.

This completes the proof of \Cref{lem:CPM-3ZK-hyb12:proof:invariant:proof:induction:preparation-claim:2}.

\end{proof}

\para{Finishing the Proof for the Induction Step.}\label{lem:CPM-3ZK-hyb12:proof:invariant:proof:induction:finish} With \Cref{lem:CPM-3ZK-hyb12:proof:invariant:proof:induction:preparation-claim,lem:CPM-3ZK-hyb12:proof:invariant:proof:induction:preparation-claim:2} at hand, we now finish the proof for the induction Step.

\subpara{Proof for \Cref{lem:CPM-3ZK-hyb12:proof:target-format:H1}.} We first establish \Cref{lem:CPM-3ZK-hyb12:proof:target-format:H1}:
\begin{align}
\ket{\phi^{(t)}} 
& = D_2 \ket{\gamma^{(t)}_0} 
\label{CPM-3ZK-hyb12:proof:invariant:proof:induction:final:phi:1} \\ 
& = D_2 
\bigg(
\ket{p_1}_{\reg{p_1}} \ket{p_2}_{\reg{p_2}} \ket{\dot{\rho}^{(t)}_{p_1,p_2}}  
+  
\sum_{(p'_1, p'_2) \ne (p_1, p_2)}  \ket{p'_1}_{\reg{p_1}} \ket{p'_2}_{\reg{p_2}}  C_{2,0} B^\dagger_{2,1} C^\dagger_{2,1}  \ket{\dot{\rho}^{(t)}_{p'_1,p'_2}} 
\bigg) 
\label{CPM-3ZK-hyb12:proof:invariant:proof:induction:final:phi:2} \\ 
& = 
\ket{p_1}_{\reg{p_1}} \ket{p_2}_{\reg{p_2}} D_{2,1} \ket{\dot{\rho}^{(t)}_{p_1,p_2}}  
+  
\sum_{(p'_1, p'_2) \ne (p_1, p_2)}  \ket{p'_1}_{\reg{p_1}} \ket{p'_2}_{\reg{p_2}}  C_{2,0} B^\dagger_{2,1} C^\dagger_{2,1}  \ket{\dot{\rho}^{(t)}_{p'_1,p'_2}} 
\label{CPM-3ZK-hyb12:proof:invariant:proof:induction:final:phi:3} \\ 
& = 
\ket{p_1}_{\reg{p_1}} \ket{p_2}_{\reg{p_2}} \ket{\rho^{(t)}_{p_1,p_2}}  
+  
\sum_{(p'_1, p'_2) \ne (p_1, p_2)}  \ket{p'_1}_{\reg{p_1}} \ket{p'_2}_{\reg{p_2}}  C_{2,0} B^\dagger_{2,1} C^\dagger_{2,1}  \ket{\rho^{(t)}_{p'_1,p'_2}} 
\label{CPM-3ZK-hyb12:proof:invariant:proof:induction:final:phi:4} 
,\end{align}
where 
\begin{itemize}
\item
\Cref{CPM-3ZK-hyb12:proof:invariant:proof:induction:final:phi:1} follows from \Cref{lem:CPM-3ZK-hyb12:proof:invariant:proof:induction:preparation-claim:eq:target:phi} in \Cref{lem:CPM-3ZK-hyb12:proof:invariant:proof:induction:preparation-claim} and \Cref{lem:CPM-3ZK-hyb12:proof:invariant:proof:induction:preparation-claim:2:eq:out:0} in \Cref{lem:CPM-3ZK-hyb12:proof:invariant:proof:induction:preparation-claim:2}; 

\item
\Cref{CPM-3ZK-hyb12:proof:invariant:proof:induction:final:phi:2} follows from \Cref{lem:CPM-3ZK-hyb12:proof:invariant:proof:induction:preparation-claim:2:eq:target:0} in \Cref{lem:CPM-3ZK-hyb12:proof:invariant:proof:induction:preparation-claim:2};

\item
\Cref{CPM-3ZK-hyb12:proof:invariant:proof:induction:final:phi:3} follows from the definition of $D_2$ (see \Cref{re-define:unitaries});

\item
\Cref{CPM-3ZK-hyb12:proof:invariant:proof:induction:final:phi:4} follows by defining
\begin{equation}
\label[Expression]{lem:CPM-3ZK-hyb12:proof:invariant:proof:induction:preparation-claim:proof:def:rho}
\ket{\rho^{(t)}_{p_1,p_2}}   \coloneqq D_{2,1} \ket{\dot{\rho}^{(t)}_{p_1,p_2}} ~~~\text{and}~~~
\ket{\rho^{(t)}_{p'_1,p'_2}}  \coloneqq \ket{\dot{\rho}^{(t)}_{p'_1,p'_2}} ~~\big(\forall (p'_1, p'_2) \in \big(\bits^\ell \times \bits^\ell\big) \setminus \Set{(p_1,p_2)}\big).
\end{equation}
\end{itemize}
Note that \Cref{CPM-3ZK-hyb12:proof:invariant:proof:induction:final:phi:4} is exactly 
\Cref{lem:CPM-3ZK-hyb12:proof:target-format:H1} in \Cref{lem:CPM-3ZK-hyb12:proof:invariant}.

\subpara{Proof for \Cref{lem:CPM-3ZK-hyb12:proof:target-format:H2}.} We next establish \Cref{lem:CPM-3ZK-hyb12:proof:target-format:H2}:
\begin{align}
\ket{\psi^{(t)}} 
& = D_2 \ket{\gamma^{(t)}_1} 
\label{CPM-3ZK-hyb12:proof:invariant:proof:induction:final:psi:1} \\ 
& = D_2 
\bigg(
\ket{p_1}_{\reg{p_1}} \ket{p_2}_{\reg{p_2}} \ket{\dot{\rho}^{(t)}_{p_1,p_2}}  
+  
\sum_{(p'_1, p'_2) \ne (p_1, p_2)}  \ket{p'_1}_{\reg{p_1}} \ket{p'_2}_{\reg{p_2}}  C_{2,0} \ket{\dot{\rho}^{(t)}_{p'_1,p'_2}} 
\bigg) 
\label{CPM-3ZK-hyb12:proof:invariant:proof:induction:final:psi:2} \\ 
& = 
\ket{p_1}_{\reg{p_1}} \ket{p_2}_{\reg{p_2}} D_{2,1} \ket{\dot{\rho}^{(t)}_{p_1,p_2}}  
+  
\sum_{(p'_1, p'_2) \ne (p_1, p_2)}  \ket{p'_1}_{\reg{p_1}} \ket{p'_2}_{\reg{p_2}}  C_{2,0}   \ket{\dot{\rho}^{(t)}_{p'_1,p'_2}} 
\label{CPM-3ZK-hyb12:proof:invariant:proof:induction:final:psi:3} \\ 
& = 
\ket{p_1}_{\reg{p_1}} \ket{p_2}_{\reg{p_2}} \ket{\rho^{(t)}_{p_1,p_2}}  
+  
\sum_{(p'_1, p'_2) \ne (p_1, p_2)}  \ket{p'_1}_{\reg{p_1}} \ket{p'_2}_{\reg{p_2}}  C_{2,0}  \ket{\rho^{(t)}_{p'_1,p'_2}} 
\label{CPM-3ZK-hyb12:proof:invariant:proof:induction:final:psi:4} 
,\end{align}
where 
\begin{itemize}
\item
\Cref{CPM-3ZK-hyb12:proof:invariant:proof:induction:final:psi:1} follows from \Cref{lem:CPM-3ZK-hyb12:proof:invariant:proof:induction:preparation-claim:eq:target:psi} in \Cref{lem:CPM-3ZK-hyb12:proof:invariant:proof:induction:preparation-claim} and \Cref{lem:CPM-3ZK-hyb12:proof:invariant:proof:induction:preparation-claim:2:eq:out:1} in \Cref{lem:CPM-3ZK-hyb12:proof:invariant:proof:induction:preparation-claim:2}; 

\item
\Cref{CPM-3ZK-hyb12:proof:invariant:proof:induction:final:psi:2} follows from \Cref{lem:CPM-3ZK-hyb12:proof:invariant:proof:induction:preparation-claim:2:eq:target:1} in \Cref{lem:CPM-3ZK-hyb12:proof:invariant:proof:induction:preparation-claim:2};

\item
\Cref{CPM-3ZK-hyb12:proof:invariant:proof:induction:final:psi:3} follows from the definition of $D_2$ (see \Cref{re-define:unitaries});

\item
\Cref{CPM-3ZK-hyb12:proof:invariant:proof:induction:final:psi:4} follows from {\em the same definitions} of $\ket{\rho^{(t)}_{p_1, p_2}}$ and $\ket{\rho^{(t)}_{p'_1, p'_2}}$ in \Cref{lem:CPM-3ZK-hyb12:proof:invariant:proof:induction:preparation-claim:proof:def:rho}.
\end{itemize}
Note that \Cref{CPM-3ZK-hyb12:proof:invariant:proof:induction:final:psi:4} is exactly 
\Cref{lem:CPM-3ZK-hyb12:proof:target-format:H2} in \Cref{lem:CPM-3ZK-hyb12:proof:invariant}. 

This finishes the proof of the induction step of \Cref{lem:CPM-3ZK-hyb12:proof:invariant}.

\vspace{1em}
Finally, we remark that our proof for the base case in \Cref{sec:lem:CPM-3ZK-hyb12:proof:invariant:proof:base:sec} and the proof for the induction step in \Cref{sec:lem:CPM-3ZK-hyb12:proof:invariant:proof:induction:sec} together finish the proof of \Cref{lem:CPM-3ZK-hyb12:proof:invariant}, which in turn finishes the proof of \Cref{lem:CPM-3ZK:H12} eventually.

\section{Proving \Cref{lem:H:IND} (Full Version)}
\label{sec:CMP:full}

In this section, we provide the proof of \Cref{lem:H:IND} for the general case (in contrast to the warm-up case $K =2$ shown in \Cref{sec:CMP:warm-up}).

Due to a similar argument as we did at the beginning of \Cref{sec:lem:CPM-3ZK:H01:proof}, we claim that: to prove \Cref{lem:H:IND}, it suffices to establish the following \Cref{lem:H:IND:proof:invariant}.
\begin{lemma}[Invariance in $H^{(J, \msf{pat})}_{k-1}$ and $H^{(J, \msf{pat})}_{k}$]
\label{lem:H:IND:proof:invariant}
For a $(J, \msf{pat})$ satisfying the requirements in \Cref{CPM:MnR:game:subnormalized} and a $k \in [K]$, assume that during the execution of $H^{(J, \msf{pat})}_{k-1}$ (and $H^{(J, \msf{pat})}_{k}$), the global counter reaches value $k$ for $T$ times in total. For each $t \in [T]$, there exist (possibly sub-normalized) pure states $\Set{\ket{\rho^{(t)}_{p'_1, \ldots, p'_k}}}_{p'_1, \ldots, p'_k \in \bits^{k\ell}}$ so that the following holds: in hybrid $H^{(J, \msf{pat})}_{k-1}$ (resp.\ $H^{(J, \msf{pat})}_{k}$), when the global counter reaches value $k$ for the $t$-th time, the overall state can be written as the $\ket{\phi^{(t)}}$ (resp.\ $\ket{\psi^{(t)}}$) defined as follows:
\begin{align}
\ket{\phi^{(t)}} 
& =  
\ket{p_1, \ldots, p_k}_{\reg{p_1\ldots p_k}} \ket{\rho^{(t)}_{p_1, \ldots, p_k}}  ~+~
   \sum_{(p'_1, \ldots, p'_k)\ne (p_1, \ldots, p_k)} \ket{p'_1, \ldots, p'_k}_{\reg{p_1} \ldots p_k}  C_{k,0} B^\dagger_{k,1} C^\dagger_{k,1} \ket{\rho^{(t)}_{p'_1, \ldots, p'_k}}, 
   \label{lem:H:IND:proof:invariant:target-format:Hk-1}\\ 
\ket{\psi^{(t)}} 
& =  
\ket{p_1, \ldots, p_k}_{\reg{p_1\ldots p_k}} \ket{\rho^{(t)}_{p_1, \ldots, p_k}}  ~+~
   \sum_{(p'_1, \ldots, p'_k)\ne (p_1, \ldots, p_k)} \ket{p'_1, \ldots, p'_k}_{\reg{p_1} \ldots p_k}  \red{C_{k,0}} \ket{\rho^{(t)}_{p'_1, \ldots, p'_k}}
 \label{lem:H:IND:proof:invariant:target-format:Hk}
,\end{align}
where the summations are taken over all $(p'_1, \ldots, p'_k) \in \bits^{k\ell} \setminus \Set{(p_1, \ldots, p_k)}$ (abbreviated as $(p'_1, \ldots, p'_k) \ne (p_1, \ldots, p_k)$ in the above), the $(p_1, \ldots, p_k)$ are defined in $\msf{pat}$, and the unitaries $B_{k,1}$, $C_{k,0}$, and $C_{k,1}$  are as defined in \Cref{re-define:unitaries}. 
\end{lemma}

\para{Proving \Cref{lem:H:IND:proof:invariant}.} Similar as the proofs for \Cref{lem:CPM-3ZK-hyb01:proof:invariant,lem:CPM-3ZK-hyb12:proof:invariant}, 
we establish \Cref{lem:H:IND:proof:invariant} through mathematical induction on the number $t \in [T]$, indicating the time at which the global counter reaches the value $k$. Throughout this proof, we will monitor the evolution of the overall states in both $H^{(J, \msf{pat})}_{k-1}$ and $H^{(J, \msf{pat})}_k$ simultaneously. This is done in \Cref{sec:lem:H:IND:proof:invariant:base,sec:lem:H:IND:proof:invariant:induction} respective.

\subsection{Base Case ($t = 1$)}
\label{sec:lem:H:IND:proof:invariant:base}

We first derive how the overall state evolves in $H^{(J, \msf{pat})}_{k-1}$. 

This base case corresponds to the very first time the global counter reaches value $k$. By definition, this is due to the $\msf{sq}(k)$ query.

We assume w.l.o.g.\ that the overall state right before this query is some pure state $\ket{\rho}$. Then, by definition, the $\msf{sq}(k)$ query corresponds to the following procedure: $\Sim$ first applies her local unitary $S$ (followed by the projector $\ketbra{\downarrow}_{\reg{u}}$ which we hide as per \Cref{rmk:CPM-3ZK:hid-proj-u}). Next, according to the notation in \Cref{MnR:game:redefine:hybrids}, 
\begin{itemize}
\item 
If $b_k = 0$, then the operator $D_kC_kB_k \ketbra{p_1, \ldots, p_k}_{\reg{p_1 \ldots p_k}} A_k U_{gc}$ will be applied;
\item 
If $b_k = 1$, then the operator $C_{k,0} \ketbra{p_1, \ldots, p_k}_{\reg{p_1 \ldots p_k}} A_k U_{gc}$ will be applied.
\end{itemize}
In the following, we show the proof for $b_k=0$ only; the other case (i.e., $b_k = 1$) can be established using the same argument.

It follows from the above discussion that the state $\ket{\phi^{(1)}}$ in $H^{(J, \msf{pat})}_1$ can be written as follows:
\begin{align*}
\ket{\phi^{(1)}} 
& =  
D_kC_kB_k \ketbra{p_1, \ldots, p_k}_{\reg{p_1 \ldots p_k}} A_k U_{gc} S   
\ket{\rho}
\\
& =  
D_kC_kB_k \ket{p_1, \ldots, p_k}_{\reg{p_1 \ldots p_k}} 
\ket{\dot{\rho}_{p_1, \ldots, p_k}}  
\numberthis \label{H:IND:proof:invariant:base:eq:1}\\
& =  
 \ket{p_1, \ldots, p_k}_{\reg{p_1 \ldots p_k}} 
D_{k,1} C_{k,1} B_{k,1} \ket{\rho_{p_1, \ldots, p_k}}  
\numberthis \label{H:IND:proof:invariant:base:eq:2}\\
& =  
 \ket{p_1, \ldots, p_k}_{\reg{p_1 \ldots p_k}} 
 \ket{\rho^{(1)}_{p_1, \ldots, p_k}}  
\numberthis \label{H:IND:proof:invariant:base:eq:3}
,\end{align*}
where
\begin{itemize}
\item
\Cref{H:IND:proof:invariant:base:eq:1} follows by defining
$
\ket{\rho_{p_1, \ldots, p_k}} 
\coloneqq 
\bra{p_1, \ldots, p_k}_{\reg{p_1 \ldots p_k}} A_k U_{gc} S \ket{\rho}  
$;

\item 
\Cref{H:IND:proof:invariant:base:eq:2} follows from the definitions of $B_k$, $C_k$, and $D_k$ (see \Cref{re-define:unitaries});

\item 
\Cref{H:IND:proof:invariant:base:eq:3} follows by defining $ \ket{\rho^{(1)}_{p_1, \ldots, p_k}}  \coloneqq D_{k,1} C_{k,1} B_{k,1} \ket{\rho_{p_1, \ldots, p_k}}$.	
\end{itemize}
It is straightforward that the $\ket{\phi^{(1)}}$ shown in \Cref{H:IND:proof:invariant:base:eq:3} satisfies the format shown in \Cref{lem:H:IND:proof:invariant:target-format:Hk-1} when $t = 1$.

Also, note that the game $H^{(J, \msf{pat})}_{k-1}$ and $H^{(J, \msf{pat})}_{k}$ are identical so far and thus $\ket{\phi^{(1)}} = \ket{\psi^{(1)}}$. Therefore, it follows from \Cref{H:IND:proof:invariant:base:eq:3} that 
$$
\ket{\psi^{(1)}} 
=  
\ket{p_1, \ldots, p_k}_{\reg{p_1 \ldots p_k}} 
 \ket{\rho^{(1)}_{p_1, \ldots, p_k}}.
$$ 
 Such a $\ket{\psi^{(1)}}$ satisfies the format shown in \Cref{lem:H:IND:proof:invariant:target-format:Hk} with $t = 1$ as well.

This finishes the proof for the base case $t =1$.

\subsection{Induction Step ($t \ge 2$)} 
\label{sec:lem:H:IND:proof:invariant:induction}

We assume $\ket{\phi^{(t-1)}}$ and $\ket{\psi^{(t-1)}}$ satisfy the requirements in \Cref{lem:H:IND:proof:invariant}, and show in the following that \Cref{lem:H:IND:proof:invariant} holds when the global counter reaches $k$ for the $t$-th time.

We establish this claim by considering the following MECE (mutually exclusive and collectively exhaustive) cases:
\begin{enumerate}
\item \label[Case]{H:IND:proof:invariant:induction:case-1}
{\bf Case 1:} The $t$-th arrival at value $k$ happens because of the following operations performed by the simulator $\Sim$: 
\begin{enumerate}
\item 
$\Sim$ first applies her local unitary $S$ and then makes\footnote{Recall that this is determined by measuring the $\reg{u}$ reigster. And further recall that this measurement is replaced with the projector $\ketbra{\downarrow}_\reg{u}$ because we are in the sub-normalized game (see \Cref{MnR:game:redefine:hybrids}).} an $\uparrow$-query, which brings the global counter to $k-1$;

\item \label[Step]{H:IND:proof:invariant:induction:case-1:2}
Next, $\Sim$ could perform any operations, as long as they keep the global counter $\le k-1$. For instance, $\Sim$ can choose to further rewind the execution by making another $\uparrow$-query, or $\Sim$ can choose to make $\uparrow$ and $\downarrow$ queries that cause the global counter to jump between $i-1$ and $i$ as long as $i \le k-1$. The only restriction is that $\Sim$ will not bring the global counter back to $k$ (which is captured by the next step).

\vspace{0.5em}
{\bf Defining Operator $\Upsilon$:} We define the operations performed by $\Sim$ as an operator $\Upsilon$.

\item
After the application of $\Upsilon$, the simulator $\Sim$ eventually makes a $\downarrow$-query that bring the global counter back to $k$ (i.e., exactly the $t$-th arrival at value $k$).
\end{enumerate}

\item \label[Case]{H:IND:proof:invariant:induction:case-2}
{\bf Case 1:} The $t$-th arrival at value $k$ happens because of the following operations performed by the simulator $\Sim$: 
\begin{enumerate}
\item \label[Step]{H:IND:proof:invariant:induction:case-2:1}
$\Sim$ first applies her local unitary $S$ and then makes a $\downarrow$-query, which brings the global counter to $k+1$;

\item \label[Step]{H:IND:proof:invariant:induction:case-2:2}
Next, $\Sim$ could perform any operations, as long as they keep the global counter $\ge k+1$. For instance, $\Sim$ can choose to make another $\downarrow$-query, or $\Sim$ can choose to make $\uparrow$ and $\downarrow$ queries that cause the global counter to jump between $i$ and $i+1$ as long as $i \ge k+1$. The only restriction is that $\Sim$ will not bring the global counter back to $k$ (which is captured by the next step).

\item \label[Step]{H:IND:proof:invariant:induction:case-2:3}
After the application of $\Lambda$, the simulator $\Sim$ eventually makes a $\uparrow$-query that bring the global counter back to $k$ (i.e., exactly the $t$-th arrival at value $k$).
\end{enumerate}
Note that \Cref{H:IND:proof:invariant:induction:case-2:1,H:IND:proof:invariant:induction:case-2:2,H:IND:proof:invariant:induction:case-2:3} described above not only transition $\ket{\phi^{(t-1)}}$ to $\ket{\phi^{(t)}}$, but they also transition $\ket{\psi^{(t-1)}}$ to $\ket{\psi^{(t)}}$. This is because these three steps maintain the global counter $\ge k$, and $H^{(J, \msf{pat})}_{k-1}$ and $H^{(J, \msf{pat})}_{k}$ behave identically by definition until the global counter returns to $k$ (see \Cref{MnR:game:redefine:hybrids}).

\vspace{0.5em}
{\bf Defining Operator $\Lambda$:} We define the operations performed in \Cref{H:IND:proof:invariant:induction:case-2:1,H:IND:proof:invariant:induction:case-2:2,H:IND:proof:invariant:induction:case-2:3} as an operator $\Lambda$.
\end{enumerate}
In the following, we show the proof for these two cases in \Cref{H:IND:proof:invariant:induction:case-1:sec,H:IND:proof:invariant:induction:case-1:sec} respectively.

\subsubsection{Proving the Induction Step: \Cref{H:IND:proof:invariant:induction:case-1}}
\label{H:IND:proof:invariant:induction:case-1:sec}

We first describe formally how $\ket{\phi^{(t-1)}}$ (resp.\ $\ket{\psi^{(t-1)}}$) evolves into $\ket{\phi^{(t)}}$ (resp.\ $\ket{\psi^{(t)}}$) in \Cref{H:IND:proof:invariant:induction:case-1}. According to the description in \Cref{H:IND:proof:invariant:induction:case-1} and our notation in \Cref{MnR:game:redefine:hybrids}, it is clear that the states $\ket{\phi^{(t)}}$ and $\ket{\psi^{(t)}}$ in this case can be written respectively as:
\begin{align}
\ket{\phi^{(t)}}
& = D_k C_k B_k A_k U_{gc} \Upsilon U^{\dagger}_{gc} A^\dagger_k B^\dagger_k C^\dagger_k D^\dagger_k S  \ket{\phi^{(t-1)}} \label{H:IND:proof:invariant:induction:case-1:eq:phi}\\ 
\ket{\psi^{(t)}}
& = D_k \ddot{C}_k A_k U_{gc} \Upsilon  U^{\dagger}_{gc} A^\dagger_k \ddot{C}^\dagger_k D^\dagger_k S \ket{\psi^{(t-1)}} \label{H:IND:proof:invariant:induction:case-1:eq:psi}
,\end{align}
where the operator $\Upsilon$ is defined in \Cref{H:IND:proof:invariant:induction:case-1:2}.

We first make a useful claim that specifies the registers on which the operator $\Upsilon$ (defined in \Cref{H:IND:proof:invariant:induction:case-1:2}) acts non-trivially. This claim can be treated as the analogue of \Cref{lem:CPM-3ZK-hyb12:proof:invariant:proof:induction:Lambda-register:claim}.
\begin{MyClaim}[Non-Trivial Registers for $\Upsilon$]\label{H:IND:proof:invariant:induction:case-1:Upsilon-register:claim}
The $\Upsilon$ defined in \Cref{H:IND:proof:invariant:induction:case-1:2} acts non-trivially only on registers $\reg{m}$, $\reg{u}$, $\reg{s}$, $\reg{p_1 \ldots p_{k-1}}$, $\reg{t_1  \ldots t_{k-1}}$, $\reg{lc}$, and $\reg{gc}$ (i.e., it works as the identity operator on the other registers). 
\end{MyClaim}
\begin{proof}[Proof of \Cref{H:IND:proof:invariant:induction:case-1:Upsilon-register:claim}]
By definition, $\Upsilon$ only contains operations that happen when the global counter is $k-1$ or smaller. 

Also, recall that we are talking about hybrids $H^{(J, \msf{pat})}_{k-1}$ and $H^{(J, \msf{pat})}_{k}$ now. Thus, all the $\downarrow$ and $\uparrow$ queries happening at global counter $i \le k-1$ will be answered by the ``dummy'' operators $\ddot{V}$ and $\ddot{V}^\dagger$ respective. (One exception is the $\tilde{V}$ corresponding $\msf{sq}(i)$. But this does not matter in the current proof---Right before the behaviors defined in \Cref{H:IND:proof:invariant:induction:case-1} happen, the global counter is already $k$, which implies that $\msf{sq}(i)$ ($\forall i \in [k]$) has already happened and thus cannot be a part of the operator $\Upsilon$.)

Therefore, according to our notation in \Cref{MnR:game:redefine:hybrids}, operator $\Upsilon$ can be written as $\Upsilon = \Gamma_z \Gamma_{z-1} \cdots \Gamma_1$ (for some integer $z$) where each $\Gamma_i$ ($i \in [z]$) comes from the set of operators $\Set{S, U_{gc}} \cup \Set{A_i}_{i \in [k-1]} \cup \Set{\ddot{C}_i}_{i \in [k-1]} \cup \Set{D_i}_{i \in [k-1]}$. It then follows from \Cref{CPM:non-trivial-registers:table} that $\Upsilon$ acts non-trivially only on registers $\reg{m}$, $\reg{u}$, $\reg{s}$, $\reg{p_1 \ldots p_{k-1}}$, $\reg{t_1  \ldots t_{k-1}}$, $\reg{lc}$, and $\reg{gc}$.

This completes the proof of \Cref{H:IND:proof:invariant:induction:case-1:Upsilon-register:claim}.

\end{proof}

\para{High-Level Idea for the Sequel.} Recall that our eventual goal is to prove that the states $\ket{\phi^{(t)}}$ and $\ket{\psi^{(t)}}$ are of the format shown in \Cref{lem:H:IND:proof:invariant:target-format:Hk-1,lem:H:IND:proof:invariant:target-format:Hk} in \Cref{lem:CPM-3ZK-hyb12:proof:invariant}. At a high level, we prove it by applying \Cref{lem:err-inv-com} to \Cref{H:IND:proof:invariant:induction:case-1:eq:phi,H:IND:proof:invariant:induction:case-1:eq:psi}. But we first need to perform some preparation work, putting \Cref{H:IND:proof:invariant:induction:case-1:eq:phi,H:IND:proof:invariant:induction:case-1:eq:psi} into a format that is more ``compatible'' with \Cref{lem:err-inv-com}. In the sequel, we first perform the preparation work in \Cref{H:IND:proof:invariant:induction:case-1:preparation-claim,H:IND:proof:invariant:induction:case-1:preparation-claim:2}. Then, we show on \Cpageref{H:IND:proof:invariant:induction:case-1:finish} how to use \Cref{H:IND:proof:invariant:induction:case-1:preparation-claim,H:IND:proof:invariant:induction:case-1:preparation-claim:2} to complete the proof for the induction step. 

The following \Cref{H:IND:proof:invariant:induction:case-1:preparation-claim} can be treated as an analogue of \Cref{lem:CPM-3ZK-hyb12:proof:invariant:proof:induction:preparation-claim}.

\begin{MyClaim}\label{H:IND:proof:invariant:induction:case-1:preparation-claim}
There exist (possibly sub-normalized) pure states $\Set{\dot{\rho}^{(t-1)}_{p'_1, \ldots, p'_k}}_{(p'_1, \ldots, p'_k) \in \bits^{k\ell}}$ so that the following holds
\begin{align*}
\ket{\phi^{(t)}}
& =  
D_k C_k B_k U_{gc} \Upsilon^{\reg{p_k}/\reg{m}} U_{gc}^\dagger   B^\dagger_k C^\dagger_k
\bigg( 
\ket{p_1, \ldots, p_k}_{\reg{p_1 \ldots p_k}} \ket{\dot{\rho}^{(t-1)}_{p_1, \ldots, p_k}}  
+ \\ 
& \hspace{12em} 
\sum_{(p'_1, \ldots, p'_k)\ne (p_1, \ldots, p_k)}  \ket{p'_1, \ldots, p'_k}_{\reg{p_1 \ldots p_k}}  C_{k,0} B^\dagger_{k,1} C^\dagger_{k,1}  \ket{\dot{\rho}^{(t-1)}_{p'_1, \ldots, p'_k}} 
\bigg) \numberthis \label{H:IND:proof:invariant:induction:case-1:preparation-claim:eq:target:phi} \\ 
\ket{\psi^{(t)}}
& = 
D_k\ddot{C}_k U_{gc}\Upsilon^{\reg{p_k}/\reg{m}}  U_{gc}^\dagger \ddot{C}^\dagger_k 
\bigg( 
\ket{p_1, \ldots, p_k}_{\reg{p_1 \ldots p_k}} \ket{\dot{\rho}^{(t-1)}_{p_1, \ldots, p_k}}  
+ \\ 
& \hspace{15em} 
\sum_{(p'_1, \ldots, p'_k)\ne (p_1, \ldots, p_k)}  \ket{p'_1, \ldots, p'_k}_{\reg{p_1 \ldots p_k}}  C_{k,0}  \ket{\dot{\rho}^{(t-1)}_{p'_1, \ldots, p'_k}}  
\bigg) 
\numberthis \label{H:IND:proof:invariant:induction:case-1:preparation-claim:eq:target:psi}
\end{align*}
where $\Upsilon^{\reg{p_k}/\reg{m}}$ is identical to the $\Upsilon$ defined in \Cref{H:IND:proof:invariant:induction:case-1:2}, except that it treats register $\reg{p_k}$ as $\reg{m}$, and the summations are taken over all $(p'_1, \ldots, p'_k) \in \bits^{k\ell}\setminus \Set{(p_1, \ldots, p_k)}$ (abbreviated as $(p'_1, \ldots, p'_k) \ne (p_1, \ldots, p_k)$ in the above).
\end{MyClaim}
\begin{proof}[Proof of \Cref{H:IND:proof:invariant:induction:case-1:preparation-claim}]
First, notice that  
\begin{align*}
 A_k U_{gc} \Upsilon U^\dagger_{gc} A^\dagger_k 
 & = 
 A_k U_{gc} \Upsilon A^\dagger_k U^\dagger_{gc}
 \numberthis \label{H:IND:proof:invariant:induction:case-1:preparation-claim:proof:eq:swap-S:1} \\ 
 & = 
 A_k U_{gc} A^\dagger_k \Upsilon^{\reg{p_k}/\reg{m}}  U^\dagger_{gc}
 \numberthis \label{H:IND:proof:invariant:induction:case-1:preparation-claim:proof:eq:swap-S:2} \\
  & = 
 A_k A^\dagger_k U_{gc} \Upsilon^{\reg{p_k}/\reg{m}}  U^\dagger_{gc}
 \numberthis \label{H:IND:proof:invariant:induction:case-1:preparation-claim:proof:eq:swap-S:3} \\
 & = 
U_{gc} \Upsilon^{\reg{p_k}/\reg{m}}  U^\dagger_{gc}
 \numberthis \label{H:IND:proof:invariant:induction:case-1:preparation-claim:proof:eq:swap-S} 
,\end{align*}
where 
\begin{itemize}
\item
\Cref{H:IND:proof:invariant:induction:case-1:preparation-claim:proof:eq:swap-S:1,H:IND:proof:invariant:induction:case-1:preparation-claim:proof:eq:swap-S:3} follows from the fact that $U_{gc}$ acts non-trivially on different registers from $A_k$;

\item
\Cref{H:IND:proof:invariant:induction:case-1:preparation-claim:proof:eq:swap-S:2} from the fact that $\Upsilon$ acts as the identity operator on $\reg{p_k}$ (see \Cref{H:IND:proof:invariant:induction:case-1:Upsilon-register:claim}) and that $A_k$ is nothing but a swap operator between $\reg{m}$ and $\reg{p_k}$ (see \Cref{CPM:V:unitary}). 
\end{itemize}

\subpara{Proving \Cref{H:IND:proof:invariant:induction:case-1:preparation-claim:eq:target:phi}.} We now show the derivation for $\ket{\phi^{(t)}}$:
\begin{align*}
\ket{\phi^{(t)}}
& =
D_k C_k B_k A_k U_{gc} \Upsilon U^{\dagger}_{gc} A^\dagger_k B^\dagger_k C^\dagger_k D^\dagger_2  S \ket{\phi^{(t-1)}}
\numberthis \label{H:IND:proof:invariant:induction:case-1:preparation-claim:proof:phi-derivation:0} \\ 
& =
D_k C_k B_k  U_{gc} \Upsilon^{\reg{p_k}/\reg{m}}  U^\dagger_{gc} B^\dagger_k C^\dagger_k D^\dagger_k  S \ket{\phi^{(t-1)}}
\numberthis \label{H:IND:proof:invariant:induction:case-1:preparation-claim:proof:phi-derivation:1} \\ 
& =
D_kC_kB_k  U_{gc} \Upsilon^{\reg{p_k}/\reg{m}}  U^\dagger_{gc} B^\dagger_k C^\dagger_k D^\dagger_k S 
\bigg(
\ket{p_1, \ldots, p_k}_{\reg{p_1 \ldots p_k}} \ket{\rho^{(t-1)}_{p_1, \ldots, p_k}}  
~+ \\ 
& \hspace{12em} 
\sum_{(p'_1, \ldots, p'_k)\ne (p_1, \ldots, p_k)}  \ket{p'_1, \ldots, p'_k}_{\reg{p_1 \ldots p_k}}  C_{k,0} B^\dagger_{k,1} C^\dagger_{k,1}  \ket{\rho^{(t-1)}_{p'_1, \ldots, p'_k}} 
\bigg)
\numberthis \label{H:IND:proof:invariant:induction:case-1:preparation-claim:proof:phi-derivation:2} \\ 
& =
D_kC_kB_k  U_{gc} \Upsilon^{\reg{p_k}/\reg{m}}  U^\dagger_{gc} B^\dagger_k C^\dagger_k D^\dagger_k  
\bigg(
\ket{p_1, \ldots, p_k}_{\reg{p_1 \ldots p_k}} S \ket{\rho^{(t-1)}_{p_1, \ldots, p_k}}  
~+ \\ 
& \hspace{11em} 
\sum_{(p'_1, \ldots, p'_k)\ne (p_1, \ldots, p_k)}  \ket{p'_1, \ldots, p'_k}_{\reg{p_1 \ldots p_k}} S C_{k,0} B^\dagger_{k,1} C^\dagger_{k,1}  \ket{\rho^{(t-1)}_{p'_1, \ldots, p'_k}}
\bigg)
\numberthis \label{H:IND:proof:invariant:induction:case-1:preparation-claim:proof:phi-derivation:3} \\ 
& =
D_kC_kB_k  U_{gc} \Upsilon^{\reg{p_k}/\reg{m}}  U^\dagger_{gc} B^\dagger_k C^\dagger_k  
\bigg(
\ket{p_1, \ldots, p_k}_{\reg{p_1 \ldots p_k}} D^\dagger_{k,1} S \ket{\rho^{(t-1)}_{p_1, \ldots, p_k}}  
~+ \\ 
& \hspace{11em} 
\sum_{(p'_1, \ldots, p'_k)\ne (p_1, \ldots, p_k)}  \ket{p'_1, \ldots, p'_k}_{\reg{p_1 \ldots p_k}} S C_{k,0} B^\dagger_{k,1} C^\dagger_{k,1}  \ket{\rho^{(t-1)}_{p'_1, \ldots, p'_k}}
\bigg)
\numberthis \label{H:IND:proof:invariant:induction:case-1:preparation-claim:proof:phi-derivation:4} \\ 
& =
D_kC_kB_k  U_{gc} \Upsilon^{\reg{p_k}/\reg{m}}  U^\dagger_{gc} B^\dagger_k C^\dagger_k  
\bigg(
\ket{p_1, \ldots, p_k}_{\reg{p_1 \ldots p_k}} D^\dagger_{k,1} S \ket{\rho^{(t-1)}_{p_1, \ldots, p_k}}  
~+ \\ 
& \hspace{10em} 
\sum_{(p'_1, \ldots, p'_k)\ne (p_1, \ldots, p_k)}  \ket{p'_1, \ldots, p'_k}_{\reg{p_1 \ldots p_k}}  C_{k,0} S^{\reg{t_k}/\reg{m}} B^\dagger_{k,1} C^\dagger_{k,1}  \ket{\rho^{(t-1)}_{p'_1, \ldots, p'_k}}
\bigg)
\numberthis \label{H:IND:proof:invariant:induction:case-1:preparation-claim:proof:phi-derivation:5} \\ 
& =
D_kC_kB_k  U_{gc} \Upsilon^{\reg{p_k}/\reg{m}}  U^\dagger_{gc} B^\dagger_k C^\dagger_k  
\bigg(
\ket{p_1, \ldots, p_k}_{\reg{p_1 \ldots p_k}} D^\dagger_{k,1} S \ket{\rho^{(t-1)}_{p_1, \ldots, p_k}}  
~+ \\ 
& \hspace{10em} 
\sum_{(p'_1, \ldots, p'_k)\ne (p_1, \ldots, p_k)}  \ket{p'_1, \ldots, p'_k}_{\reg{p_1 \ldots p_k}}  C_{k,0}  B^\dagger_{k,1} C^\dagger_{k,1} S^{\reg{t_k}/\reg{m}} \ket{\rho^{(t-1)}_{p'_1, \ldots, p'_k}}
\bigg)
\numberthis \label{H:IND:proof:invariant:induction:case-1:preparation-claim:proof:phi-derivation:6} \\ 
& =
D_kC_kB_k  U_{gc} \Upsilon^{\reg{p_k}/\reg{m}}  U^\dagger_{gc} B^\dagger_k C^\dagger_k  
\bigg(
\ket{p_1, \ldots, p_k}_{\reg{p_1 \ldots p_k}} \ket{\dot{\rho}^{(t-1)}_{p_1, \ldots, p_k}}  
~+ \\ 
& \hspace{12em} 
\sum_{(p'_1, \ldots, p'_k)\ne (p_1, \ldots, p_k)}  \ket{p'_1, \ldots, p'_k}_{\reg{p_1 \ldots p_k}}  C_{k,0}  B^\dagger_{k,1} C^\dagger_{k,1} \ket{\dot{\rho}^{(t-1)}_{p'_1, \ldots, p'_k}}
\bigg)
\numberthis \label{H:IND:proof:invariant:induction:case-1:preparation-claim:proof:phi-derivation:7} 
,\end{align*}
where 
\begin{itemize}
\item
\Cref{H:IND:proof:invariant:induction:case-1:preparation-claim:proof:phi-derivation:0} follows from \Cref{H:IND:proof:invariant:induction:case-1:eq:phi};

\item
\Cref{H:IND:proof:invariant:induction:case-1:preparation-claim:proof:phi-derivation:1} follows from \Cref{H:IND:proof:invariant:induction:case-1:preparation-claim:proof:eq:swap-S};

\item
\Cref{H:IND:proof:invariant:induction:case-1:preparation-claim:proof:phi-derivation:2} follows from our induction assumption;

\item
\Cref{H:IND:proof:invariant:induction:case-1:preparation-claim:proof:phi-derivation:3} from  from the fact that $S$ acts as the identity operator on registers $\reg{p_1 \ldots p_k}$ (see \Cref{CPM:non-trivial-registers:table});

\item
\Cref{H:IND:proof:invariant:induction:case-1:preparation-claim:proof:phi-derivation:4} from the definition of $D_k$ (see \Cref{re-define:unitaries});

\item
\Cref{H:IND:proof:invariant:induction:case-1:preparation-claim:proof:phi-derivation:5} from the fact that $S$ acts as the identity operator on $\reg{t_k}$ (see \Cref{CPM:non-trivial-registers:table}) and $C_{k,0}$ is nothing but a swap operator between $\reg{t_k}$ and $\reg{m}$ (see \Cref{re-define:unitaries}); (Note that $S^{\reg{t_k}/\reg{m}}$ is defined to be an operator that is identical to $S$ except that it treats $\reg{t_k}$ as $\reg{m}$.)

\item
\Cref{H:IND:proof:invariant:induction:case-1:preparation-claim:proof:phi-derivation:6} follows from the fact that $S^{\reg{t_k}/\reg{m}}$ acts non-trivially on different registers from $B_{k,1}$ and $C_{k,1}$ (see \Cref{CPM:non-trivial-registers:table});

\item
\Cref{H:IND:proof:invariant:induction:case-1:preparation-claim:proof:phi-derivation:7} follows by defining 
\begin{equation}
\label[Expression]{H:IND:proof:invariant:induction:case-1:preparation-claim:proof:def:rho-dot}
\begin{cases}
\ket{\dot{\rho}^{(t-1)}_{p_1, \ldots, p_k}}   \coloneqq D^\dagger_{k,1} S \ket{\rho^{(t-1)}_{p_1, \ldots, p_k}} & 
\vspace{0.3em} \\
\ket{\dot{\rho}^{(t-1)}_{p'_1, \ldots, p'_k}}  \coloneqq S^{\reg{t_k}/\reg{m}} \ket{\rho^{(t-1)}_{p'_1, \ldots, p'_k}} & ~~\forall (p'_1, \ldots, p'_k) \in \big(\bits^{k\ell} \setminus \Set{(p_1, \ldots, p_k)}\big).
\end{cases}
\end{equation}
\end{itemize}
\Cref{H:IND:proof:invariant:induction:case-1:preparation-claim:proof:phi-derivation:7} finishes the proof of \Cref{H:IND:proof:invariant:induction:case-1:preparation-claim:eq:target:phi} in \Cref{H:IND:proof:invariant:induction:case-1:preparation-claim}.

\subpara{Proving \Cref{H:IND:proof:invariant:induction:case-1:preparation-claim:eq:target:psi}.} We now show the derivation for $\ket{\psi^{(t)}}$. This is almost identical to the above proof for \Cref{H:IND:proof:invariant:induction:case-1:preparation-claim:eq:target:phi}. Nevertheless, we present it for the sake of completeness.
\begin{align*}
\ket{\psi^{(t)}}
& =
D_k \ddot{C}_k A_kU_{gc} \Upsilon U^{\dagger}_{gc} A^\dagger_k \ddot{C}^\dagger_k D^\dagger_k  S \ket{\psi^{(t-1)}}
\numberthis \label{H:IND:proof:invariant:induction:case-1:preparation-claim:proof:psi-derivation:0} \\ 
& =
D_k \ddot{C}_k  U_{gc} \Upsilon^{\reg{p_k}/\reg{m}}  U^\dagger_{gc}  \ddot{C}^\dagger_k D^\dagger_k  S \ket{\psi^{(t-1)}}
\numberthis \label{H:IND:proof:invariant:induction:case-1:preparation-claim:proof:psi-derivation:1} \\ 
& =
D_k \ddot{C}_k  U_{gc} \Upsilon^{\reg{p_k}/\reg{m}}  U^\dagger_{gc}  \ddot{C}^\dagger_k D^\dagger_k  S  
\bigg(
\ket{p_1, \ldots, p_k}_{\reg{p_1 \ldots p_k}} \ket{\rho^{(t-1)}_{p_1, \ldots, p_k}}  
~+ \\ 
& \hspace{12em} 
\sum_{(p'_1, \ldots, p'_k)\ne (p_1, \ldots, p_k)}  \ket{p'_1, \ldots, p'_k}_{\reg{p_1 \ldots p_k}}  C_{k,0}  \ket{\rho^{(t-1)}_{p'_1, \ldots, p'_k}} 
\bigg)
\numberthis \label{H:IND:proof:invariant:induction:case-1:preparation-claim:proof:psi-derivation:2} \\ 
& =
D_k \ddot{C}_k  U_{gc} \Upsilon^{\reg{p_k}/\reg{m}}  U^\dagger_{gc}  \ddot{C}^\dagger_k   
\bigg(
\ket{p_1, \ldots, p_k}_{\reg{p_1 \ldots p_k}} D^\dagger_{k,1}  S \ket{\rho^{(t-1)}_{p_1, \ldots, p_k}}  
~+ \\ 
& \hspace{12em} 
\sum_{(p'_1, \ldots, p'_k)\ne (p_1, \ldots, p_k)}  \ket{p'_1, \ldots, p'_k}_{\reg{p_1 \ldots p_k}}  C_{k,0} S^{\reg{t_k}/\reg{m}} \ket{\rho^{(t-1)}_{p'_1, \ldots, p'_k}} 
\bigg)
\numberthis \label{H:IND:proof:invariant:induction:case-1:preparation-claim:proof:psi-derivation:6} \\ 
& =
D_k \ddot{C}_k  U_{gc} \Upsilon^{\reg{p_k}/\reg{m}}  U^\dagger_{gc}  \ddot{C}^\dagger_k 
\bigg(
\ket{p_1, \ldots, p_k}_{\reg{p_1 \ldots p_k}} \ket{\dot{\rho}^{(t-1)}_{p_1, \ldots, p_k}}  
~+ \\ 
& \hspace{12em} 
\sum_{(p'_1, \ldots, p'_k)\ne (p_1, \ldots, p_k)}  \ket{p'_1, \ldots, p'_k}_{\reg{p_1 \ldots p_k}}  C_{k,0} \ket{\dot{\rho}^{(t-1)}_{p'_1, \ldots, p'_k}}
\bigg)
\numberthis \label{H:IND:proof:invariant:induction:case-1:preparation-claim:proof:psi-derivation:7} 
,\end{align*}
where 
\begin{itemize}
\item
\Cref{H:IND:proof:invariant:induction:case-1:preparation-claim:proof:psi-derivation:0} follows from \Cref{H:IND:proof:invariant:induction:case-1:eq:psi};

\item
\Cref{H:IND:proof:invariant:induction:case-1:preparation-claim:proof:psi-derivation:1} follows from \Cref{H:IND:proof:invariant:induction:case-1:preparation-claim:proof:eq:swap-S};

\item
\Cref{H:IND:proof:invariant:induction:case-1:preparation-claim:proof:psi-derivation:2}  follows from our induction assumption;

\item
\Cref{H:IND:proof:invariant:induction:case-1:preparation-claim:proof:psi-derivation:6} follows from a similar argument as we did to derive \Cref{H:IND:proof:invariant:induction:case-1:preparation-claim:proof:phi-derivation:6} from \Cref{H:IND:proof:invariant:induction:case-1:preparation-claim:proof:phi-derivation:2};

\item
\Cref{H:IND:proof:invariant:induction:case-1:preparation-claim:proof:psi-derivation:7} follows from {\em the same definitions} of $\ket{\dot{\rho}^{(t-1)}_{p_1, \ldots, p_k}}$ and $\ket{\dot{\rho}^{(t-1)}_{p'_1, \ldots, p'_k}}$ as shown in \Cref{H:IND:proof:invariant:induction:case-1:preparation-claim:proof:def:rho-dot}.
\end{itemize}
\Cref{H:IND:proof:invariant:induction:case-1:preparation-claim:proof:psi-derivation:7} finishes the proof of \Cref{H:IND:proof:invariant:induction:case-1:preparation-claim:eq:target:psi} in \Cref{H:IND:proof:invariant:induction:case-1:preparation-claim}.

\vspace{1em}
This completes the proof of \Cref{H:IND:proof:invariant:induction:case-1:preparation-claim}.

\end{proof}

The following \Cref{H:IND:proof:invariant:induction:case-1:preparation-claim:2} can be treated as an analogue of \Cref{lem:CPM-3ZK-hyb12:proof:invariant:proof:induction:preparation-claim:2}. 
\begin{MyClaim}\label{H:IND:proof:invariant:induction:case-1:preparation-claim:2}
Let $\Upsilon^{\reg{p_k}/\reg{m}}$ and $\Set{\dot{\rho}^{(t-1)}_{p'_1, \ldots, p'_k}}_{(p'_1, \ldots, p'_k ) \in \bits^{k\ell}}$ be as defined in \Cref{H:IND:proof:invariant:induction:case-1:preparation-claim}. Let
\begin{align}
\ket{\gamma^{(t-1)}_0} 
& \coloneqq 
\ket{p_1, \ldots, p_k}_{\reg{p_1 \ldots p_k}} \ket{\dot{\rho}^{(t-1)}_{p_1, \ldots, p_k}}  
+ 
\sum_{(p'_1, \ldots, p'_k)\ne (p_1, \ldots, p_k)}  \ket{p'_1, \ldots, p'_k}_{\reg{p_1 \ldots p_k}}  C_{k,0} B^\dagger_{k,1} C^\dagger_{k,1}  \ket{\dot{\rho}^{(t-1)}_{p'_1, \ldots, p'_k}} 
\label{H:IND:proof:invariant:induction:case-1:preparation-claim:2:eq:in:0} \\ 
\ket{\gamma^{(t-1)}_1} 
& \coloneqq 
\ket{p_1, \ldots, p_k}_{\reg{p_1 \ldots p_k}} \ket{\dot{\rho}^{(t-1)}_{p_1, \ldots, p_k}}  
+ 
\sum_{(p'_1, \ldots, p'_k)\ne (p_1, \ldots, p_k)}  \ket{p'_1, \ldots, p'_k}_{\reg{p_1 \ldots p_k}}  C_{k,0}  \ket{\dot{\rho}^{(t-1)}_{p'_1, \ldots, p'_k}}  
\label{H:IND:proof:invariant:induction:case-1:preparation-claim:2:eq:in:1} \\ 
\ket{\gamma^{(t)}_0} 
& \coloneqq 
 C_k B_k U_{gc} \Upsilon^{\reg{p_k}/\reg{m}} U_{gc}^\dagger   B^\dagger_k C^\dagger_k \ket{\gamma^{(t-1)}_0} 
\label{H:IND:proof:invariant:induction:case-1:preparation-claim:2:eq:out:0} \\ 
\ket{\gamma^{(t)}_1}
& \coloneqq 
\ddot{C}_k U_{gc}\Upsilon^{\reg{p_k}/\reg{m}}  U_{gc}^\dagger \ddot{C}^\dagger_k 
 \ket{\gamma^{(t-1)}_1}
\label{H:IND:proof:invariant:induction:case-1:preparation-claim:2:eq:out:1} 
.\end{align}
Then, there exist (possibly sub-normalized) pure states $\Set{\dot{\rho}^{(t)}_{p'_1, \ldots, p'_k}}_{(p'_1, \ldots, p'_k ) \in \bits^{k\ell}}$ so that the following holds:
\begin{align}
\ket{\gamma^{(t)}_0} 
& =
\ket{p_1, \ldots, p_k}_{\reg{p_1 \ldots p_k}} \ket{\dot{\rho}^{(t)}_{p_1, \ldots, p_k}}  
+ 
\sum_{(p'_1, \ldots, p'_k)\ne (p_1, \ldots, p_k)}  \ket{p'_1, \ldots, p'_k}_{\reg{p_1 \ldots p_k}}  C_{k,0} B^\dagger_{k,1} C^\dagger_{k,1}  \ket{\dot{\rho}^{(t)}_{p'_1, \ldots, p'_k}}  
\label{H:IND:proof:invariant:induction:case-1:preparation-claim:2:eq:target:0} \\ 
\ket{\gamma^{(t)}_1} 
& = 
\ket{p_1, \ldots, p_k}_{\reg{p_1 \ldots p_k}} \ket{\dot{\rho}^{(t)}_{p_1, \ldots, p_k}}  
+ 
\sum_{(p'_1, \ldots, p'_k)\ne (p_1, \ldots, p_k)}  \ket{p'_1, \ldots, p'_k}_{\reg{p_1 \ldots p_k}}  C_{k,0}  \ket{\dot{\rho}^{(t)}_{p'_1, \ldots, p'_k}} 
\label{H:IND:proof:invariant:induction:case-1:preparation-claim:2:eq:target:1} 
.\end{align}
\end{MyClaim}
\begin{proof}[Proof of \Cref{H:IND:proof:invariant:induction:case-1:preparation-claim:2}]
This claim follows from an application of \Cref{lem:err-inv-com}, with the notation correspondence listed in \Cref{H:IND:proof:invariant:induction:case-1:cor-table}. We provide a detailed explanation below.
\begin{table}[!htb]
\centering
\caption{Notation Correspondence between \Cref{lem:err-inv-com} and \Cref{H:IND:proof:invariant:induction:case-1:preparation-claim:2}}
\label{H:IND:proof:invariant:induction:case-1:cor-table}
\vspace{0.5em}
\begin{tabular}{ C{50pt} C{90pt} c C{50pt} C{70pt} c C{50pt} C{80pt} }
\toprule
 \multicolumn{2}{c}{Registers}   & \phantom{abc}   & \multicolumn{2}{c}{Operators}   &
\phantom{abc}   & \multicolumn{2}{c}{Random Variables}  \\
\cmidrule{1-2} \cmidrule{4-5} \cmidrule{7-8}
In \Cref{lem:err-inv-com} & In \Cref{H:IND:proof:invariant:induction:case-1:preparation-claim:2} & & In \Cref{lem:err-inv-com} & In \Cref{H:IND:proof:invariant:induction:case-1:preparation-claim:2} & & In \Cref{lem:err-inv-com} & In \Cref{H:IND:proof:invariant:induction:case-1:preparation-claim:2}  \\ 
\midrule
\addlinespace[0.3em]
$\reg{a}$    & $\reg{p_1 \ldots p_k}$  & & $W_1$ & $C_{k,1}$ & & $\ket{a}_{\reg{a}}$    & $\ket{p_1, \ldots, p_k}_{\reg{p_1 \ldots p_k}}$  \\ 
\addlinespace[0.3em]
$\reg{m}$    & $\reg{m}$ & & $W_0$ & $C_{k,0}$ & & $\ket{a'}_{\reg{a}}$    & $\ket{p'_1, \ldots,p'_k}_{\reg{p_1 \ldots p_k}}$  \\
\addlinespace[0.3em]
$\reg{t}$    & $\reg{t_k}$ & & $W$   & $C_{k}$ & & $\ket{\rho^{(\msf{in})}_a}_{\reg{mtso}}$    & $\ket{\dot{\rho}^{(t-1)}_{p_1, \ldots, p_k}}$  \\ 
\addlinespace[0.3em]
$\reg{s}$    & $\reg{u}$, $\reg{s}$, $\reg{t_1 \ldots t_{k-1}}$, $\reg{gc}$, $\reg{lc}$ & & $\tilde{W}$ & $\ddot{C}_k$ & & $\ket{\rho^{(\msf{in})}_{a'}}_{\reg{mtso}}$    & $\ket{\dot{\rho}^{(t-1)}_{p'_1, \ldots, p'_k}}$ \\ 
\addlinespace[0.3em]
$\reg{o}$    & other registers & & $U_1$ & $B_{k,1}$ & &   $\ket{\eta^{(\msf{in})}_0}$ & $\ket{\gamma^{(t-1)}_0}$ \\ 
\addlinespace[0.3em]
   & & & $U$   & $B_k$ & & $\ket{\eta^{(\msf{in})}_1}$ & $\ket{\gamma^{(t-1)}_1}$  \\ 
\addlinespace[0.3em]
 &  & &   $S$   & $U_{gc}\Upsilon^{\reg{p_k}/\reg{m}}U^\dagger_{gc}$ & &  $\ket{\rho^{(\msf{out})}_a}_\reg{mtso}$    & $\ket{\dot{\rho}^{(t)}_{p_1, \ldots, p_k}}$  \\ 
\addlinespace[0.3em]
  &  & &   &  & &  $\ket{\rho^{(\msf{out})}_{a'}}_\reg{mtso}$     & $\ket{\dot{\rho}^{(t)}_{p'_1, \ldots, p'_k}}$  \\  
  \addlinespace[0.3em]
 &  & & & & &  $\ket{\eta^{(\msf{out})}_0}$    & $\ket{\gamma^{(t)}_0}$  \\
\addlinespace[0.3em] 
  &  & &   &  & &  $\ket{\eta^{(\msf{out})}_1}$    & $\ket{\gamma^{(t)}_1}$  \\   
\bottomrule
\end{tabular}
\end{table}

First, we argue that the premises in \Cref{lem:err-inv-com} are satisfied with the notation listed in \Cref{H:IND:proof:invariant:induction:case-1:cor-table}:
\begin{itemize}
\item
\Cref{lem:err-inv-com} requires that $W_1$ should act as the identity operator on register $\reg{s}$. In terms of the \Cref{H:IND:proof:invariant:induction:case-1:preparation-claim:2} notation, this is satisfied by $C_{k,1}$ (playing the role of $W_1$) who works as identity on registers $\reg{u}$, $\reg{s}$, $\reg{t_1\ldots t_{k-1}}$, $\reg{gc}$, and $\reg{lc}$ (playing the role of registers $\reg{s}$). (Recall $C_{k,1}$ from \Cref{CPM:non-trivial-registers:table}.)

\item
\Cref{lem:err-inv-com} requires that $W_0$ should be the swap operator between $\reg{m}$ and $\reg{t}$. In terms of the \Cref{H:IND:proof:invariant:induction:case-1:preparation-claim:2} notation, this is satisfied by $C_{k,0}$ (playing the role of $W_0$), who is the swap operator between registers $\reg{m}$ and $\reg{t_k}$ (playing the role of $\reg{m}$ and $\reg{t}$ respectively). (Recall $C_{k,0}$ from \Cref{re-define:unitaries}.)

\item
\Cref{lem:err-inv-com} requires that $\tilde{W}$ is the identity operator on branch $\ket{a}_{\reg{a}}$ and is identical to $W_0$ on branches $\ket{a'}_{\reg{a}}$ with $a' \ne a$. In terms of the \Cref{H:IND:proof:invariant:induction:case-1:preparation-claim:2} notation, this is satisfied by $\ddot{C}_k$ (playing the role of $\tilde{W}$), who is the identity operator on branch $\ket{p_1, \ldots, p_k}_{\reg{p_1\ldots p_k}}$ (playing the role of $\ket{a}_{\reg{a}}$) and is identical to $C_{k,0}$ (playing the role of $W_0$) on branches $\ket{p'_1, \ldots, p'_k}_{\reg{p_1\ldots p_k}}$ (playing the role of $\ket{a'}_{\reg{a}}$) with $(p'_1, \ldots, p'_k) \ne (p_1, \ldots, p_k)$. (Recall $\ddot{C}_k$ from \Cref{re-define:unitaries})

\item
 \Cref{lem:err-inv-com} requires that $U_1$ should work as identity on register $\reg{s}$. In terms of the \Cref{H:IND:proof:invariant:induction:case-1:preparation-claim:2} notation, this is satisfied by $B_{k,1}$ (playing the role of $U_1$), who works as identity on registers $\reg{u}$, $\reg{s}$, $\reg{t_1\ldots t_{k-1}}$, $\reg{gc}$, and $\reg{lc}$ (playing the role of register $\reg{s}$). (Recall $B_{k,1}$ from \Cref{CPM:non-trivial-registers:table}.)

\item
 \Cref{lem:err-inv-com} requires that $S$ should act non-trivially {\em only} on registers $\reg{a}$ and $\reg{s}$. In terms of the \Cref{H:IND:proof:invariant:induction:case-1:preparation-claim:2} notation, this is satisfied by $U_{gc}\Upsilon^{\reg{p_k}/\reg{m}}U^\dagger_{gc}$ (playing the role of $S$)---recall from \Cref{H:IND:proof:invariant:induction:case-1:Upsilon-register:claim} that $\Upsilon$ acts non-trivially on registers $\reg{m}$, $\reg{u}$, $\reg{s}$, $\reg{p_1\ldots p_{k-1}}$, $\reg{t_1 \ldots t_{k-1}}$, $\reg{lc}$, and $\reg{gc}$. Thus, $U_{gc}\Upsilon^{\reg{p_k}/\reg{m}}U^\dagger_{gc}$ acts non-trivially on registers $\reg{p_k}$, $\reg{u}$, $\reg{s}$, $\reg{p_1 \ldots p_{k-1}}$, $\reg{t_1 \ldots t_{k-1}}$, $\reg{lc}$, and $\reg{gc}$, which constitute registers $\reg{a}$ and $\reg{s}$ (see \Cref{H:IND:proof:invariant:induction:case-1:cor-table}). 
\end{itemize}

Finally, we apply \Cref{lem:err-inv-com} (with the notation in \Cref{H:IND:proof:invariant:induction:case-1:cor-table}) to the $\ket{\gamma^{(t-1)}_0}$ and $\ket{\gamma^{(t-1)}_1}$ defined in \Cref{H:IND:proof:invariant:induction:case-1:preparation-claim:2:eq:in:0,H:IND:proof:invariant:induction:case-1:preparation-claim:2:eq:in:1}  (playing the role of $\ket{\eta^{(\msf{in})}_0}$ and $\ket{\eta^{(\msf{in})}_1}$ in \Cref{lem:err-inv-com}). This implies the existence of (possibly sub-normalized) pure states $\Set{\ket{\dot{\rho}^{(t)}_{p'_1, \ldots, p'_k}}}_{(p'_1, \ldots, p'_k)\in\bits^{k\ell}}$  (playing the role of $\Set{\ket{\rho^{(\msf{out})}_{a'}}_{\reg{mtso}}}_{a'\in\bits^\ell }$ in \Cref{lem:err-inv-com}) such that the following holds
\begin{align*}
\ket{\gamma^{(t)}_0} 
& =
\ket{p_1, \ldots, p_k}_{\reg{p_1\ldots p_k}}  \ket{\dot{\rho}^{(t)}_{p_1, \ldots, p_k}}  
+  
\sum_{(p'_1, \ldots, p'_k)\ne (p_1, \ldots, p_k)}  \ket{p'_1, \ldots, p'_k}_{\reg{p_1\ldots p_k}}  C_{k,0} B^\dagger_{k,1} C^\dagger_{k,1}   \ket{\dot{\rho}^{(t)}_{p'_1, \ldots, p'_k}} \\ 
\ket{\gamma^{(t)}_1} 
& = 
\ket{p_1, \ldots, p_k}_{\reg{p_1\ldots p_k}}  \ket{\dot{\rho}^{(t)}_{p_1, \ldots, p_k}}  
+  
\sum_{(p'_1, \ldots, p'_k)\ne (p_1, \ldots, p_k)}  \ket{p'_1, \ldots, p'_k}_{\reg{p_1\ldots p_k}}  C_{k,0}  \ket{\dot{\rho}^{(t)}_{p'_1, \ldots, p'_k}}   
,\end{align*}
which are exactly \Cref{H:IND:proof:invariant:induction:case-1:preparation-claim:2:eq:target:0,H:IND:proof:invariant:induction:case-1:preparation-claim:2:eq:target:1} in  \Cref{H:IND:proof:invariant:induction:case-1:preparation-claim:2}.

This completes the proof of \Cref{H:IND:proof:invariant:induction:case-1:preparation-claim:2}.

\end{proof}

\para{Finishing the Proof for the Induction Step.}\label{H:IND:proof:invariant:induction:case-1:finish} With \Cref{H:IND:proof:invariant:induction:case-1:preparation-claim,H:IND:proof:invariant:induction:case-1:preparation-claim:2} at hand, we now finish the proof for the induction Step.

\subpara{Proof for \Cref{lem:H:IND:proof:invariant:target-format:Hk-1}.} We first establish \Cref{lem:H:IND:proof:invariant:target-format:Hk-1}:
\begin{align}
\ket{\phi^{(t)}} 
& = D_k \ket{\gamma^{(t)}_0} 
\label{H:IND:proof:invariant:induction:case-1:final:phi:1} \\ 
& = D_k 
\bigg(
\ket{p_1, \ldots, p_k}_{\reg{p_1 \ldots p_k}} \ket{\dot{\rho}^{(t)}_{p_1, \ldots, p_k}}  
+ 
\sum_{(p'_1, \ldots, p'_k)\ne (p_1, \ldots, p_k)}  \ket{p'_1, \ldots, p'_k}_{\reg{p_1 \ldots p_k}}  C_{k,0} B^\dagger_{k,1} C^\dagger_{k,1}  \ket{\dot{\rho}^{(t)}_{p'_1, \ldots, p'_k}}   
\bigg) 
\label{H:IND:proof:invariant:induction:case-1:final:phi:2} \\ 
& = 
\ket{p_1, \ldots, p_k}_{\reg{p_1 \ldots p_k}} D_{k,1} \ket{\dot{\rho}^{(t)}_{p_1, \ldots, p_k}}  
+ 
\sum_{(p'_1, \ldots, p'_k)\ne (p_1, \ldots, p_k)}  \ket{p'_1, \ldots, p'_k}_{\reg{p_1 \ldots p_k}}  C_{k,0} B^\dagger_{k,1} C^\dagger_{k,1}  \ket{\dot{\rho}^{(t)}_{p'_1, \ldots, p'_k}}  
\label{H:IND:proof:invariant:induction:case-1:final:phi:3} \\ 
& = 
\ket{p_1, \ldots, p_k}_{\reg{p_1 \ldots p_k}} \ket{\rho^{(t)}_{p_1, \ldots, p_k}}  
+ 
\sum_{(p'_1, \ldots, p'_k)\ne (p_1, \ldots, p_k)}  \ket{p'_1, \ldots, p'_k}_{\reg{p_1 \ldots p_k}}  C_{k,0} B^\dagger_{k,1} C^\dagger_{k,1}  \ket{\rho^{(t)}_{p'_1, \ldots, p'_k}}   
\label{H:IND:proof:invariant:induction:case-1:final:phi:4} 
,\end{align}
where 
\begin{itemize}
\item
\Cref{H:IND:proof:invariant:induction:case-1:final:phi:1} follows from \Cref{H:IND:proof:invariant:induction:case-1:preparation-claim:eq:target:phi} in \Cref{H:IND:proof:invariant:induction:case-1:preparation-claim} and \Cref{H:IND:proof:invariant:induction:case-1:preparation-claim:2:eq:out:0} in \Cref{H:IND:proof:invariant:induction:case-1:preparation-claim:2}; 

\item
\Cref{H:IND:proof:invariant:induction:case-1:final:phi:2} follows from \Cref{H:IND:proof:invariant:induction:case-1:preparation-claim:2:eq:target:0} in \Cref{H:IND:proof:invariant:induction:case-1:preparation-claim:2};

\item
\Cref{H:IND:proof:invariant:induction:case-1:final:phi:3} follows from the definition of $D_k$ (see \Cref{re-define:unitaries});

\item
\Cref{H:IND:proof:invariant:induction:case-1:final:phi:4} follows by defining
\begin{equation}
\label[Expression]{H:IND:proof:invariant:induction:case-1:preparation-claim:proof:def:rho}
\begin{cases}
\ket{\rho^{(t)}_{p_1,\ldots,p_k}}   
\coloneqq 
D_{k,1} \ket{\dot{\rho}^{(t)}_{p_1,\ldots,p_k}} & 
\vspace{0.3em}\\
\ket{\rho^{(t)}_{p'_1, \ldots, p'_k}}  
\coloneqq 
\ket{\dot{\rho}^{(t)}_{p'_1, \ldots, p'_k}} & \forall (p'_1, \ldots, p'_k) \in \big(\bits^{k\ell} \setminus \Set{(p_1, \ldots, p_k)}\big).
\end{cases}
\end{equation}
\end{itemize}
Note that \Cref{H:IND:proof:invariant:induction:case-1:final:phi:4} is exactly 
\Cref{lem:H:IND:proof:invariant:target-format:Hk-1} in \Cref{lem:H:IND:proof:invariant}.

\subpara{Proof for \Cref{lem:H:IND:proof:invariant:target-format:Hk}.} Next, we establish \Cref{lem:H:IND:proof:invariant:target-format:Hk}:
\begin{align}
\ket{\psi^{(t)}} 
& = D_k \ket{\gamma^{(t)}_1} 
\label{H:IND:proof:invariant:induction:case-1:final:psi:1} \\ 
& = D_k 
\bigg(
\ket{p_1, \ldots, p_k}_{\reg{p_1 \ldots p_k}} \ket{\dot{\rho}^{(t)}_{p_1, \ldots, p_k}}  
+ 
\sum_{(p'_1, \ldots, p'_k)\ne (p_1, \ldots, p_k)}  \ket{p'_1, \ldots, p'_k}_{\reg{p_1 \ldots p_k}}  C_{k,0}  \ket{\dot{\rho}^{(t)}_{p'_1, \ldots, p'_k}}   
\bigg) 
\label{H:IND:proof:invariant:induction:case-1:final:psi:2} \\ 
& = 
\ket{p_1, \ldots, p_k}_{\reg{p_1 \ldots p_k}} D_{k,1} \ket{\dot{\rho}^{(t)}_{p_1, \ldots, p_k}}  
+ 
\sum_{(p'_1, \ldots, p'_k)\ne (p_1, \ldots, p_k)}  \ket{p'_1, \ldots, p'_k}_{\reg{p_1 \ldots p_k}}  C_{k,0}   \ket{\dot{\rho}^{(t)}_{p'_1, \ldots, p'_k}}  
\label{H:IND:proof:invariant:induction:case-1:final:psi:3} \\ 
& = 
\ket{p_1, \ldots, p_k}_{\reg{p_1 \ldots p_k}} \ket{\rho^{(t)}_{p_1, \ldots, p_k}}  
+ 
\sum_{(p'_1, \ldots, p'_k)\ne (p_1, \ldots, p_k)}  \ket{p'_1, \ldots, p'_k}_{\reg{p_1 \ldots p_k}}  C_{k,0}   \ket{\rho^{(t)}_{p'_1, \ldots, p'_k}}   
\label{H:IND:proof:invariant:induction:case-1:final:psi:4} 
,\end{align}
where 
\begin{itemize}
\item
\Cref{H:IND:proof:invariant:induction:case-1:final:psi:1} follows from \Cref{H:IND:proof:invariant:induction:case-1:preparation-claim:eq:target:psi} in \Cref{H:IND:proof:invariant:induction:case-1:preparation-claim} and \Cref{H:IND:proof:invariant:induction:case-1:preparation-claim:2:eq:out:1} in \Cref{H:IND:proof:invariant:induction:case-1:preparation-claim:2}; 

\item
\Cref{H:IND:proof:invariant:induction:case-1:final:psi:2} follows from \Cref{H:IND:proof:invariant:induction:case-1:preparation-claim:2:eq:target:1} in \Cref{H:IND:proof:invariant:induction:case-1:preparation-claim:2};

\item
\Cref{H:IND:proof:invariant:induction:case-1:final:psi:3} follows from the definition of $D_k$ (see \Cref{re-define:unitaries});

\item
\Cref{H:IND:proof:invariant:induction:case-1:final:psi:4} follows from {\em the same definitions} of $\ket{\rho^{(t)}_{p_1,\ldots,p_k}}$ and $\ket{\rho^{(t)}_{p'_1,\ldots,p'_k}}$ shown in \Cref{H:IND:proof:invariant:induction:case-1:preparation-claim:proof:def:rho}.
\end{itemize}
Note that \Cref{H:IND:proof:invariant:induction:case-1:final:psi:4} is exactly 
\Cref{lem:H:IND:proof:invariant:target-format:Hk} in \Cref{lem:H:IND:proof:invariant}. 

\vspace{1em}
This completes the proof of the induction step of \Cref{lem:H:IND:proof:invariant} for \Cref{H:IND:proof:invariant:induction:case-1}.

\subsubsection{Proving the Induction Step: \Cref{H:IND:proof:invariant:induction:case-2}}
\label{H:IND:proof:invariant:induction:case-2:sec}

According to the description of \Cref{H:IND:proof:invariant:induction:case-2}, it holds that
\begin{align}
\ket{\phi^{(t)}} 
& = 
\Lambda \ket{\phi^{(t-1)}} 
\label{H:IND:proof:invariant:induction:case-2:initial:phi} \\ 
\ket{\psi^{(t)}} 
& = 
\Lambda \ket{\psi^{(t-1)}}
\label{H:IND:proof:invariant:induction:case-2:initial:psi}
,\end{align} 
where recall that the operator $\Lambda$ was defined toward the end of the description of \Cref{H:IND:proof:invariant:induction:case-2}.

\para{Structure of $\Lambda$.} While the exact format of $\Lambda$ will not be substantial, our proof of \Cref{H:IND:proof:invariant:induction:case-2} will rely on certain properties of $\Lambda$. We formalize these useful properties in the following \Cref{H:IND:proof:invariant:induction:case-2:claim:operator-commute}, which can be treated as an analogue of \Cref{lem:CPM-3ZK-hyb01:proof:case-2-1:claim:operator-commute}.
\begin{MyClaim}\label{H:IND:proof:invariant:induction:case-2:claim:operator-commute}
For the operator $\Lambda$ defined above, there exist two operators $\Lambda_0$ and $\Lambda_1$ so that
\begin{itemize}
\item
both  $\Lambda_1$ and $\Lambda_0$ act as the identity operator on $\reg{p_1\ldots p_k}$;
\item
$\Lambda_0$ acts as the identity operator on $\reg{t_k}$;
\end{itemize}
and the following holds
\begin{align}
& 
\Lambda = \ketbra{p_1, \ldots,p_k}_{\reg{p_1\ldots p_k}} \tensor \Lambda_1 + \sum_{(p'_1,\ldots,p'_k) \ne (p_1, \ldots, p_k)}\ketbra{p'_1, \ldots,p'_k}_{\reg{p_1\ldots p_k}} \tensor \Lambda_0, \label{H:IND:proof:invariant:induction:case-2:claim:operator-commute:eq:1} \\ 
& 
\Lambda_0 C_{k,0}  = C_{k,0} \Lambda^{\reg{t_k}/\reg{m}}_0, \label{H:IND:proof:invariant:induction:case-2:claim:operator-commute:eq:2} \\ 
& 
\Lambda_0 C_{k,0}  B^\dagger_{k,1} C^\dagger_{k,1} = C_{k,0}  B^\dagger_{k,1} C^\dagger_{k,1} \Lambda^{\reg{t_k}/\reg{m}}_0 \label{H:IND:proof:invariant:induction:case-2:claim:operator-commute:eq:3} 
,\end{align}
where $\Lambda^{\reg{t_k}/\reg{m}}_0$ is identical to $\Lambda_0$ except that it treats $\reg{t_k}$ as $\reg{m}$, and the summation is taken over all $(p'_1, \ldots, p'_k) \in \bits^{k\ell}\setminus \Set{(p_1, \ldots, p_k)}$ (abbreviated as $(p'_1, \ldots, p'_k) \ne (p_1, \ldots, p_k)$ in the above).  
\end{MyClaim}
\begin{proof}[Proof of \Cref{H:IND:proof:invariant:induction:case-2:claim:operator-commute}]

First, note that $\Lambda$ can be written as $\Lambda = \Gamma_z \Gamma_{z-1} \cdots \Gamma_1$ (for some integer $z$), where each $\Gamma_i$ ($i \in [z]$) comes from the following set 
\begin{equation}\label[Expression]{H:IND:proof:invariant:induction:case-2:claim:operator-commute:set}
\Set{S,  U_{gc}} \cup \Set{(A_i, B_i, C_i, D_i)}_{i \in \Set{k+1, k+2, \ldots, K}} \cup \Set{\ketbra{p_1, \ldots, p_i}_{\reg{p_1\ldots p_i}}}_{i \in \Set{k+1, k+2, \ldots, K}}.
\end{equation} 
This is because $\Lambda$ only corresponds to the operations that happen when the global counter ``jumps'' between some $i$ and $i+1$ with $i \ge k$ (see \Cref{MnR:game:redefine:hybrids}). We remark that $\Sim$ may also apply projectors on $\reg{u}$, but we consider it as a part of $S$ as per \Cref{rmk:CPM-3ZK:hid-proj-u}.

Therefore, to prove \Cref{H:IND:proof:invariant:induction:case-2:claim:operator-commute}, it suffices to show that for each operator $\Gamma$ in the set shown in \Cref{H:IND:proof:invariant:induction:case-2:claim:operator-commute:set}, there exist two operators $\Gamma_0$ and $\Gamma_1$ so that
\begin{itemize}
\item
both  $\Gamma_1$ and $\Gamma_0$ act as the identity operator on $\reg{p_1 \ldots p_k}$;
\item
$\Gamma_0$ acts as the identity operator on $\reg{t_k}$;
\end{itemize}
and the following holds
\begin{align*}
& 
\Gamma = \ketbra{p_1, \ldots,p_k}_{\reg{p_1\ldots p_k}} \tensor \Gamma_1 + \sum_{(p'_1,\ldots,p'_k) \ne (p_1, \ldots, p_k)}\ketbra{p'_1, \ldots,p'_k}_{\reg{p_1\ldots p_k}} \tensor \Gamma_0,\\ 
& 
\Gamma_0 C_{k,0}  = C_{k,0} \Gamma^{\reg{t_k}/\reg{m}}_0, \\ 
& 
\Gamma_0 C_{k,0}  B^\dagger_{k,1} C^\dagger_{k,1} = C_{k,0}  B^\dagger_{k,1} C^\dagger_{k,1} \Gamma^{\reg{t_k}/\reg{m}}_0 
.\end{align*}
In the following, we prove it for each possible $\Gamma$.

First, notice that the above is true for $\Gamma =\ketbra{p_1, \ldots, p_i}_{\reg{p_1 \ldots p_i}}$ for each $i \in \Set{k+1, k+2, \ldots, K}$, simply because such a $\Gamma$ can be written in the following format:
\begin{align*}
\Gamma 
& = 
\ketbra{p_1, \ldots, p_i}_{\reg{p_1 \ldots p_i}} \\ 
& = 
\ketbra{p_1, \ldots, p_k}_{\reg{p_1 \ldots p_k}} \tensor \Gamma_1  
+ 
\sum_{(p'_1,\ldots,p'_k) \ne (p_1, \ldots, p_k)}\ketbra{p'_1, \ldots,p'_k}_{\reg{p_1\ldots p_k}} \tensor \Gamma_0,
\end{align*}
with $\Gamma_1  \coloneqq \ketbra{p_{k+1}, p_{k+2} \ldots, p_i}_{\reg{p_{k+1} \ldots p_i}}$ and $\Gamma_0  \coloneqq I$, and such a $\Gamma_0$ vacuously satisfies the requires $\Gamma_0 C_{k,0}  = C_{k,0} \Gamma^{\reg{t_k}/\reg{m}}_0$ and $\Gamma_0 C_{k,0}  B^\dagger_{k,1} C^\dagger_{k,1} = C_{k,0}  B^\dagger_{k,1} C^\dagger_{k,1} \Gamma^{\reg{t_k}/\reg{m}}_0$.

The above is true for $\Gamma = U_{gc}$ as well, because $U_{gc}$ acts non-trivially only on register $\reg{gc}$.

The above is also true for $\Gamma = S$ because (1) $S$ does not act on $\reg{p_1 \ldots p_k}$, (2) $S$ does not act on $\reg{t_k}$ and $C_{k,0}$ is nothing but a swap operator between $\reg{m}$ and $\reg{t_k}$, and (3) $S^{\reg{t_k}/\reg{m}}$ acts non-trivially on different registers from $B_{k,1}$ and $C_{k,1}$ (see \Cref{CPM:non-trivial-registers:table}).

The above is true for $\Gamma \in \Set{A_i}_{i \in \Set{k+1, k+2, \ldots, K}}$ as well, because for each $i\in \Set{k+1, k+2, \ldots, K}$, $A_i$ is nothing but the swap operator between registers $\reg{p_i}$ and $\reg{m}$ (see \Cref{CPM:V:unitary}).

The only  cases left are $\Gamma \in \Set{(B_i, C_i, D_i)}_{i \in \Set{k+1, k+2, \ldots, K}}$. The proof for these cases are (almost) identical. In the following, we only present the proof for $\Gamma = C_i$ $(\forall i \in \Set{k+1, k+2, \ldots, K})$ as a representative example:
\begin{align*}
\Gamma 
& = 
C_i 
\\
& = 
\ketbra{p_1, \ldots, p_i}_{\reg{p_1 \ldots p_i}} \tensor C_{i,1} 
+ 
\sum_{(p'_1,\ldots,p'_i) \ne (p_1, \ldots, p_i)} \ketbra{p'_1, \ldots,p'_i}_{\reg{p_1\ldots p_i}} \tensor C_{i,0}
\numberthis \label{H:IND:proof:invariant:induction:case-2:claim:operator-commute:proof:C2:1} \\ 
& =
\ketbra{p_1, \ldots, p_k}_{\reg{p_1 \ldots p_k}} \tensor \ketbra{p_{k+1}, \ldots, p_i}_{\reg{p_{k+1} \ldots p_i}} \tensor C_{i,1}
~ +  \\ 
& \hspace{2em}
\sum_{(p'_1,\ldots,p'_i) \ne (p_1, \ldots, p_i)}  \ketbra{p'_1, \ldots, p'_k}_{\reg{p_1 \ldots p_k}} \tensor \ketbra{p'_{k+1}, \ldots, p'_i}_{\reg{p_{k+1} \ldots p_i}} \tensor C_{i,0} 
\\ 
& =
\ketbra{p_1, \ldots, p_k}_{\reg{p_1 \ldots p_k}} \tensor \ketbra{p_{k+1}, \ldots, p_i}_{\reg{p_{k+1} \ldots p_i}} \tensor C_{i,1}
~ +  \\ 
& \hspace{1em}
 \ketbra{p_1, \ldots, p_k}_{\reg{p_1 \ldots p_k}} \tensor 
 \bigg( 
 \sum_{(p'_{k+1},\ldots,p'_i) \ne (p_{k+1}, \ldots, p_i)} 
 \ketbra{p'_{k+1}, \ldots, p'_i}_{\reg{p_{k+1} \ldots p_i}}
 \bigg) 
\tensor C_{i,0} 
~ + \\ 
& \hspace{3em}
\sum_{(p'_1,\ldots,p'_k) \ne (p_1, \ldots, p_k)}  \ketbra{p'_1, \ldots, p'_k}_{\reg{p_1 \ldots p_k}} 
\tensor \\ 
& \hspace{10em}
 \bigg( 
 \sum_{(p'_{k+1},\ldots,p'_i) \in \bits^{(i - k)\cdot \ell} } 
 \ketbra{p'_{k+1}, \ldots, p'_i}_{\reg{p_{k+1} \ldots p_i}}
 \bigg) 
\tensor C_{i,0} 
\\ 
& =
\ketbra{p_1, \ldots, p_k}_{\reg{p_1 \ldots p_k}} \tensor \Gamma_1  
+ 
\sum_{(p'_1,\ldots,p'_k) \ne (p_1, \ldots, p_k)}\ketbra{p'_1, \ldots,p'_k}_{\reg{p_1\ldots p_k}} \tensor \Gamma_0
 \numberthis \label{H:IND:proof:invariant:induction:case-2:claim:operator-commute:proof:C2:2}
,\end{align*}
where \Cref{H:IND:proof:invariant:induction:case-2:claim:operator-commute:proof:C2:1} follows from the definition of $C_i$ (see \Cref{re-define:unitaries}), and \Cref{H:IND:proof:invariant:induction:case-2:claim:operator-commute:proof:C2:2} follows by defining $\Gamma_0$ and $\Gamma_1$ as follows 
\begin{align*}
 \Gamma_1 
 & \coloneqq 
 \ketbra{p_{k+1}, \ldots, p_i}_{\reg{p_{k+1} \ldots p_i}} \tensor C_{i,1}
~ + \\ 
& \hspace{6em}
\bigg( 
 \sum_{(p'_{k+1},\ldots,p'_i) \ne (p_{k+1}, \ldots, p_i)} 
 \ketbra{p'_{k+1}, \ldots, p'_i}_{\reg{p_{k+1} \ldots p_i}}
 \bigg) 
\tensor C_{i,0} 
\\ 
 \Gamma_0 
 & \coloneqq 
  \bigg( 
 \sum_{(p'_{k+1},\ldots,p'_i) \in \bits^{(i - k)\cdot \ell} } 
 \ketbra{p'_{k+1}, \ldots, p'_i}_{\reg{p_{k+1} \ldots p_i}}
 \bigg) 
\tensor C_{i,0}.
\end{align*} 
Clearly, such a $\Gamma_0$ satisfies the requires $\Gamma_0 C_{k,0}  = C_{k,0} \Gamma^{\reg{t_k}/\reg{m}}_0$ and $\Gamma_0 C_{k,0}  B^\dagger_{k,1} C^\dagger_{k,1} = C_{k,0}  B^\dagger_{k,1} C^\dagger_{k,1} \Gamma^{\reg{t_k}/\reg{m}}_0$, because (1) we are in the case $i \in \Set{k+1, k+2, \ldots, K}$, (2) $C_{i,0}$ is nothing but a swap operator between $\reg{m}$ and $\reg{t_i}$, and (3) $C_{i,0}^{\reg{t_k}/\reg{m}}$ acts non-trivially on different registers from $B_{k,1}$ and $ C_{k,1}$ (see \Cref{CPM:non-trivial-registers:table}).

\vspace{1em}
This finishes the proof of \Cref{H:IND:proof:invariant:induction:case-2:claim:operator-commute}.   

\end{proof}

\para{Finishing the Proof for \Cref{H:IND:proof:invariant:induction:case-2}.} With \Cref{H:IND:proof:invariant:induction:case-2:claim:operator-commute} in hand, we now show how to finish the proof for \Cref{H:IND:proof:invariant:induction:case-2}.

\subpara{Proof of \Cref{lem:H:IND:proof:invariant:target-format:Hk-1}.} First, we establish \Cref{lem:H:IND:proof:invariant:target-format:Hk-1}:
\begin{align*}
 \ket{\phi^{(t)}} 
 & = 
 \Lambda  \ket{\phi^{(t-1)}}  
 \numberthis \label{H:IND:proof:invariant:induction:case-2:final:derivation:eq:0}\\ 
& = 
\Lambda  
  \bigg(
  \ket{p_1, \ldots, p_k}_{\reg{p_1\ldots p_k}} \ket{\rho^{(t-1)}_{p_1, \ldots, p_k}}  
   ~+~ \\ 
   & \hspace{10em}
   \sum_{(p'_1, \ldots, p'_k)\ne (p_1, \ldots, p_k)} \ket{p'_1, \ldots, p'_k}_{\reg{p_1} \ldots p_k}  C_{k,0} B^\dagger_{k,1} C^\dagger_{k,1} \ket{\rho^{(t-1)}_{p'_1, \ldots, p'_k}}
  \bigg)  
  \numberthis \label{H:IND:proof:invariant:induction:case-2:final:derivation:eq:1}\\ 
  & =  
  \ket{p_1, \ldots, p_k}_{\reg{p_1\ldots p_k}} \Lambda_1  \ket{\rho^{(t-1)}_{p_1, \ldots, p_k}}  
   ~+~ \\ 
   & \hspace{10em}
   \sum_{(p'_1, \ldots, p'_k)\ne (p_1, \ldots, p_k)} \ket{p'_1, \ldots, p'_k}_{\reg{p_1} \ldots p_k} \Lambda_0  C_{k,0} B^\dagger_{k,1} C^\dagger_{k,1} \ket{\rho^{(t-1)}_{p'_1, \ldots, p'_k}}
  \numberthis \label{H:IND:proof:invariant:induction:case-2:final:derivation:eq:2}\\ 
  & =  
    \ket{p_1, \ldots, p_k}_{\reg{p_1\ldots p_k}} \Lambda_1  \ket{\rho^{(t-1)}_{p_1, \ldots, p_k}}  
    ~+~ \\ 
   & \hspace{10em}
   \sum_{(p'_1, \ldots, p'_k)\ne (p_1, \ldots, p_k)} \ket{p'_1, \ldots, p'_k}_{\reg{p_1} \ldots p_k}   C_{k,0} B^\dagger_{k,1} C^\dagger_{k,1} \Lambda^{\reg{t_k}/\reg{m}}_0 \ket{\rho^{(t-1)}_{p'_1, \ldots, p'_k}}
  \numberthis \label{H:IND:proof:invariant:induction:case-2:final:derivation:eq:3}\\ 
& =  
  \ket{p_1, \ldots, p_k}_{\reg{p_1\ldots p_k}} \ket{\rho^{(t)}_{p_1, \ldots, p_k}}  
   ~+~ 
   \sum_{(p'_1, \ldots, p'_k)\ne (p_1, \ldots, p_k)} \ket{p'_1, \ldots, p'_k}_{\reg{p_1} \ldots p_k}  C_{k,0} B^\dagger_{k,1} C^\dagger_{k,1} \ket{\rho^{(t)}_{p'_1, \ldots, p'_k}} 
  \numberthis \label{H:IND:proof:invariant:induction:case-2:final:derivation:eq:4}
,\end{align*}
where  
\begin{itemize}
\item

\Cref{H:IND:proof:invariant:induction:case-2:final:derivation:eq:0} follows from \Cref{H:IND:proof:invariant:induction:case-2:initial:phi}; 

\item
\Cref{H:IND:proof:invariant:induction:case-2:final:derivation:eq:1} follows from our induction assumption; 

\item
\Cref{H:IND:proof:invariant:induction:case-2:final:derivation:eq:2}  follows from \Cref{H:IND:proof:invariant:induction:case-2:claim:operator-commute:eq:1} in \Cref{H:IND:proof:invariant:induction:case-2:claim:operator-commute}; 

\item
\Cref{H:IND:proof:invariant:induction:case-2:final:derivation:eq:3} follows from \Cref{H:IND:proof:invariant:induction:case-2:claim:operator-commute:eq:3} in \Cref{H:IND:proof:invariant:induction:case-2:claim:operator-commute}; 

\item 
\Cref{H:IND:proof:invariant:induction:case-2:final:derivation:eq:4} follows by defining
\begin{equation}\label[Expression]{H:IND:proof:invariant:induction:case-2:final:derivation:def:rho-t}
\begin{cases}
\ket{\rho^{(t)}_{p_1, \ldots, p_k}}
\coloneqq 
 \Lambda_1 \ket{\rho^{(t-1)}_{p_1, \ldots, p_k}} & 
 \vspace{0.3em}
 \\ 
\ket{\rho^{(t)}_{p'_1, \ldots, p'_k}}
\coloneqq 
\Lambda^{\reg{t_k}/\reg{m}}_0 \ket{\rho^{(t-1)}_{p'_1, \ldots, p'_k}} 
& ~~\forall (p'_1, \ldots, p'_k) \in \big(\bits^{k\ell} \setminus \Set{(p_1, \ldots,p_k)}\big). 
\end{cases}
\end{equation}
\end{itemize}
Clearly, \Cref{H:IND:proof:invariant:induction:case-2:final:derivation:eq:4} is of the same format as \Cref{lem:H:IND:proof:invariant:target-format:Hk-1} in \Cref{lem:H:IND:proof:invariant}.

\subpara{Proof of \Cref{lem:H:IND:proof:invariant:target-format:Hk}.} Next, we present the derivation for \Cref{lem:H:IND:proof:invariant:target-format:Hk}:
\begin{align*}
 \ket{\psi^{(t)}} 
 & = 
 \Lambda  \ket{\psi^{(t-1)}}  
 \numberthis \label{H:IND:proof:invariant:induction:case-2:final:derivation:eq:psi:0}\\ 
& = 
\Lambda  
  \bigg(
  \ket{p_1, \ldots, p_k}_{\reg{p_1\ldots p_k}} \ket{\rho^{(t-1)}_{p_1, \ldots, p_k}}  
   ~+~ \\ 
   & \hspace{10em}
   \sum_{(p'_1, \ldots, p'_k)\ne (p_1, \ldots, p_k)} \ket{p'_1, \ldots, p'_k}_{\reg{p_1} \ldots p_k}  C_{k,0}  \ket{\rho^{(t-1)}_{p'_1, \ldots, p'_k}}
  \bigg)  
  \numberthis \label{H:IND:proof:invariant:induction:case-2:final:derivation:eq:psi:1}\\ 
  & =  
  \ket{p_1, \ldots, p_k}_{\reg{p_1\ldots p_k}} \Lambda_1  \ket{\rho^{(t-1)}_{p_1, \ldots, p_k}}  
   ~+~ \\ 
   & \hspace{10em}
   \sum_{(p'_1, \ldots, p'_k)\ne (p_1, \ldots, p_k)} \ket{p'_1, \ldots, p'_k}_{\reg{p_1} \ldots p_k} \Lambda_0  C_{k,0} \ket{\rho^{(t-1)}_{p'_1, \ldots, p'_k}}
  \numberthis \label{H:IND:proof:invariant:induction:case-2:final:derivation:eq:psi:2}\\ 
  & =  
    \ket{p_1, \ldots, p_k}_{\reg{p_1\ldots p_k}} \Lambda_1  \ket{\rho^{(t-1)}_{p_1, \ldots, p_k}}  
    ~+~ \\ 
   & \hspace{10em}
   \sum_{(p'_1, \ldots, p'_k)\ne (p_1, \ldots, p_k)} \ket{p'_1, \ldots, p'_k}_{\reg{p_1} \ldots p_k}   C_{k,0} \Lambda^{\reg{t_k}/\reg{m}}_0 \ket{\rho^{(t-1)}_{p'_1, \ldots, p'_k}}
  \numberthis \label{H:IND:proof:invariant:induction:case-2:final:derivation:eq:psi:3}\\ 
& =  
  \ket{p_1, \ldots, p_k}_{\reg{p_1\ldots p_k}} \ket{\rho^{(t)}_{p_1, \ldots, p_k}}  
   ~+~ 
   \sum_{(p'_1, \ldots, p'_k)\ne (p_1, \ldots, p_k)} \ket{p'_1, \ldots, p'_k}_{\reg{p_1} \ldots p_k}  C_{k,0} \ket{\rho^{(t)}_{p'_1, \ldots, p'_k}} 
  \numberthis \label{H:IND:proof:invariant:induction:case-2:final:derivation:eq:psi:4}
,\end{align*}
where  
\begin{itemize}
\item

\Cref{H:IND:proof:invariant:induction:case-2:final:derivation:eq:psi:0} follows from \Cref{H:IND:proof:invariant:induction:case-2:initial:psi}; 

\item
\Cref{H:IND:proof:invariant:induction:case-2:final:derivation:eq:psi:1} follows from our induction assumption; 

\item
\Cref{H:IND:proof:invariant:induction:case-2:final:derivation:eq:psi:2}  follows from \Cref{H:IND:proof:invariant:induction:case-2:claim:operator-commute:eq:1} in \Cref{H:IND:proof:invariant:induction:case-2:claim:operator-commute}; 

\item
\Cref{H:IND:proof:invariant:induction:case-2:final:derivation:eq:psi:3} follows from \Cref{H:IND:proof:invariant:induction:case-2:claim:operator-commute:eq:2} in \Cref{H:IND:proof:invariant:induction:case-2:claim:operator-commute}; 

\item 
\Cref{H:IND:proof:invariant:induction:case-2:final:derivation:eq:psi:4} follows from {\em the same definitions of} $\ket{\rho^{(t)}_{p_1, \ldots, p_k}}$ and $\ket{\rho^{(t)}_{p'_1, \ldots, p'_k}}$ as shown in \Cref{H:IND:proof:invariant:induction:case-2:final:derivation:def:rho-t}.
\end{itemize}
Clearly, \Cref{H:IND:proof:invariant:induction:case-2:final:derivation:eq:psi:4} is of the same format as \Cref{lem:H:IND:proof:invariant:target-format:Hk} in \Cref{lem:H:IND:proof:invariant}.

\vspace{1em}
This completes the proof of the induction step of \Cref{lem:H:IND:proof:invariant} for \Cref{H:IND:proof:invariant:induction:case-2}.

\vspace{1em}
Finally, we remark that our proof for the base case in \Cref{sec:lem:H:IND:proof:invariant:base} and the proof for the induction step in \Cref{sec:lem:H:IND:proof:invariant:induction} together finish the proof of \Cref{lem:H:IND:proof:invariant}, which in turn finishes the proof of \Cref{lem:H:IND} eventually.

\section{On Expected-Polynomial Time Simulator and Efficient Verifier}
\label{sec:expected_sim_efficient_verifier}

In this section, we will fully show how the result \Cref{thm:impossibility:QZK} also works with expected polynomial-time simulator and how to make the malicious verifier designed in \Cref{sec:CPM:Vtil:unitaries} efficient.

We first describe the high-level ideas and intuition.

\para{Expected Polynomial-Time Simulator.} 
To show that our lower bound holds for expected polynomial time simulator, we follow the idea in \cite{FOCS:CCLY21}.
The idea is to design a modified random-terminating malicious verifier $\Vexp$  based on $\tilde{V}$ designed in \Cref{sec:CPM:Vtil:unitaries}.
If there exists any simulator $\Sexp$ that runs in expected polynomial time and $\Sexp^{\Vexp}$ outputs an accepting view, then we can turn $\Sexp$ into a strict-polynomial time simulator $S$, which would violate our already proven lower bound.

The new verifier $\Vexp$ is as follows. $\Vexp$ operates an honest verifier $V$ and a random-terminating verifier $\tilde{V}$ designed in \Cref{sec:CPM:Vtil:unitaries} in superposition,  by using a control-qubit in an additional register  $\mathbf{Cont}$: 
\begin{align*}
    \ket{\tilde{\psi}}_{\Cont, \reg{aux}} = \frac{1}{\sqrt{2}}(\ket{0}_{\Cont} +\ket{1}_{\Cont}) \otimes \ket{\psi_\epsilon}_{\reg{aux}},
    \end{align*}
where $\ket{\psi_\epsilon}_{\reg{aux}}$ is defined as:
\begin{align*}
    \ket{\psi_\epsilon}_{\reg{aux}}:= \sum_{H \in \cH_\epsilon} \sqrt{D(H)} \ket{H}_{\reg{aux}},
\end{align*}
where $H\in \cH_\epsilon$ 
is the function family defined in \Cref{def:H-epsilon} and D is the density function corresponding to $\cH_\epsilon$.

The hope is that if we have any expected-polynomial time simulator $\Sexp$, we run it up to a fixed $q$ number of steps and simply stop-- the final output view of this ''truncated'' simulator $S$ with $\Vexp$ is still an accepting one, with inverse polynomial probability. More importantly, we not only want any accepting view, we also want the random-terminating branch of the verifier $\Vexp$ to be accepting in this case, because otherwise we would not be able to transform the simulator-verifier interaction $S^{\Vexp}$ into the BQP decider we designed in \Cref{sec:BQP-decider}. Here comes how we use the control-qubit:
at the end of the protocol's execution between $S$ and $\Vexp$, we show that the state in register $\Cont$ has a large enough weight on the $\ket{1}$ component. Therefore, if we measure $\Cont$ in the end, there is a large probability that we fall into the branch where we use the random-terminating $\tilde{V}$.

A major technical task we deal with is to ensure that the state in $\Cont$ is a pure state with a specific format at the end of the protocol execution, for the rest of the arguments in \cite{FOCS:CCLY21} to go through.
To make sure that the final state in the $\Cont$ register is pure, \cite{FOCS:CCLY21} uses an ''adjusting'' unitary when the verifier $\Vexp$ operates on the $\ket{0}$ (i.e. honest verifier branch) at the end of the protocol execution, to disentangle the $\reg{aux}$ register and the $\Cont$ register. Our adjusting unitary has the  same design as in \cite{FOCS:CCLY21}, but the analysis on why this procedure helps us achieve the above goal deviates from the analysis of \cite{FOCS:CCLY21}. In \cite{FOCS:CCLY21}, the proof considers the verifier to take input a fixed classical randomness, which we don't use in the quantum setting. But we make the observation that the use of a fixed classical randomness is not necessary for the proof.  We will prove the corresponding lemma in our scenario and connect it with the rest of the proofs. 

\jiahui{expanded this part}

\para{Efficient Malicious Verifier.} 
Note that the state $\ket{\psi_\epsilon}_{\reg{aux}}$ is exponentially large, but we will use the same method as in \cite{FOCS:CCLY21}
to make our verifier efficient, by modifying a $2q$-independent hash function to simulate the functins in the family $\cH_\epsilon$.

\subsection{Expected Polynomial-Time Simulator}
We first cite a lemma from \cite{FOCS:CCLY21}. This lemma helps us make sure that the qubit in register $\Cont$ remains as a pure state for the rest of the arguments to go through.

To achieve this, we must let the verifier $\Vexp$ apply an "adjusting" unitary when operating under the $\ket{0}$ (i.e. honest verifier) branch,
The following unitary will be used at the end of the honest verifier brancha and performs the following task: for any $K$-round prover message $\bp$, it adjusts the $\aux$ register to have support only on the $H$'s that satisfies $H(\bp) = \vecone$.

\begin{lemma}[Lemma 3.8 \cite{FOCS:CCLY21}]\label{lem:uncomputing_random}
    For any $\bp =(p_1,...,p_k)\in \cM^k$, let $S_{\bp}\subseteq \cH_\epsilon: \cM^{\leq K} \to\bits$ be the subset consisting of all $H$ such that $H(p_1,...,p_i)=1$ for all $i\in[K]$.  
    There exists a  unitary $U_{\bp}$ such that
    \[
    U_{\bp}\sum_{H\in \cH_\epsilon}\sqrt{D(H)}\ket{H}_{\reg{aux}}=\sum_{H\in S_{\bp}}\sqrt{\frac{D(H)} {\epsilon^K}}\ket{H}_{\reg{aux}}.
    \]
\end{lemma}

The proof is the same as in \cite{FOCS:CCLY21} and we omit it here.

Next, we describe how the new malicious verifier $\Vexp$ works.

$\Vexp$ uses the control-qubit $\Cont$ to control whether to apply an honest verifier unitary $V_{\mathsf{hon}}$ or a random-terminating verifier unitary $\tilde{V}$ defined in \Cref{sec:CPM:Vtil:unitaries}.

\begin{align*}
    & \Vexp(\ket{0}_{\Cont}\ket{\mathbf{other}_0}_{\mathbf{other}} + \ket{1}_{\Cont}\ket{\mathbf{other}_1}_{\mathbf{other}} ) \\
    = & \ket{0}_{\Cont}\Vhon\ket{\mathbf{other}_0}_{\mathbf{other}} + \ket{1}_{\Cont}\tilde{V} \ket{\mathbf{other}_1}_{\mathbf{other}}
\end{align*}
where the $\mathbf{other}$ register refers to all the other registers the verifier will ever act on. 

Additionally, the honest unitary $V_{\mathsf{hon}}$ applies the unitary $U_{\bp}$ from the \Cref{lem:uncomputing_random} to register $\Cont$ at the end of the protocol. More specifically, $\Vhon$ operates as follows:

\para{Honest $\Vhon$:} it non-trivially acts on registers $\reg{ins}, \reg{gc, lc}, \reg{p_1}, \cdots,\reg{p_K}, \reg{v_1}, \cdots,\reg{v_K}, \reg{t_1}, \cdots,\reg{t_K},\reg{m}, \reg{w}, \reg{aux}$ defined in \Cref{sec:CPM:Vtil:registers}:
\begin{enumerate}
    \item It reads the value $j$ in $\reg{gc}$ and increments it to $i = j + 1 \mod C$.\footnote{$C = 2^\lambda$ defined in \Cref{sec:CPM:Vtil:registers}.} It swaps $\reg{m}$
and $\reg{p_i}$.

\item If the value in the global counter $\reg{gc}$ is $i < K$, it then applies the honest verifier's unitary at round $i$ to registers $\reg{ins}, \reg{p_1}, \cdots,\reg{p_K}, \reg{v_1}, \cdots,\reg{v_K}, \reg{m}, \reg{w}$. Note that register $\reg{aux}$ is not used by the honest verifier because it only contains the function used for determining random termination.

\item If the value in the global counter $\reg{gc}$ is $K$ 
, then it first applies the honest verifier's unitary in the corresponding round, then applies the adjusting unitary $U_{\bp}$ to $\reg{p_1}, \cdots, \reg{p_K}$ and $\reg{aux}$:
\begin{align*}
    \ket{p_1, \cdots, p_K, H} \to U_{\bp} \ket{p_1, \cdots, p_K, H}
\end{align*}

\end{enumerate}

Next, we show that by running the above new malicious verifier $\Vexp$ with an honest prover $P$ on some instance $x \in L$, the final state in register $\Cont$ will be negligibly close to a pure state that contains $\frac{1}{\epsilon^K+1}$ fraction of weight on the $\ket{0}$ component and $\frac{\epsilon^K}{\epsilon^K+1}$ fraction of weight on the $\ket{1}$ component.

\begin{lemma}\label{lem:final_state_pure}
For any $x\in L\cap \bits^\lambda$ and $w\in R_L(x)$, suppose that we run $\langle P(w), \Vexp(\ket{\widetilde{\psi}_\epsilon}) \rangle (x)$ and measure the final output register and obtain outcome $1$. Then the resulting state in $\Cont$ (tracing out other registers) is negligibly close to $\ket{\phi_\epsilon}_{\Cont}:=\sqrt{\frac{1}{1+\epsilon^K}}\ket{0}_{\Cont}+\sqrt{\frac{\epsilon^K}{1+\epsilon^K}}\ket{1}_{\Cont}$.

\end{lemma}

\begin{proof}

We adapt the proof in \cite{FOCS:CCLY21} to be of use in our setting.
By the completeness of the honestly executed protocol $\Pi$, the completeness error is $\negl(\lambda)$. 

Let $\ket{\eta}$ be the final state of the internal register of $\Vexp$ after executing $\langle P(w), \Vexp(\ket{\widetilde{\psi}_\epsilon}) \rangle (x)$. We additionally note that the final output register that stores $\Vexp$'s decision bit is called  $\reg{d}$ (defined in \Cref{sec:def:BBQZK}). We denote $\acc$ 
as a final measurement on the $\reg{d}$ register in the computational basis.

For $\beta\in \bits$, let $\ket{\eta_\beta}$ be the final state of the internal register of $\Vexp$ after executing $\langle P(w), \Vexp(\ket{\widetilde{\psi}_\epsilon}) \rangle (x)$.
where $\ket{\widetilde{\psi}_\epsilon^{(\beta)}}_{\Cont,\aux}:=
\ket{\beta}_{\Cont}\otimes \ket{\psi_\epsilon}_{\aux}
$.
 Since $\Vexp$ only uses $\Cont$ as a control register
and $\ket{\widetilde{\psi}_\epsilon}_{\Cont,\aux}=\frac{1}{\sqrt{2}}\left(\ket{\widetilde{\psi}_\epsilon^{(0)}}_{\Cont,\aux}+\ket{\widetilde{\psi}_\epsilon^{(1)}}_{\Cont,\aux}\right)$
, it is easy to see that we have 
\begin{align}  \label{eq:eta}
    \ket{\eta}=\frac{1}{\sqrt{2}}\left(\ket{\eta_0}+\ket{\eta_1}\right).
\end{align}
In the following, when we consider summations over  $H$ (they are over all $H\in \cH_\epsilon$, respectively, unless otherwise specified) and prover messages $\bp = (p_1, \cdots, p_K)$. 

By the definition of $\Vexp$ and \Cref{lem:uncomputing_random}, we have  
\xiao{While the following is obviously in CCLY, we need to argue why this is the case! Also, our verifier has a working space and other registers. It's not as simple as what you wrote here.}
\xiao{summation should also be over $p$'s}
\xiao{Moreover, explain the meaning of $\reg{d}$ register.}
\xiao{same issue for \Cref{eq:eta_one} as well.}
\jiahui{fixed}
\begin{align*}
 \ket{\eta_0}
 &= U_{\bp}  \ket{0}_{\Cont}\ket{x}_{\reg{ins}}\ket{K}_{\reg{gc}} 
 \otimes \left( \sum_{\bp} \alpha_\bp \left( \sum_{H} \sqrt{D(H)}\ket{H}_{\aux}  \right) 
\ket{\bp}_{\reg{p_1},...,\reg{p_K}} \ket{\mathbf{other}_\bp}_{\reg{other}} \ket{b_{\bp}}_{\reg{d}} \right)\\
 &= \ket{0}_{\Cont}\ket{x}_{\reg{ins}}\ket{K}_{\reg{gc}} 
\otimes \left( \sum_{\bp} \alpha_\bp \left(\sum_{ H\in S_{\bp}} \sqrt{\frac{D(H)}{\epsilon^K}}\ket{H}_{\aux} \right)  \ket{\bp}_{\reg{p_1},...,\reg{p_K}} \ket{\mathbf{other}_\bp}_{\reg{other}}\ket{b_{\bp}}_{\reg{d}}\right)
\end{align*}
where 
$\bp$ is prover's messages and the register $\reg{other}$ refers to the concatenation of the registers $\reg{w}$, $\reg{m}$, $\reg{lc}$, $\reg{v_1 \cdots v_K}$, and $\reg{t_1, \cdots, t_K}$. The $\ket{\bother_{\bp}}_{\reg{other}}$ is some sub-normalized state.
The verifier's final output decision state $\ket{b_{\bp}}$ in register $\reg{d}$ is dependent on the contents in register $\reg{other}$.

Since $\Vhon$ is an honest verifier's unitary except applying $U_\bp$ at the very end: the registers $\reg{p_1 \cdots p_k}$, $\reg{other}$, and $\reg{d}$ will evolve to exactly what they will be at the end of an honestly executed protocol before $U_{\bp}$ is applied. After we apply $U_{\bp}$, the only change is the weight on the $\aux$ register for each $\bp$, according to \Cref{lem:uncomputing_random}. 
Also note that since $P$ is an honest prover, the register $\reg{lc}$ and $\reg{gc}$ will always have the same value throughout the protocol and the registers $\reg{t_1, \cdots, t_K}$ are not used.
While for each $\bp$, the corresponding $\reg{other}$ register may hold a mixed state, the analysis assuming $\reg{other}$  holding a pure state is without loss of generality.


By the completeness of $\Pi$, we have 
$\Pr[\acc( \ket{\eta_0})=1]=1-\negl(\lambda)$. 
This implies 
\begin{align} \label{eq:eta_zero}
 \ket{\eta_0}
\approx \ket{0}_{\Cont}\ket{x}_{\reg{ins}}\ket{K}_{\reg{gc}} 
\otimes \left( \sum_{\bp} \alpha_{\bp,1} \left(\sum_{ H\in S_{\bp}} \sqrt{\frac{D(H)}{\epsilon^k}}\ket{H}_{\aux} \right)   \ket{\bp}_{\reg{p_1},...,\reg{p_k}} \ket{\mathbf{other}_{\bp, 1}}_{\reg{other}} \right)\ket{1}_{\reg{d}}
\end{align}
where $\approx$ means that the trace distance between both sides is $\negl(\lambda)$. The register $\reg{p_1, \cdots, p_k, other}$ now contains a state
such that the final output $b_{\bp} = 1$. 

On the other hand, by the definition of $\Vexp$, the value in $\reg{d}$ of $\ket{\eta_1}$ can be $1$ only if $H\in S_{\bp}$ for the transcript $\bp$. 
Therefore we have 
\begin{align*}
 \ket{\eta_1}=   \ket{1}_{\Cont}\ket{x}_{\reg{ins}}\ket{K}_{\reg{gc}}
 \otimes
 \left(
 \begin{array}{ll}
 &
 \sum_{\bp} \alpha_{\bp,1} \sum_{H\in S_{\bp}}\left(\sqrt{D(H)}\ket{H}_{\aux} \otimes  \ket{\bp}_{\reg{p_1},...,\reg{p_k}} 
\ket{\mathbf{other}_{\bp,1}}_{\reg{other}} \otimes {\ket{1}_{\reg{d}}}
 \right)\\
+& \ket{\sf garbage}_{\aux,\reg{p_1},...,\reg{p_k, other}}
 \otimes \ket{0}_{\reg{d}}
 \end{array}
 \right)
\end{align*}
for some (sub-normalized) state $\ket{\sf garbage}_{\aux,\reg{p_1},...,\reg{p_k, other}}
 \otimes \ket{0}_{\reg{d}}$. By our design of the random-terminating adversary $\tilde{V}$,
 conditioned on that $H \in S_{\bp}$, $\tilde{V}$ performs the same operations as the honest verifier does on the registers $\reg{p_1, \cdots, p_k, other}$. $\tilde{V}$ can also easily uncompute the registers $\reg{t_1, \cdots, t_K}$ used to store aborting information for each round, which are not used in $\ket{\eta_0}$. 
 Therefore, if we have $b = 1$ in register $\reg{d}$,   the state over the registers $\reg{p_1, \cdots, p_k, other}$ is exactly the same as the state over registers $\reg{p_1, \cdots, p_k, other}$  in $\ket{\eta_0}$ (conditioned on $b = 1$ in $\reg{d}$). 

By a similar argument to that for $\ket{\eta_0}$, we have 
\begin{align}  \label{eq:eta_one}
 \ket{\eta_1}\approx   \ket{1}_{\Cont}\ket{x}_{\reg{ins}}\ket{K}_{\reg{gc}}
 \otimes
 \Biggl(
 \begin{array}{ll}
 &
\sum_{\bp} \alpha_{\bp,1} \sum_{H\in S_{\bp}}\left(\sqrt{D(H)}\ket{H}_{\aux} \right)  \otimes  \ket{\bp}_{\reg{p_1},...,\reg{p_k}}
\ket{\mathbf{other}_{\bp,1}}_{\reg{other}}\otimes
{\ket{1}_{\reg{d}}}\\
+& 
 \ket{\sf garbage}_{\aux,\reg{p_1},...,\reg{p_k},\reg{other}}
\ket{0}_{\reg{d}} 
 \end{array}
 \Biggr)
\end{align}


By \Cref{eq:eta,eq:eta_zero,eq:eta_one}, we have 
\begin{align*}
(\ket{1}\bra{1})_{\reg{d}}\ket{\eta}
& \approx\frac{1}{\sqrt{2}}\left(
\sqrt{\frac{1}{\epsilon^K}} 
\ket{0}_{\Cont}+\ket{1}_{\Cont}\right)
\otimes
\ket{x}_{\reg{ins}}\ket{K}_{\reg{gc}}
\\
 & \hspace{8em}
 \otimes \sum_{\bp} \alpha_{\bp,1} \sum_{H\in S_{\bp}}\left(\sqrt{D(H)}\ket{H}_{\aux}\right) \ket{\bp}_{\reg{p_1},...,\reg{p_k}} 
\ket{\mathbf{other}_{\bp,1}}_{\reg{other}} \otimes
 \ket{1}_{\reg{d}}.
\end{align*}

 We omit the identity operator on registers other than $\reg{d}$ and  $(\ket 1 \bra 1)_\reg{d}$  simply means the projection onto states whose values in $\reg{d}$ is $0$.

By normalization, we can see that the final state in $\Cont$ conditioned on the measurement outcome of $\reg{d}$ is $1$ is negligibly close to $\ket{\phi_\epsilon}_\Cont$.

\end{proof}


The rest of the proof follows from \cite{FOCS:CCLY21} Section 3.2. We describe the ideas here. 
Suppose that there is a quantum black-box simulator $\Sexp$ for the
protocol $\Pi$ whose expected number of queries is at most $q/2 = \poly(\lambda)$ that works for all possibly
inefficient verifiers, then we can truncate it into a strict polynomial time simulator $S$ that makes only $q$ queries based on the following arguments:
\begin{enumerate}
    \item The probability that $\Vexp$ accepts and the number of $\Sexp$’s queries is at most $q$ is at least $1/4 -\negl(\lambda)$ (See \cite{FOCS:CCLY21} Lemma 3.10).

    \item The final state in $\Cont$ after the execution of $\Sexp^{\Vexp(\ket{\widetilde{\psi}_\epsilon})}$ conditioned on the above event is negligibly close in trace distance to $\ket{\phi_\epsilon}$ defined in \Cref{lem:final_state_pure} (See \cite{FOCS:CCLY21} Lemma 3.11).
\end{enumerate}

In the end, we can argue that the truncated simulator $S$ interacting with $\Vexp(\ket{\widetilde{\psi}_\epsilon})$ will lead to an accepting view with probability $\frac{\epsilon^k}{4} - \negl(\lambda)$ on any $x \in L \cap \bits^\lambda$.

We refer the readers to \cite{FOCS:CCLY21} Section 3.2 for the full proof.

\subsection{Expected Polynomial-Time Simulator with Efficient Malicious Verifier}

Now we describe how to make the above malicious verifier $\Vexp$ efficient. 

The following lemma states that we can replace the inefficient function (represented by a truth table) $H \in \cH_\epsilon$ used for random-termination in the previous sections by a function with an efficient description. Moreover, there exists an efficient implementation of the unitary in \Cref{lem:uncomputing_random}.
\begin{lemma}[\cite{FOCS:CCLY21} Lemma 3.13]\label{lem:efficient_verifier}
Let $\epsilon\in [0,1]$ be a rational number expressed as $\epsilon=\frac{B}{A}$ for some $A,B\in \mathbb{N}$ such that $\log A=\poly(\secpar)$ and $\log B=\poly(\secpar)$.\footnote{Note that $\epsilon$ is also a function of $\secpar$, but we omit to explicitly write the dependence on $\epsilon$ for simplicity.} 
For any  $Q=\poly(\secpar)$,  there exists a family $\tilde{\cH}_\epsilon=\{\tilde{\cH}_\kappa:\cM^{\leq k}\to \bits\}_{\kappa\in \mathcal{K}}$ of classical polynomial-time computable functions that satisfies the following properties.
\begin{enumerate}
    \item For any algorithm $\A$ that 
    makes at most $q$ quantum queries and any quantum input $\rho$, we have 
    \[
    \Pr_{H\gets \mathcal{H}_\epsilon}\left[\A^{H}(\rho)=1\right]
    =
      \Pr_{\kappa\gets\mathcal{K}}\left[\A^{\tilde{\cH}_\kappa}(\rho)=1\right].
    \]
    \item 
    For any $\bp=(p_1,...,p_k)\in \cM^k$, let $S_{\bp}\subseteq \mathcal{K}$ be the subset consisting of all $\kappa$ such that $\tilde{H}_\kappa(p_1,...,p_i)=1$ for all $i\in[k]$.  
    There exists a  unitary $U_{\bp}^{(q)}$ such that
    \[
    U_{\bp}^{(q)}\sqrt{\frac{1}{|\mathcal{K}|}}\sum_{\kappa\in \keyspace}\ket{\kappa}=\sqrt{\frac{1}{|S_{\bp}|}}\sum_{\kappa\in S_{\bp}}\ket{\kappa}.
    \]
    $U_{\bp}$ can be implemented by a quantum circuit of size $\poly(\secpar)$. 
\end{enumerate}
\end{lemma}
The proof follows from \cite{FOCS:CCLY21}. Therefore we present the statement and intuition  and refer the readers to \cite{FOCS:CCLY21} for the full proof.

The idea of is to construct a family of $2q$-wise independent hash function that have output distribution the same as $\cH_\epsilon$, from a family of $2q$-wise independent hash functions that have almost uniform-random output distributions. We first make use of a $2q$-independent hash function $H'$. We take the key $\kappa$ to be $k$ values of random $\{a_i\}_{i \in [k]}, a_i \gets [A]$. When computing $\tilde{H}_\kappa(p_1, \cdots, p_i)$: we add $a_i$ to the outcome of $H'(p_1, \cdots, p_i)$ and check if the sum satisfies $H'(p_1, \cdots, p_i) + a_i \mod A \leq B$; output 1 if yes and 0 otherwise.

\bibliographystyle{alpha}
\bibliography{cryptobib/abbrev0,cryptobib/crypto,additionalRef}
\addcontentsline{toc}{section}{References}

\newpage
\appendix

\end{document}